\documentclass{aa}

\newcommand{\iso}       {\mbox{$ \rm sech^{2} $}}
\newcommand{\Rmax}      {\mbox{$  R_{\rm max} $}}

\newcommand{\zo}        {\mbox{$  z_{0}       $}}

\renewcommand{\~}       {\mbox{$ \hspace{0.4mm} $}}

\newcommand{\A}         {\mbox{$ ^{\rm a)}    $}}
\newcommand{\B}         {\mbox{$ ^{\rm b)}    $}}

\usepackage{graphicx}
\begin{document}

   \thesaurus{03         % A&A Section 11: Galaxies
              (11.05.2;  % Galaxies: evolution
               11.07.1;  % Galaxies: general,
               11.09.2;  % Galaxies: interactions,
               11.19.2;  % Galaxies: spiral,
               11.19.6;  % Galaxies: structure,
               11.19.7)} % Galaxies: statistics.

   \title{Properties of tidally-triggered vertical disk perturbations
          \thanks{Based on observations obtained at the European Southern Observatory
          (ESO, La Silla, Chile), Calar Alto Observatory operated by the MPIA (DSAZ, Spain),
          Lowell Observatory (Flagstaff,AZ, USA), and Hoher List Observatory (Germany).}
         }

%   \subtitle{}

   \authorrunning{U. Schwarzkopf \& R.-J. Dettmar}

   \titlerunning{Properties of tidally-triggered vertical disk perturbations.}

   \author{U. Schwarzkopf \hspace{-0.8mm} \inst{1,}\inst{2,}\thanks
          {Lynen Fellow at Steward Observatory}
          \and
           R.-J. Dettmar \inst{2}}

   \offprints{U. Schwarzkopf \\ (schwarz@as.arizona.edu)}

   \institute{Steward Observatory, University of Arizona,
              933 North Cherry Avenue, Tucson, Arizona 85721, USA
             \and
              Astronomisches Institut, Ruhr-Universit\"at Bochum,
              Universit\"atsstra{\ss}e 150, 44780 Bochum, Germany}

   \date{Received 15 July 2000 / Accepted 12 April 2001}

   \maketitle

   \begin{abstract}

We present a detailed analysis of the properties of warps and tidally-triggered
perturbations perpendicular to the plane of 47 interacting/merging edge-on
spiral galaxies. The derived parameters are compared with those obtained for a sample
of 61 non-interacting edge-on spirals. The entire optical ($R$-band) sample used for
this study was presented in two previous papers.
We find that the scale height of disks in the interacting/merging sample is
characterized by perturbations on both large ($\simeq$ disk cut-off radius) and short
($\simeq \zo$) scales, with amplitudes of the order of 280\~pc and 130\~pc on average,
respectively.
The size of these large (short) -scale instabilities corresponds to 14\% (6\%) of the
mean disk scale height. This is a factor of 2 (1.5) larger than the value found for
non-interacting galaxies.
A hallmark of nearly all tidally distorted disks is a scale height that increases
systematically with radial distance. The frequent occurrence and the significantly
larger size of these gradients indicate that disk asymmetries on large scales
are a common and persistent phenomenon, while local disturbances and bending instabilities
decline on shorter timescales.
Nearly all (93\%) of the interacting/merging and 45\% of the non-interacting galaxies studied
are noticeably warped. Warps of interacting/merging galaxies are $\approx 2.5$ times larger on
average than those observed in the non-interacting sample, with sizes of the order of
340\~pc and 140\~pc, respectively. This indicates that tidal distortions do considerably
contribute to the formation and size of warps. However, they cannot entirely explain the
frequent occurrence of warped disks.


   \end{abstract}

%
%________________________________________________________________

\section{Introduction}

The vertical structure of galactic disks seen nearly edge-on and, in particular, the behaviour
of the disk scale height with galactocentric distance is a major source of information on disk
properties, such as their global and local stability, the secular evolution of their stellar
population as well as the vertical heating process. Until recently most observational results
indicated that the constancy of the scale height over the whole radial extent is a fundamental
property of galactic disks (van der Kruit \& Searle \cite{kruit1981a},\cite{kruit1981b},
\cite{kruit1982}; Wainscoat et al. \cite{wainscoat1989}; Barnaby \& Thronson \cite{barnaby1992}).
This result was consistent with the theoretical picture of continuously heated disk stars by
massive molecular clouds (Spitzer \& Schwarzschild \cite{spitzer1951}; Binney \cite{binney1981})
and with the assumption that at all times the star formation rate is proportional to the number
density of molecular clouds (Fall \cite{fall1980}). Hence, also the vertical velocity dispersion
of the disk stars should exponentially decrease with radial distance.

Although the overall constancy of the scale height of galactic disks was confirmed thereafter
by several studies in the optical and near-infrared, the derived constancy levels differ
significantly from each other.
While Shaw \& Gilmore (\cite{shaw1990}) found that the intrinsic variation of the disk
scale height along the major axis of a galaxy is typically within 3\% of its mean value,
de Grijs \& van der Kruit (\cite{grijs1996}) obtained a value between $10\% - 15\%$ for Sc and
later type galaxies. Reshetnikov \& Combes (\cite{reshetnikov1996}, \cite{reshetnikov1997}) have 
split their galaxy sample into two subsamples and found a dispersion of 7\% for non-interacting
galaxies, while the heavily disturbed disks of interacting pairs of galaxies (equal mass) possess
a dispersion of 11\%. Although this is almost twice compared to non-interacting galaxies, their
values are still consistent with the above mentioned level found for normal galaxies by
de Grijs \& van der Kruit (\cite{grijs1996}). Furthermore, the results of the latter two
studies indicate that there is a correlation between the radial behaviour of the scale height
and environment.

At present the existing uncertainties cannot be constrained by statistical studies based on
a significantly larger and thus more reliable galaxy sample. Hence the large differences between
individual values make it difficult to give a coherent interpretation. Considering that small
variations of the scale height are difficult to measure it may be argued that the observed
differences are mainly due to a combination of several biasing effects such as selection
criteria, the size of the galaxy sample used, or due to different disk models applied
for fitting the scale height. On the other hand, there is some evidence for a physically
motivated explanation: as a result of a detailed study of vertical surface brightness profiles
of edge-on disk galaxies de Grijs \& Peletier (\cite{grijs1997}) found that the exponential
scale height increases with distance along the major axis. The effect is strongest for early-type
galaxies (increase by 50\% for $T=-2$), but negligible for the latest-type galaxies ($T=8$).
Therefore, they concluded that this effect might be closely related to the presence of an
underlying thick disk, with a scale length larger than that of the thin disk (with the possibility
that thick disks were formed by accretion of small satellites or gas clouds).

Another frequently observed feature of vertical distortions are warped and bended disks.
Various theories have been proposed to explain the existence of warps, considering either
tidal interactions with neighbouring galaxies (e.g. Schwarz \cite{schwarz1985};
Zaritsky \& Rix \cite{zaritsky1997}), gravitational action of the halo on the disk
(Sparke \cite{sparke1984}; Sparke \& Casertano \cite{sparke1988}) or interaction
with or accretion of extragalactic gas (Binney \cite{binney1992}). Other possible
explanations for gentle warps and bended disks are intergalactic magnetic fields
(Battaner et al. \cite{battaner1990}) self-gravitation (Pfenniger et al. \cite{pfenniger1994}),
or discrete oscillations and bending modes (Sellwood \cite{sellwood1996}).

The results of recently conducted statistical studies (Sanchez-Saavedra et al. \cite{sanchez1990};
Reshetnikov \& Combes \cite{reshetnikov1998a}) enable to limit the number of different explanations
given for warped disks. They conclude that at least 50\% (up to 80\% after correction effects)
of all observed galaxies exhibit warped distortions. Furthermore, they find a strong positive correlation
of observed warps with the environment, suggesting that tidal interactions have a large influence in
creating or re-enforcing warped deformations. In spite of selection biases or identification problems
(Sanchez-Saavedra et al. \cite{sanchez1990}) the fraction of galaxies possessing warps, particularly
that of non-interacting galaxies, is unexpectedly high.

In two previous papers (Schwarzkopf \& Dettmar \cite{schwarzkopf2000a},\cite{schwarzkopf2000b},
hereafter SD~I+II) we studied the global properties of disks in 110 edge-on spiral galaxies.
The sample was divided into two subsamples of interacting/merging and non-interacting galaxies.
We found that galactic disks recently affected by tidal interactions and minor mergers of the
order of $M_{\rm sat}/M_{\rm disk} \approx 0.1$ statistically possess a 50\% -- 60\% larger
scale height. However, it is still ambiguous whether this thickening is due to a locally increased
disk scale height as a result of the vertical perturbations, or caused by global disk thickening.
Moreover, the properties of other tidally triggered features in the disks such as ``warps'' and
``flares'' have not yet been investigated.

Therefore, we use the galaxy sample described in SD~I+II in order to study the scale height of
both non-interacting and interacting/merging galaxies in greater detail. The main questions are:

\begin{itemize}

\item How constant is the disk scale height of interacting/merging and non-interacting galaxies?

\smallskip

\item Is there a systematic dependence of disk thickness with radial distance (scale height
 gradients)?

\smallskip

\item Do the tidally induced disk perturbations possess a typical size and lifetime?

\smallskip

\item Are warped disks restricted to interacting/merging galaxies, and what is the typical size
 of these warps?

\end{itemize}

In Sect.\~2 of this paper the sample and approach will be briefly reviewed. In Sect.\~3 we present
the results of a detailed analysis of the vertical disk structure for both non-interacting and
interacting/merging galaxies. Finally, the results of this paper will be discussed in Sect.\~4
and summarized in Sect.\~5.

%__________________________________________________________________

\section{Sample and disk fitting procedure}

\subsection{The sample}

The galaxy sample studied in this paper consists of two subsamples of 47 interacting/merging
and 61 non-interacting disk galaxies with morphological types ranging from $-1 \le T \le 9$.
All galaxies of the optical data set (passbands $R$ and $r$) presented here are nearly edge-on:
85\% with inclinations $ i \geq 88\degr$, and 15\% with $ i \ge 85\degr$.
Since the whole project (Schwarzkopf \cite{schwarzkopf1999} and SD~I+II) is focused on the
influence of interactions and minor mergers on the disk structure of spiral galaxies only
merging candidates in the mass range $M_{\rm sat}/M_{\rm disk} \approx 0.05 - 0.2$ were
included in the sample (mean ratio $0.08 \pm 0.035$). Both subsamples were selected in such
a way that selection biases, i.e. an uncertain classification of galaxies belonging
to the interacting or non-interacting sample, are largely avoided (SD\~I, Sect.\~2).
We have also shown that both the distribution of morphological galaxy types and distances
are statistically indistinguishable for the two subsamples (Sect.\~2 of SD\~I+II, resp.).
A detailed description of the selection criteria, observations, and data reduction as well
as a complete list of the parameters of all sample galaxies and their disk components was
given in SD\~I+II.

%__________________________________________________________________

\subsection{Approach}

First, the global disk parameters for each individual galaxy in the sample were roughly
estimated using the semi-automated disk fitting procedure described in Sect.\~4 of SD\~I.
These disk parameters are: inclination angle $i$, central luminosity density $L_{0}$, cut-off
radius $\Rmax$, scale length $h$, scale height $\zo$, and one of the 3 possible vertical disk
profiles $f(z) \propto$ exp, sech, and $\iso$. Since both the scale height and the choice
of the optimum vertical disk profile were still subject of further independent investigation,
we used the previously derived mean value of the disk scale height as a starting point for
the following vertical fitting procedure. The obtained raidal disk parameters were kept
fixed. This is a valid approximation since the effects of small ($\approx 10\%$) variations of
vertical disk parameters on radial disk profiles are negligible (SD\~I, Sect.\~4). Thus, the
only disk parameter that was changed according to the variable scale height was the central
luminosity density $L_{0}$.

Using these start parameters a fully automatical least-square fitting procedure was applied
in order to fit the disk scale height as a function of the galactocentric distance, $\zo(R)$.
The fit was performed for all 108 galaxies of the total sample, and for each vertical disk model
$f(z)$. Due to their importance for the subsequent analysis of the vertical disk structure the
main features of this fitting procedure and its output parameters (listed in Table~\ref{parameters})
are described in the following (for an illustration see examples in Figs.\~2-4 and 6):

\begin{itemize}

\item The scale height was fitted to the vertical disk profiles, starting in a region as close as
 possible to the galactic plane but outside the strongest dust absorption, out to $\approx 2\~\zo$.
 The radial fit region covers both sides of the major axis out to the disk cut-off, and outside the
 bulge-contaminated region. The selection of the fit regions (Table\~\ref{regions}) depends on
 the S/N ratio of the individual image as well as on several other factors (dust, foreground stars,
 position of companions, warps, etc.).

\smallskip

\item The mean disk scale height $(\zo)_{\rm mean}$, given in pc, and its standard deviation
 $(\zo)_{\rm std}$, given both in pc and in percent of $(\zo)_{\rm mean}$, were estimated.
 These values (average from both disk sides) are listed in Table\~\ref{parameters},
 columns\~(4) and (6), respectively.

\smallskip

\item A possible gradient (or constancy) of the disk scale height along the major axis, in the
 following referred to as $(\zo)_{\rm grad}$, was identified by using a 1st order fit (average
 of both sides) on the radial distribution of disk scale height. Thus the value $(\zo)_{\rm grad}$
 is a reliable estimate of the behaviour of disk scale height on large ($\approx \Rmax$) scales
 (column\~(5) in Table\~\ref{parameters}).

\smallskip

\item The variations of $\zo$ around the gradient $(\zo)_{\rm grad}$, i.e. an estimate of
 vertical disk perturbations on short scales and hereafter $(\zo)_{\rm std1}$, were defined
 as the standard deviation of the first order fit to the disk scale height:
 $(\zo)_{\rm std1} = \rm std [(\zo)_{\rm grad} - \zo(R)]$. The value $(\zo)_{\rm std1}$
 (average of both disk sides) is given in pc and in percent of $(\zo)_{\rm mean}$ (column\~(7)
 in Table\~\ref{parameters}).

\smallskip

\item The standard deviation around the position found for the mean galactic plane, named as
 ``Warping'' in column\~(8) of Table\~\ref{parameters}, was used as an estimate for the mean
 amplitude of stellar warps. It was calculated using the center of symmetric vertical disk
 profiles.

\smallskip

\item The complete vertical fit, $\zo(R)$, for all galaxies in the sample was performed for each
 individual luminosity profile $f(z) \propto$ exp, sech, $\iso$. The two quantitative best disk
 models, listed in column\~(3) of Table\~\ref{parameters}, were selected using the goodness-of-fit
 parameter $Q$ of the least-square test, which is given in Table\~\ref{parameters}, column\~(9).

\end{itemize}

The software used for the fit was designed primarily to characterize the vertical disk structure
of a large number of galaxies in a statistically systematical and homogeneous way. Therefore we
decided to use mean values and standard deviations as output parameters for quantifying the
properties of the disks.
This makes the applied method very robust against ``outliers''. The obtained results
(Tables\~\ref{perturb} - \ref{parameters}) are thus statistically more reliable than
measurements emphasizing individual amplitudes of vertical disk perturbations.

The first part of the paper provides a small sample of spiral galaxies that were found to be
``prototypes'' for particular disk features. They were included in order to illustrate
the behaviour of the scale height. A summary of all measured disk parameters
is given in Table\~\ref{parameters}. The appendix shows the behaviour of disk scale height
and mean galactic plane for the complete sample of 108 objects. The variations of disk
thickness (upper panels) and galactic plane (lower panels) are drawn to the same scale.
The sample of interacting/merging galaxies is shown in Appendix A, and Appendix B contains
the non-interacting galaxies. The corresponding catalog of contour maps can be found in
Figs.\~4 and 5 of SD\~I, respectively.

%%%%%%%%%%%%%%%%%%%%%%%%%%%%%%%%%%%%%%%%%%%%%%%%%%%%%%%%%%%%%%%%%%%

%__________________________________________________________________
%
% Table 1 - Regions used for radial \& vertical fit.
%
%__________________________________________________________________
%

\tabcolsep1.25mm

  \begin{table}[t]
  \caption[ ]{Regions adopted for fitting the disk scale height $\zo$.
   Columns: (1) Sample used for the statistics: total= all galaxies ($n=108$);
   non-int.= non-interacting galaxies ($n=61$); int./merg.= interacting/merging
   galaxies ($n=47$); (2) Radial region used for the fit (avg. values, given in
   units of disk cut-off radius $\Rmax$ and disk scale length $h$); (3) Vertical
   region used for the fit (avg. values, given in units of disk scale height $\zo$).}
  \label{regions}
  \begin{flushleft}
  \begin{tabular}{lcccc}
  \cline{1-5}
  \hline\hline\noalign{\smallskip}
  \multicolumn{1}{c}{Sample}  &
  \multicolumn{3}{c}{\underline{ \hspace{1mm} Radial region of fit \hspace{1mm} }}    &
  \multicolumn{1}{c}{\underline{Vertical region of fit}} \\
  \noalign{\smallskip}
  \multicolumn{1}{c}{}  & \multicolumn{1}{c}{$[\Rmax]$}  && \multicolumn{1}{c}{$[h]$} &
  \multicolumn{1}{c}{$[\zo]$} \\
  \noalign{\smallskip}
  \multicolumn{1}{c}{(1)} & \multicolumn{3}{c}{(2)} & \multicolumn{1}{c}{(3)} \\
  \noalign{\smallskip}
  \hline\noalign{\smallskip}
   total        &  0.23 -- 0.83  &&  0.76 -- 2.77   &  0.11 -- 1.81  \\
   non-int.     &  0.24 -- 0.82  &&  0.79 -- 2.69   &  0.08 -- 1.95  \\
   int./merg.   &  0.21 -- 0.83  &&  0.72 -- 2.85   &  0.15 -- 1.67  \\
  \noalign{\smallskip}
  \hline\noalign{\smallskip}
  \end{tabular}
  \end{flushleft}

  \end{table}

%%%%%%%%%%%%%%%%%%%%%%%%%%%%%%%%%%%%%%%%%%%%%%%%%%%%%%%%%%%%%%%%%%%

%__________________________________________________________________

%%%%%%%%%%%%%%%%%%%%%%%%%%%%%%%%%%%%%%%%%%%%%%%%%%%%%%%%%%%%%%%%%%%

\begin{figure*}[t]

\vspace*{76mm}

\begin{minipage}[b]{8.8cm}
\begin{picture}(8.8,7.8)
{\includegraphics[angle=90,viewport=10 55 650 735,clip,width=90mm]{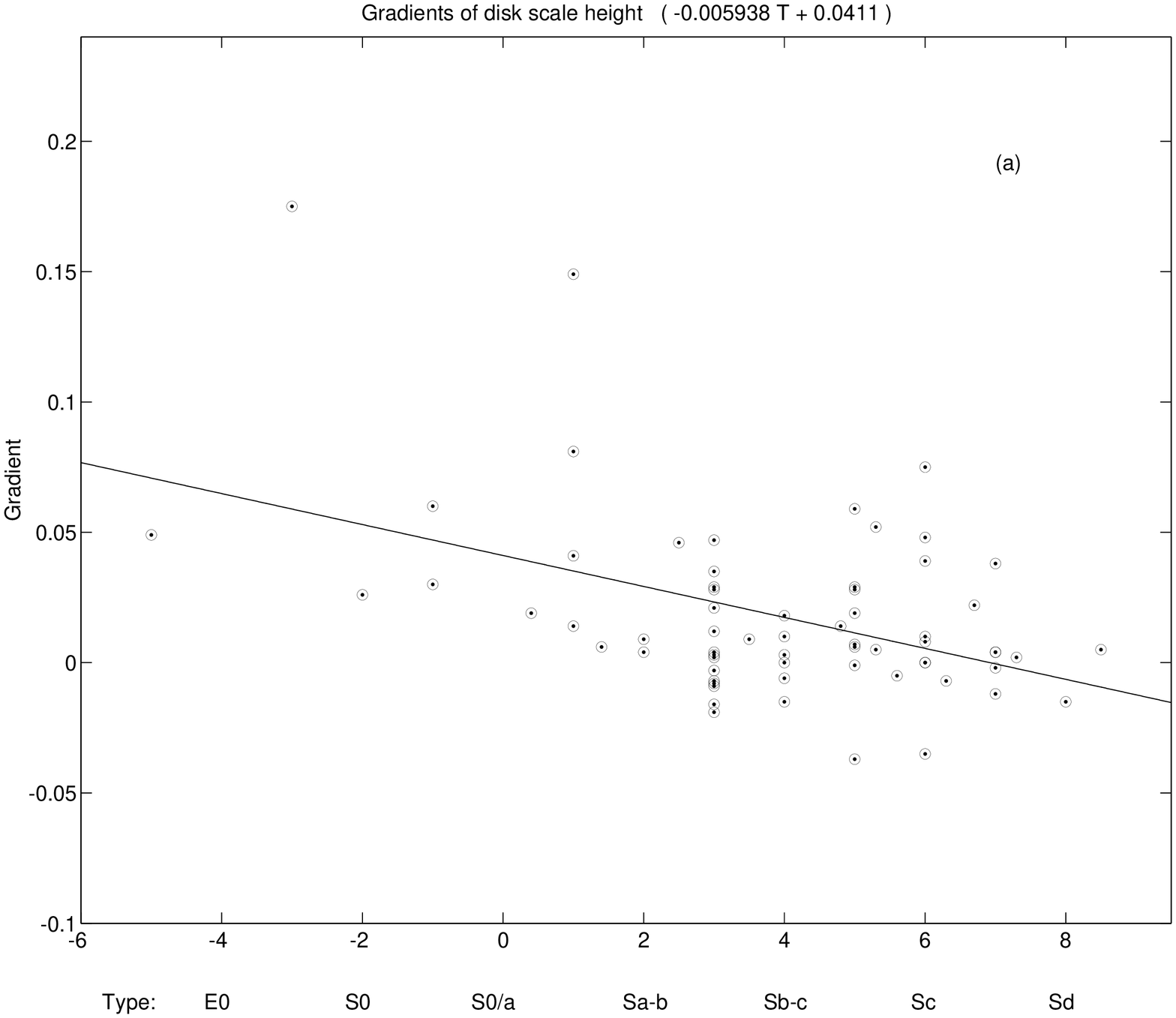}}
\end{picture}
\end{minipage}
\hfill
\begin{minipage}[b]{8.8cm}
\begin{picture}(8.8,7.6)
{\includegraphics[angle=90,viewport=10 55 650 735,clip,width=90mm]{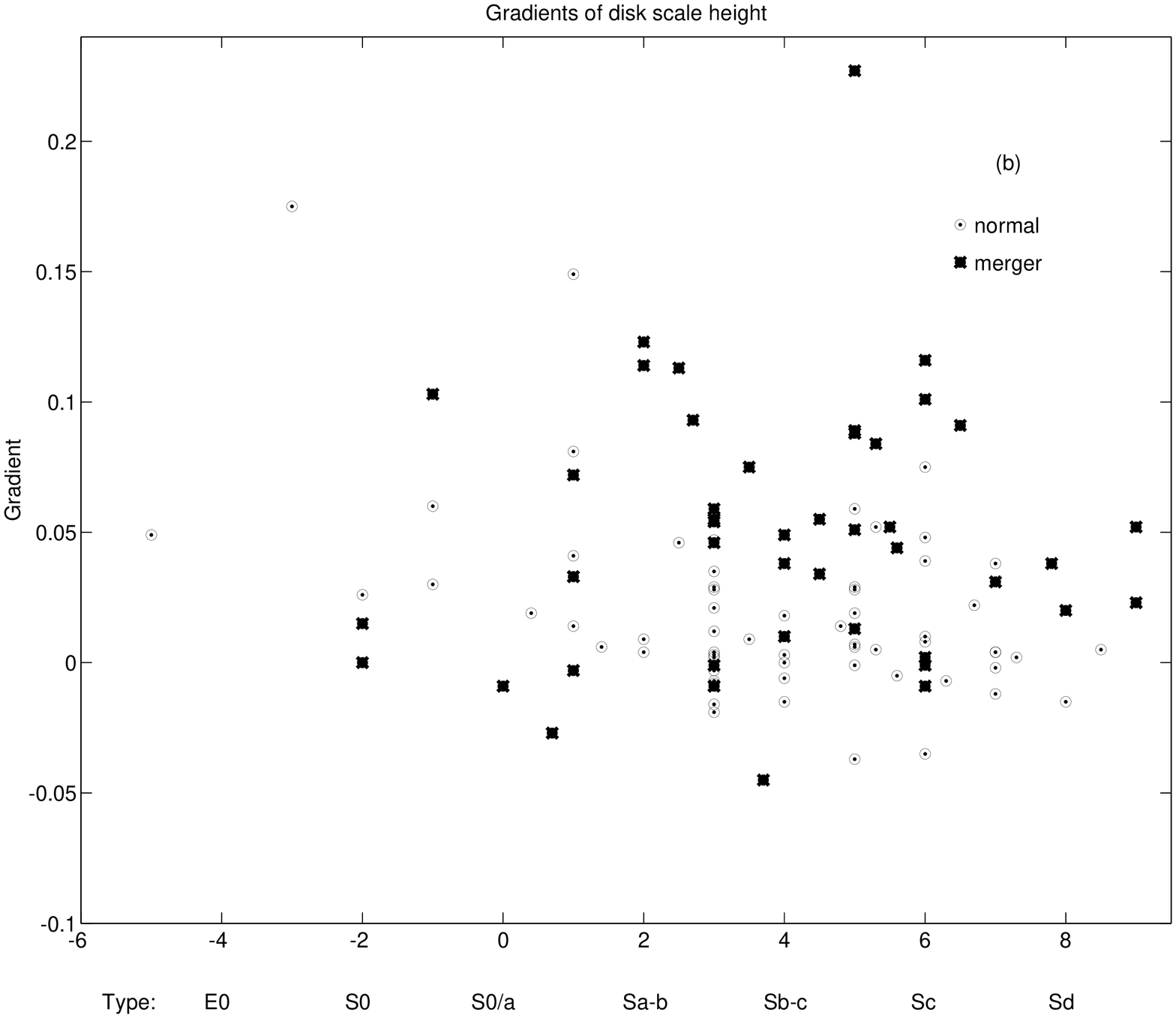}}
\end{picture}
\end{minipage}

\vspace{2mm}

{\bf \noindent Fig.\~1.} Gradients $(\zo)_{\rm grad}$ of disk scale height
(taken from Table\~\ref{parameters}, column\~(5)) versus morphological galaxy type.
Shown are {\bf (a)} the sample of non-interacting galaxies, {\bf (b)} both
samples of non-interacting and interacting/merging galaxies.

\end{figure*}

%%%%%%%%%%%%%%%%%%%%%%%%%%%%%%%%%%%%%%%%%%%%%%%%%%%%%%%%%%%%%%%%%%%

\section{Results}

\subsection{The radial behaviour of the disk scale height}

The behaviour of the disk scale height as a function of galactocentric distance, $\zo(R)$, is
studied for the total galaxy sample. A short description of the applied fitting procedure and
their output parameters was given in Sect.\~2. The complete set of parameters obtained from
an analysis of the vertical disk structure is listed in Table\~\ref{parameters}.

In order to derive reliable fit parameters it is important to use similar regions for both the radial
and vertical fit, independent of the subsample investigated. As shown in Table\~\ref{regions}, the
radial and vertical fit regions used for the two galaxy samples are typically between $0.76 - 2.77$
disk scale lengths and $0.11 - 1.81$ scale heights and thus almost identical. These regions nearly
cover the whole disk outside the bulge-contaminated region (named as ``bulge'' in the plots of
the disk scale height). All galaxies in the sample are nearly edge-on (Sect.\~2), and the vertical
profiles were fitted out to $\approx 2$ scale heights (Table\~\ref{regions}). As a result the
influence of dust on the shape of the vertical profiles and thus on the measured scale heights
is negligible. This is confirmed by comparable results obtained after fitting $\approx 50\%$ of
the sample galaxies in the near-infrared (Table\~3 and Sect.\~3.3 in SD\~II).
All images used for this study were binned radially by $\approx 2''$ in order to
increase the S/N ratio. In order to retain the full vertical resolution no binning was applied in
that direction. For the majority of galaxies within a distance of 10\~Mpc -- 50\~Mpc this gives
a radial and vertical spatial resolution between 100\~pc -- 600\~pc and 10\~pc -- 50\~pc, respectively.
This resolution is sufficient to detect even small deviations and perturbations in the vertical
disk structure.

%__________________________________________________________________

\subsubsection{The mean value of the disk scale height}

The mean value $(\zo)_{\rm mean}$ of the disk scale height and its error $(\zo)_{\rm std}$, measured
for all galaxies of the two subsamples, are listed in columns\~(4) and (6) of Table\~\ref{parameters},
respectively.
Since many disks show medium or even large gradients in the scale height (i.e. their scale height
increases significantly with radial distance, see next section) the obtained errors for the total
sample are spread in a wide range around an average value of $(\zo)_{\rm std} \approx 12\%$
(Table\~\ref{perturb}, column\~(2)). Within this error the measured scale heights $(\zo)_{\rm mean}$
are consistent with those obtained using an independent, semi-automated fitting procedure
(described in Sect.\~4 of SD\~I, values listed in Fig.\~5 and Table\~5 of SD\~II).
Both methods also independently deliver the same difference between the mean scale height of
both subsamples, which was interpreted as tidally-triggered vertical heating in SD\~II.
As a result, disks of interacting/merging galaxies were found systematically $\approx 50\%$
thicker than those of non-interacting galaxies. A detailed discussion of the results and
the derived conclusions can be found in SD\~I+II.

%__________________________________________________________________

\subsubsection{The increase of scale height (gradients)}

The behaviour of the scale height over the whole radial extent of the disk, i.e. its
gradient $(\zo)_{\rm grad}$, is listed in column\~(5) of Table\~\ref{parameters}. In the
following we will use the term ``gradient'' independently whether there is an observed increase or
decrease of the scale height ($\zo_{\rm grad} \neq 0 $) or not ($\zo_{\rm grad}$ = 0). In Fig.\~1
the measured gradients are plotted against the morphological type of the galaxies (both subsamples
are shown). Although the correlation is weak, Fig.\~1a indicates that the scale height of
non-interacting, early-type ($T \le 2$) galaxies increases towards the outer parts of the disk.
While the obtained values are relatively large ($\zo_{\rm grad} > 0.04$) almost no
gradients were found for non-interacting galaxies with $T \ge 3$.

In order to illustrate the different radial behaviour of disk scale heights Fig.\~2 shows, as an
example, the results of vertical fits obtained for 4 galaxies of the non-interacting sample with
morphological types ranging from $-1.0 \le T \le 7.0$. The used disk parameters as well as the
corresponding scale height gradients, $(\zo)_{\rm grad}$, are indicated.
The reason for the relatively weak correlation with type is the high intrinsic scatter. This scatter
is due to a number of disks that show considerably low or high gradients compared to the mean level
for galaxies with about the same morphological type.

Although many of the interacting/merging spirals in our sample possess considerable gradients
(Fig.\~1b) no systematic correlation between these gradients and morphological type was found.
The size and substructure of scale height gradients within the sample of interacting/merging
galaxies will be investigated in greater detail in the next section, together with an analysis
of accompanying environmental effects like disk ``warping'' and ``flaring''.

%%%%%%%%%%%%%%%%%%%%%%%%%%%%%%%%%%%%%%%%%%%%%%%%%%%%%%%%%%%%%%%%%%%

\begin{figure*}[t]

\vspace*{60mm}

\hspace*{26mm}
\begin{minipage}[b]{6.0cm}
\begin{picture}(6.0,6.0)
{\includegraphics[angle=0,viewport=40 220 540 420,clip,width=128mm]{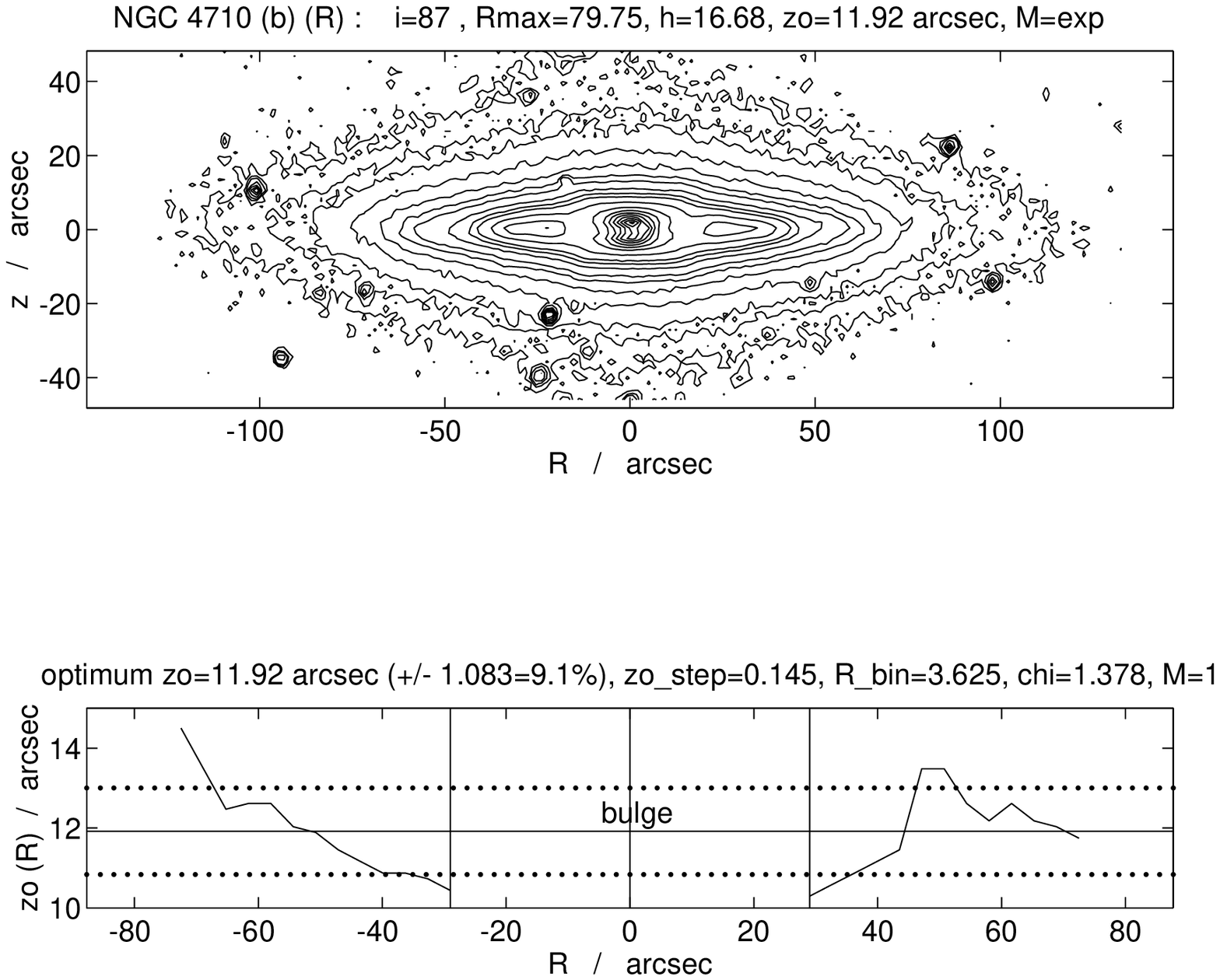}}
\end{picture}
\end{minipage}

\vspace*{37mm}

\hspace*{26mm}
\begin{minipage}[b]{6.0cm}
\begin{picture}(6.0,6.0)
{\includegraphics[angle=0,viewport=40 10 540 150,clip,width=128mm]{ngc4710.ps}}
\end{picture}
\end{minipage}

\vspace*{66mm}

\hspace*{26mm}
\begin{minipage}[b]{6.0cm}
\begin{picture}(6.0,6.0)
{\includegraphics[angle=0,viewport=40 220 540 420,clip,width=128mm]{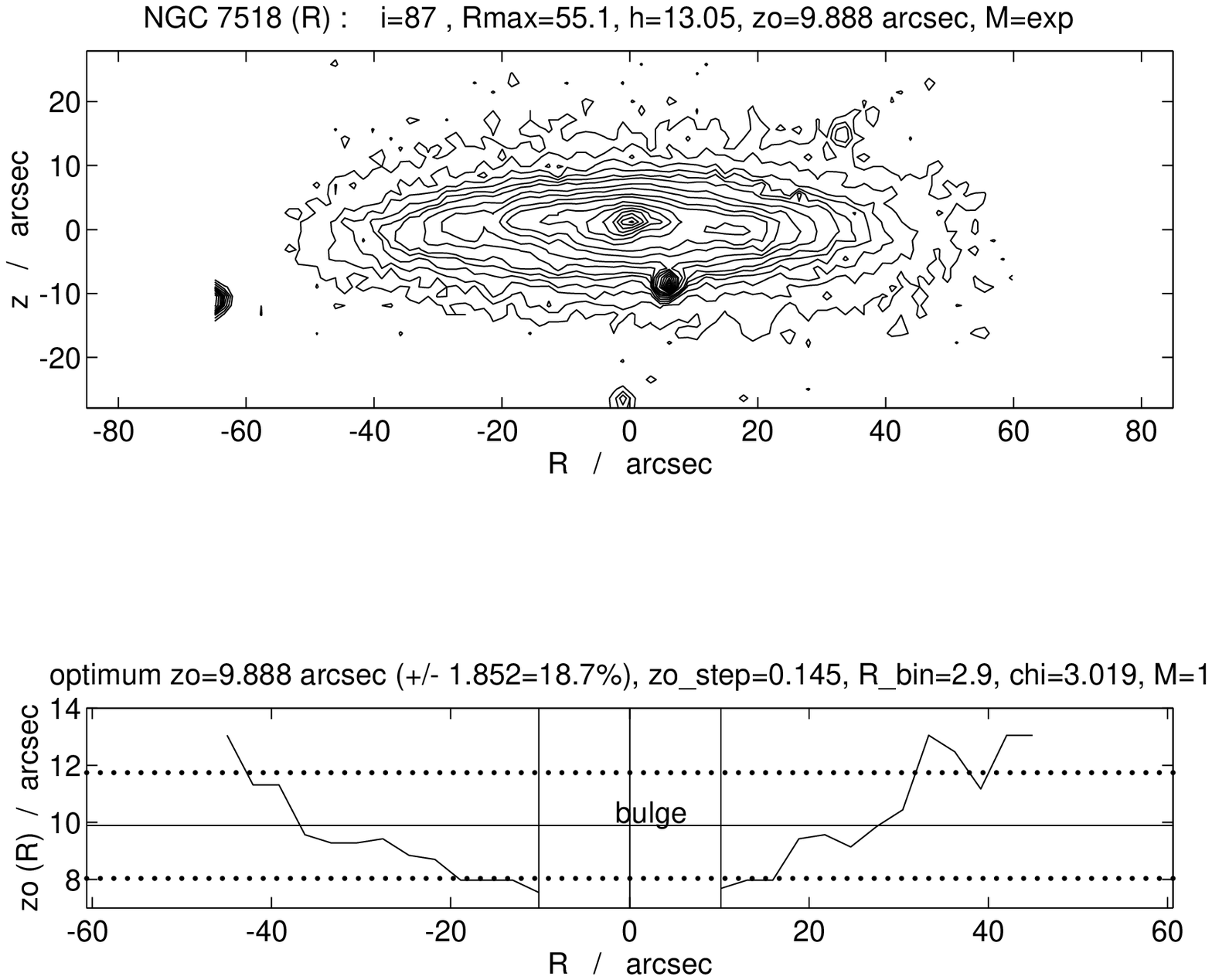}}
\end{picture}
\end{minipage}

\vspace*{37mm}

\hspace*{26mm}
\begin{minipage}[b]{160cm}
\begin{picture}(6.0,6.0)
{\includegraphics[angle=0,viewport=40 10 540 150,clip,width=128mm]{ngc7518.ps}}
\end{picture}
\end{minipage}

\vspace{5mm}

{\bf \noindent Fig.\~2.} Examples of non-interacting galaxies with morphological types
between $ -1 \le T \le 6$ showing different levels of variation in their scale height
(corresponding lower panels). The galaxies, their type and the measured scale height
gradients are: \\
NGC 4710, $T=-1.0$, $(\zo)_{\rm grad}=0.06\pm0.009$ and NGC 7518, $T=1.0$,
$(\zo)_{\rm grad}=0.149\pm0.008$.

\end{figure*}

%%%%%%%%%%%%%%%%%%%%%%%%%%%%%%%%%%%%%%%%%%%%%%%%%%%%%%%%%%%%%%%%%%%

%%%%%%%%%%%%%%%%%%%%%%%%%%%%%%%%%%%%%%%%%%%%%%%%%%%%%%%%%%%%%%%%%%%

\begin{figure*}[t]

\vspace*{78mm}

\hspace*{26mm}
\begin{minipage}[b]{6.0cm}
\begin{picture}(6.0,6.0)
{\includegraphics[angle=90,viewport=230 50 590 725,clip,width=128mm]{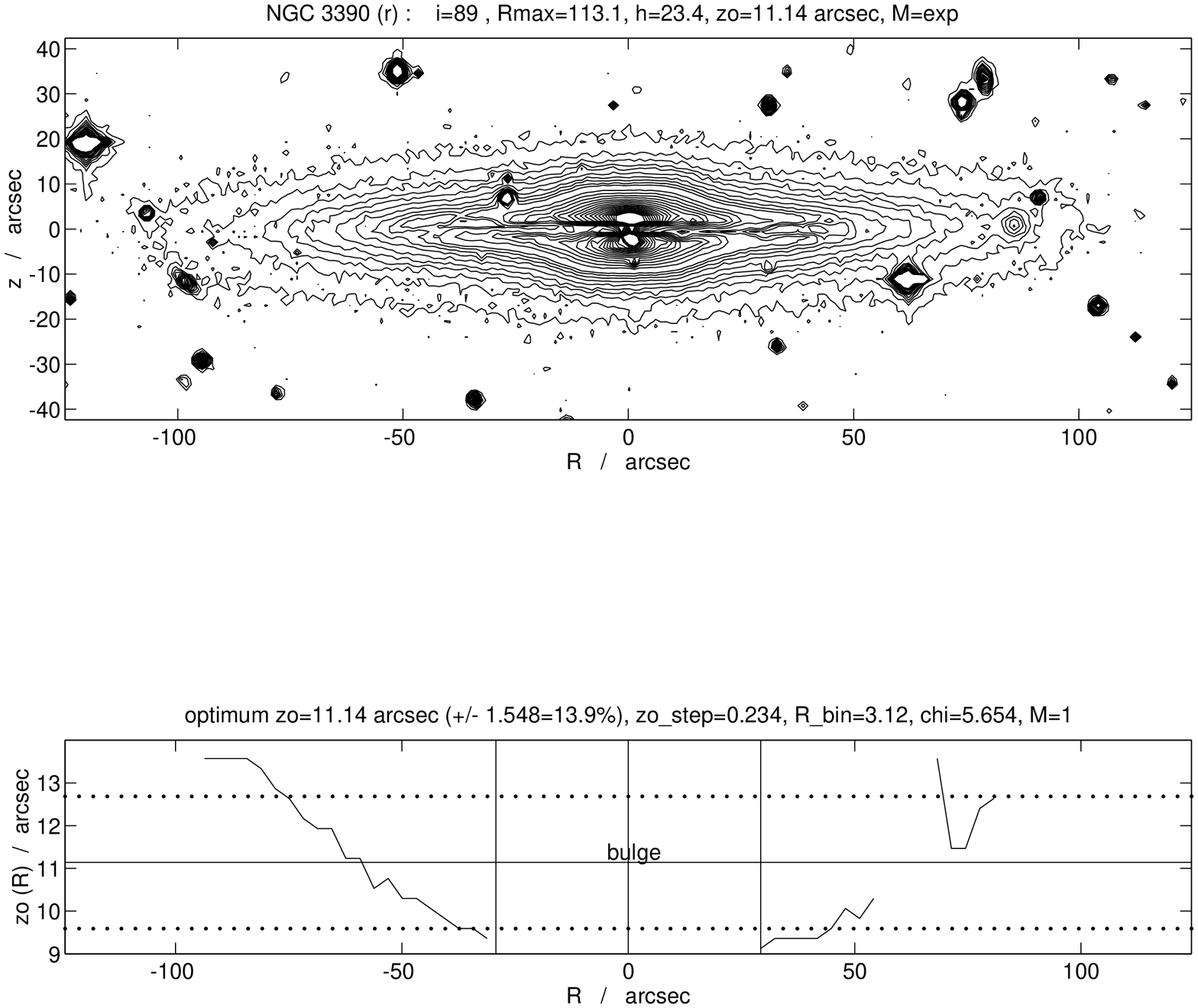}}
\end{picture}
\end{minipage}

\vspace*{20mm}

\hspace*{26mm}
\begin{minipage}[b]{6.0cm}
\begin{picture}(6.0,6.0)
{\includegraphics[angle=90,viewport=20 50 210 725,clip,width=128mm]{ngc3390.ps}}
\end{picture}
\end{minipage}

\vspace*{85mm}

\hspace*{26mm}
\begin{minipage}[b]{6.0cm}
\begin{picture}(6.0,6.0)
{\includegraphics[angle=90,viewport=230 50 590 725,clip,width=128mm]{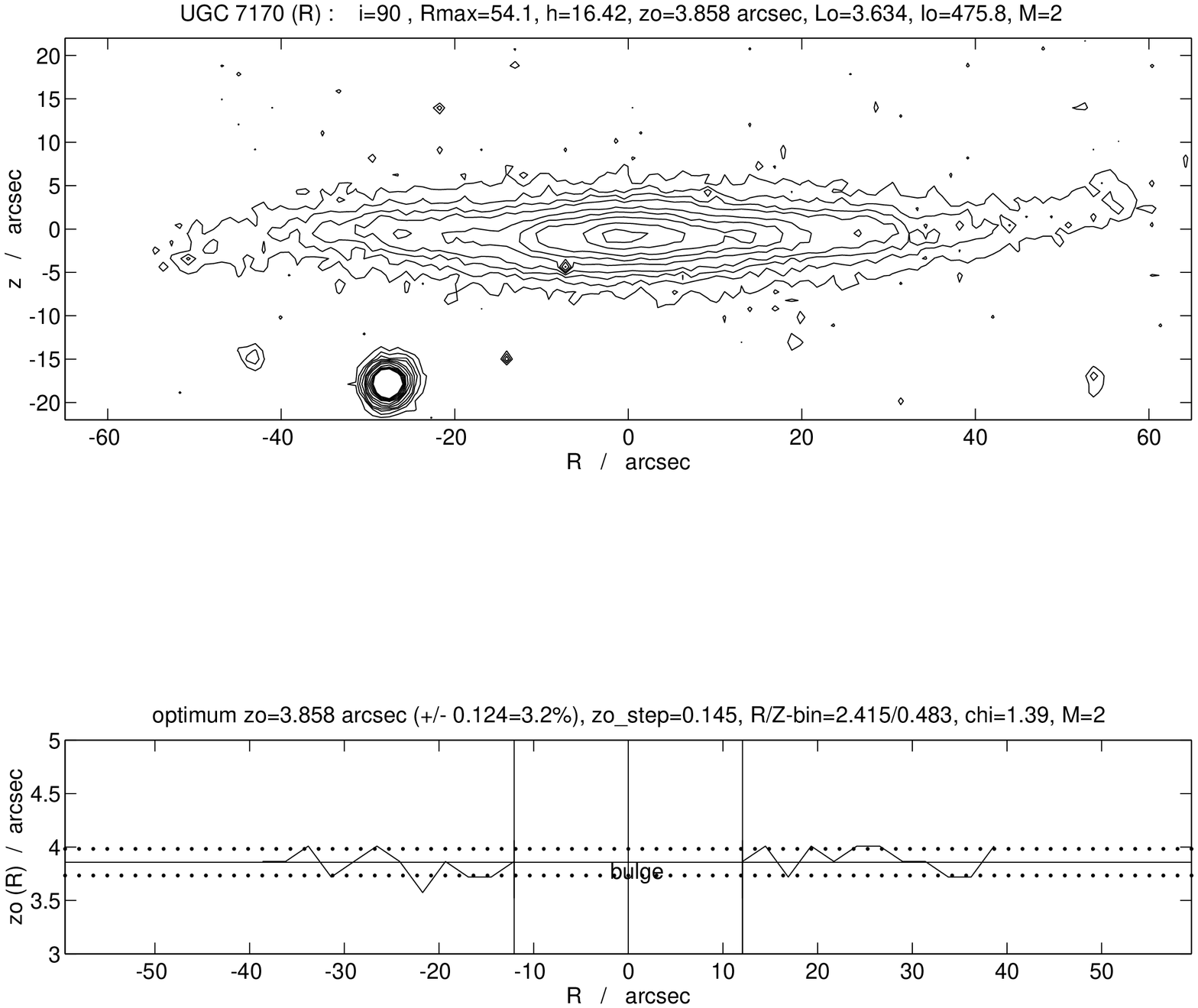}}
\end{picture}
\end{minipage}

\vspace*{20mm}

\hspace*{26mm}
\begin{minipage}[b]{6.0cm}
\begin{picture}(6.0,6.0)
{\includegraphics[angle=90,viewport=20 50 210 725,clip,width=128mm]{ugc7170.ps}}
\end{picture}
\end{minipage}

\vspace{6mm}

{\bf \noindent Fig.\~2.} (continued) NGC 3390, $T=3.0$, $(\zo)_{\rm grad}=0.03\pm0.009$ and
UGC 7170, $T=6.0$, $(\zo)_{\rm grad}=0\pm0.003$.

\end{figure*}

%%%%%%%%%%%%%%%%%%%%%%%%%%%%%%%%%%%%%%%%%%%%%%%%%%%%%%%%%%%%%%%%%%%

%%%%%%%%%%%%%%%%%%%%%%%%%%%%%%%%%%%%%%%%%%%%%%%%%%%%%%%%%%%%%%%%%%%

\begin{figure*}[t]

\vspace*{78mm}

\hspace*{26mm}
\begin{minipage}[b]{6.0cm}
\begin{picture}(6.0,6.0)
{\includegraphics[angle=90,viewport=220 50 585 725,clip,width=128mm]{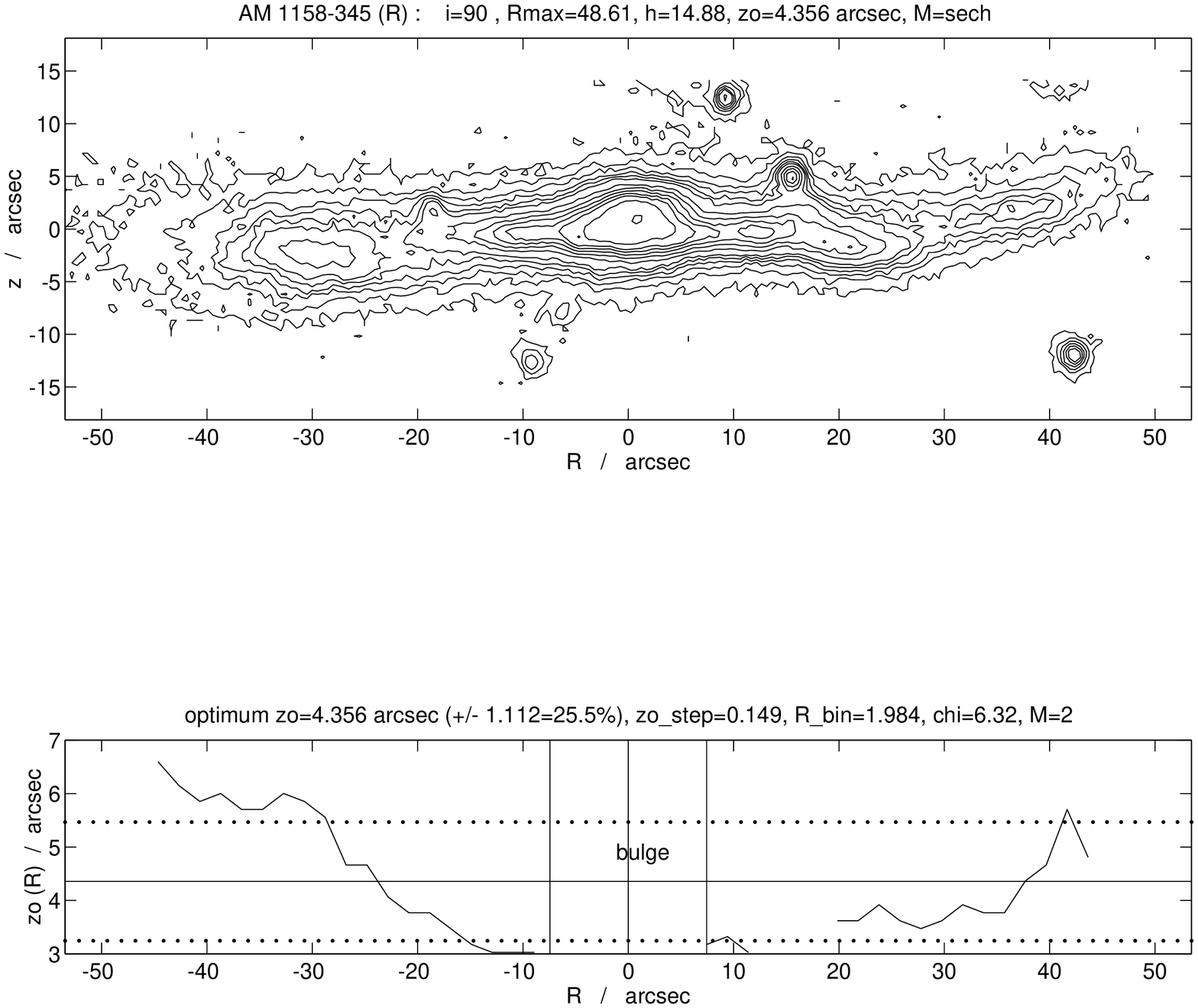}}
\end{picture}
\end{minipage}

\vspace*{20mm}

\hspace*{26mm}
\begin{minipage}[b]{6.0cm}
\begin{picture}(6.0,6.0)
{\includegraphics[angle=90,viewport=20 50 220 725,clip,width=128mm]{eso379_20.ps}}
\end{picture}
\end{minipage}

\vspace*{80mm}

\hspace*{26mm}
\begin{minipage}[b]{6.0cm}
\begin{picture}(6.0,6.0)
{\includegraphics[angle=90,viewport=220 50 585 725,clip,width=128mm]{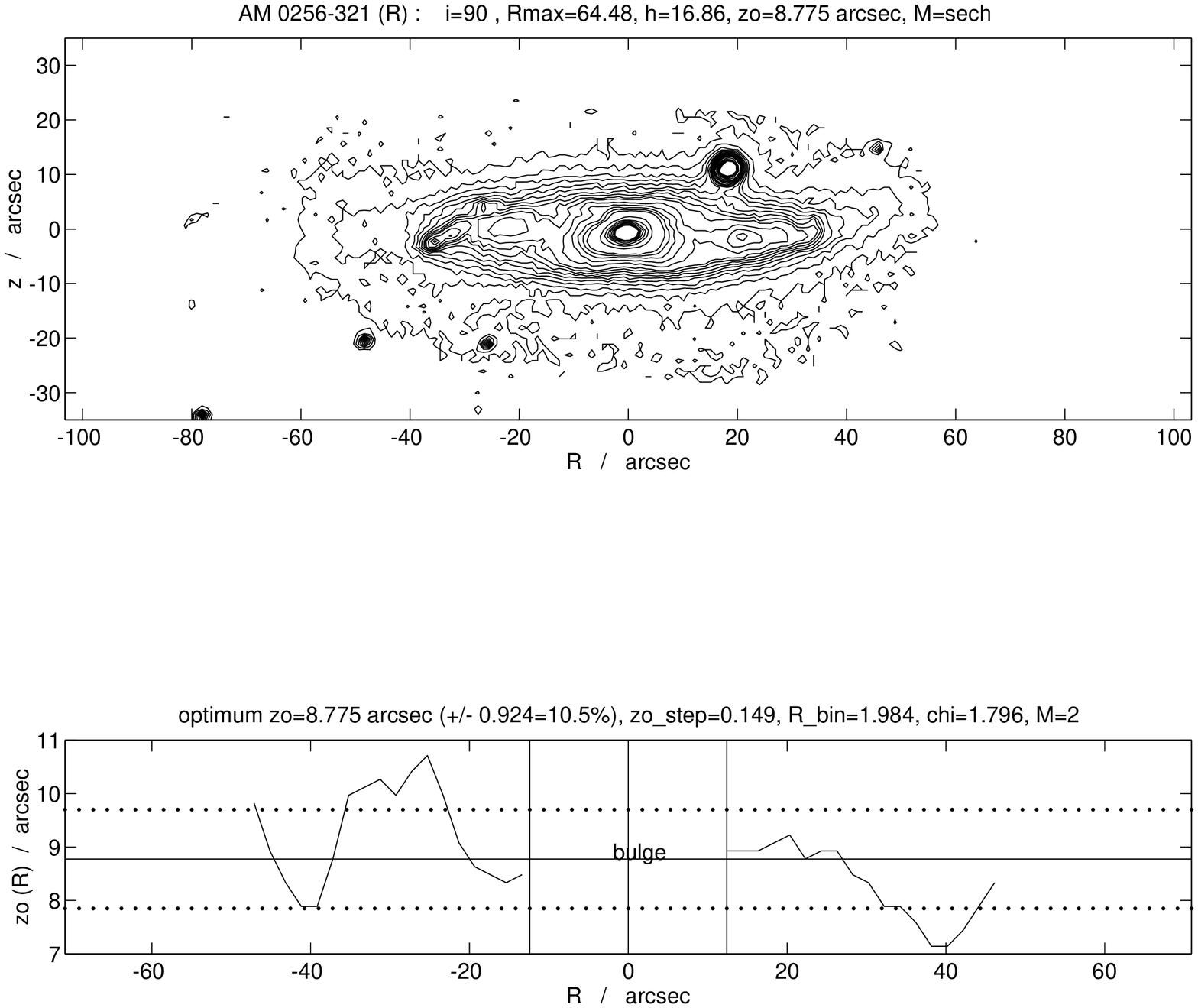}}
\end{picture}
\end{minipage}

\vspace*{20mm}

\hspace*{26mm}
\begin{minipage}[b]{6.0cm}
\begin{picture}(6.0,6.0)
{\includegraphics[angle=90,viewport=20 50 220 725,clip,width=128mm]{eso417_08.ps}}
\end{picture}
\end{minipage}

\vspace{6mm}

{\bf \noindent Fig.\~3.} Two examples of galaxies seen during a phase of an ongoing
minor merger. The small companion and/or material of debris are still visible.
The corresponding lower panels show variations of the scale height over the whole
radial extent of the disk:  a large scale height gradient for AM\~1158-345 (above)
and variations on short scales for AM\~0256-321 (below).

\end{figure*}

%%%%%%%%%%%%%%%%%%%%%%%%%%%%%%%%%%%%%%%%%%%%%%%%%%%%%%%%%%%%%%%%%%%

%%%%%%%%%%%%%%%%%%%%%%%%%%%%%%%%%%%%%%%%%%%%%%%%%%%%%%%%%%%%%%%%%%%

\begin{figure*}[t]

\vspace*{80mm}

\hspace*{26mm}
\begin{minipage}[b]{6.0cm}
\begin{picture}(6.0,6.0)
{\includegraphics[angle=90,viewport=220 50 585 725,clip,width=128mm]{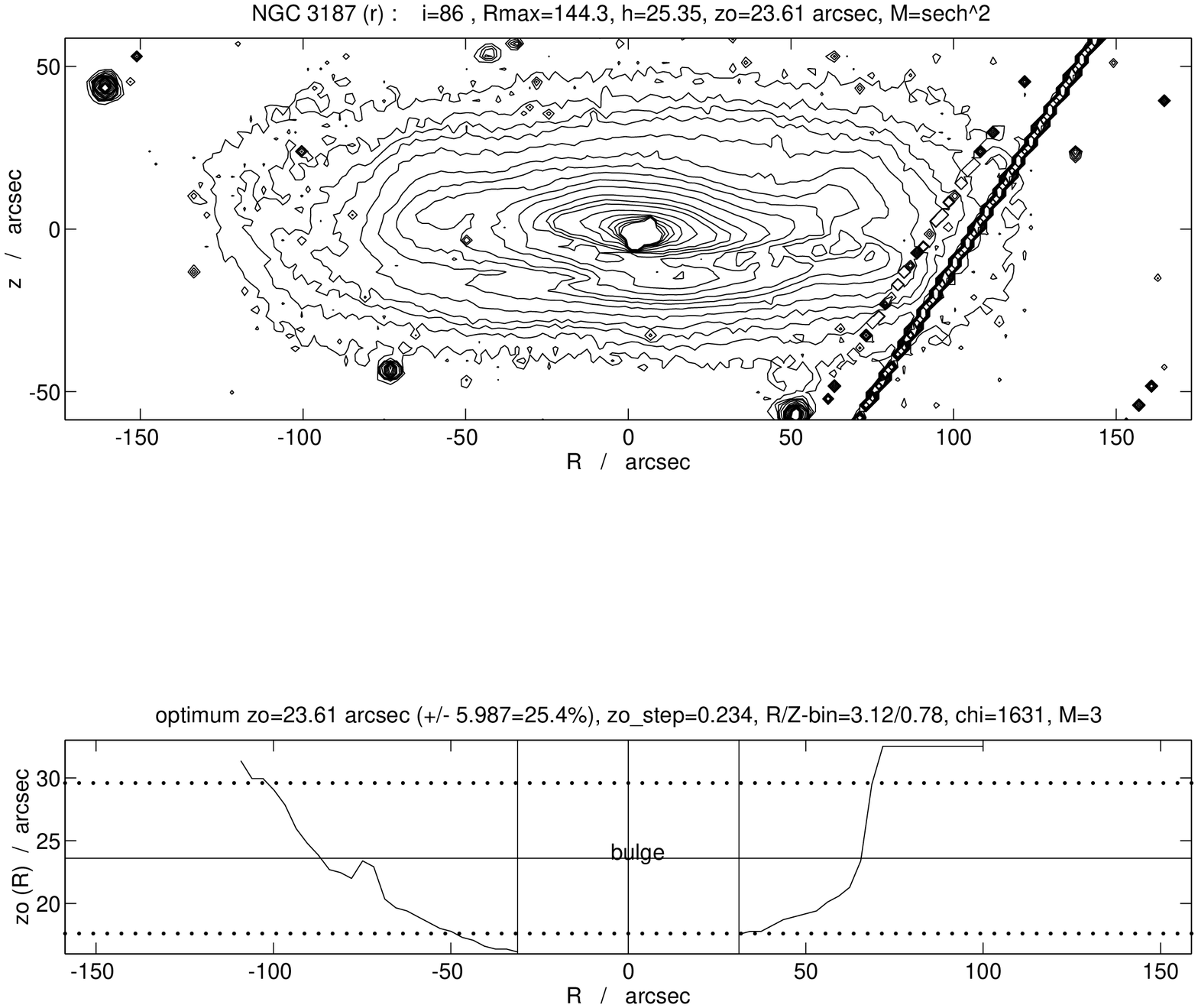}}
\end{picture}
\end{minipage}

\vspace*{13mm}

\hspace*{26mm}
\begin{minipage}[b]{6.0cm}
\begin{picture}(6.0,6.0)
{\includegraphics[angle=90,viewport=20 50 220 725,clip,width=128mm]{ngc3187.ps}}
\end{picture}
\end{minipage}

\vspace*{88mm}

\hspace*{26mm}
\begin{minipage}[b]{6.0cm}
\begin{picture}(6.0,6.0)
{\includegraphics[angle=90,viewport=220 50 585 725,clip,width=128mm]{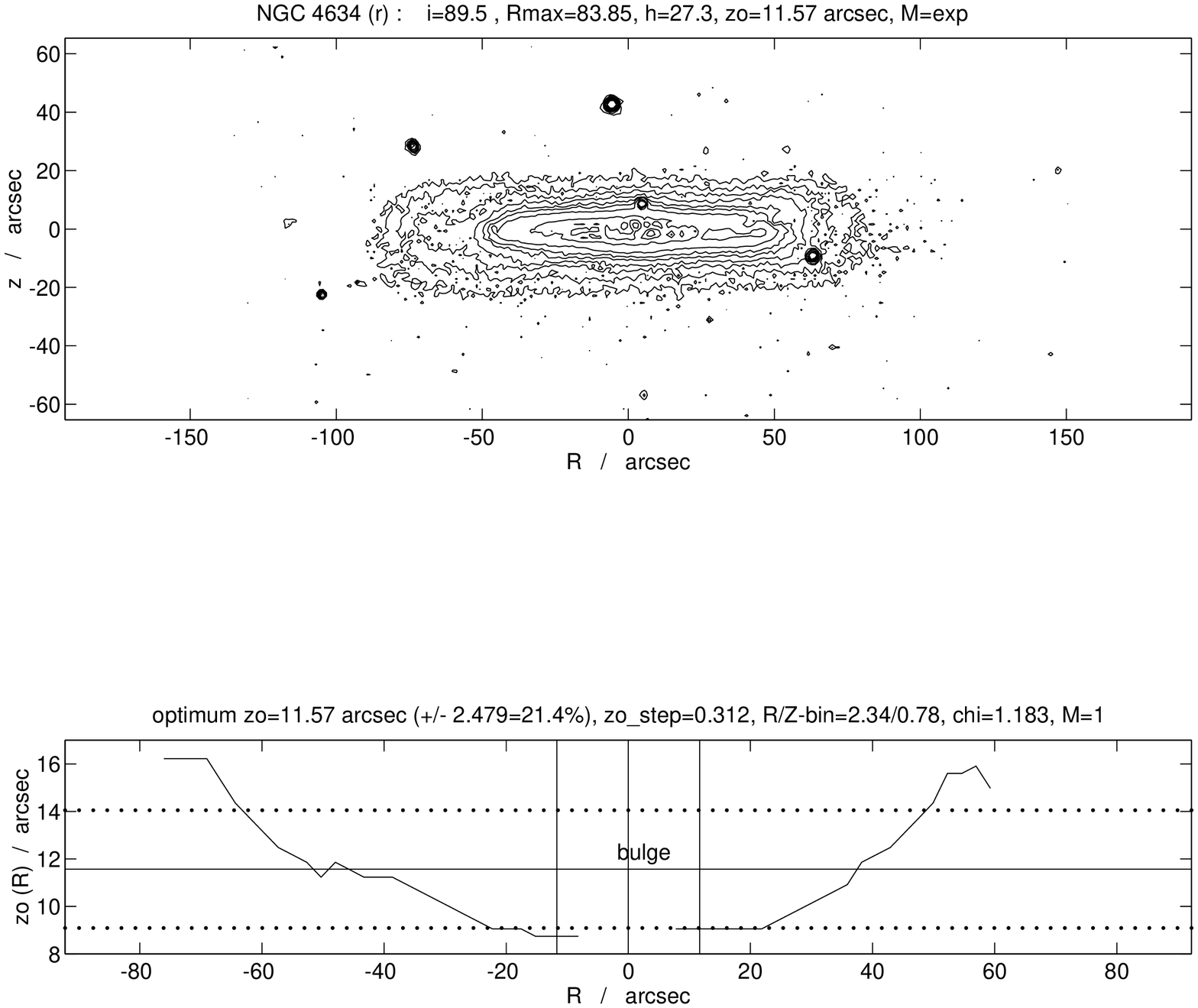}}
\end{picture}
\end{minipage}

\vspace*{13mm}

\hspace*{26mm}
\begin{minipage}[b]{6.0cm}
\begin{picture}(6.0,6.0)
{\includegraphics[angle=90,viewport=20 50 220 725,clip,width=128mm]{ngc4634.ps}}
\end{picture}
\end{minipage}

\vspace{6mm}

{\bf \noindent Fig.\~4.} Two examples of late-type ($T= 5$ and 6) edge-on spirals
possessing a considerable increase of scale height with radial distance and hence
large scale height gradients (between 0.12 and 0.23). The outer parts of their
disks show ``fringed'' isophotes. Both galaxies are members of compact groups and
have been affected by tidal interactions in the past.

\end{figure*}

%%%%%%%%%%%%%%%%%%%%%%%%%%%%%%%%%%%%%%%%%%%%%%%%%%%%%%%%%%%%%%%%%%%

%__________________________________________________________________

\subsubsection{Tidally-triggered vertical disk perturbations}

In this section we investigate the typical size of tidally induced vertical perturbations
in the disk and compare them with values found for non-interacting galaxies. In order to
quantify the perturbations both subsamples of non-interacting and interacting/merging
galaxies were compared using the disk parameters $(\zo)_{\rm std}$ and $(\zo)_{\rm std1}$.
The approach was described in Sect.\~2. Due to their importance for the forthcomming analysis
of the vertical disk structure in the following we give a brief interpretation of these
parameters (listed in columns\~(6) and (7) of Table\~\ref{parameters}).

The standard deviation $(\zo)_{\rm std}$ of the mean scale height characterizes the {\em sum}
of all those vertical variations that are due to both perturbations on short ($\approx \zo$)
scales (e.g. ``flaring'') {\em and} a possibly systematical trend (scale height gradients).
By contrast, the standard deviation $(\zo)_{\rm std1}$ of the first-order fit of the scale
height considers a large-scale gradient and only gives an estimate of vertical disk perturbations
on short scales. The values $(\zo)_{\rm std}$ and $(\zo)_{\rm std1}$ are listed in columns\~(2)
and (3) of Table\~\ref{perturb}, respectively (averaged for the total galaxy sample and for both
subsamples).

%__________________________________________________________________
%
% Table 2 - Vertical perturbation and warping of galaxy disks.
%
%__________________________________________________________________
%

\tabcolsep1.1mm

  \begin{table}
  \caption[ ]{Size of vertical disk perturbations and warps.\\
   Columns: (1) Galaxy sample used for the statistics: total= all; non-int.= non-interacting;
   int./merg.= interacting/merging; (2) and (3) Scale of vertical disk perturbations, i.e.
   (2) sum of (large+short) and (3) short-scale perturbations only, given in percent of
   mean disk scale height $(\zo)_{\rm mean}$ and in pc (average values, derived from
   $(\zo)_{\rm std}$ and $(\zo)_{\rm std1}$ in Table \ref{parameters}, resp.);
   (4) Disk warping, given in percent of $(\zo)_{\rm mean}$ and in pc (average values,
   derived from column\~(8) in Table\~\ref{parameters}).}
  \label{perturb}
  \begin{flushleft}
  \begin{tabular}{lccccccccc}
  \cline{1-10}
  \hline\hline\noalign{\smallskip}
  \multicolumn{1}{c}{Sample}     &&  \multicolumn{5}{c}{Scale of vertical perturbation}  &&
  \multicolumn{2}{c}{Disk}    \\
  \noalign{\smallskip}
   &&  \multicolumn{2}{c}{large + short}  &&  \multicolumn{2}{c}{short-scale}  &&
  \multicolumn{2}{c}{Warping} \\
  \noalign{\smallskip}
  \cline{3-4}
  \cline{6-7}
  \cline{9-10}
  \noalign{\smallskip}
  \multicolumn{1}{c}{}     &&  \multicolumn{1}{c}{[\%]}  &  \multicolumn{1}{c}{[pc]}  &&
  \multicolumn{1}{c}{[\%]} &   \multicolumn{1}{c}{[pc]}  && \multicolumn{1}{c}{[\%]}  &
  \multicolumn{1}{c}{[pc]} \\
  \noalign{\smallskip}
  \multicolumn{1}{c}{(1)}  &&  \multicolumn{2}{c}{(2)}   && \multicolumn{2}{c}{(3)}   &&
  \multicolumn{2}{c}{(4)}  \\
  \noalign{\smallskip}
  \hline\noalign{\smallskip}
   total            &&  12      &  210  &&  6      &  100  &&  14       &  220  \\
   non-int.         &&  10      &  140  &&  6      &  80   &&  11       &  140  \\
   int./merg.$\A$   &&  14 (22) &  280  &&  6 (9)  &  130  &&  17 (26)  &  340  \\
  \noalign{\smallskip}
  \hline\noalign{\smallskip}
  \end{tabular}
  \begin{list}{}{}
  \item[$\A$] Values in parenthesis are normalized to the disk thick- ness
              of non-int. galaxies, i.e. corrected by a factor of 1.5.\\
  \end{list}
  \end{flushleft}
  \end{table}

%__________________________________________________________________

According to Table\~\ref{perturb} the obtained results can be summarized as follows: the typical
size of vertical disk perturbations in interacting/merging galaxies -- including both gradients
on large scales as well as perturbations on short scales -- is $\approx 14\%$ of the mean scale
height. The corresponding value found for non-interacting spirals is $\approx 10\%$. In order
to compare both values it must be considered that disks of interacting/merging galaxies were
found to be systematically thicker by $\approx 50\%-60\%$ on average (Sect.\~3 of SD\~II and
previous section of this paper). Applying this correction the relative size of vertical disk
perturbations in interacting/merging and non-interacting galaxies is of the order of
22\% and 10\%, respectively (normalized to the scale height of non-interacting galaxies).
This factor of two is confirmed by the absolute size of 280\~pc and 140\~pc measured for
the perturbations (Table\~\ref{perturb}, column\~(2)).

The size of disk perturbations on short scales is, although statistically significant, considerably
smaller and typically of the order of 130\~pc and 80\~pc for the sample of interacting/merging and
non-interacting galaxies, respectively. This fact is likely due to a combination of two
different properties of such perturbations on short scales, namely their transient character
and a systematically smaller amplitude of the induced distortions itself. As a result of
the first of these effects the amplitude of the perturbations, and thus the velocity dispersion
of disk stars within this region, would subside. Since the size of the affected region
($\approx$\~kpc) is small compared to the radial extend of the disk the vertical motion excess
could be absorbed by neighbouring disk stars. This would contribute to an overall increase
of vertical velocity dispersion and thus to a larger value for the scale height (SD\~II).
However, the existing measurements (Table\~\ref{perturb}, columns\~(2) and (3)) do not allow
us to disentangle the individual effects nor to draw conclusions about their significance.
We therefore infer that tidally-triggered, vertical disk asymmetries on large ($\approx \Rmax$)
scales are a common and persistent phenomenon, while perturbations on small ($\approx \zo$)
scales are systematically smaller and decline on shorter timescales.

%__________________________________________________________________

\subsubsection{Properties of tidally distorted disks}

In addition to the statistical differences found between both galaxy samples we also studied
the properties of vertical, tidally-triggered disk perturbations in individual galaxies.
This also allows a better illustration of some of the previously obtained results. We
therefore selected 4 galaxies in different stages of the interacting/merging process that
clearly display some of the above mentioned properties of galactic disks. The upper panels of
Figs.\~3 and 4 show these galaxies together with the adopted disk parameters. The corresponding
lower panels display the radial behaviour of their scale height.

Figure~3 shows two galaxies in a transient phase of the merging process, accreting a low-mass
satellite. While the small companion and its tidal debris are still clearly visible
(left part of the disk in both images) the measured radial behaviour of scale height shows
significant variations on short scales. The amplitude of these variations is $\approx 1.5$
times the mean scale height of the corresponding object. In the case of AM\~1158-345
the lower panel clearly shows that the scale height increases substantially towards
the outer parts of the disk while showing significant perturbations on small scales at the same
time. The first of these two effects is responsible for the large scale height gradient of the
order of $(\zo)_{\rm grad} \approx 0.05$ (depending on the vertical disk model used for the fit;
for comparison see Fig.\~1). It also causes the large deviation of $\approx 25\%-30\%$ of
$(\zo)_{\rm mean}$ that was found for the scale height of this disk. The disk of AM\~0256-321
shows perturbations on short scales with large amplitudes of the order of 10\%. There is no
evidence of a significant gradient of the scale height. This behaviour was found
to be typical for those galaxy disks observed in a transition phase of a still ongoing
interacting/merging process.

Figure\~4 shows two examples of late-type $(T=5,6)$ edge-on spirals that are, suggested by
their location in compact groups of galaxies (Verdes-Montenegro et al. \cite{verdes1998};
Zepf \cite{zepf1993}) and by the typical features detected in their isophotes, in a more
progressed phase of the interaction/merging process. Although there is evidence that
these galaxies were strongly affected by such events in the recent past it is not clear
whether their disk properties do result from accreting a small companion or from tidal
interactions without a merger. However, the behaviour of the scale height of these disks
is characteristic for many galaxies in the interacting/merging sample: they show
substantial scale height gradients between 0.1 and 0.25, caused by a $1.5 - 2.5$ times
larger scale height in a region between 0.5 and 4 disk scale lengths (exceeding slightly
the upper limit of the fit region used in the right part of the NGC~3187-panel). Furthermore,
the outer isophotes of NGC 3187 and NGC 4634 both show significantly ``fringed'' disks.
The deviations $(\zo)_{\rm std}$ from their mean scale heights were found to be between
22\% and 26\%. Considering their morphological types ($T= 5-6$) and the derived ratio
between radial and vertical scale parameters ($1 \le h/\zo \le 2$) their disks are
$\approx 4$ times thicker than those of typical non-interacting galaxies of the same
type (Figs.\~6 and 7 in SD\~II).

Despite the described variety of the radial behaviour of the disk scale height within
the sample of interacting/merging galaxies the obtained measurements suggest that perturbations
on large scales (i.e. gradients) are a widely common feature, while perturbations on
small scales seem to be typical only during an ongoing phase of accretion or interaction
(Table\~\ref{perturb}, columns\~(2) and (3)).
Furthermore, our data indicate that galaxy disks possessing a considerable scale height
gradient do also show a smoother trend of the measured scale height with radial distance,
i.e. a smaller level of short-scale perturbations. The detection of features like scale height
gradients and perturbations on both short and large scales as described in this section is
consistent with results of N-body simulations analyzing the stability of galaxy disks as a
result of minor mergers (T$\acute{\rm o}$th \& Ostriker \cite{toth1992}; Mihos et al. \cite{mihos1995})
and with studies of the stability of galactic disks in general (Griv \& Peter \cite{griv1996c}).

\subsection{Size, type, and frequency of stellar warps}

The majority of galaxies in our interacting/merging sample as well as some of the non-interacting
galaxies show substantial geometrical distortions. Apart from vertical perturbations studied
in the previous section the most frequently observed kind of distortions are variations
of the geometrical center of vertical disk profiles around a mean galactic plane, i.e.
``warps''. In the following we investigate their relative and absolute sizes, their types and
the frequency of occurrence for both galaxy samples. The method used for quantifying warped
disks was described in Sect.\~2.

Due to the sensitivity of the applied fitting procedure we are able to detect vertical deviations
from the mean galactic plane of the order of 10\~pc -- 50\~pc, depending on the distance of galaxies
(Sect.\~3.1).
In order to evaluate the influence of dust extinction, misaligned warps, and the appearance of
spiral structures on the detection of warps we refer to a series of studies performed
by Reshetnikov \& Combes (\cite{reshetnikov1998a}, \cite{reshetnikov1998b}), in which these effects
were statistically studied in great detail. According to their results, in particular the simulations
presented in section\~4 and Fig.\~10 (same section) of the first paper, the probability of false
detections of $S$-shaped warps is around 30\% for disks with inclinations $i \approx 85\degr$,
and drops to 0\% for $i \approx 90\degr$. The simulations were performed for 405 flat model
galaxies with a $B/D = 0.3 - 0.5$, axis ratios $a/b > 7$ (average $a/b \approx 9.1$), and with a
nice contrasted spiral structure (the latter two properties both overestimate the simulated
projection effects with respect to the sample studied in this paper). 
According to Fig.\~10 in Reshetnikov \& Combes (\cite{reshetnikov1998a}) we therefore estimate
that the percentage of false $S$ ($U$)-shaped warps due to projection effects is smaller than
10\% (5\%) for our highly inclined (85\% $\geq 88\degr$ and 15\% $\ge 85\degr$) galaxy sample.
This statistical uncertainty is smaller than the error introduced by the applied visual
classification of warp types ($\approx 10\%-15\%$) and does therefore not affect the results
and conclusions of this study (Table\~\ref{warp}).

%%%%%%%%%%%%%%%%%%%%%%%%%%%%%%%%%%%%%%%%%%%%%%%%%%%%%%%%%%%%%%%%%%%

\begin{figure*}[t]

\vspace*{55mm}

\hspace*{26mm}
\begin{minipage}[b]{18.0cm}
\begin{picture}(18.0,10.3)
{\includegraphics[angle=90,viewport=320 50 590 725,clip,width=128mm]{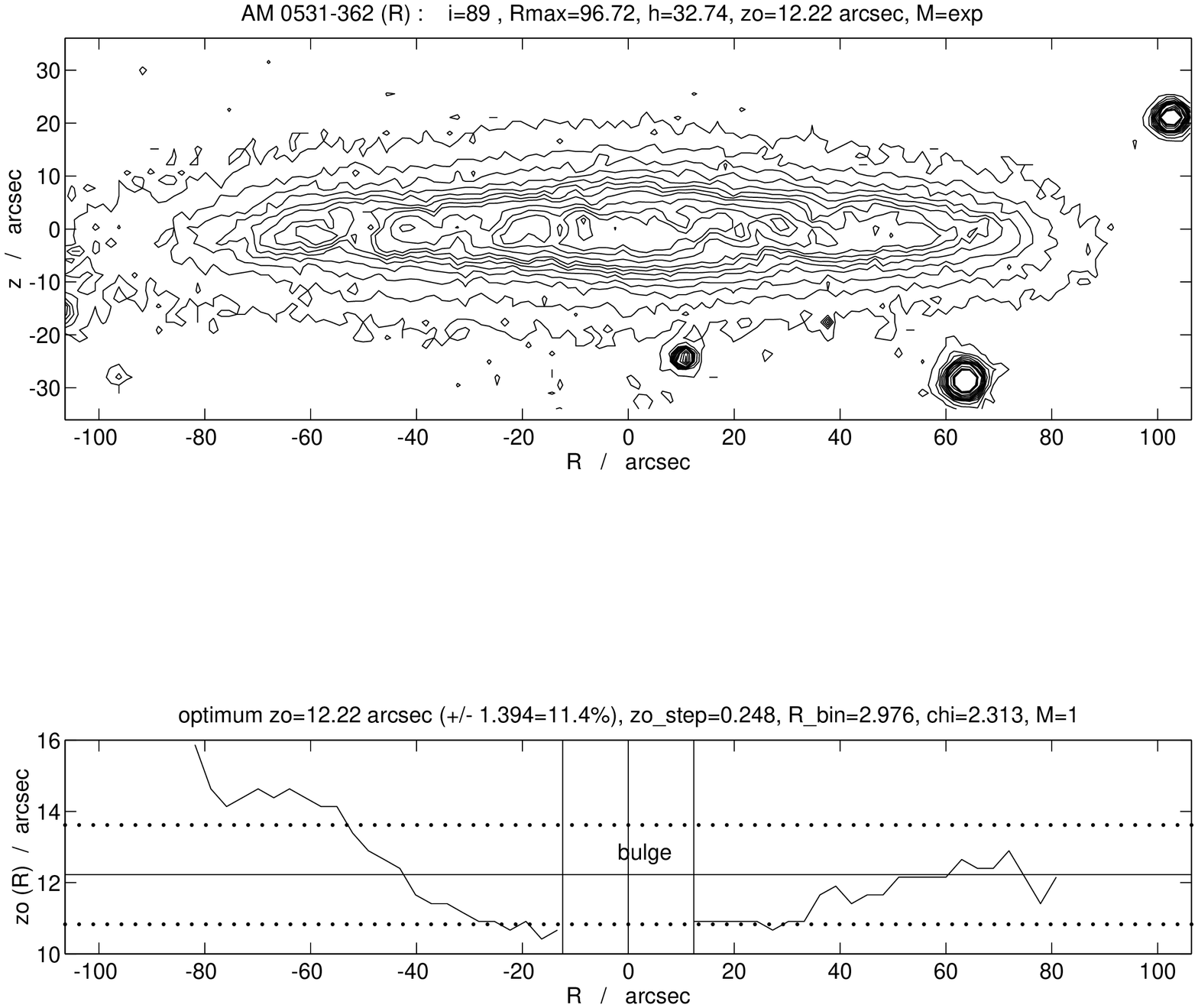}}
\end{picture}
\end{minipage}

\vspace*{42mm}

\hspace*{26mm}
\begin{minipage}[b]{18.0cm}
\begin{picture}(18.0,10.3)
{\includegraphics[angle=90,viewport=20 50 210 725,clip,width=128mm]{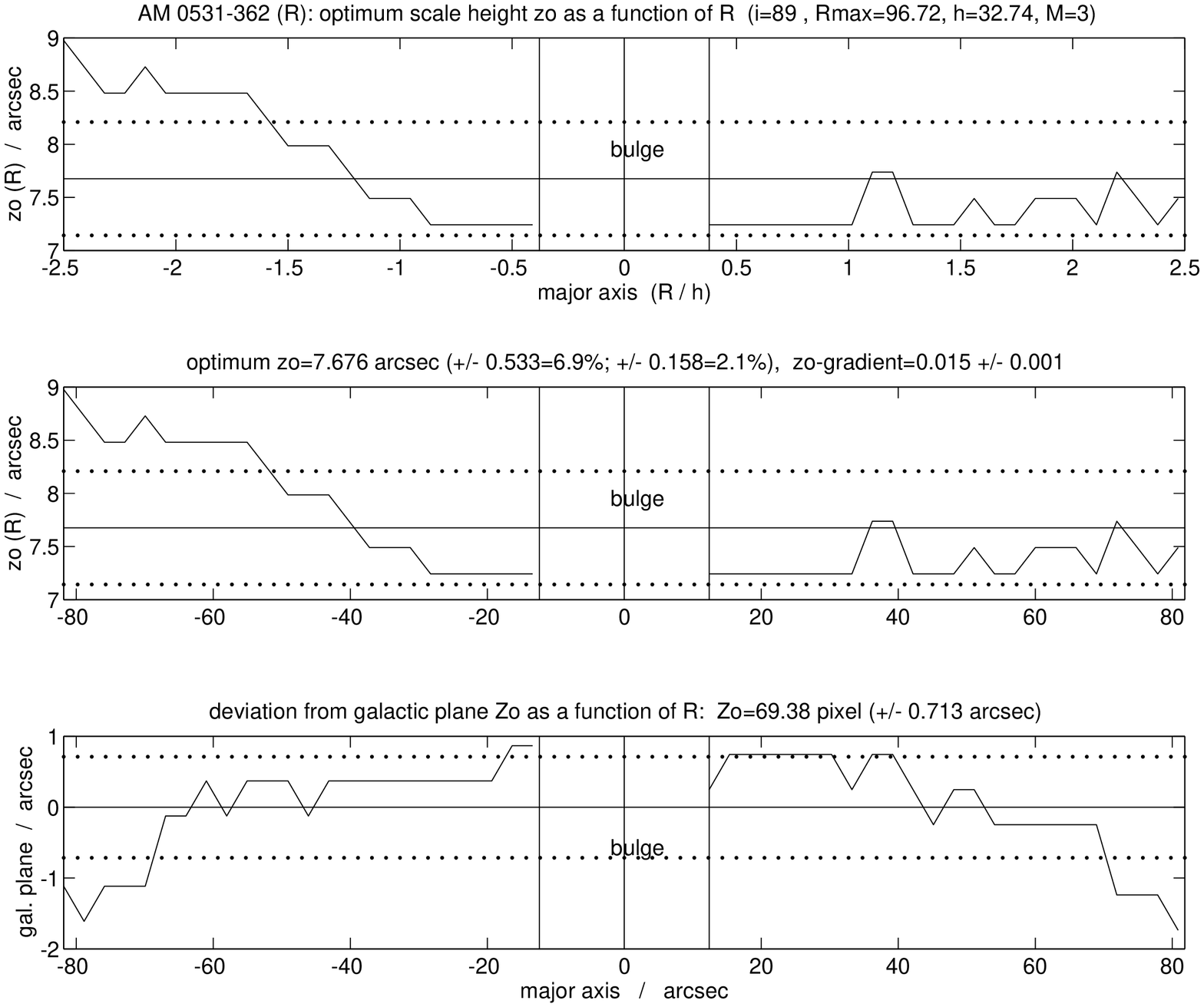}}
\end{picture}
\end{minipage}

\vspace*{62mm}

\hspace*{26mm}
\begin{minipage}[b]{18.0cm}
\begin{picture}(18.0,11.0)
{\includegraphics[angle=90,viewport=320 50 590 725,clip,width=128mm]{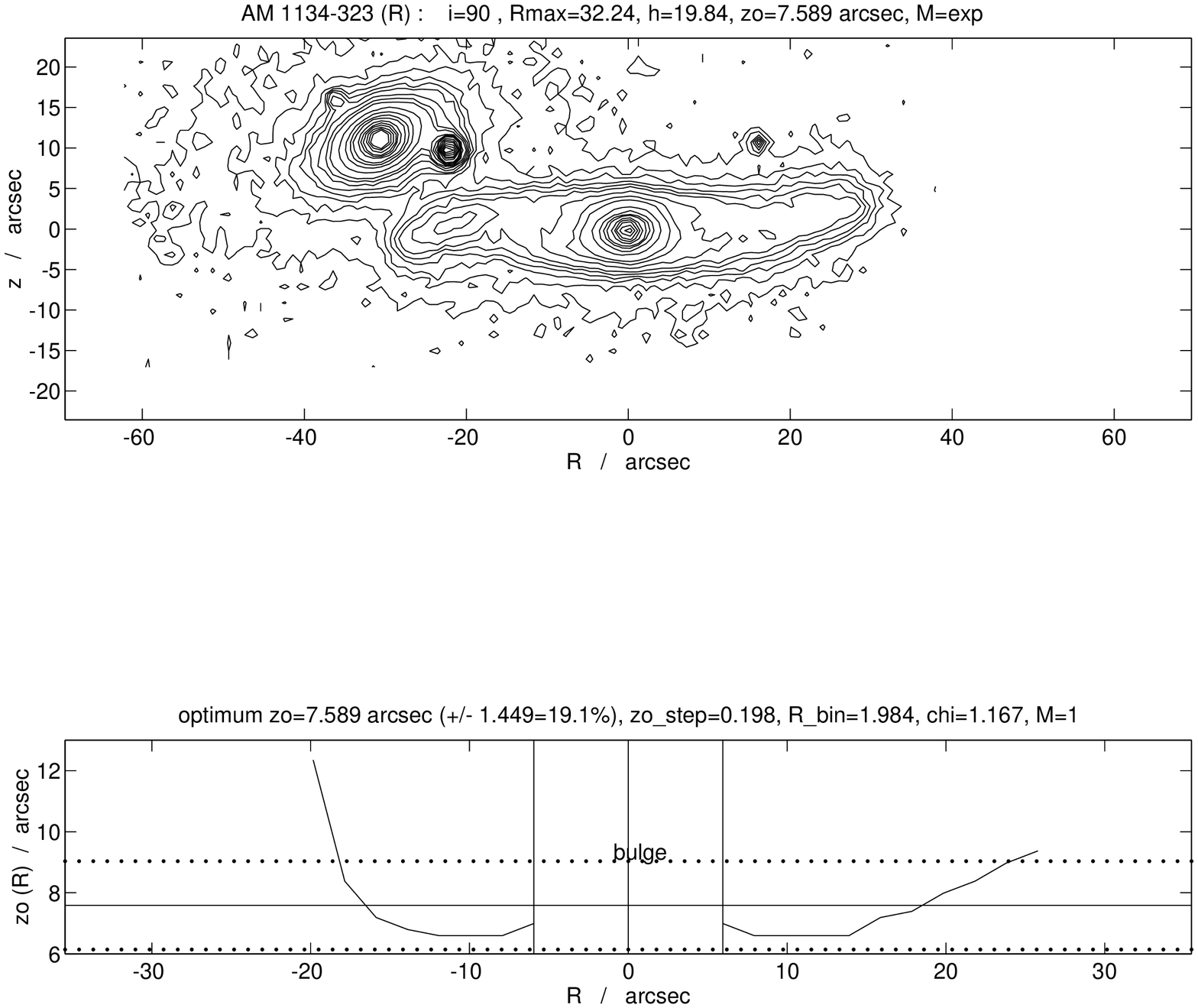}}
\end{picture}
\vspace{1mm}
\end{minipage}

\vspace*{42mm}

\hspace*{26mm}
\begin{minipage}[b]{18.0cm}
\begin{picture}(18.0,10.3)
{\includegraphics[angle=90,viewport=20 50 210 725,clip,width=128mm]{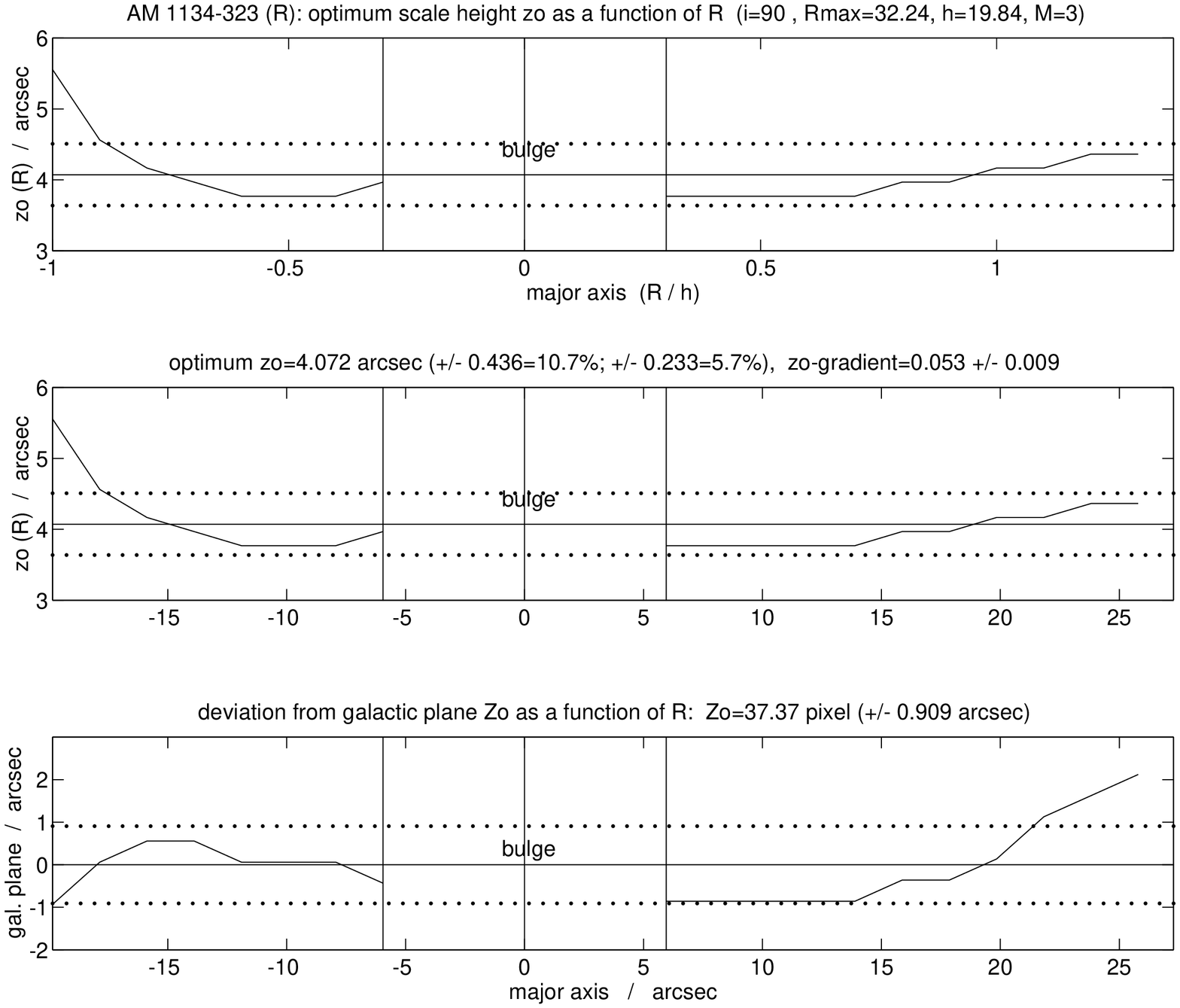}}
\end{picture}
\end{minipage}

\vspace*{2mm}

{\bf \noindent Fig.\~6.} Examples of different types of warps, shown together with
the corresponding variations around the mean galactic plane:
$U$-shaped warp (upper two panels) and $S$-shaped warp (lower two panels).
\end{figure*}

%%%%%%%%%%%%%%%%%%%%%%%%%%%%%%%%%%%%%%%%%%%%%%%%%%%%%%%%%%%%%%%%%%%

%%%%%%%%%%%%%%%%%%%%%%%%%%%%%%%%%%%%%%%%%%%%%%%%%%%%%%%%%%%%%%%%%%%

\begin{figure*}[t]

\vspace*{55mm}

\hspace*{26mm}
\begin{minipage}[b]{18.0cm}
\begin{picture}(18.0,10.3)
{\includegraphics[angle=90,viewport=320 50 590 725,clip,width=128mm]{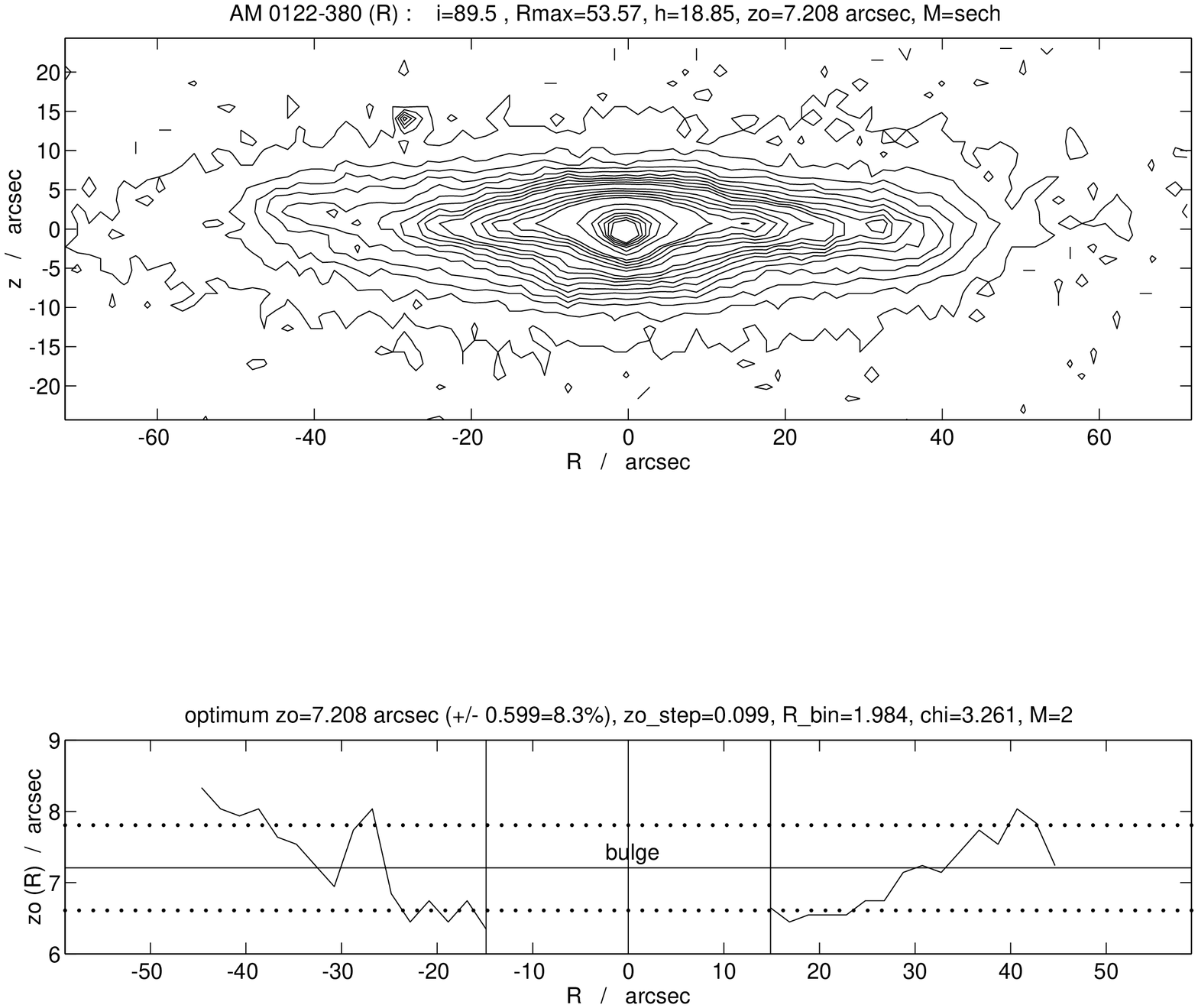}}
\end{picture}
\end{minipage}

\vspace*{42mm}

\hspace*{26mm}
\begin{minipage}[b]{18.0cm}
\begin{picture}(18.0,10.3)
{\includegraphics[angle=90,viewport=20 50 210 725,clip,width=128mm]{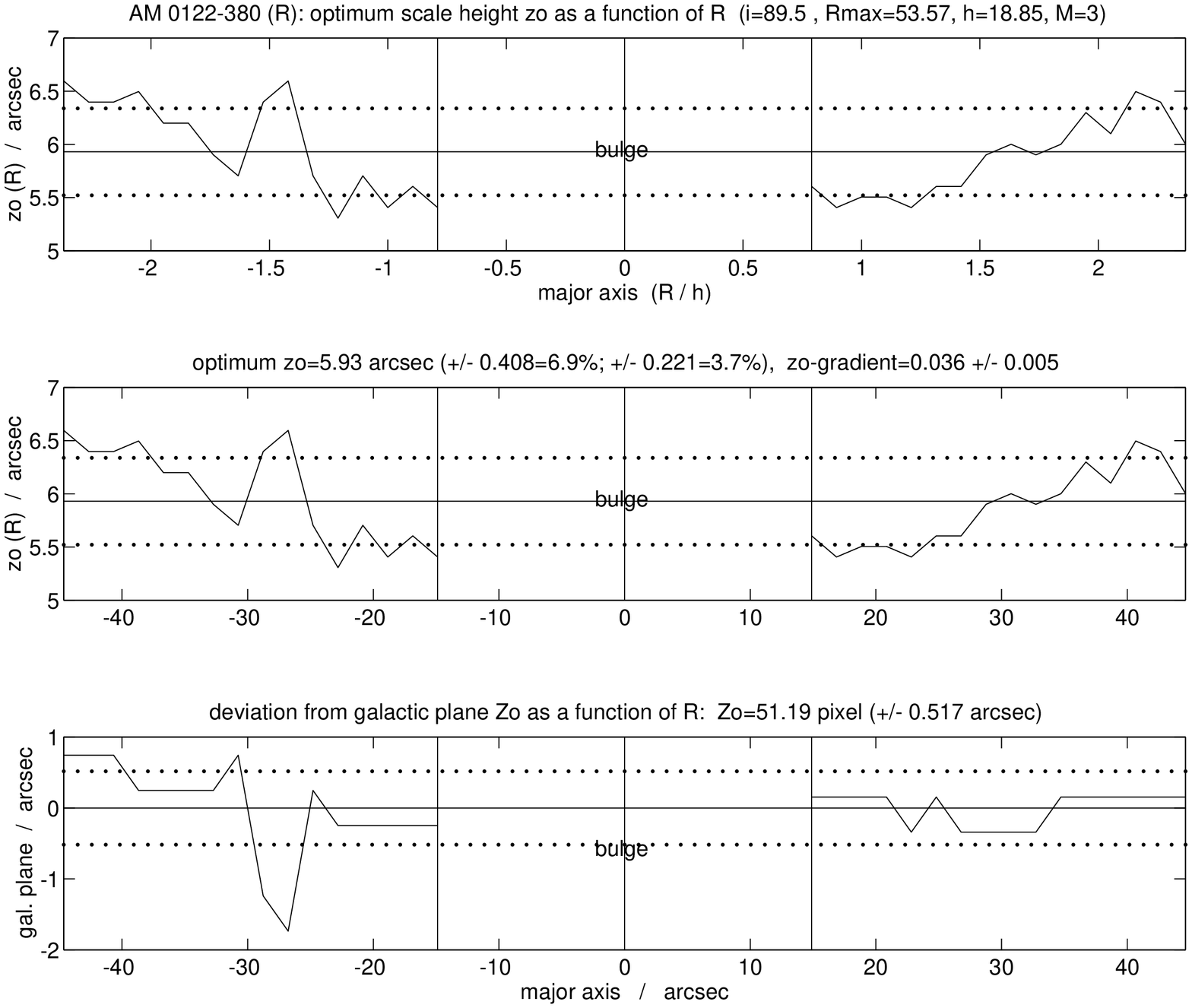}}
\end{picture}
\end{minipage}

\vspace*{62mm}

\hspace*{26mm}
\begin{minipage}[b]{18.0cm}
\begin{picture}(18.0,11.0)
{\includegraphics[angle=90,viewport=320 50 590 725,clip,width=128mm]{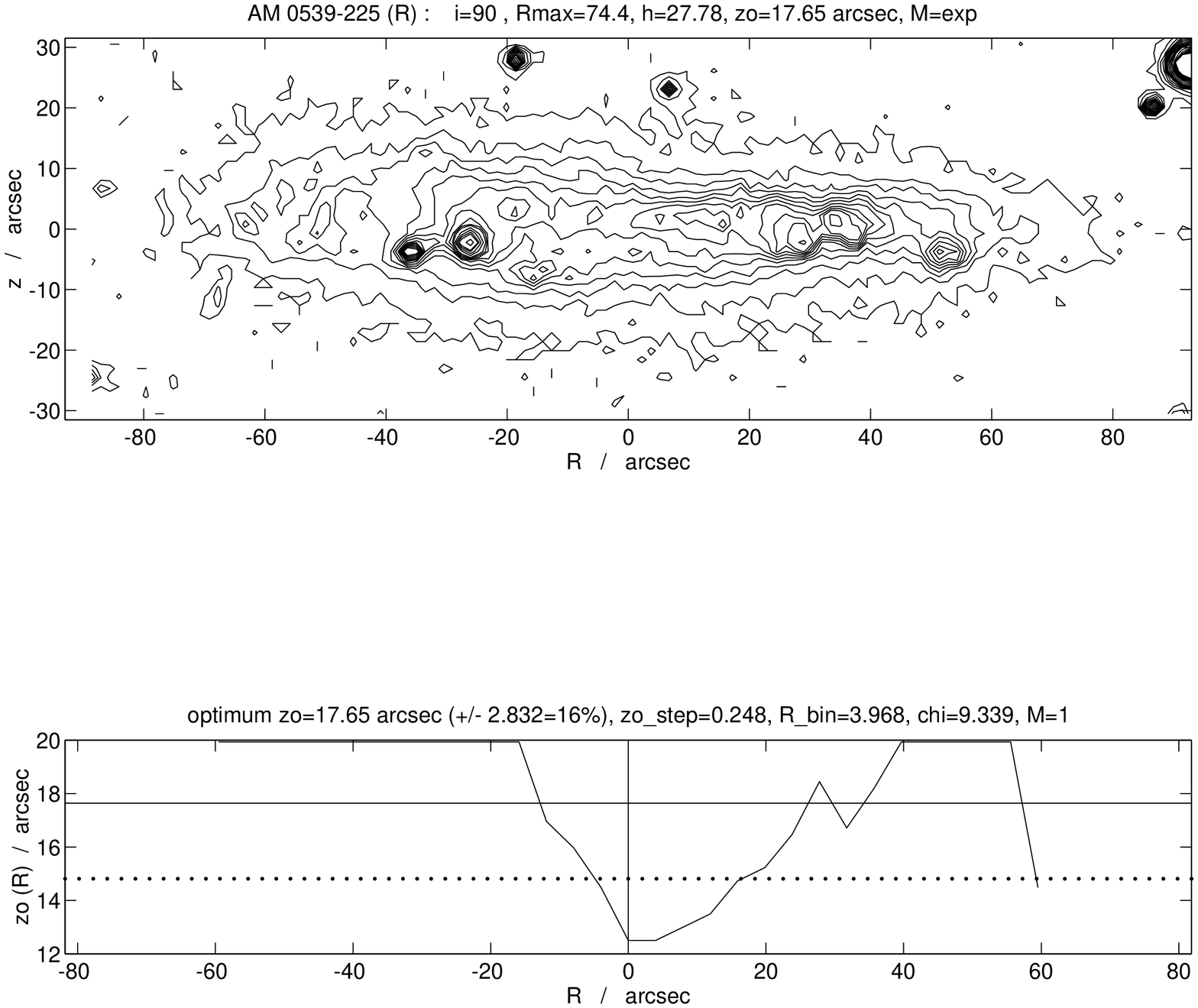}}
\end{picture}
\vspace{1mm}
\end{minipage}

\vspace*{42mm}

\hspace*{26mm}
\begin{minipage}[b]{18.0cm}
\begin{picture}(18.0,10.3)
{\includegraphics[angle=90,viewport=20 50 210 725,clip,width=128mm]{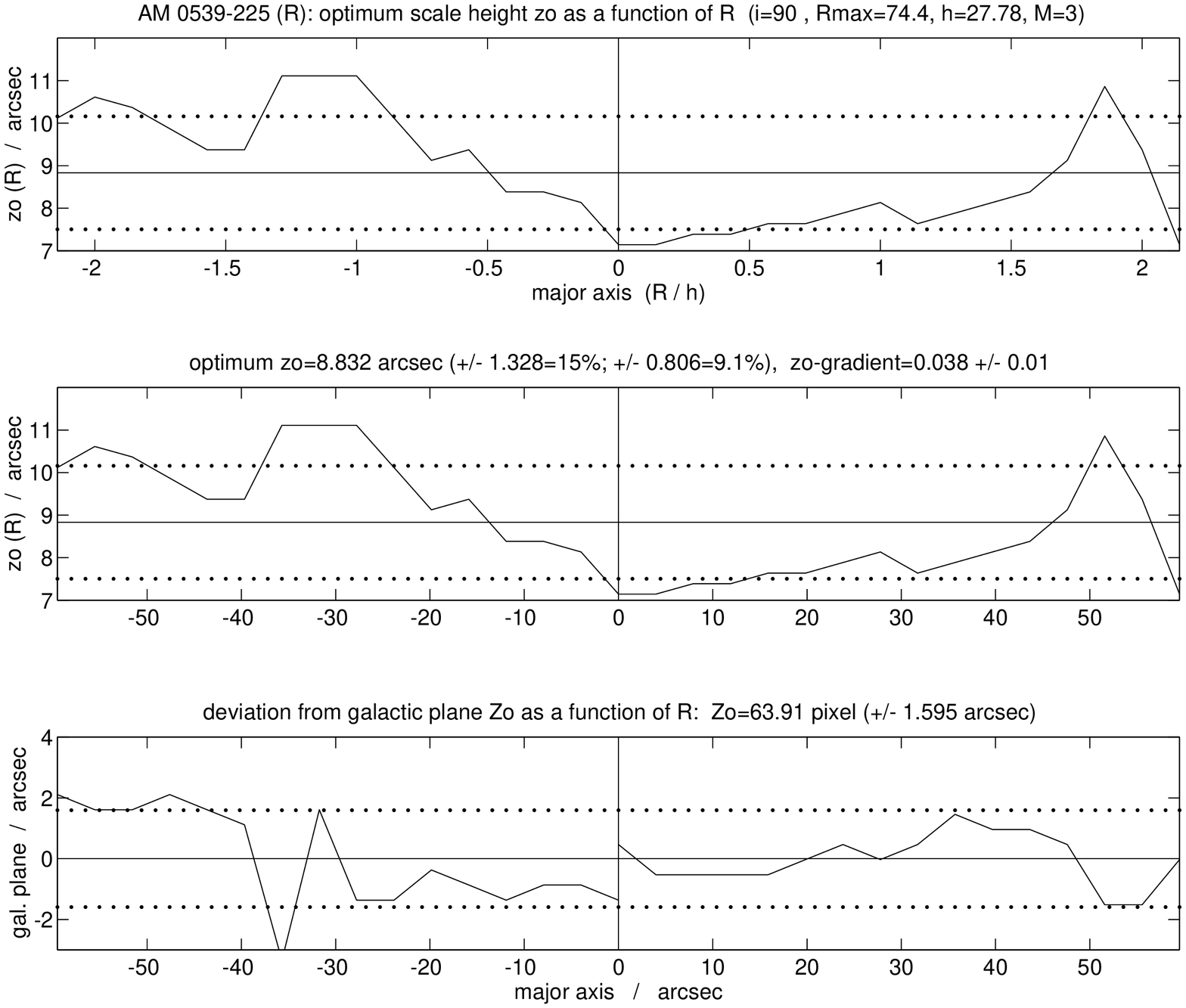}}
\end{picture}
\end{minipage}

\vspace*{2mm}

{\bf \noindent Fig.\~6.} (continued)
$One-side$ warp (upper two panels) and $Irr$ warp (lower two panels). 
\end{figure*}

%%%%%%%%%%%%%%%%%%%%%%%%%%%%%%%%%%%%%%%%%%%%%%%%%%%%%%%%%%%%%%%%%%%

%%%%%%%%%%%%%%%%%%%%%%%%%%%%%%%%%%%%%%%%%%%%%%%%%%%%%%%%%%%%%%%%%%%

\begin{figure*}[t]

\vspace*{55mm}

\hspace*{26mm}
\begin{minipage}[b]{18.0cm}
\begin{picture}(18.0,10.8)
{\includegraphics[angle=0,viewport=40 220 540 540,clip,width=128mm]{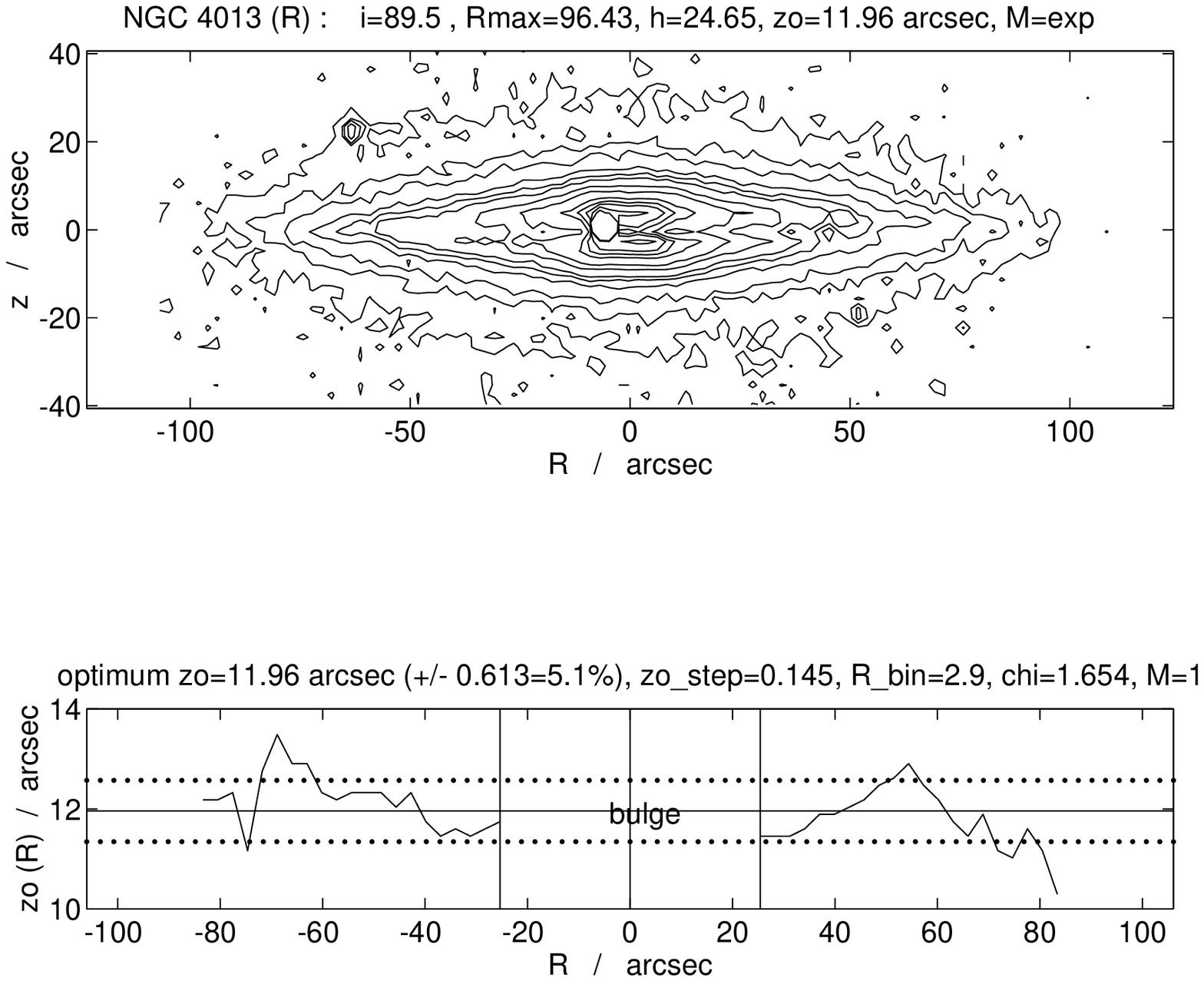}} % ugc5341 !!
\end{picture}
\end{minipage}

\vspace*{46mm}

\hspace*{26mm}
\begin{minipage}[b]{18.0cm}
\begin{picture}(18.0,10.3)
{\includegraphics[angle=0,viewport=40 0 540 150,clip,width=128mm]{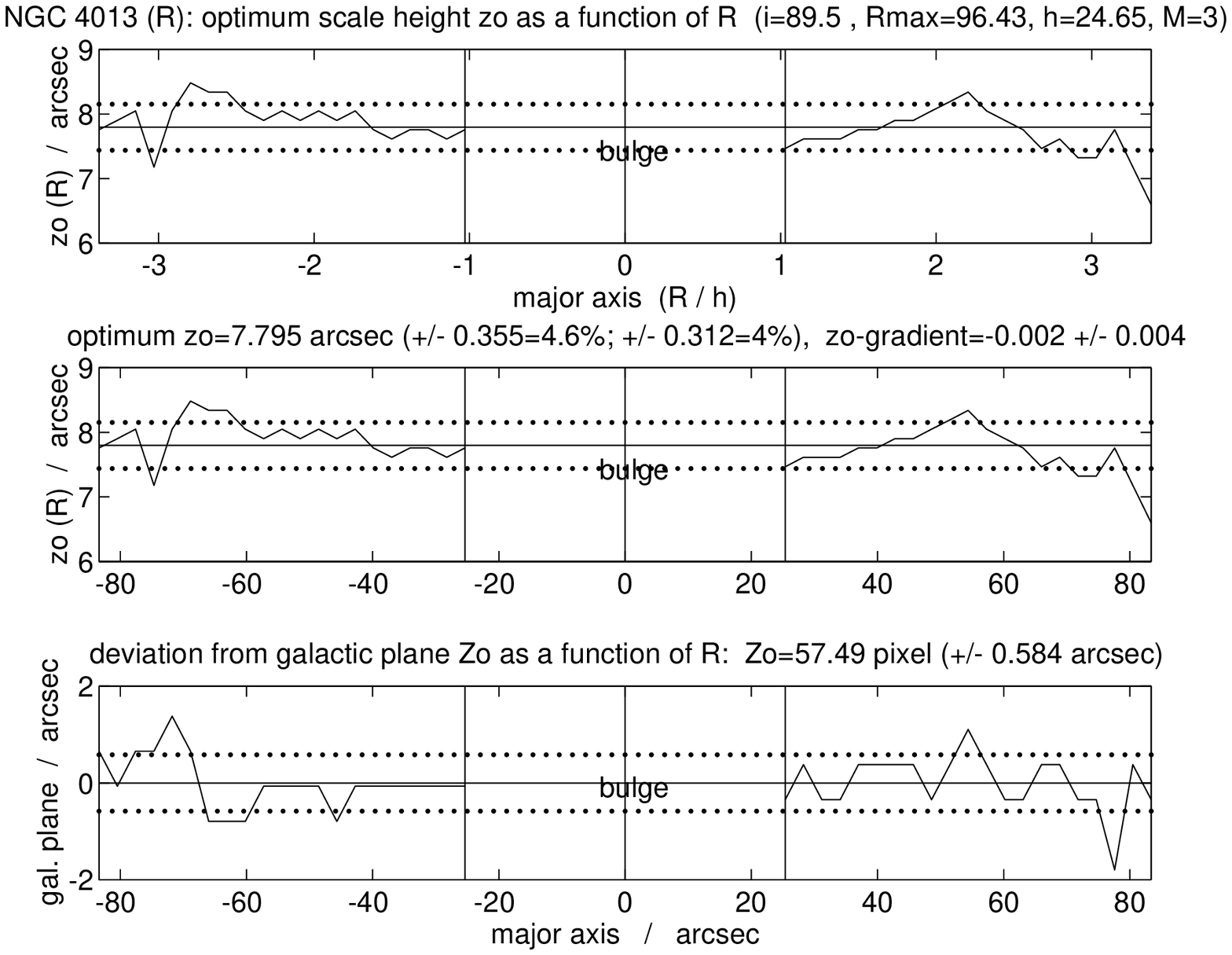}}
\end{picture}
\end{minipage}

\vspace*{2mm}

{\bf \noindent Fig.\~6.} (continued) Galaxy disk without significant warps.
\end{figure*}

%%%%%%%%%%%%%%%%%%%%%%%%%%%%%%%%%%%%%%%%%%%%%%%%%%%%%%%%%%%%%%%%%%%

%%%%%%%%%%%%%%%%%%%%%%%%%%%%%%%%%%%%%%%%%%%%%%%%%%%%%%%%%%%%%%%%%%%

\begin{figure}[t]

\vspace*{74mm}

\begin{minipage}[b]{8.8cm}
\begin{picture}(8.8,7.5)
{\includegraphics[angle=90,viewport=320 430 570 720,clip,width=88mm]{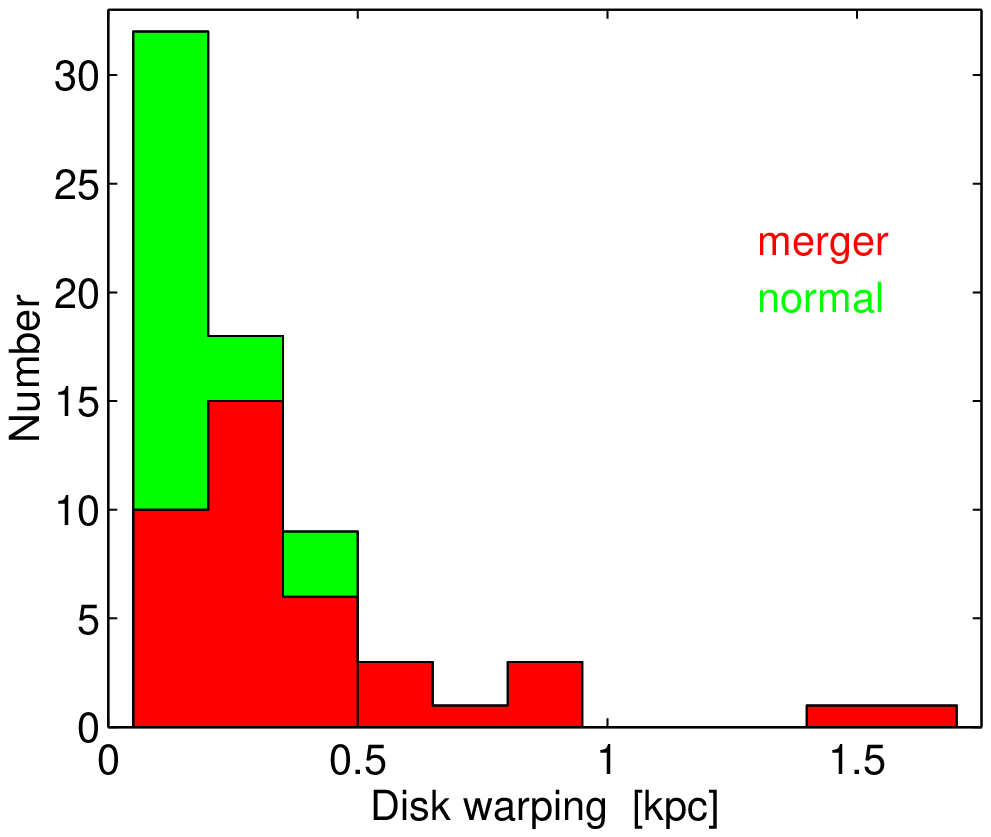}}
\end{picture}

\vspace{1mm}

{\bf \noindent Fig.\~5.} The size of warps in galactic disks (average values),
shown for the sample of non-interacting and interacting/merging galaxies.
\end{minipage}

\end{figure}

%%%%%%%%%%%%%%%%%%%%%%%%%%%%%%%%%%%%%%%%%%%%%%%%%%%%%%%%%%%%%%%%%%%

%__________________________________________________________________
%
% Table 3 - Types and frequency of optical stellar warps.
%
%__________________________________________________________________
%

\tabcolsep1.5mm

  \begin{table}[t]

  \vspace{2mm}

  \caption[ ]{Types and frequency of optical stellar warps. \\
   Columns: (1) Galaxy sample used for the statistics: total= all; non-int.= non-interacting;
   int./merg.= interacting/merging; (2) Fraction of galaxies (in percent of the investigated
   sample) with warped disks, for explanation of types see Sect.\~3.2.}
  \label{warp}
  \begin{flushleft}
  \begin{tabular}{lccccc}
  \cline{1-6}
  \hline\hline\noalign{\smallskip}
  \multicolumn{1}{c}{Sample}           &  \multicolumn{5}{c}{Frequency of warp type [\%]}  \\
  \noalign{\smallskip}
  \cline{2-6}
  \noalign{\smallskip}
   & \multicolumn{1}{c}{U-shaped$\A$}  &  \multicolumn{1}{c}{S-shaped$\A$} &
  \multicolumn{1}{c}{One-side}         &  \multicolumn{1}{c}{Irr}  &  \multicolumn{1}{c}{None}  \\
  \noalign{\smallskip}
  \multicolumn{1}{c}{(1)}              &  \multicolumn{5}{c}{(2)}  \\
  \noalign{\smallskip}
  \hline\noalign{\smallskip}
%
%                      U     S   one-side  irr   none
%
   total           &  19  &  24  &  20   &  3  &  34  \\
   non-int.        &   8  &  19  &  16   &  2  &  55  \\
   int./merg.      &  33  &  31  &  25   &  4  &   7  \\
  \noalign{\smallskip}
  \hline\noalign{\smallskip}
  \end{tabular}
  \begin{list}{}{}
  \item[$\A$] Types according to Reshetnikov \& Combes (\cite{reshetnikov1998a}). \\

\vspace{1mm}

  \end{list}
  \end{flushleft}
  \end{table}

%__________________________________________________________________

One of the most unexpected results derived from the standard deviation around the
mean galactic plane (named as ``Warping'' in column\~(8) of Table\~\ref{parameters}) was
that {\em all} galaxy disks investigated are warped on some level. However, clear 
differences in the size of warps between both subsamples can be seen (Fig.\~5 and
Table\~\ref{perturb}). Figure\~5 shows the distribution of absolute sizes of warps
for both subsamples of non-interacting and interacting/merging galaxies. Most of the
non-interacting galaxies possess disks with warps that have amplitudes around 140\~pc
on average, corresponding to $\approx 11\%$ of the mean scale height of these disks. In
contrast the majority of interacting/merging galaxies shows considerably warped disks with
amplitudes up to $\approx 1$\~kpc, and with an average size of 340\~pc. This value corresponds
to $\approx 17\%$ of their mean scale height and $\approx 26\%$ of the scale height of
non-interacting galaxies (column\~(4) in Table\~\ref{perturb}).

In addition to the size of warps their different types and their frequency of occurrence
were investigated. In order to classify the types we visually inspected the results of the
above described vertical fitting procedure, i.e. the behaviour of vertical variations around
the mean galactic plane along the major axis of the disk (shown in the corresponding lower
panels of Fig.\~6). For our classification we distinguished four different types of warps
(the plots in Fig.\~6 show a typical example for each of these types): \\
\noindent
$U$-shaped warp, i.e. both sides of the disk are warped in the same direction;
$S$-shaped warp with an integral sign-shaped disk plane; $One-side$ warp, i.e. deformations
are visible only at one side of the disk; $Irr$ for irregular warp, i.e. completely asymmetric.
While the types $S$ and $U$  were described in detail by Reshetnikov \& Combes
(\cite{reshetnikov1998b}), it was necessary to introduce One-sided and Irr-shaped warps to
enable an adequate description of the observed variety of warped distortions.

The results of this study are summarized in Table\~\ref{warp}, distinguishing between the
total, non-interacting, and interacting/merging galaxy sample. Accordingly, nearly all
($\approx 93\%$) of the interacting/merging spirals are warped. One third of them shows
either $U$- or $S$-shaped warps, and nearly another third is warped only at one side.
The frequency of warped disks in isolated, non-interacting galaxies is with $\approx 45\%$
unexpectedly high. These galaxies either have $S$-shaped ($\approx 19\%$) or One-side
($\approx 16\%$) warps, and only 8\% of their disks are $U$-shaped. A correlation between
size or type of the warps and morphological type of galaxies was not found.

Finally, it should be stressed that the $\approx 34\%$ of galaxies in the total sample
classified as ``not warped'' are not in contradiction to the previously made conclusion
that all galaxy disks investigated are warped (column\~(4) in Table\~\ref{perturb}).
This is due to the larger tolerances used in the visual inspection of deviations
from galactic plane. The results of the classification scheme in Table\~\ref{warp} are
therefore qualitative and less sensitive than the measurements listed in column\~(4)
of Table\~\ref{perturb}.

%__________________________________________________________________

\section{Discussion}

The correlation between disk scale height gradients and morphological type of non-interacting
galaxies found in this study (gradient $-0.006 \pm 0.002$ in $R$-band, Sect.\~3.1 and Fig.\~1)
is consistent with results obtained by de Grijs \& Pelletier (\cite{grijs1997}). They found
a gradient of $-0.0086 \pm 0.0012$ ($I$-band) for spiral galaxies with Hubble types between
$-2 \le T \le 8$. Furthermore, we observe a significant
number of galaxies possessing no or very small bulge components but having a large scale height
gradient (e.g. NGC 7518). Therefore we can rule out a correlation between the existence of a
bulge component and thickened disks. This also supports the results obtained in the two-dimensional
study of bulge-disk decomposition by de Grijs \& Pelletier (\cite{grijs1997}). Investigating
galaxies with considerable scale height gradients they find that the contribution of the bulge
to the total light can be assumed to be negligible (less than 5\%).
We can also discard an influence of warps on the measured thickness of disks -- as proposed
by van der Kruit \& Searle (\cite{kruit1981a}) -- since this effect was considered in the
fitting procedure (Sect.\~2).

A more plausible explanation for an increase of disk scale height with radial distance
might be either the presence of an underlying thick disk with a scale length comparable
or larger than the thin disk (Burstein \cite{burstein1979}; de Grijs et al.
\cite{grijsetal1997}; de Grijs \& Pelletier \cite{grijs1997}) or recent accretion of material
onto the outer disk parts (T$\acute{\rm o}$th \& Ostriker \cite{toth1992}; Quinn et al.
\cite{quinn1993}; Zaritsky \cite{zaritsky1995}).
However, our observations do not sufficiently support the first of these explanations,
i.e. the presence of a thick disk. The studied sample only contains a few galaxies with vertical
disk profiles weakly indicating the existence of an underlying thick disk. Hence, accretion of
small satellites continues to be a promising mechanism in order to explain the origin of gradients
in disk scale height.

The size and frequency of tidally-triggered vertical disk perturbations found in this study
(Sect.\~3 and Table\~\ref{perturb}) both indicate that such perturbations are typical during a
transient phase of close interactions or minor mergers. They likely result from large-scale
asymmetries of the gravitational potential, caused by the proximity of the companion
(Reshetnikov \& Combes \cite{reshetnikov1997}). According to results of N-body simulations
of minor mergers (T$\acute{\rm o}$th \& Ostriker \cite{toth1992}; Walker et al. \cite{walker1996})
the vertical structure of affected disks is mainly characterized by a systematical increase
of the scale height towards the outer regions (``slanting'' disks) as well as by local
instabilities and disk ``flaring''. Our observations (Figs.\~3 and 4) confirm the qualitative
predictions of these studies, and the derived statistical results (Table\~\ref{perturb})
give a quantitative estimate for the size of tidally-triggered perturbations.

The simulations made by T$\acute{\rm o}$th \& Ostriker (\cite{toth1992}) and Walker et al.
(\cite{walker1996}) also make some predictions on the lifetime of the evoked perturbations.
They conclude that after the encounter the tidally induced perturbations do rapidly decline
and the vertically heated disk settles into a new, nearly axisymmetric equilibrium
which is stable on longer timescales ($>1$\~Gyr).
This is consistent with the results of this study (columns\~(2) and (3) in
Table\~\ref{perturb} and Sect.\~3), indicating that tidally-triggered perturbations
on large ($\approx \Rmax$) scales, i.e. gradients in the disk scale height, are common
and persistent phenomena if compared to short-lived perturbations on small ($\approx \zo$) scales.
Together with the value of $\approx 1.5$ found for vertical disk thickening due to interactions
and minor mergers (SD\~I+II) this is one of the main conclusions of this paper.

Summarizing the results obtained on the size and frequency of stellar warps the high fraction
($\approx 45\%$) of non-interacting galaxies possessing warped distortions is in good agreement
with the results of Sanchez-Saavedra et al. (\cite{sanchez1990}) and Reshetnikov \& Combes
(\cite{reshetnikov1998b}, Table\~1). They found that the fraction of isolated galaxies with warped
disks is at least 50\% and $(42 \pm 6)\%$, respectively. Assuming a hierarchical galaxy formation
scenario it is unrealistic to explain this high fraction of warped disks entirely as a result
of tidal interactions in the distant past. Hence we have to consider alternative processes
that are able to form new or reinforce existent warps. Possible mechanisms were mentioned
in the introduction of this paper. On the other hand, the measured differences in their
size --  warps of interacting/merging galaxies were found to be at least two times larger than
those of non-interacting galaxies on average -- indicate that such processes are not capable
to produce warps of comparable size. Hence tidal interactions/minor mergers as studied in this
paper do at least considerably contribute to the formation and size of warps. According to
the results found for our total galaxy sample, the fraction of warped spirals
($\approx 66\%$) is also consistent with a study of stellar warps made by de Grijs
(\cite{grijs1997a}), who detected warped disks for $\approx 64\%$ of his objects.

%__________________________________________________________________

\section{Summary and conclusions}

A detailed study of the vertical disk structure of 108 highly-inclined/edge-on spiral
galaxies was presented. The sample consists of two subsamples of 61 non-interacting and 47
interacting/merging galaxies. In particular, we analyzed the behaviour of the scale height
of all disks as a function of galactocentric distance and investigated the properties
of vertical, tidally-triggered disk perturbations. Furthermore, the size and frequency
of occurrence of warped distortions and bends in galactic disks were studied.
The main conlusions are:

\begin{enumerate}

\item The scale height of disks of interacting/merging galaxies is characterized by
 perturbations on both large ($\simeq \Rmax$) and short ($\simeq \zo$) scales.
 The size of these perturbations are of the order of 280\~pc and 130\~pc, or 14\%
 and 6\% of the corresponding mean scale height, respectively.

\smallskip

\item The scale height of non-interacting galaxies shows similar variations on large
 (short) scales, with amplitudes of the order of 140\~pc (80\~pc). This is only about
 half the size of the perturbations measured for interacting/merging galaxies, and
 equivalent to 10\% and 6\% (6\% and 4\% if corrected for thickened disks) of the
 corresponding mean scale height, respectively.

\smallskip

\item A hallmark of nearly all disks in the interacting/merging sample is a scale
 height that increases systematically with radial distance (gradient). The size of these
 gradients is typically 14\% (22\% if corrected for thickened disks) of the mean scale height.
 This is more than twice the value found for non-interacting galaxies.

\smallskip

\item This indicates that tidally-triggered, large-scale disk asymmetries such as scale
 height gradients are common and persistent phenomena, while local disturbances and
 bending instabilities decline on shorter timescales.

\smallskip

\item The existence of large scale height gradients for non-interacting, early-type
 $(T \le 2)$ spirals, weakly dependent on Hubble type, is not associated with the
 presence of an underlying thick disk. Since many interacting/merging galaxies show
 the same large gradients, this effect might be due to previous accretion of material
 or a minor merger.

\smallskip

\item Nearly all (93\%) of the interacting/merging and 45\% of the non-interacting galaxies
 investigated are noticeably warped. Down to small amplitudes ($\le 80$\~pc) all galaxy
 disks possess warped distortions.

\smallskip

\item Warps of interacting/merging galaxies are $\approx 2.5$ times larger on average than
 those observed in non-interacting galaxies, with a mean amplitude of 340\~pc and 140\~pc,
 respectively.

\smallskip

\item The results indicate that tidal interactions/minor mergers considerably contribute
 to the formation and size of warps. However, these processess cannot entirely explain the
 observed frequent occurrence of warped disks.

\end{enumerate}

%__________________________________________________________________

\begin{acknowledgements}

The authors would like to thank the referee of this paper, Dr. J.A. Sellwood,
for his useful comments and suggestions, and M. R. Corbin for his careful
reading of the manuscript. 
\\
This work was supported by a {\em Lynen} fellowship of the {\em Alexander von
Humboldt Foundation} and by the {\em Deutsche Forschungsgemeinschaft} (DFG)
under grants no. De 385 and GRK 118.
\\
This research has made use of the NASA/IPAC Extragalactic Database (NED) which is
operated by the Jet Propulsion Laboratory, California Institute of Technology, under
contract with the National Aeronautics and Space Administration. 

\end{acknowledgements}

%__________________________________________________________________

%__________________________________________________________________
%
% Table 4 - Vertical disk parameters of the optical galaxy samples.
%
%__________________________________________________________________
%

\tabcolsep1.75mm

\begin{center}
  \begin{table*}
  \caption[ ]{Parameters obtained by an analysis of the vertical disk structure
   (based on $R$-band data of SD\~I+II, Tables\~4 and 5).\\
   Columns: (1) Serial number; (2) Galaxy name; (3) Best-fitting / Alternative vertical disk model:
   $1= \rm exp, 2= \rm sech, 3= \iso$; (4) Mean value of disk scale height $\zo$; (5) Gradient of
   the first-order scale height fit (average from both disk sides), and its error; (6) -- (8):
   Standard deviation (in pc and in percent of its mean value $(\zo)_{\rm mean}$) of: (6) ... mean
   disk scale height; (7) ... first-order scale height fit; (8) ... average galactic plane, i.e. disk
   warping; (9) Goodness-of-fit parameter, defined as the difference between observations and fit
   (for a detailed explanation of fit parameters see Sect.\~2).}
  \label{parameters}
  \begin{flushleft}
  \begin{tabular}{rlcccrccrcrrrrrrrrrrcr}
  \cline{1-22}
  \hline\hline
  \noalign{\smallskip}
  \multicolumn{1}{r}{No.}                  & \multicolumn{1}{c}{Galaxy}  &
  \multicolumn{2}{c}{Vertical Fit}        && \multicolumn{2}{c}{$(\zo)_{\rm mean}$}  &&
  \multicolumn{3}{c}{$(\zo)_{\rm grad}$}  && \multicolumn{2}{c}{$(\zo)_{\rm std}$}   &&
  \multicolumn{2}{c}{$(\zo)_{\rm std1}$}  && \multicolumn{2}{c}{Warping}             &&
  \multicolumn{1}{c}{$Q$}  \\
  \noalign{\smallskip}
  \cline{3-4}
  \cline{6-7}
  \cline{9-11}
  \cline{13-14}
  \cline{16-17}
  \cline{19-20}
  \noalign{\smallskip}
   &  & \multicolumn{1}{c}{Best}  & \multicolumn{1}{c}{Alt.}        &&  \multicolumn{1}{c}{[$''$]} &
   \multicolumn{1}{c}{[kpc]}     && \multicolumn{3}{c}{[$10^{-3}$]} &&  \multicolumn{1}{c}{[pc]}   &
   \multicolumn{1}{c}{[\%]}      && \multicolumn{1}{c}{[pc]}        &   \multicolumn{1}{c}{[\%]}   &&
   \multicolumn{1}{c}{[pc]}       & \multicolumn{1}{c}{[\%]}        &&   \\
  \noalign{\smallskip}
  \multicolumn{1}{r}{(1)}         & \multicolumn{1}{c}{(2)}          &  \multicolumn{2}{c}{(3)} &&
  \multicolumn{2}{c}{(4)}        && \multicolumn{3}{c}{(5)}         &&  \multicolumn{2}{c}{(6)} &&
  \multicolumn{2}{c}{(7)}        && \multicolumn{2}{c}{(8)}         &&  \multicolumn{1}{c}{(9)} \\
  \noalign{\smallskip}
  \hline
%
%--------------------------------------------------------------------------------------------------------
%
%              Vertical Disk Model
% No.  Galaxy  Best & Alternative  zo_mean   zo_grad+err   zo_std(0.)      zo_std(1.)    warping       Q
%        
%                            ['']  [kpc]                  [pc]   [%]      [pc]   [%]     [pc]  [%]
%
% (1)   (2)          (3)         (4)           (5)              (6)            (7)            (8)     (9)
%---------- merger --------------------------------------------------------------------------------------
%
  \noalign{\medskip}
  \multicolumn{22}{c}{\bf Interacting / Merging} \\
  \noalign{\medskip}
  \hline
  \noalign{\medskip}
%--------------------------------------------------------------------------------------------------------
  1 & NGC 7       & 3 &2 &&  8.4 & 0.81 &&  55 &$\pm$& 5 &&  76 &  9.5 &&  35 & 4.3 && 117 & 14.4 && 1.23 \\
  2 & UGC 260     & 2 &--&& 10.0 & 1.54 &&$-$9 &$\pm$& 2 &&  70 &  4.5 &&  60 & 3.9 && 145 &  9.4 && 3.32 \\
  3 & NGC 128     & 2 &1 && 11.5 & 3.34 &&  15 &$\pm$& 4 && 690 & 20.7 &&  62 & 1.9 && 568 & 17.0 && 2.92 \\
  4 & AM 0107-375 & 2 &--&&  3.1 & ---  &&  75 &$\pm$&11 && --- & 22.1 && --- & 6.4 && --- & 13.2 && 1.14 \\
  5 & ESO 296-G17 & 2 &--&&  6.9 & 2.80 &&  55 &$\pm$& 5 && 242 &  8.6 && 116 & 4.1 && 209 &  7.5 && 3.26 \\
  6 & ESO 354-G05 & 2 &--&&  3.1 & ---  &&  38 &$\pm$& 7 && --- & 10.4 && --- & 5.7 && --- & 10.4 && 1.46 \\
  7 & ESO 245-G10 & 2 &--&&  8.3 & 3.20 &&  46 &$\pm$& 4 && 267 &  8.3 &&  89 & 2.8 && 317 &  9.9 && 6.45 \\
  8 & ESO 417-G08 & 2 &--&&  8.5 & 2.80 &&$-$27&$\pm$& 4 && 302 & 10.8 && 218 & 7.8 && 377 & 13.5 && 1.79 \\
  9 & ESO 199-G12 & 2 &--&&  4.0 & 1.87 &&  20 &$\pm$& 6 && 230 & 12.3 && 117 & 6.3 && 373 & 19.9 && 2.76 \\
 10 & ESO 357-G16 & 2 &--&&  5.7 & 0.50 &&  28 &$\pm$& 5 &&  40 &  8.0 &&  23 & 4.6 &&  50 &  9.9 && 1.71 \\
%
%--------------------------------------------------------------------------------------------------------
%
 11 & ESO 357-G26 & 1 &--&& 19.0 & 1.69 && 103 &$\pm$& 6 && 229 & 13.6 &&  61 & 3.6 && 160 &  9.5 && 4.89 \\
 12 & ESO 418-G15 & 2 &--&& 10.2 & 0.68 &&  10 &$\pm$& 3 &&  36 &  5.3 &&  12 & 1.8 && 131 & 19.3 && 3.85 \\
 13 & NGC 1531/32 & 1 &2 && 64.4 & 4.99 &&  93 &$\pm$&14 && 930 & 18.6 && 674 &13.5 && 791 & 15.9 && 6.91 \\
 14 & ESO 202-G04 & 2 &--&&  8.0 & 0.57 && 114 &$\pm$& 5 &&  66 & 11.6 &&  21 & 3.7 &&  51 &  9.0 && 1.92 \\
 15 & ESO 362-G11 & 1 &--&& 10.7 & 0.99 &&  49 &$\pm$&14 && 156 & 15.8 &&  74 & 7.5 && 115 & 11.6 && 1.61 \\
 16 & NGC 1888    & 2 &--&& 13.6 & 2.10 &&  88 &$\pm$& 8 && 453 & 21.6 && 148 & 7.0 && 585 & 27.8 && 9.60 \\
 17 & ESO 363-G07 & 1 &--&& 12.1 & 1.10 &&  52 &$\pm$& 3 && 128 & 11.6 &&  37 & 3.4 &&  65 &  5.9 && 2.31 \\
 18 & ESO 487-G35 & 3 &2 &&  8.6 & 0.98 &&  38 &$\pm$&10 && 152 & 15.5 &&  92 & 9.4 && 182 & 18.6 && 4.68 \\
 19 & NGC 2188    & 1 &--&& 21.1 & 1.18 &&  52 &$\pm$&10 && 131 & 11.1 &&  82 & 6.9 && 244 & 20.7 && 5.89 \\
 20 & UGC 3697    & 2 &--&&  3.3 & 0.69 &&   4 &$\pm$& 3 &&  77 & 11.2 &&  22 & 3.2 && 215 & 31.1 && 1.81 \\
 21 & ESO 060-G24 & 2 &--&&  9.0 & 2.51 && 113 &$\pm$&15 && 500 & 19.9 && 236 & 9.4 &&   8 &  0.3 && 2.52 \\
%
%--------------------------------------------------------------------------------------------------------
%
 22 & ESO 497-G14 & 2 &--&&  6.8 & 1.50 &&  54 &$\pm$&11 && 166 & 11.1 &&  89 & 5.9 && 134 &  8.9 && 2.54 \\
 23 & NGC 3044    & 1 &--&& 11.3 & 1.17 &&  13 &$\pm$&14 && 183 & 15.6 && 110 & 9.4 && 243 & 20.7 && 4.83 \\
 24 & NGC 3187    & 3 &1 && 23.5 & 2.85 && 227 &$\pm$&11 && 727 & 25.5 && 223 & 7.8 && 561 & 19.7 && 3.30 \\
 25 & ESO 317-G29 & 1 &2 && 18.9 & 3.16 &&$-$3 &$\pm$&34 && 232 &  7.3 && 121 & 3.8 &&  83 &  2.6 && 2.73 \\
 26 & ESO 264-G29 & 2 &--&&  6.0 & 1.33 &&  44 &$\pm$& 2 && 493 & 37.1 && 129 & 9.7 && 780 & 58.7 && 5.29 \\
 27 & NGC 3432    & 1 &--&& 17.6 & 0.90 &&  23 &$\pm$& 7 && 125 & 13.9 &&  43 & 4.8 && 324 & 36.0 && 2.67 \\
 28 & NGC 3628    & 2 &--&& 55.8 & 1.30 &&  59 &$\pm$& 5 && 105 &  8.1 &&  47 & 3.6 && 157 & 12.1 && 8.00 \\
 29 & ESO 378-G13 & 2 &--&&  4.7 & ---  &&  72 &$\pm$&16 && --- & 13.2 && --- & 7.9 && --- & 19.3 && 0.82 \\
 30 & ESO 379-G20 & 3 &2 &&  3.0 & 2.82 &&  33 &$\pm$& 9 && 726 & 25.7 && 197 & 7.0 &&1479 & 52.5 && 6.30 \\
 31 & NGC 4183    & 2 &--&&  8.1 & 0.69 &&  70 &$\pm$&20 &&  66 &  9.6 &&  46 & 6.7 &&  74 & 10.8 && 6.47 \\
%
%--------------------------------------------------------------------------------------------------------
%
 32 & NGC 4631    & 1 &--&& 34.1 & 1.32 &&  31 &$\pm$&22 && 299 & 22.7 && 130 & 9.8 && 178 & 13.5 && 6.95 \\
 33 & NGC 4634    & 1 &--&& 11.4 & 0.88 && 116 &$\pm$& 5 && 190 & 21.6 &&  55 & 6.3 &&  98 & 11.2 && 1.18 \\
 34 & NGC 4747    & 1 &--&& 16.1 & 1.75 && 101 &$\pm$& 4 && 232 & 13.3 &&  62 & 3.5 && 254 & 14.5 && 1.59 \\
 35 & NGC 4762    & 3 &2 && 28.1 & 1.59 &&  46 &$\pm$&38 && 364 & 22.9 && 299 &18.8 && 311 & 19.5 && 4.10 \\
 36 & ESO 443-G21 & 2 &--&&  5.1 & 0.95 &&   2 &$\pm$& 4 && 103 & 10.8 && 101 &10.6 &&  83 &  8.8 && 2.43 \\
 37 & NGC 5126    & 2 &--&&  6.7 & 2.09 &&$-$9 &$\pm$& 7 && 183 &  8.8 && 155 & 7.4 && 125 &  6.0 && 0.90 \\
 38 & ESO 324-G23 & 2 &1 && 14.3 & 1.85 &&  91 &$\pm$& 8 && 372 & 20.1 &&  97 & 5.2 && 550 & 29.7 && 6.49 \\
 39 & ESO 383-G05 & 1 &--&& 15.4 & 3.73 &&$-$45&$\pm$& 2 && 518 & 13.9 && 191 & 5.1 && 383 & 10.3 && 2.40 \\
 40 & NGC 5297    & 3 &2 &&  9.0 & 1.65 &&  34 &$\pm$& 8 && 103 &  6.2 &&  67 & 4.1 && 212 & 12.9 && 2.13 \\
 41 & ESO 445-G63 & 2 &--&&  7.6 & ---  &&  84 &$\pm$&25 && --- & 26.6 && --- & 9.7 && --- &  7.6 && 2.02 \\
 42 & NGC 5529    & 1 &--&&  8.0 & 1.43 &&  51 &$\pm$& 3 && 227 & 15.9 &&  81 & 5.7 && 254 & 17.7 && 3.32 \\
%
%--------------------------------------------------------------------------------------------------------
%
 43 & NGC 5965    & 1 &--&&  9.1 & 2.02 &&$-$1 &$\pm$& 6 && 203 & 10.0 && 132 & 6.5 && 296 & 14.6 && 1.12 \\
 44 & NGC 6045    & 2 &1 &&  4.5 & 2.88 &&  89 &$\pm$&13 && 476 & 16.5 && 271 & 9.4 &&1388 & 48.2 && 0.80 \\
 45 & NGC 6361    & 3 &1 &&  9.7 & 2.49 &&$-9$ &$\pm$& 2 && 117 &  4.7 && 101 & 4.1 && 242 &  9.7 && 2.51 \\
 46 & Arp 121     & 2 &--&&  7.9 & 2.89 && 123 &$\pm$&14 && 466 & 16.1 && 197 & 6.8 && 804 & 27.8 && 3.99 \\
 47 & IC 4991     & 3 &--&&  2.5 & 0.95 &&   0 &$\pm$& 8 && 103 & 10.8 &&  73 & 7.7 && 182 & 19.2 && 4.43 \\
%
%--------------------------------------------------------------------------------------------------------
%
  \noalign{\smallskip}
  \hline
  \end{tabular}
  \end{flushleft}
  \end{table*}
\end{center}

\tabcolsep1.7mm

\begin{center}
  \begin{table*}
 {\bf Table 4.} (continued) \\
  \begin{flushleft}
  \begin{tabular}{rlcccrccrcrrrrrrrrrrcr}
  \cline{1-22}
  \hline\hline
  \noalign{\smallskip}
  \multicolumn{1}{r}{No.}                  & \multicolumn{1}{c}{Galaxy}  &
  \multicolumn{2}{c}{Vertical Fit}        && \multicolumn{2}{c}{$(\zo)_{\rm mean}$}  &&
  \multicolumn{3}{c}{$(\zo)_{\rm grad}$}  && \multicolumn{2}{c}{$(\zo)_{\rm std}$}   &&
  \multicolumn{2}{c}{$(\zo)_{\rm std1}$}  && \multicolumn{2}{c}{Warping}             &&
  \multicolumn{1}{c}{$Q$}  \\
  \noalign{\smallskip}
  \cline{3-4}
  \cline{6-7}
  \cline{9-11}
  \cline{13-14}
  \cline{16-17}
  \cline{19-20}
  \noalign{\smallskip}
   &  & \multicolumn{1}{c}{Best}  & \multicolumn{1}{c}{Alt.}        &&  \multicolumn{1}{c}{[$''$]} &
   \multicolumn{1}{c}{[kpc]}     && \multicolumn{3}{c}{[$10^{-3}$]} &&  \multicolumn{1}{c}{[pc]}   &
   \multicolumn{1}{c}{[\%]}      && \multicolumn{1}{c}{[pc]}        &   \multicolumn{1}{c}{[\%]}   &&
   \multicolumn{1}{c}{[pc]}       & \multicolumn{1}{c}{[\%]}        &&   \\
  \noalign{\smallskip}
  \multicolumn{1}{r}{(1)}         & \multicolumn{1}{c}{(2)}          &  \multicolumn{2}{c}{(3)} &&
  \multicolumn{2}{c}{(4)}        && \multicolumn{3}{c}{(5)}         &&  \multicolumn{2}{c}{(6)} &&
  \multicolumn{2}{c}{(7)}        && \multicolumn{2}{c}{(8)}         &&  \multicolumn{1}{c}{(9)} \\
  \noalign{\smallskip}
  \hline
%
%--------------------------------------------------------------------------------------------------------
%
%              Vertical Disk Model
% No.  Galaxy  Best & Alternative  zo_mean   zo_grad+err   zo_std(0.)      zo_std(1.)    warping       Q
%        
%                            ['']  [kpc]                  [pc]   [%]      [pc]   [%]     [pc]  [%]
%
% (1)   (2)          (3)         (4)           (5)              (6)            (7)            (8)     (9)
%--------------------------------------------------------------------------------------------------------
%
  \noalign{\medskip}
  \multicolumn{22}{c}{\bf Non -- Interacting} \\
  \noalign{\medskip}
  \hline
  \noalign{\medskip}
%
%------------ superthin & andere ------------------------------------------------------------------------
%
  1 & UGC 231     & 2 &3 &&  8.6 & 0.57 &&  10 &$\pm$& 5 &&  41 &  7.2 &&  36 & 6.3 &&  31 &  5.5 && 1.57 \\
  2 & ESO 150-G07 & 1 &2 &&  5.4 & ---  &&  41 &$\pm$&13 && --- &  9.5 && --- & 3.5 && --- &  6.4 && 1.26 \\
% (!)
  3 & ESO 112-G04$\A$&2&--&& 2.1 & ---  &&$-$5 &$\pm$& 7 && --- & 24.2 && --- &18.8 && --- & 17.3 && 0.99 \\
  4 & ESO 150-G14$\A$&2&--&& 4.9 & 2.73 &&  19 &$\pm$& 8 && 300 & 11.0 && 235 & 8.6 && 357 & 13.1 && 1.04 \\
  5 & UGC 711     & 2 &--&&  7.0 & 0.92 &&  22 &$\pm$& 7 && 116 & 12.6 &&  46 & 5.0 &&  67 &  7.3 && 1.20 \\
  6 & ESO 244-G48 & 1 &--&&  6.3 & ---  &&   3 &$\pm$& 3 && --- &  1.7 && --- & 1.3 && --- &  3.7 && 0.75 \\
% (!)
  7 & UGC 1839    & 2 &--&&  8.0 & 0.80 &&   2 &$\pm$&15 && 110 & 13.8 &&  95 &11.9 && 117 & 14.6 && 3.00 \\
  8 & NGC 891     & 2 &--&& 23.8 & 1.10 &&$-$8 &$\pm$& 9 &&  79 &  7.2 &&  56 & 5.1 &&  84 &  7.6 && 2.61 \\
  9 & ESO 416-G25$\A$&3&2&&  4.7 & 1.56 &&  21 &$\pm$& 7 && 296 & 19.0 && 135 & 8.7 &&  93 &  6.0 && 4.71 \\
 10 & UGC 2411    & 2 &--&&  6.2 & 1.13 &&   5 &$\pm$&20 && 455 & 40.3 && 343 &30.4 && 217 & 19.2 && 1.91 \\
%
%--------------------------------------------------------------------------------------------------------
%
 11 & IC 1877     & 2 &3 &&  2.1 & ---  &&$-$9 &$\pm$& 5 && --- & 11.6 && --- & 8.0 && --- & 10.7 && 4.63 \\
% bei AM0302-504 = ao0012
 12 & ESO 201-G22 & 2 &--&&  4.2 & 1.17 &&$-$1 &$\pm$& 3 && 105 &  9.0 &&  53 & 4.5 && 120 & 10.3 && 4.16 \\
 13 & NGC 1886    & 2 &--&&  5.8 & 0.67 &&   9 &$\pm$& 4 &&  73 & 10.9 &&  18 & 2.7 &&  58 &  8.6 && 2.99 \\
 14 & UGC 3474    & 2 &--&&  4.2 & 1.03 &&$-$35&$\pm$& 7 && 124 & 12.0 &&  50 & 4.9 &&  70 &  6.8 && 0.74 \\
 15 & NGC 2310    & 3 &2 &&  9.3 & 0.83 &&  26 &$\pm$& 4 &&  77 &  9.3 &&  34 & 4.1 && 158 & 19.0 && 3.39 \\
 16 & UGC 4278    & 2 &--&&  5.8 & 0.31 &&   4 &$\pm$& 2 &&  13 &  4.2 &&   7 & 2.3 && 102 & 32.9 && 3.87 \\
 17 & ESO 564-G27$\A$ &3&2&& 5.2 & 0.83 &&$-$2 &$\pm$& 6 && 189 & 22.8 && 118 &14.2 &&  59 &  7.1 && 1.92 \\
 18 & UGC 4943    & 2 &--&&  2.1 & 0.34 &&  12 &$\pm$& 5 &&  36 & 10.6 &&  20 & 5.9 &&  47 & 13.8 && 1.21 \\
 19 & IC 2469     & 1 &--&& 16.2 & 1.75 &&   9 &$\pm$& 6 && 132 &  7.5 && 118 & 6.7 && 397 & 22.2 && 3.40 \\
%
%--------------------------------------------------------------------------------------------------------
%
 20 & UGC 5341    & 2 &--&&  4.0 & 1.87 &&   0 &$\pm$& 4 && 188 & 10.1 && 115 & 6.1 && 257 & 13.7 && 1.39 \\
 21 & IC 2531     & 3 &2 && 10.4 & 1.67 &&  52 &$\pm$&24 && 544 & 32.6 && 469 &28.1 && 381 & 22.8 && 8.61 \\
% 22 & NGC 3115    & 1 &3 && 69.4 & 3.70 && 175 &$\pm$& 6 && 237 &  6.4 &&  70 & 1.9 && 186 &  8.5 && 6.85 \\
 22 & NGC 3390    & 1 &--&& 11.0 & 2.02 &&  29 &$\pm$& 9 && 285 & 14.1 &&  78 & 3.9 && 162 &  8.0 && 5.65 \\
 23 & ESO 319-G26$\A$&3&--&& 2.4 & 0.48 &&   5 &$\pm$&11 &&  79 & 16.5 &&  60 &12.5 && 135 & 28.2 && 2.09 \\
 24 & NGC 3957    & 1 &--&&  9.3 & 1.33 &&  30 &$\pm$& 8 && 132 &  9.9 &&  94 & 7.1 && 108 &  8.1 && 4.28 \\
 25 & NGC 4013    & 1 &--&& 11.7 & 0.68 &&   2 &$\pm$& 7 &&  35 &  5.1 &&  31 & 4.6 &&  34 &  5.0 && 1.65 \\
 26 & ESO 572-G44 & 1 &--&&  7.0 & 3.03 &&  47 &$\pm$& 3 && 221 &  7.3 &&  85 & 2.8 && 159 &  5.3 && 0.41 \\
% (!)
 27 & UGC 7170    & 2 &--&&  3.0 & 0.43 &&   0 &$\pm$& 3 &&  20 &  4.7 &&  18 & 4.2 &&  42 &  9.8 && 1.61 \\
 28 & ESO 321-G10$\A$&1&--&& 4.9 & 1.00 &&   6 &$\pm$& 3 &&  49 &  4.9 &&  43 & 4.3 &&  60 &  6.0 && 1.36 \\
%
%--------------------------------------------------------------------------------------------------------
%
 29 & NGC 4217    & 2 &--&& 21.7 & 1.53 &&  35 &$\pm$& 5 && 148 &  9.7 &&  56 & 3.7 && 145 &  9.5 && 3.63 \\
 30 & NGC 4244    & 2 &--&& 22.1 & 0.41 &&  48 &$\pm$&22 &&  71 & 17.3 &&  50 &12.2 &&  67 & 16.4 && 8.16 \\
 31 & UGC 7321    & 2 &--&&  4.9 & 0.38 &&$-$2 &$\pm$& 4 &&  28 &  7.4 &&  24 & 6.3 &&  44 & 11.7 && 2.72 \\
 32 & NGC 4302    & 2 &--&&  8.1 & 0.62 &&  39 &$\pm$& 3 &&  52 &  8.4 &&  22 & 3.5 &&  72 & 11.6 && 0.81 \\
 33 & NGC 4330    & 1 &--&& 16.6 & 1.27 &&  75 &$\pm$& 9 && 176 & 13.9 &&  86 & 6.8 && 110 &  8.6 && 1.71 \\
 34 & NGC 4565    & 1 &--&& 10.7 & 0.52 &&$-$7 &$\pm$& 2 &&  36 &  6.9 &&  24 & 4.6 &&  53 & 10.2 && 2.04 \\
 35 & NGC 4710    & 1 &2 && 11.6 & 0.81 &&  60 &$\pm$& 9 &&  76 &  9.4 &&  42 & 5.2 &&  60 &  7.4 && 1.38 \\
% 37 & NGC 5126    & 2 &--&& 12.1 & 3.00 &&  49 &$\pm$& 7 && 158 &  5.3 &&  39 & 1.3 && 231 &  7.7 && 1.70 \\
% bei NGC 5126 = ao0013 
 36 & NGC 5170    & 1 &--&& 22.1 & 1.69 &&$-37$&$\pm$&26 && 219 & 13.0 && 121 & 7.2 && 247 & 14.6 && 2.39 \\
%
%--------------------------------------------------------------------------------------------------------
%
 37 & ESO 510-G18 & 2 &--&&  1.0 & ---  &&  14 &$\pm$& 1 && --- & 11.3 && --- & 1.2 && --- & 10.4 && 0.54 \\
% (!)
 38 & UGC 9242    & 1 &--&&  5.7 & 0.70 &&$-$12&$\pm$&11 && 173 & 24.7 &&  65 & 9.3 &&  40 &  5.7 && 1.81 \\
 39 & NGC 5775    & 1 &--&& 13.5 & 1.90 &&  29 &$\pm$&10 && 218 & 11.5 && 142 & 7.5 &&  87 &  4.6 && 2.95 \\
 40 & NGC 5907    & 2 &--&& 18.2 & 0.97 &&   7 &$\pm$& 7 && 106 & 10.9 &&  60 & 6.2 &&  94 &  9.7 && 4.74 \\
 41 & NGC 5908    & 2 &--&&  8.9 & 1.92 &&$-$19&$\pm$& 9 && 192 & 10.0 && 156 & 8.1 && 126 &  6.6 && 1.39 \\
 42 & ESO 583-G08 & 2 &--&&  1.7 & 0.84 &&  10 &$\pm$& 2 &&  66 &  7.9 &&  34 & 4.0 && 117 & 14.0 && 0.53 \\
 43 & NGC 6181    & 2 &--&&  5.2 & 0.91 &&  59 &$\pm$& 6 &&  88 &  9.7 &&  36 & 4.0 &&  70 &  7.7 && 1.53 \\
 44 & ESO 230-G11 & 3 &--&&  3.8 & 1.28 && $-$6&$\pm$& 5 &&  96 &  7.5 &&  90 & 7.0 && 348 & 27.2 && 1.53 \\
% (!)
 45 & NGC 6722$\A$& 1 &--&&  7.9 & 2.53 &&$-$16&$\pm$& 9 && 186 &  7.4 &&  78 & 3.1 && 291 & 11.5 && 2.03 \\
 46 & ESO 461-G06 & 1 &--&&  3.0 & ---  &&$-$27&$\pm$&13 && --- & 11.8 && --- & 5.4 && --- &  4.8 && 1.43 \\
 47 & ESO 339-G16$\A$ &2&--&&4.5 & ---  &&  81 &$\pm$& 2 && --- & 18.2 && --- & 3.3 && --- &  6.9 && 0.29 \\
 48 & IC 4937$\A$ & 3 &2 &&  5.4 & 1.66 &&$-$3 &$\pm$& 2 && 101 &  6.1 &&  79 & 4.8 && 200 & 12.1 && 1.04 \\
 49 & ESO 187-G08 & 2 &3 &&  3.3 & 1.02 &&   8 &$\pm$& 3 && 139 & 13.6 && 105 &10.3 && 182 & 17.8 && 0.68 \\
 50 & IC 5052     & 1 &--&& 19.3 & 0.75 &&  38 &$\pm$&10 && 117 & 15.6 &&  86 &11.5 && 101 & 13.5 && 1.96 \\
 51 & IC 5096$\B$ & 1 &--&&  6.5 & 1.46 && --- &&  ---   && --- &  --- && --- & --- && --- & ---  && ---  \\
%
%--------------------------------------------------------------------------------------------------------
%
  \noalign{\smallskip}
  \hline
  \end{tabular}
  \end{flushleft}
  \end{table*}
\end{center}

\begin{center}
  \begin{table*}
 {\bf Table 4.} (continued) \\
  \begin{flushleft}
  \begin{tabular}{rlcccrccrcrrrrrrrrrrcr}
  \cline{1-22}
  \hline\hline
  \noalign{\smallskip}
  \multicolumn{1}{r}{No.}                  & \multicolumn{1}{c}{Galaxy}  &
  \multicolumn{2}{c}{Vertical Fit}        && \multicolumn{2}{c}{$(\zo)_{\rm mean}$}  &&
  \multicolumn{3}{c}{$(\zo)_{\rm grad}$}  && \multicolumn{2}{c}{$(\zo)_{\rm std}$}   &&
  \multicolumn{2}{c}{$(\zo)_{\rm std1}$}  && \multicolumn{2}{c}{Warping}             &&
  \multicolumn{1}{c}{$Q$}  \\
  \noalign{\smallskip}
  \cline{3-4}
  \cline{6-7}
  \cline{9-11}
  \cline{13-14}
  \cline{16-17}
  \cline{19-20}
  \noalign{\smallskip}
   &  & \multicolumn{1}{c}{Best}  & \multicolumn{1}{c}{Alt.}        &&  \multicolumn{1}{c}{[$''$]} &
   \multicolumn{1}{c}{[kpc]}     && \multicolumn{3}{c}{[$10^{-3}$]} &&  \multicolumn{1}{c}{[pc]}   &
   \multicolumn{1}{c}{[\%]}      && \multicolumn{1}{c}{[pc]}        &   \multicolumn{1}{c}{[\%]}   &&
   \multicolumn{1}{c}{[pc]}       & \multicolumn{1}{c}{[\%]}        &&   \\
  \noalign{\smallskip}
  \multicolumn{1}{r}{(1)}         & \multicolumn{1}{c}{(2)}          &  \multicolumn{2}{c}{(3)} &&
  \multicolumn{2}{c}{(4)}        && \multicolumn{3}{c}{(5)}         &&  \multicolumn{2}{c}{(6)} &&
  \multicolumn{2}{c}{(7)}        && \multicolumn{2}{c}{(8)}         &&  \multicolumn{1}{c}{(9)} \\
  \noalign{\smallskip}
  \hline
  \noalign{\medskip}
%
%--------------------------------------------------------------------------------------------------------
%
%              Vertical Disk Model
% No.  Galaxy  Best & Alternative  zo_mean   zo_grad+err   zo_std(0.)      zo_std(1.)    warping       Q
%        
%                            ['']  [kpc]                  [pc]   [%]      [pc]   [%]     [pc]  [%]
%
% (1)   (2)          (3)         (4)           (5)              (6)            (7)            (8)     (9)
%--------------------------------------------------------------------------------------------------------
%
% (!)
 52 & ESO 466-G01$\A$&1&--&& 6.8 & 3.24 &&   4 &$\pm$& 4 && 319 &  9.8 &&  95 & 2.9 && 333 & 10.3 && 0.42 \\
% (!)
 53 & ESO 189-G12 & 1 &--&& 3.60 & 2.03 &&  19 &$\pm$&12 && 270 & 11.7 && 191 & 8.3 && 330 & 16.2 && 1.82 \\
% (!)
 54 & UGC 11859$\A$&1 &--&&  1.5 & 0.32 &&$-$15&$\pm$& 2 &&  50 & 15.6 &&  29 & 9.1 &&  57 & 17.8 && 0.45 \\
 55 & ESO 533-G04 & 1 &--&&  7.6 & 1.40 &&  41 &$\pm$& 6 && 158 & 11.3 &&  86 & 6.1 && 146 & 10.5 && 1.25 \\
 56 & IC 5199     & 1 &--&&  5.0 & 1.72 &&  20 &$\pm$& 5 &&  70 &  4.1 &&  62 & 3.6 && 145 &  8.4 && 2.44 \\
%
%--------------------------------------------------------------------------------------------------------
%
 57 & UGC 11994   & 2 &3 &&  4.0 & 1.37 &&  18 &$\pm$& 2 && 116 &  8.5 &&  79 & 5.8 && 208 & 15.2 && 1.23 \\
 58 & UGC 12281   & 2 &--&&  5.7 & 1.09 &&$-$15&$\pm$& 6 && 107 &  9.8 &&  54 & 5.0 && 174 & 16.0 && 3.83 \\
 59 & UGC 12423   & 1 &--&&  7.9 & 2.69 &&  28 &$\pm$&16 && 796 & 29.6 && 277 &10.3 && 229 &  8.5 && 1.22 \\
 60 & NGC 7518    & 1 &--&&  9.6 & 2.44 && 149 &$\pm$& 8 && 472 & 19.3 && 171 & 7.0 && 296 & 12.1 && 3.02 \\
 61 & ESO 604-G06 & 2 &--&&  3.4 & 1.74 &&   3 &$\pm$& 2 && 150 &  8.6 && 138 & 7.9 && 147 &  8.5 && 1.08 \\
%
%--------------------------------------------------------------------------------------------------------
%
  \noalign{\smallskip}
  \hline
  \end{tabular}
  \begin{list}{}{}
  \item[$\A$] Supplementary data from Barteldrees \& Dettmar (\cite{barteldrees1994}). \\
  \item[$\B$] No fit was performed. Disk parameters were taken from Table\~5 in SD\~II. \\
  \end{list}
  \end{flushleft}
  \end{table*}
\end{center}

%__________________________________________________________________

%\end{document}
%\graphicspath{{/d0/schwarz/publications/aa/99_3/appendix/}{/ps/}}

%%%%%%%%%%%%%%%%%%%%%%%%%%%%%  1  %%%%%%%%%%%%%%%%%%%%%%%%%%%%%%%%%

\begin{figure*}[t]
\vspace*{3mm}
\hspace*{5mm}
\begin{minipage}[b]{5.5cm}
\begin{picture}(3.0,3.0)
{\includegraphics[angle=180,viewport=40 50 400 730,clip,width=52mm]{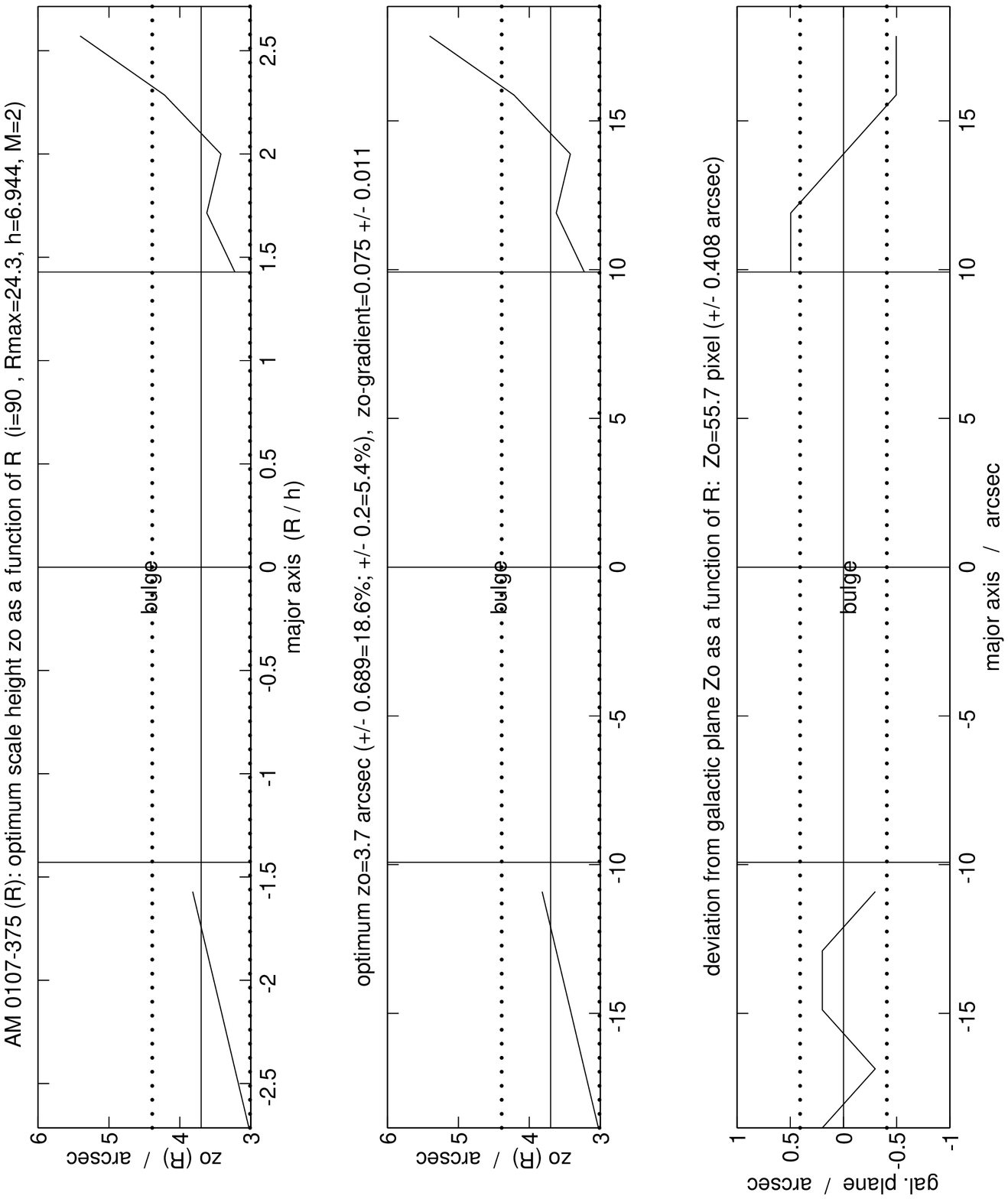}}
\end{picture}
\end{minipage}
\hfill
\begin{minipage}[b]{5.5cm}
\begin{picture}(3.0,3.0)
{\includegraphics[angle=180,viewport=40 50 400 730,clip,width=52mm]{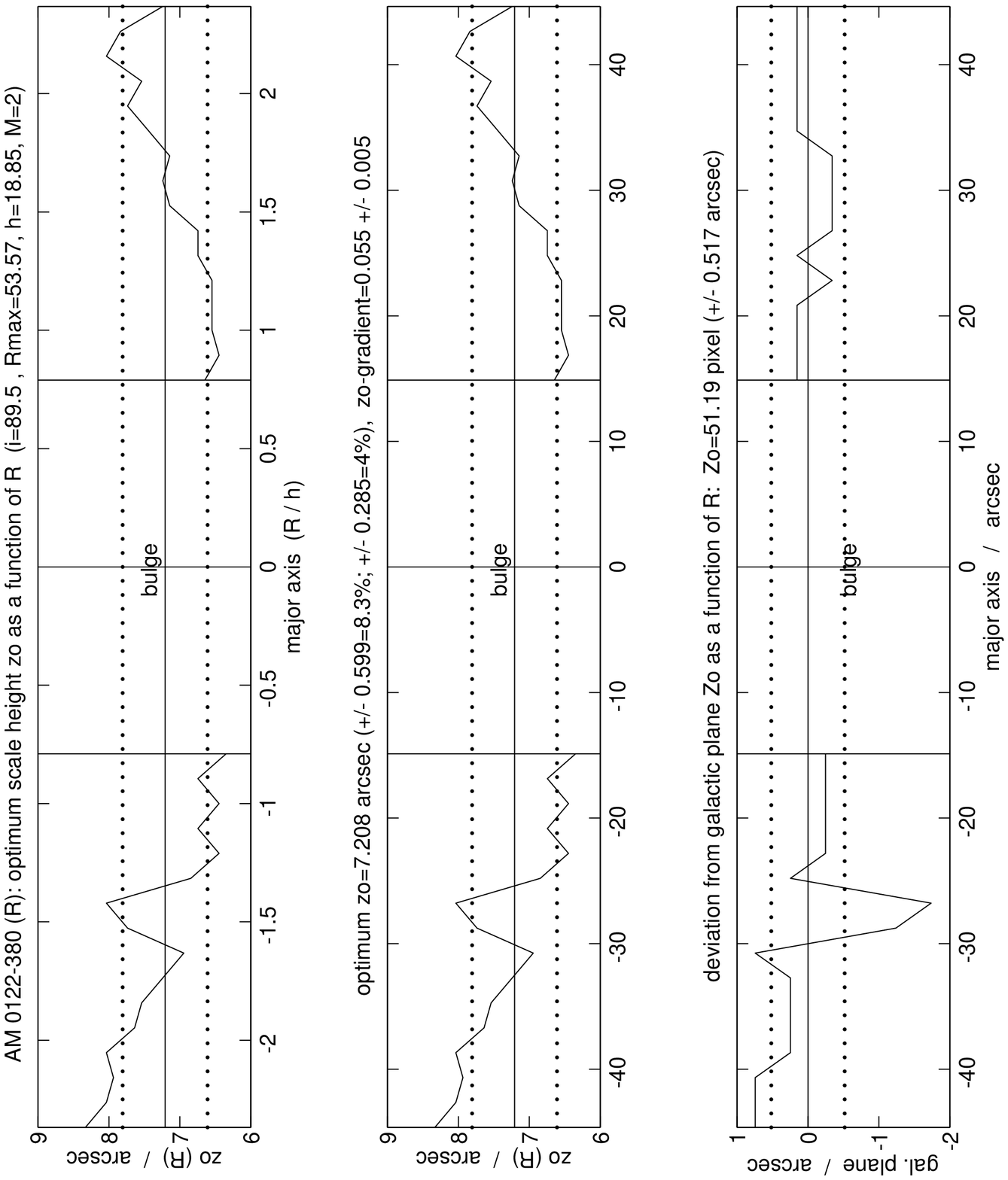}}
\end{picture}
\end{minipage}
\hfill
\begin{minipage}[b]{5.5cm}
\begin{picture}(3.0,3.0)
{\includegraphics[angle=180,viewport=40 50 400 730,clip,width=52mm]{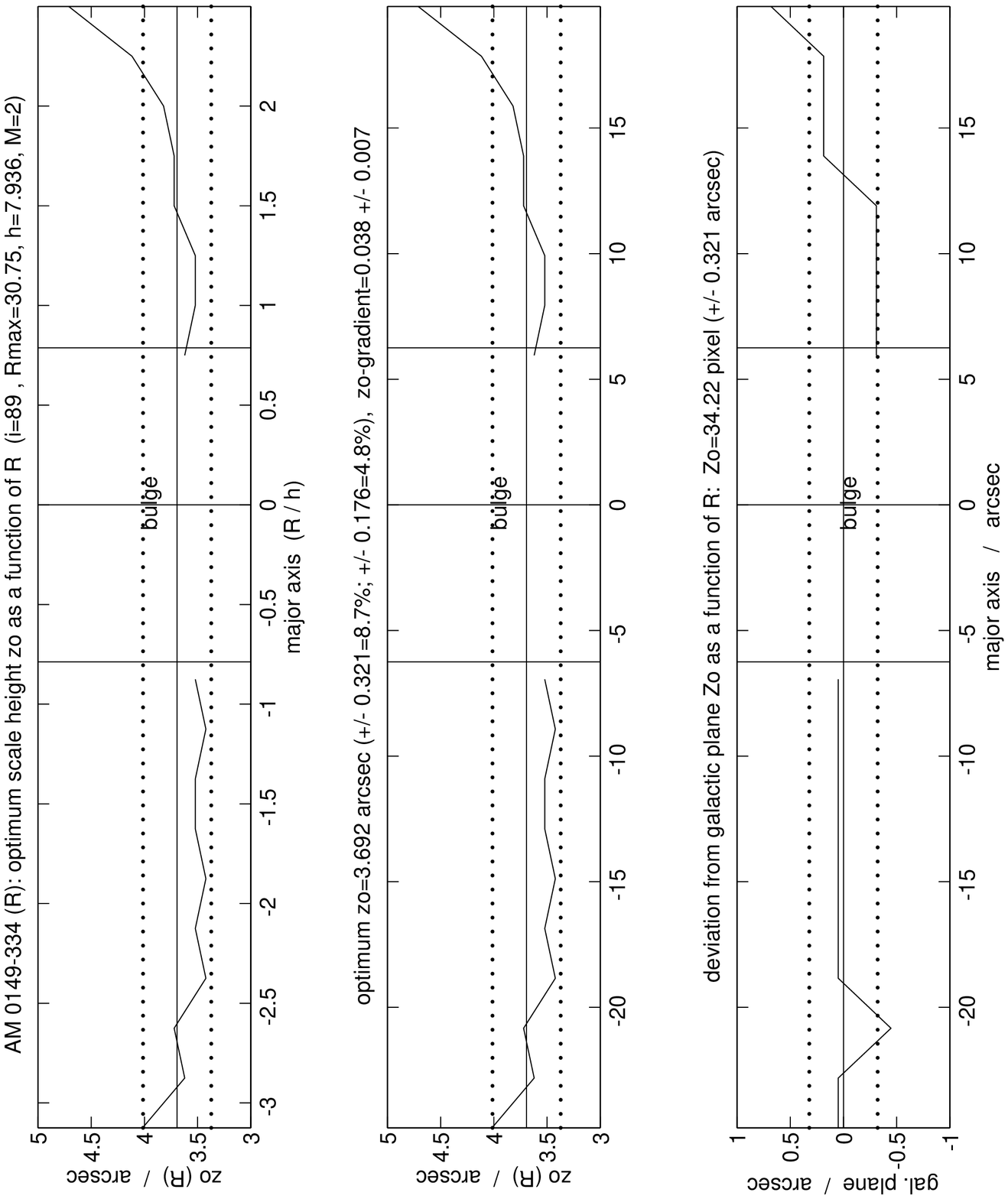}}
\end{picture}
\end{minipage}

\vspace*{98mm}

\hspace*{18mm}\parbox{165mm}{AM\~0107-375  \hspace{40mm}  ESO 296-G17  \hspace{40mm}  ESO 354-G05}

\vspace*{5mm}

\hspace*{5mm}
\begin{minipage}[b]{5.5cm}
\begin{picture}(3.0,3.0)
{\includegraphics[angle=180,viewport=40 50 400 730,clip,width=52mm]{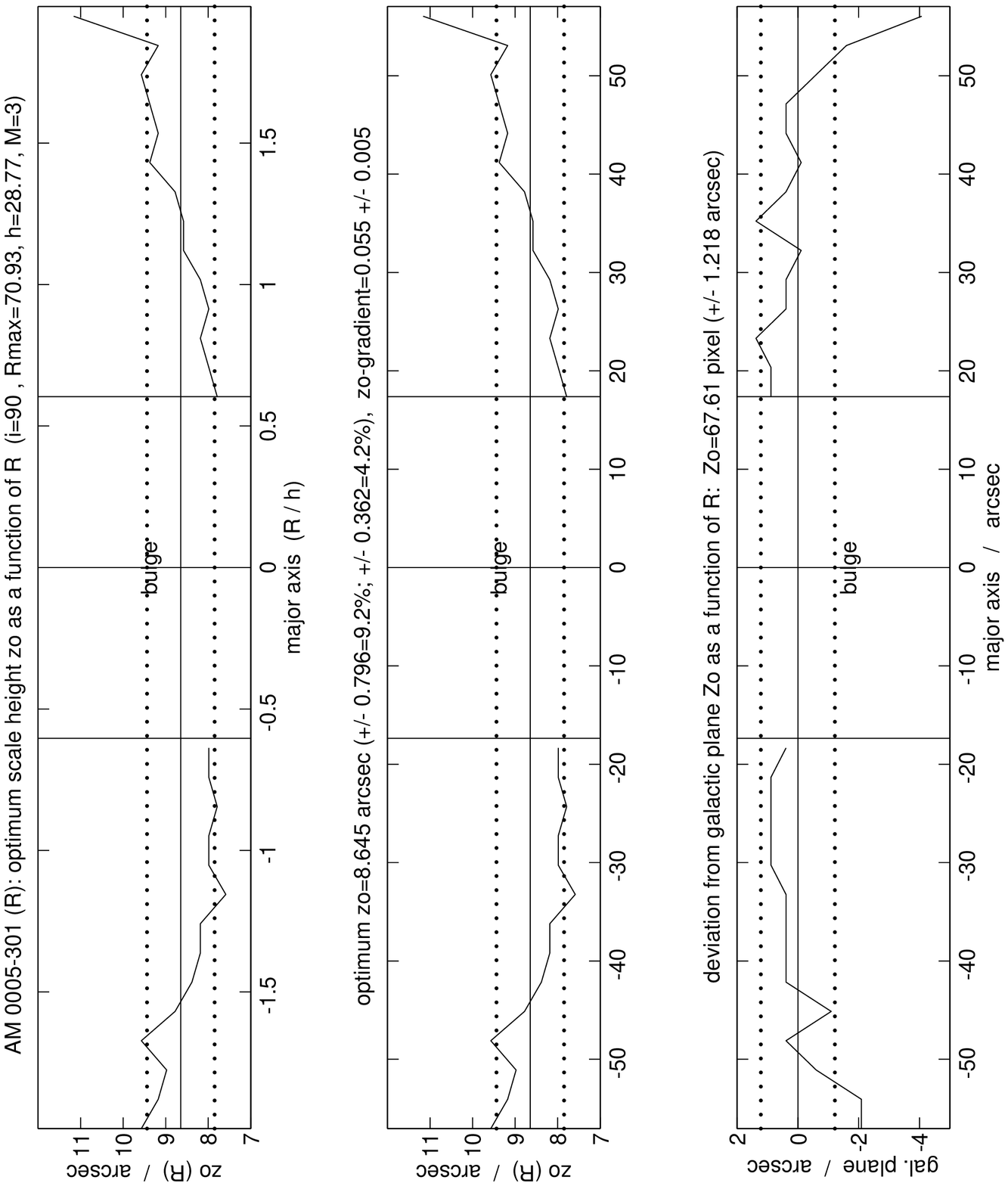}}
\end{picture}
\end{minipage}
\hfill
\begin{minipage}[b]{5.5cm}
\begin{picture}(3.0,3.0)
{\includegraphics[angle=180,viewport=40 50 400 730,clip,width=52mm]{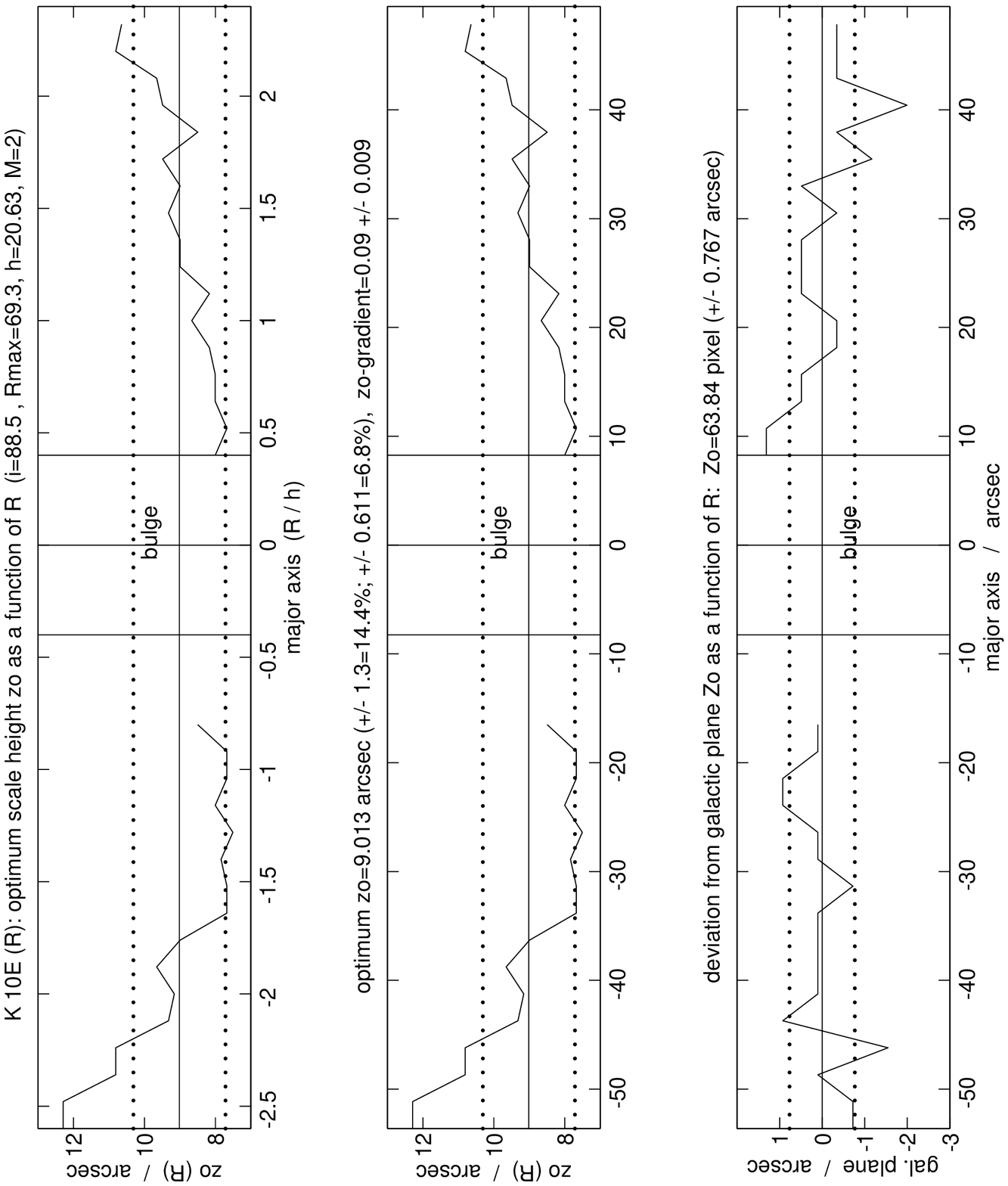}}
\end{picture}
\end{minipage}
\hfill
\begin{minipage}[b]{5.5cm}
\begin{picture}(3.0,3.0)
{\includegraphics[angle=180,viewport=40 50 400 730,clip,width=52mm]{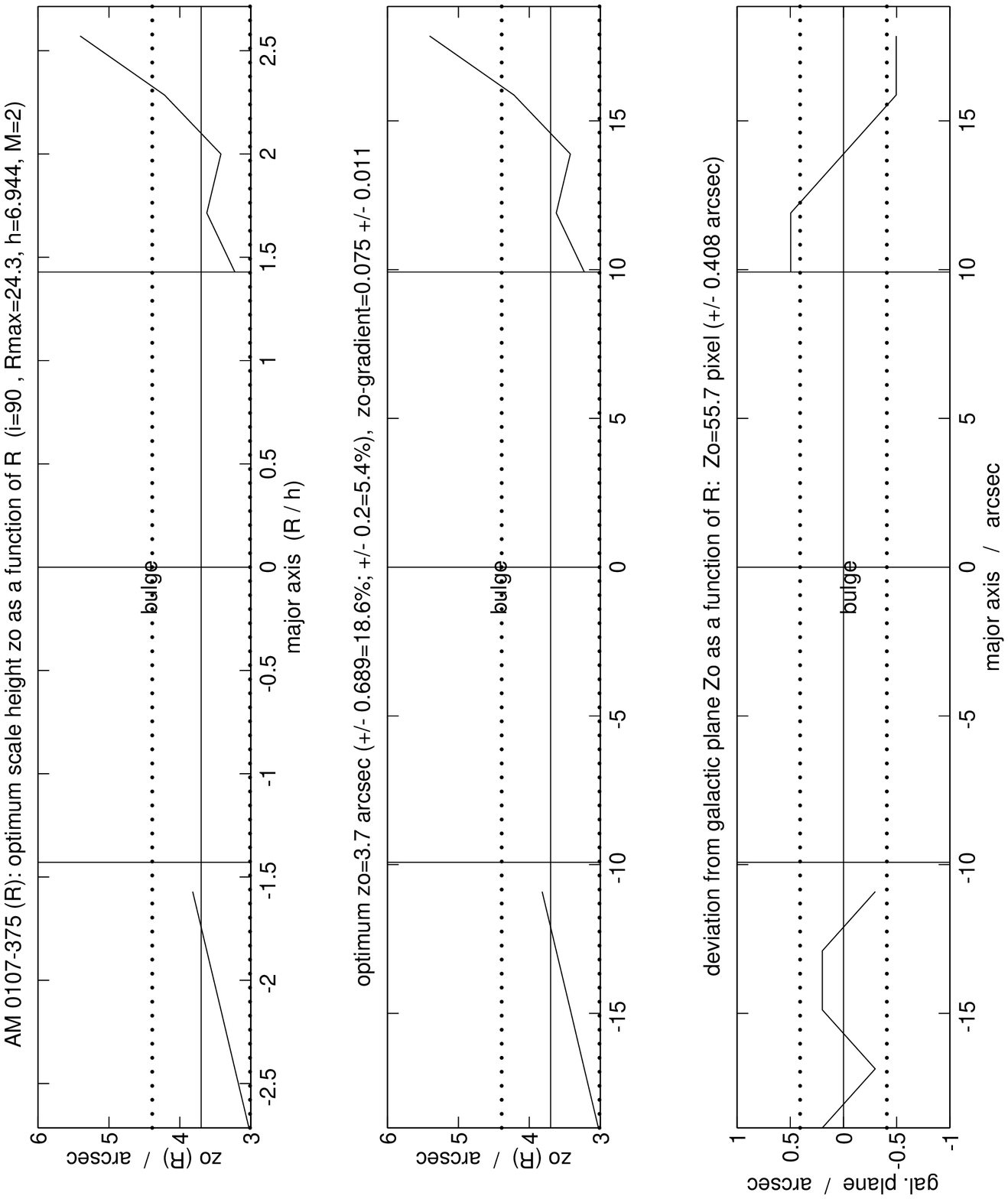}}
\end{picture}
\end{minipage}

\vspace*{98mm}

\hspace*{25mm}\parbox{165mm}{NGC 7  \hspace{46mm}  UGC 260  \hspace{44mm}  NGC 128}

\vspace*{8mm}

\hspace*{8mm}\parbox{165mm}{
{\bf \noindent Appendix A.} The measured behaviour of scale height (upper panels) and
mean galactic plane (lower panels) along the major axis of the disk, shown for each
galaxy in the interacting/merging sample. The main output parameters are indicated. For
details see Sect.\~3 and Table\~4 in this paper. Contour maps were given in Paper\~I,
Figs.\~4 and 5.
}
\end{figure*}

%%%%%%%%%%%%%%%%%%%%%%%%%%%%%%%%%%%%%%%%%%%%%%%%%%%%%%%%%%%%%%%%%%%

\clearpage

%%%%%%%%%%%%%%%%%%%%%%%%%%%%%  2  %%%%%%%%%%%%%%%%%%%%%%%%%%%%%%%%%

\begin{figure*}[t]
\vspace*{3mm}
\hspace*{5mm}
\begin{minipage}[b]{5.5cm}
\begin{picture}(3.0,3.0)
{\includegraphics[angle=180,viewport=40 50 400 730,clip,width=52mm]{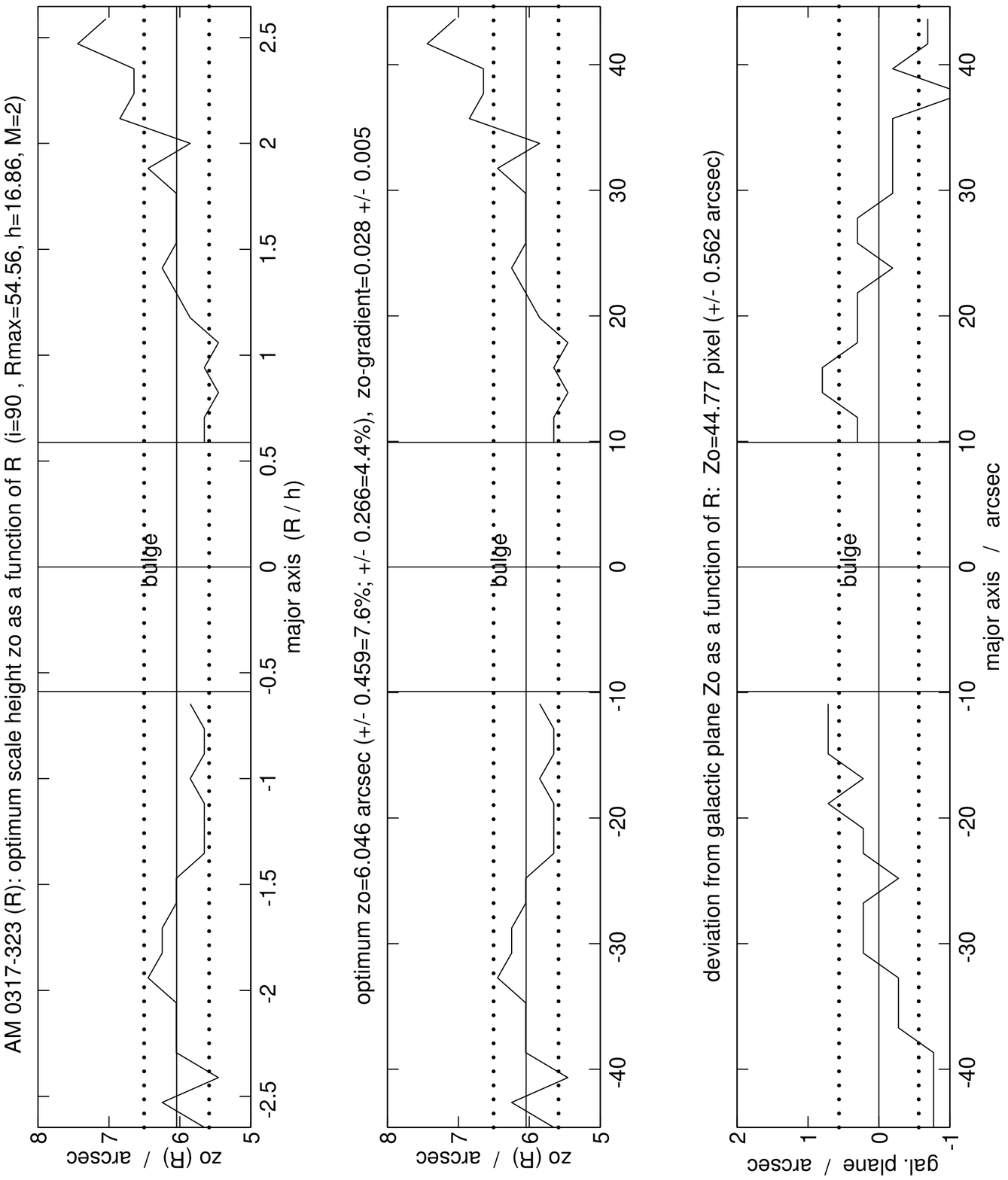}}
\end{picture}
\end{minipage}
\hfill
\begin{minipage}[b]{5.5cm}
\begin{picture}(3.0,3.0)
{\includegraphics[angle=180,viewport=40 50 400 730,clip,width=52mm]{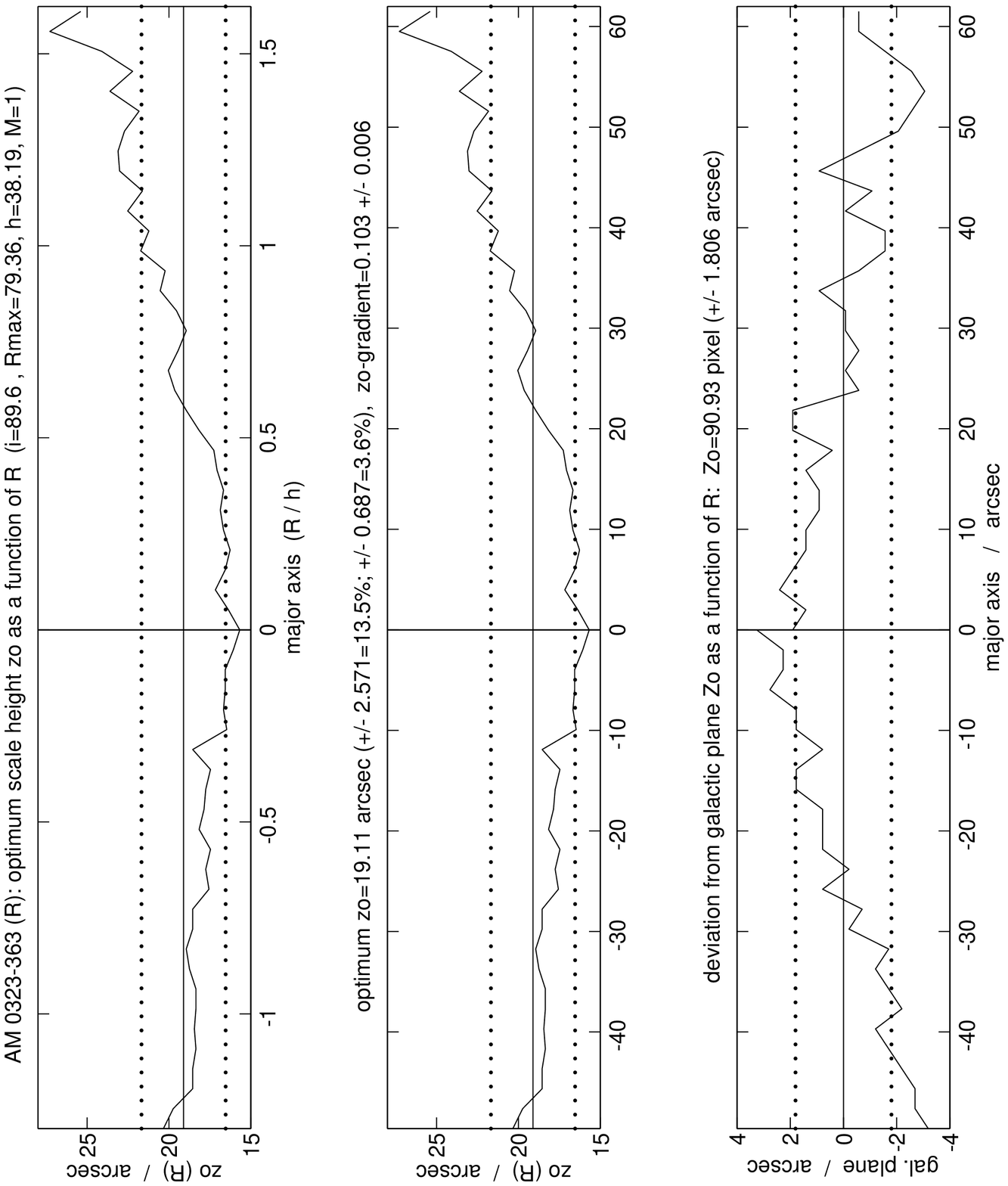}}
\end{picture}
\end{minipage}
\hfill
\begin{minipage}[b]{5.5cm}
\begin{picture}(3.0,3.0)
{\includegraphics[angle=180,viewport=40 50 400 730,clip,width=52mm]{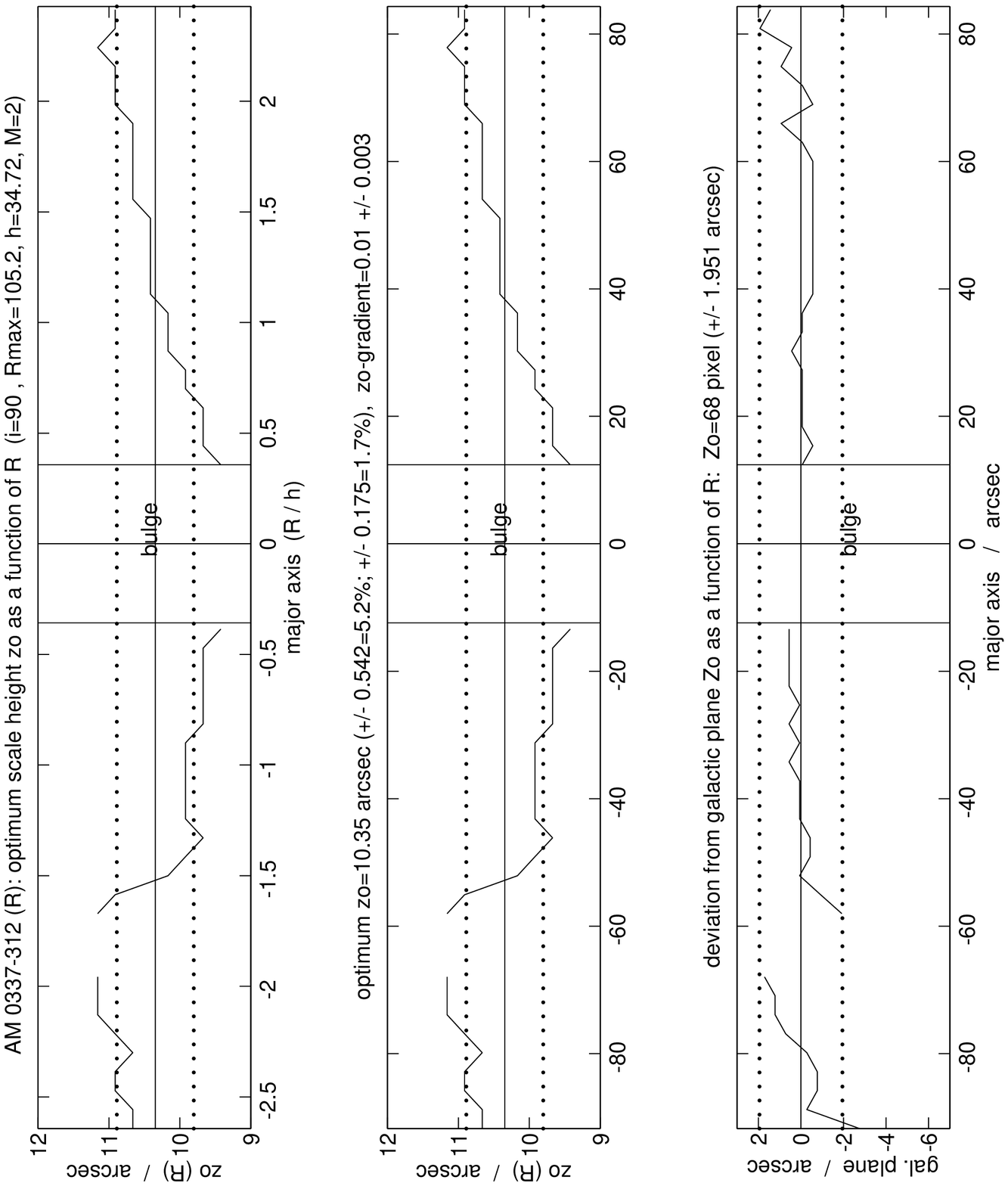}}
\end{picture}
\end{minipage}

\vspace*{98mm}

\hspace*{18mm}\parbox{165mm}{ESO 357-G16  \hspace{40mm}  ESO 357-G26  \hspace{40mm}  ESO 418-G15}

\vspace*{5mm}

\hspace*{5mm}
\begin{minipage}[b]{5.5cm}
\begin{picture}(3.0,3.0)
{\includegraphics[angle=180,viewport=40 50 400 730,clip,width=52mm]{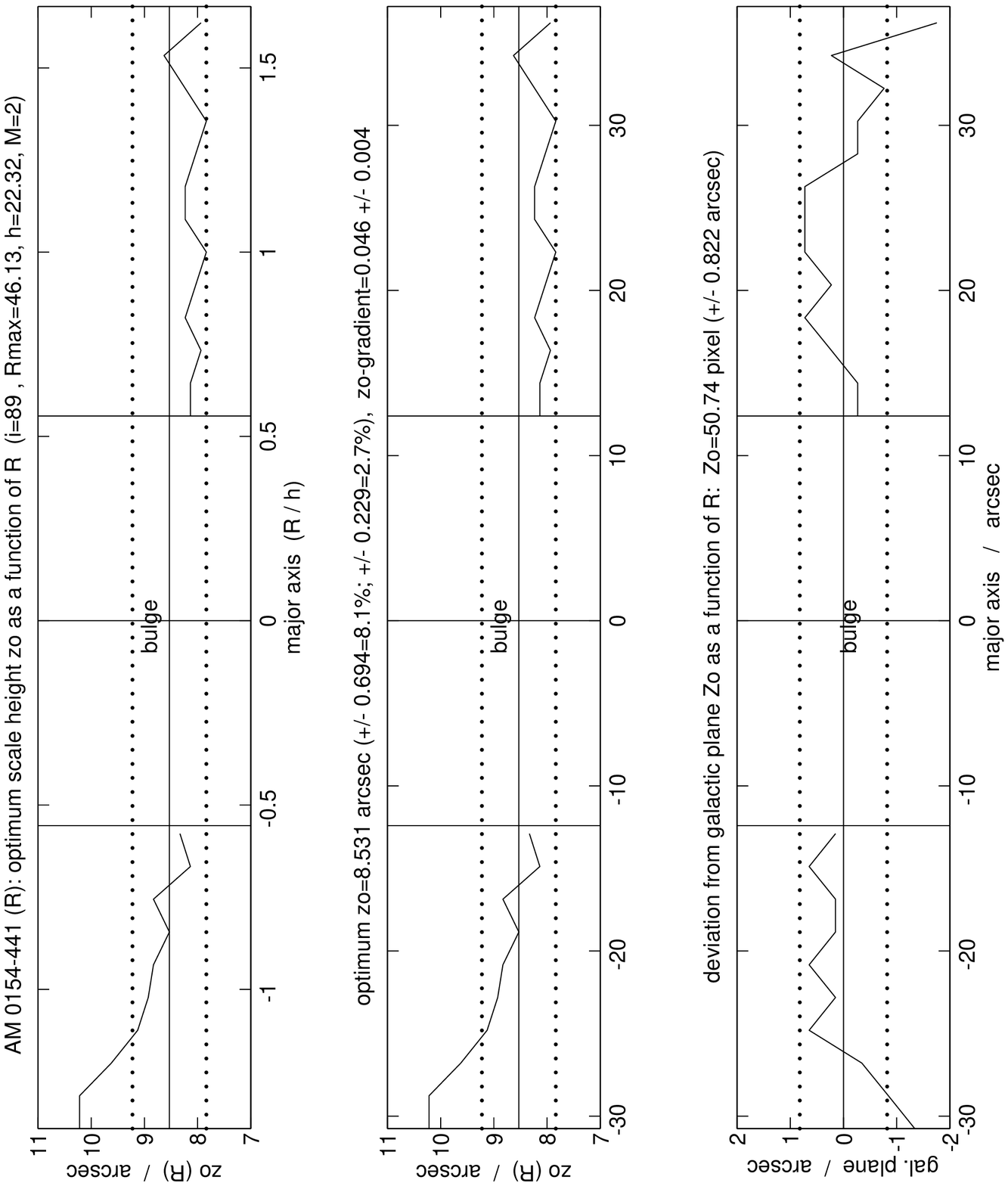}}
\end{picture}
\end{minipage}
\hfill
\begin{minipage}[b]{5.5cm}
\begin{picture}(3.0,3.0)
{\includegraphics[angle=180,viewport=40 50 400 730,clip,width=52mm]{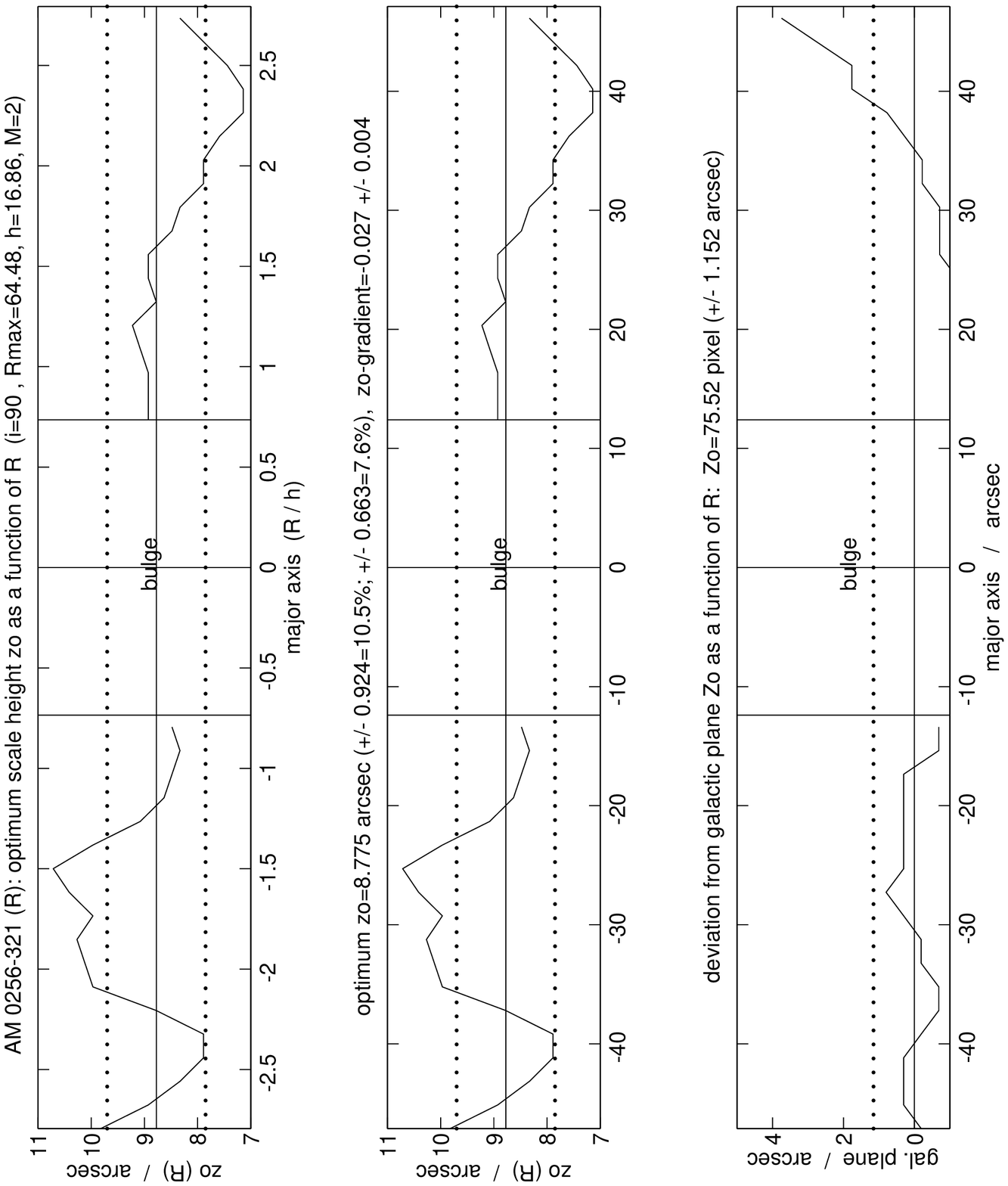}}
\end{picture}
\end{minipage}
\hfill
\begin{minipage}[b]{5.5cm}
\begin{picture}(3.0,3.0)
{\includegraphics[angle=180,viewport=40 50 400 730,clip,width=52mm]{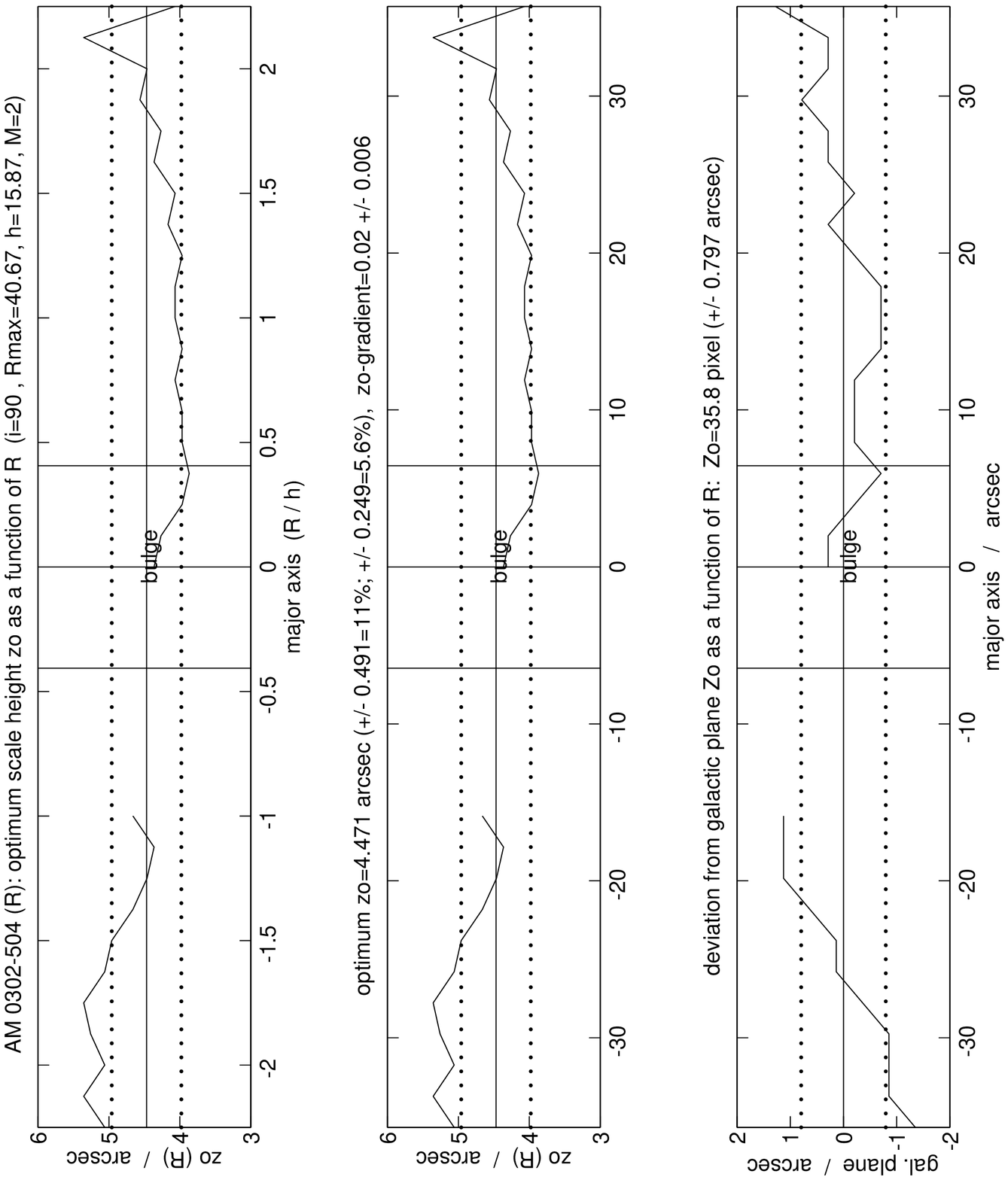}}
\end{picture}
\end{minipage}

\vspace*{98mm}

\hspace*{18mm}\parbox{165mm}{ESO 245-G10  \hspace{40mm}  ESO 417-G08  \hspace{40mm}  ESO 199-G12}

\vspace*{8mm}

\hspace*{8mm}\parbox{165mm}{
{\bf \noindent Appendix A.} (continued)
}
\end{figure*}

%%%%%%%%%%%%%%%%%%%%%%%%%%%%%%%%%%%%%%%%%%%%%%%%%%%%%%%%%%%%%%%%%%%

\clearpage

%%%%%%%%%%%%%%%%%%%%%%%%%%%%%  3  %%%%%%%%%%%%%%%%%%%%%%%%%%%%%%%%%

\begin{figure*}[t]
\vspace*{3mm}
\hspace*{5mm}
\begin{minipage}[b]{5.5cm}
\begin{picture}(3.0,3.0)
{\includegraphics[angle=180,viewport=40 50 400 730,clip,width=52mm]{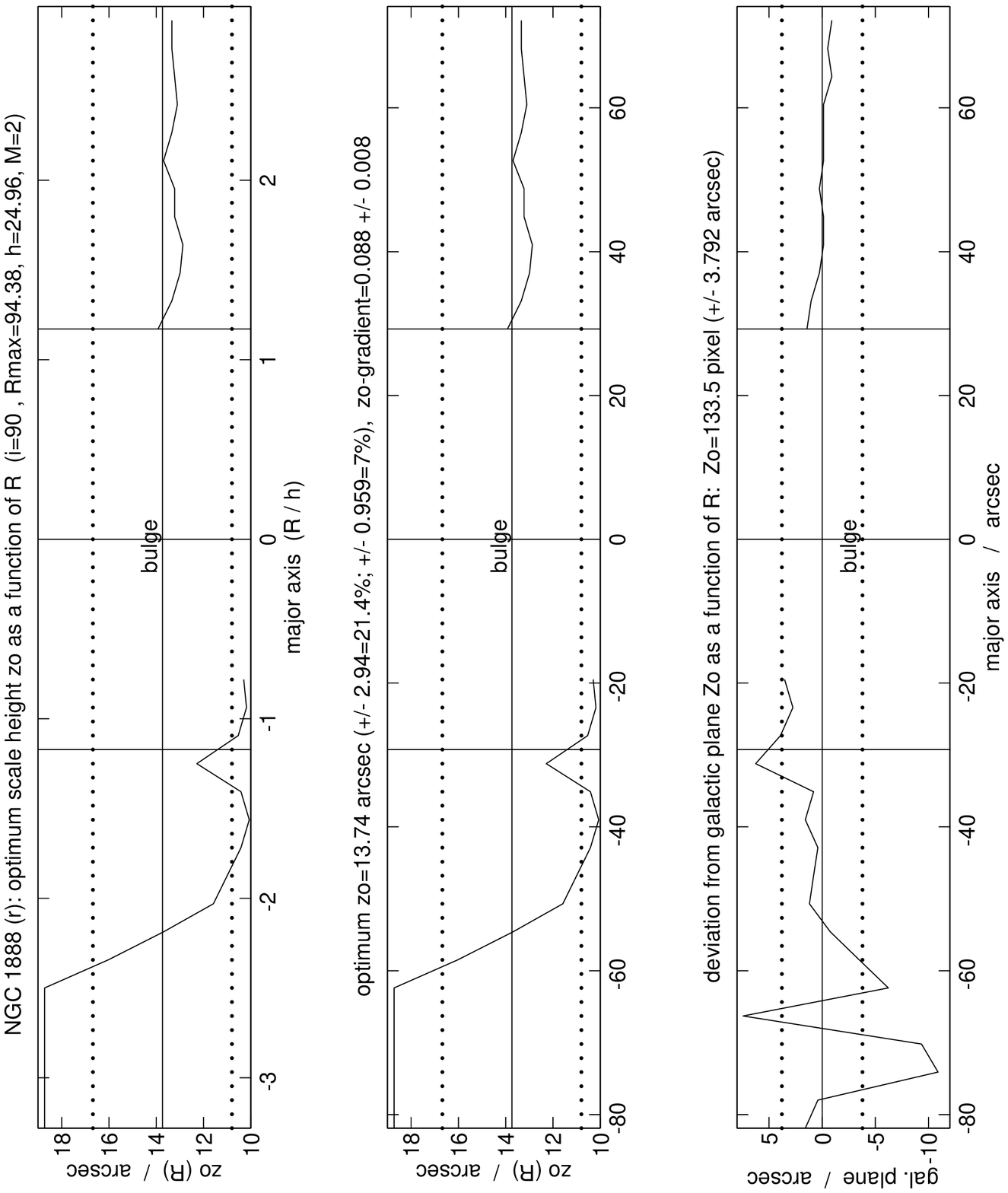}}
\end{picture}
\end{minipage}
\hfill
\begin{minipage}[b]{5.5cm}
\begin{picture}(3.0,3.0)
{\includegraphics[angle=180,viewport=40 50 400 730,clip,width=52mm]{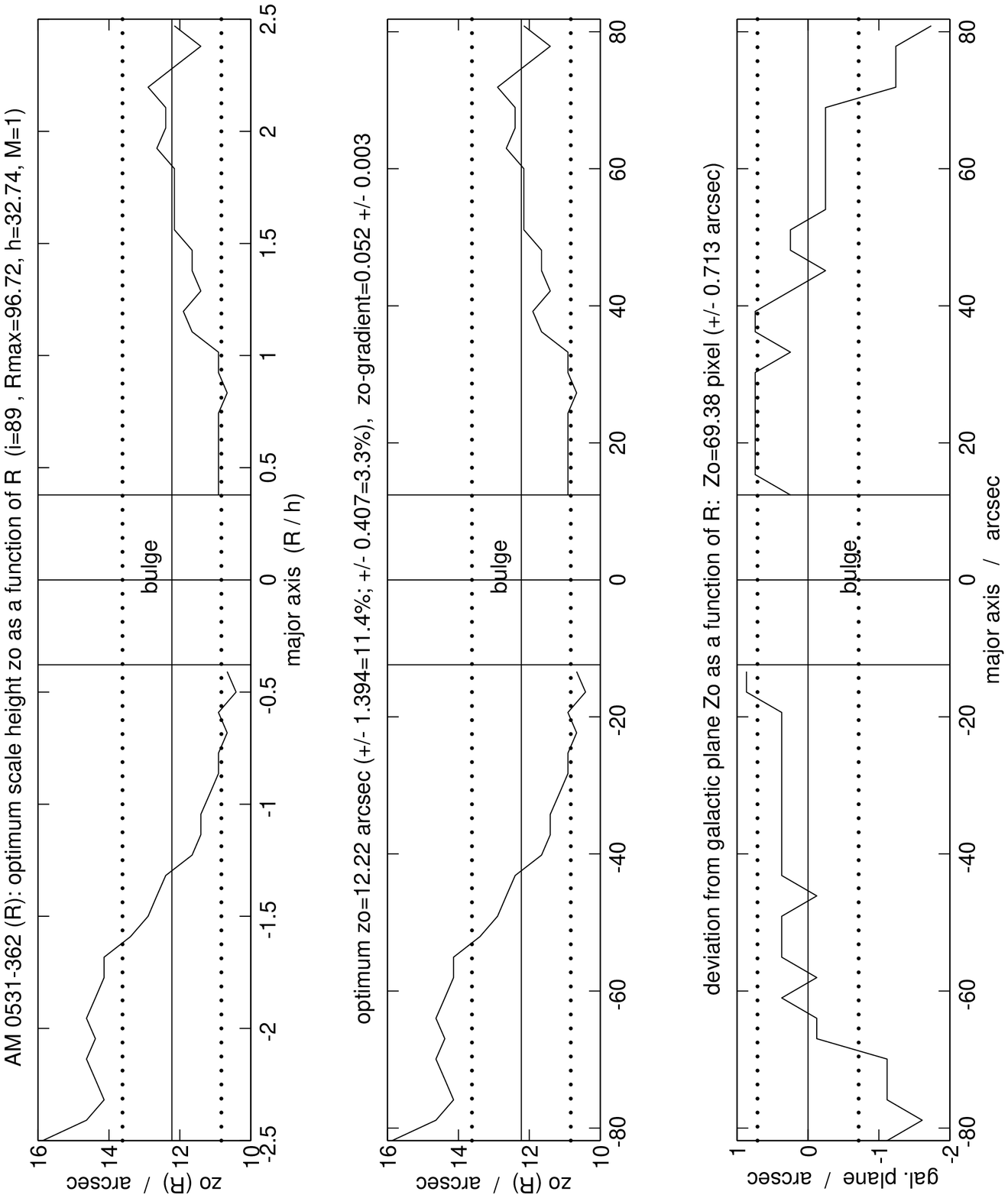}}
\end{picture}
\end{minipage}
\hfill
\begin{minipage}[b]{5.5cm}
\begin{picture}(3.0,3.0)
{\includegraphics[angle=180,viewport=40 50 400 730,clip,width=52mm]{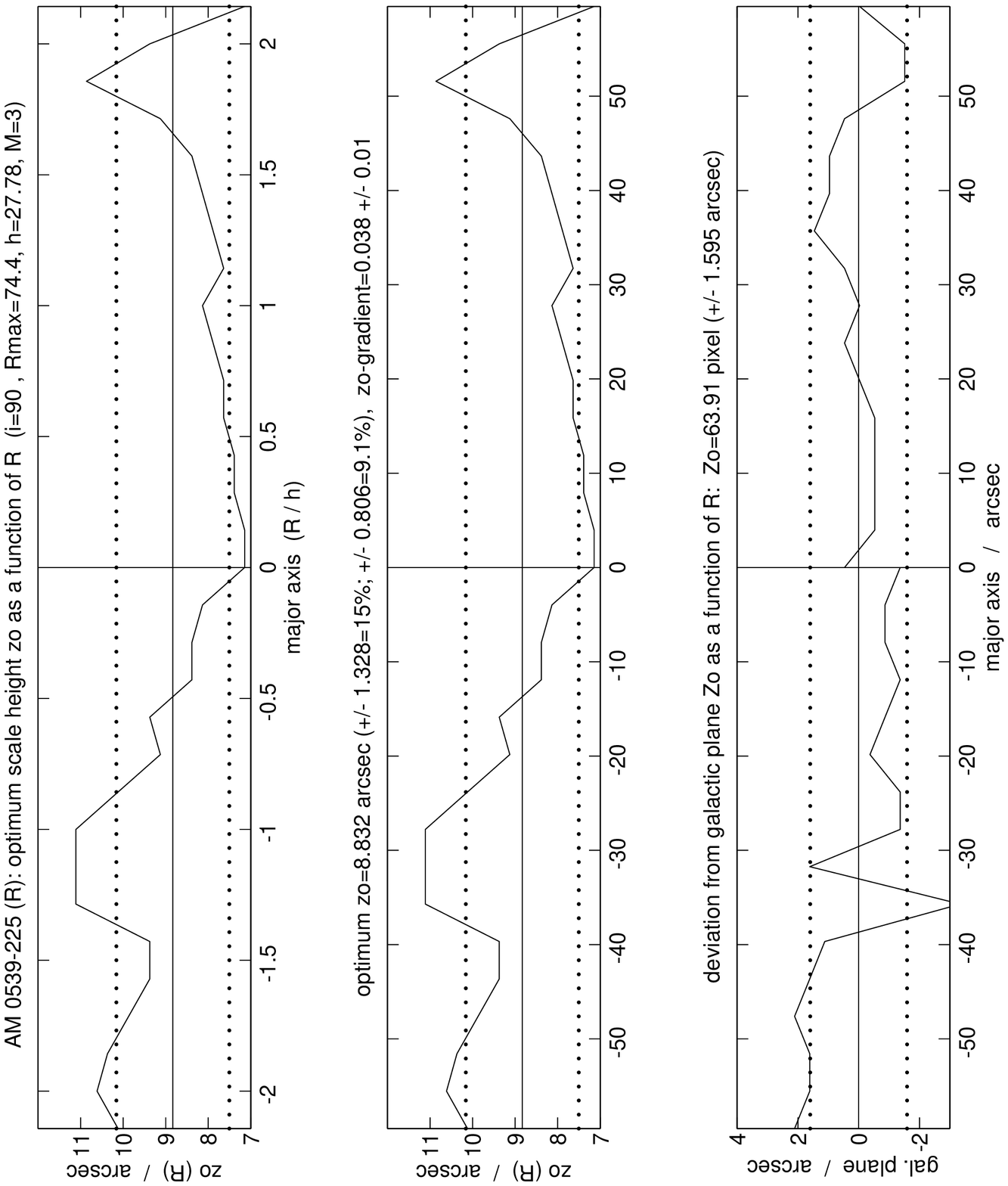}}
\end{picture}
\end{minipage}

\vspace*{98mm}

\hspace*{18mm}\parbox{165mm}{NGC 1888  \hspace{40mm}  ESO 363-G07  \hspace{40mm}  ESO 487-G35}

\vspace*{5mm}

\hspace*{5mm}
\begin{minipage}[b]{5.5cm}
\begin{picture}(3.0,3.0)
{\includegraphics[angle=180,viewport=40 50 400 730,clip,width=52mm]{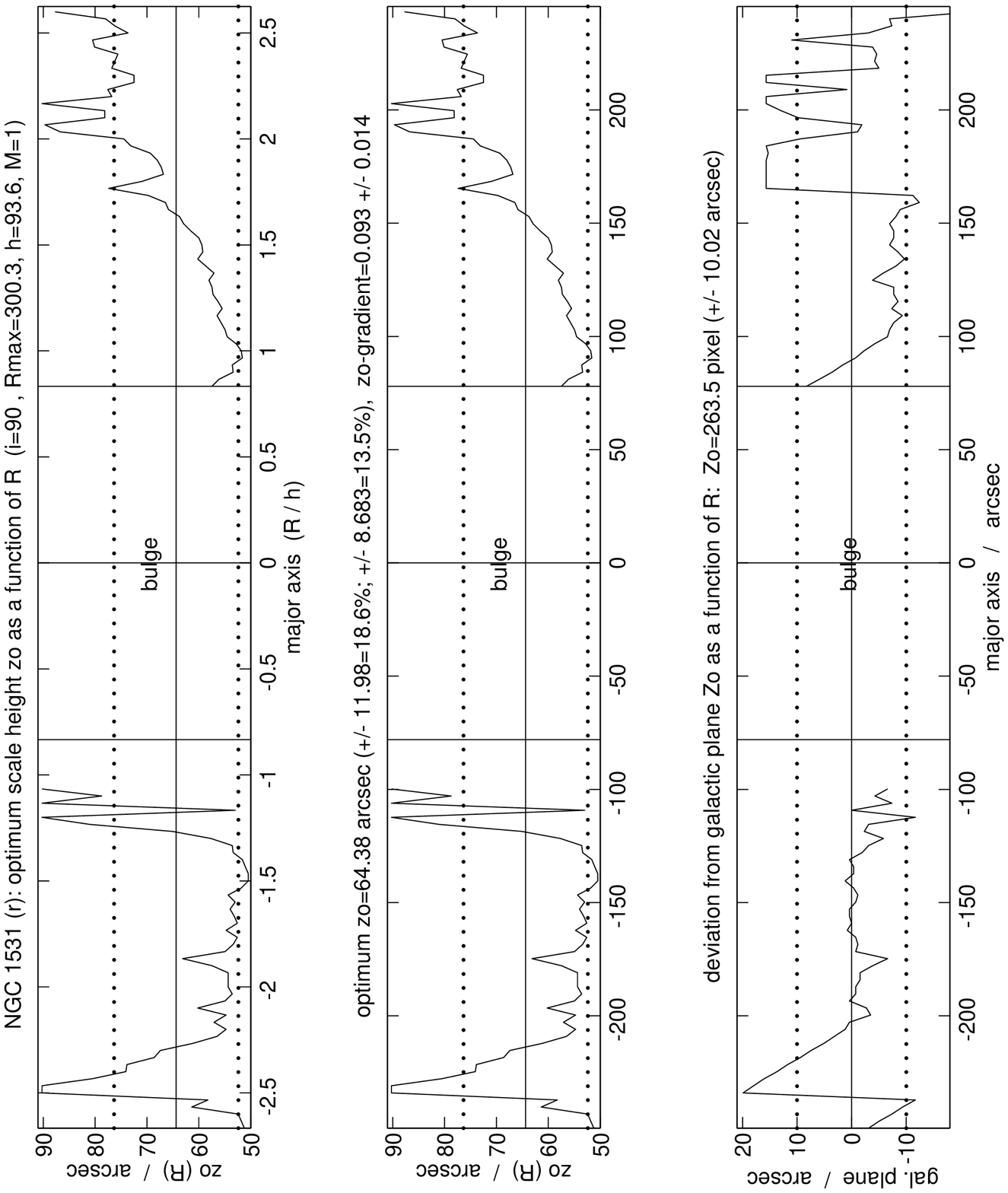}}
\end{picture}
\end{minipage}
\hfill
\begin{minipage}[b]{5.5cm}
\begin{picture}(3.0,3.0)
{\includegraphics[angle=180,viewport=40 50 400 730,clip,width=52mm]{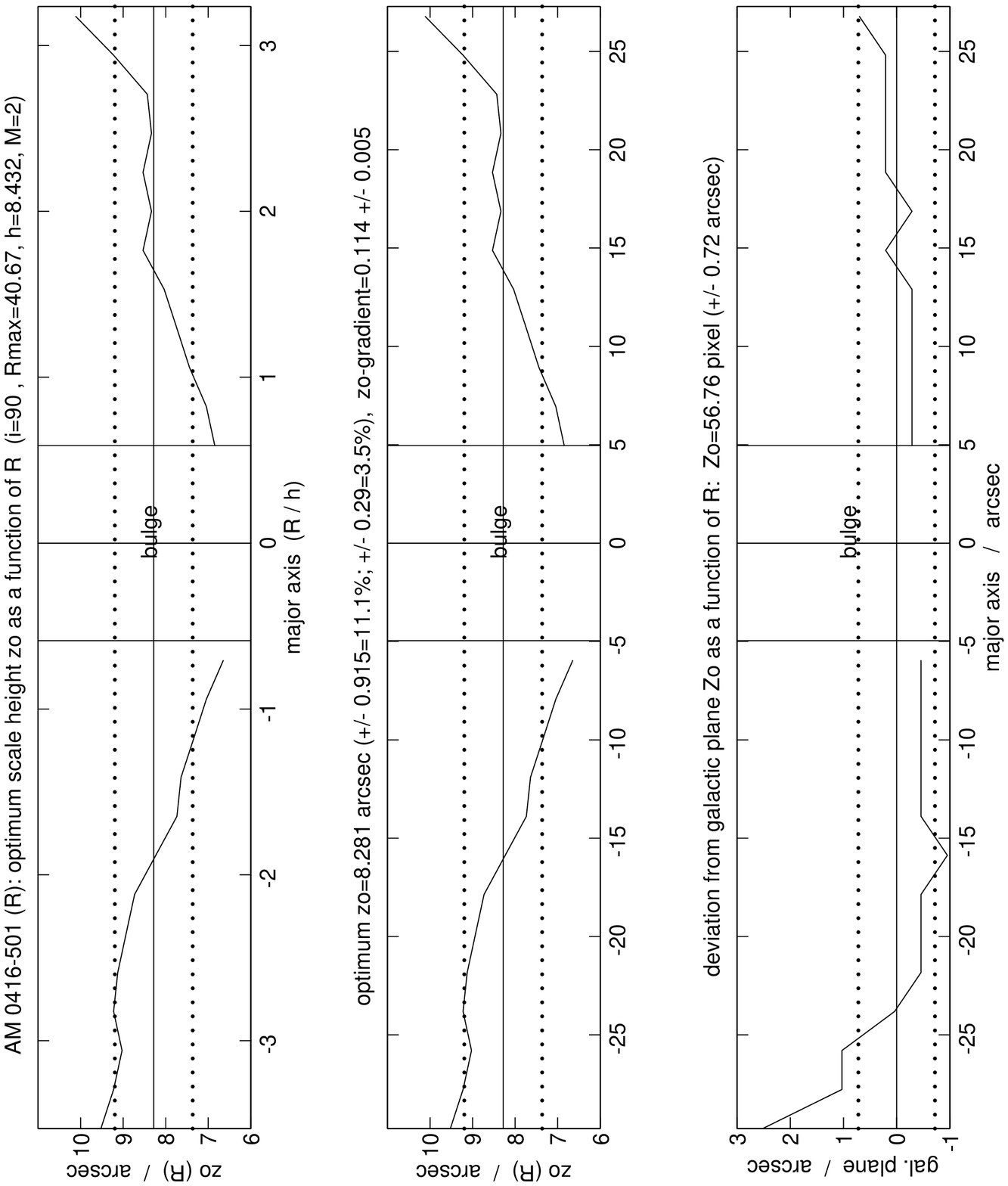}}
\end{picture}
\end{minipage}
\hfill
\begin{minipage}[b]{5.5cm}
\begin{picture}(3.0,3.0)
{\includegraphics[angle=180,viewport=40 50 400 730,clip,width=52mm]{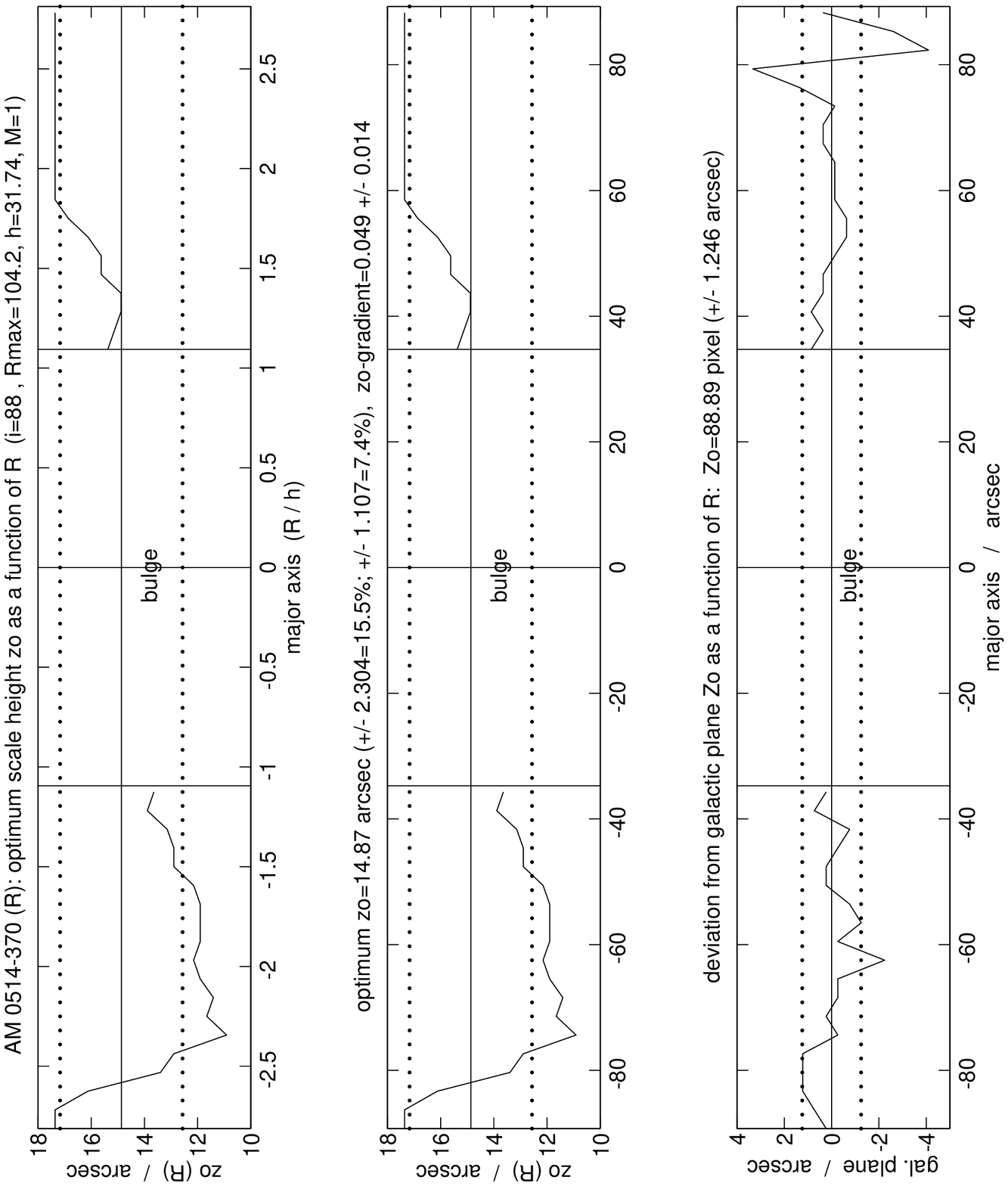}}
\end{picture}
\end{minipage}

\vspace*{98mm}

\hspace*{18mm}\parbox{165mm}{NGC 1531  \hspace{40mm}  ESO 202-G04  \hspace{40mm}  ESO 362-G11}

\vspace*{8mm}

\hspace*{8mm}\parbox{165mm}{
{\bf \noindent Appendix A.} (continued)
}
\end{figure*}

%%%%%%%%%%%%%%%%%%%%%%%%%%%%%%%%%%%%%%%%%%%%%%%%%%%%%%%%%%%%%%%%%%%

\clearpage

%%%%%%%%%%%%%%%%%%%%%%%%%%%%%  4  %%%%%%%%%%%%%%%%%%%%%%%%%%%%%%%%%

\begin{figure*}[t]
\vspace*{5mm}
\hspace*{6mm}
\begin{minipage}[b]{5.5cm}
\begin{picture}(3.0,3.0)
{\includegraphics[angle=180,viewport=15 -15 342 630,clip,width=49mm]{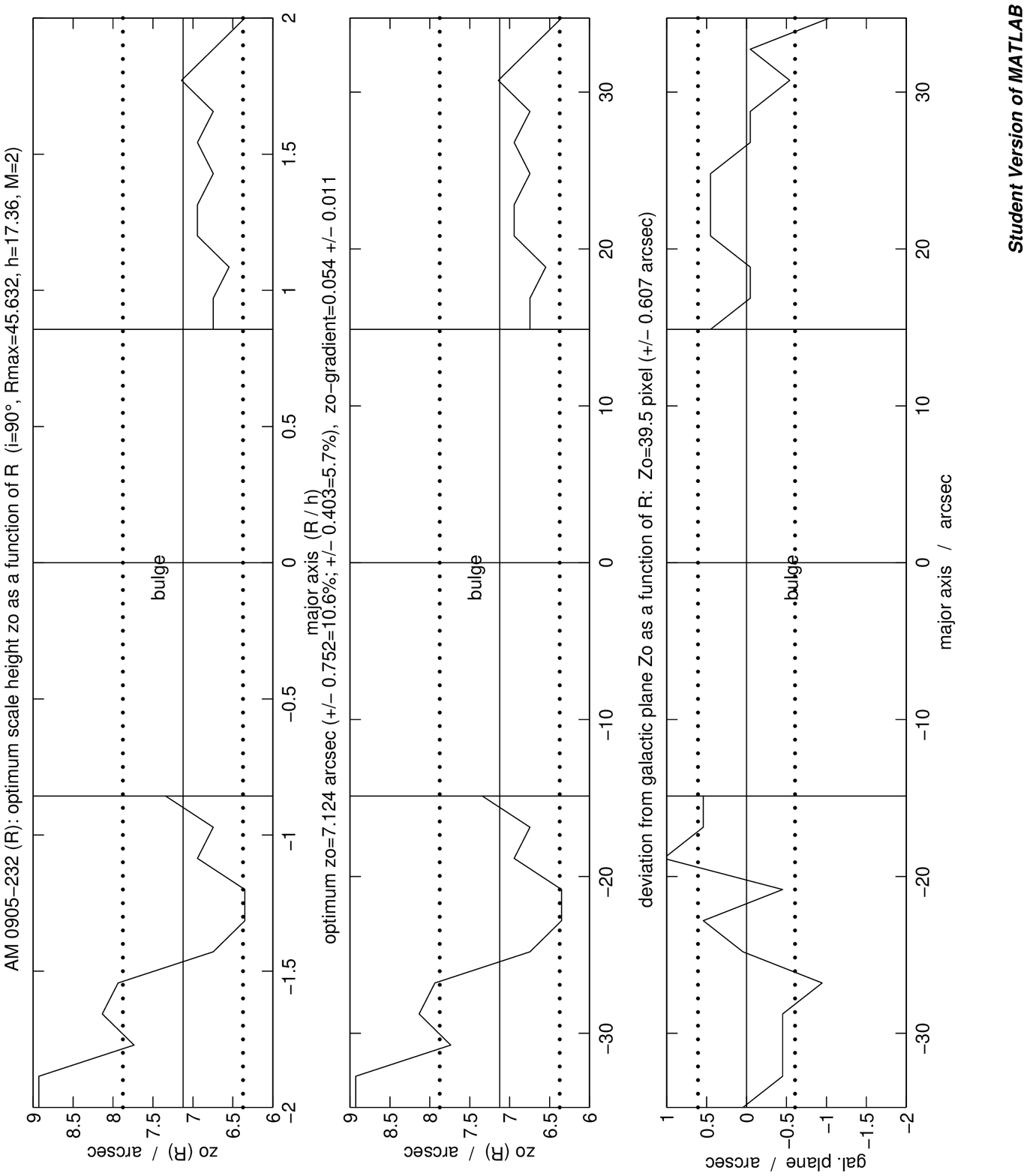}}
\end{picture}
\end{minipage}
\hfill
\begin{minipage}[b]{5.5cm}
\begin{picture}(3.0,3.0)
{\includegraphics[angle=180,viewport=40 50 400 730,clip,width=52mm]{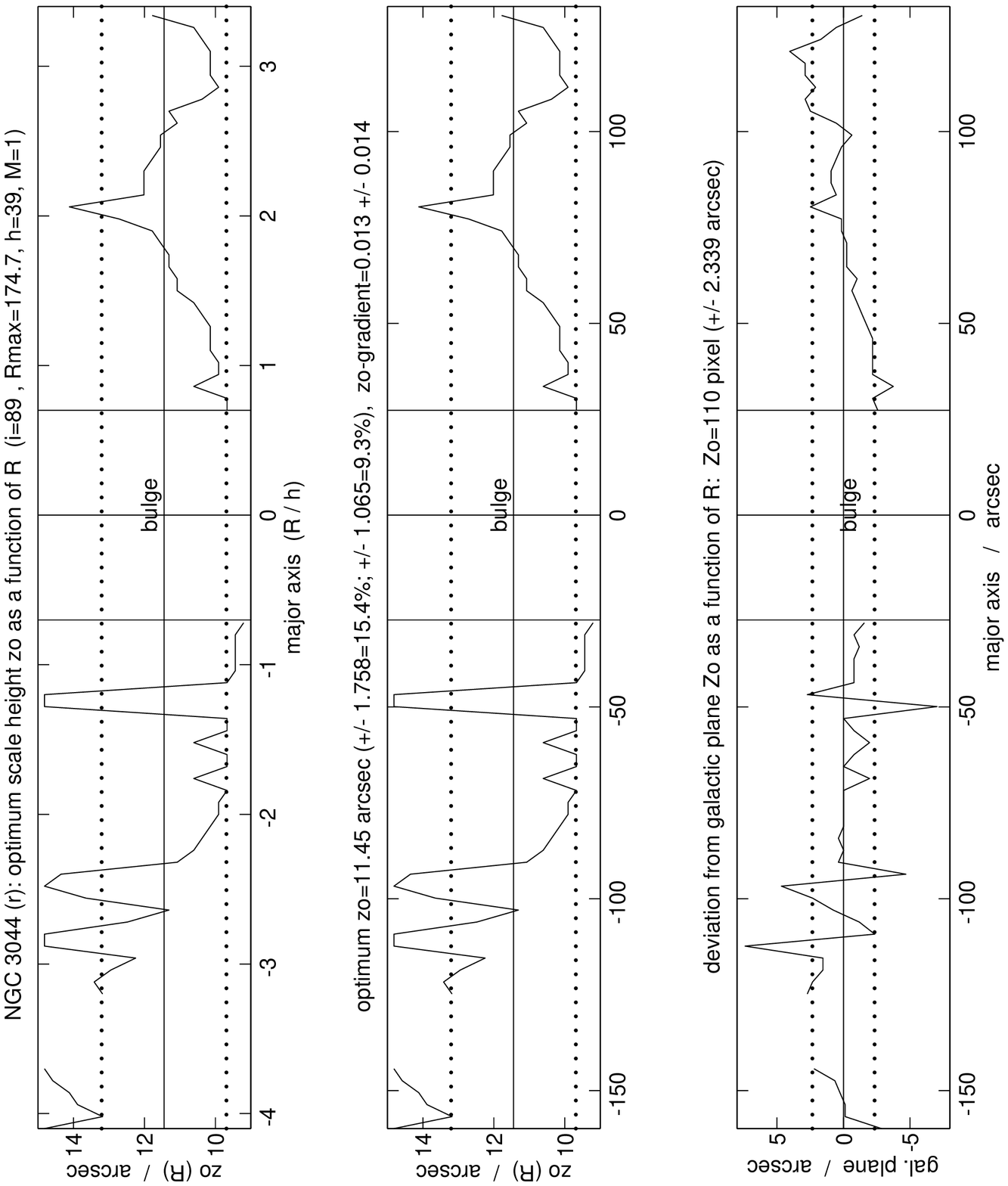}}
\end{picture}
\end{minipage}
\hfill
\begin{minipage}[b]{5.5cm}
\begin{picture}(3.0,3.0)
{\includegraphics[angle=180,viewport=40 50 400 730,clip,width=52mm]{ngc3187.ps}}
\end{picture}
\end{minipage}

\vspace*{98mm}

\hspace*{18mm}\parbox{165mm}{ESO 497-G14  \hspace{40mm}  NGC 3044  \hspace{40mm}  NGC 3187}

\vspace*{5mm}

\hspace*{5mm}
\begin{minipage}[b]{5.5cm}
\begin{picture}(3.0,3.0)
{\includegraphics[angle=180,viewport=40 50 400 730,clip,width=52mm]{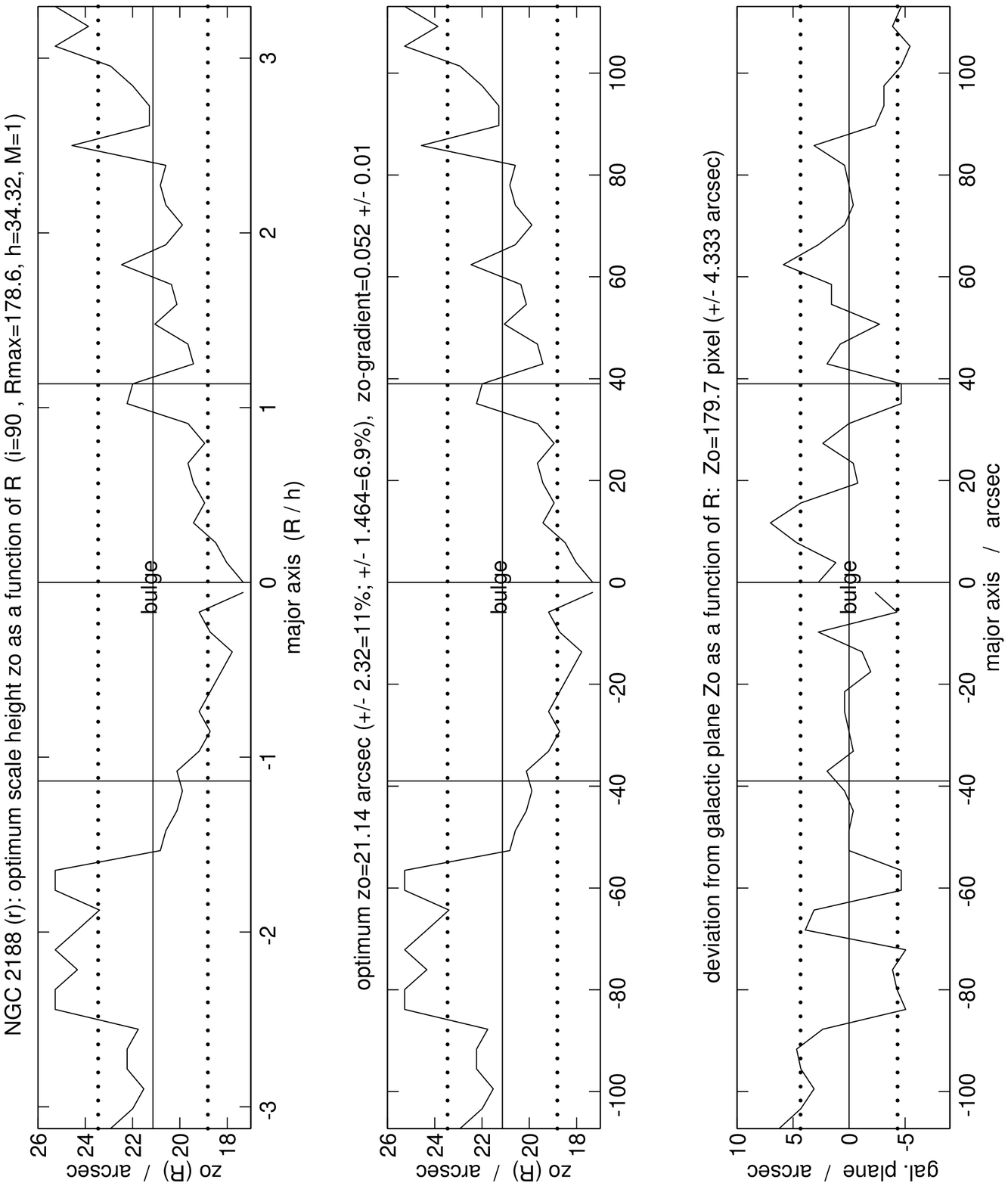}}
\end{picture}
\end{minipage}
\hfill
\begin{minipage}[b]{5.5cm}
\begin{picture}(3.0,3.0)
{\includegraphics[angle=180,viewport=00 -30 342 730,clip,width=50.5mm]{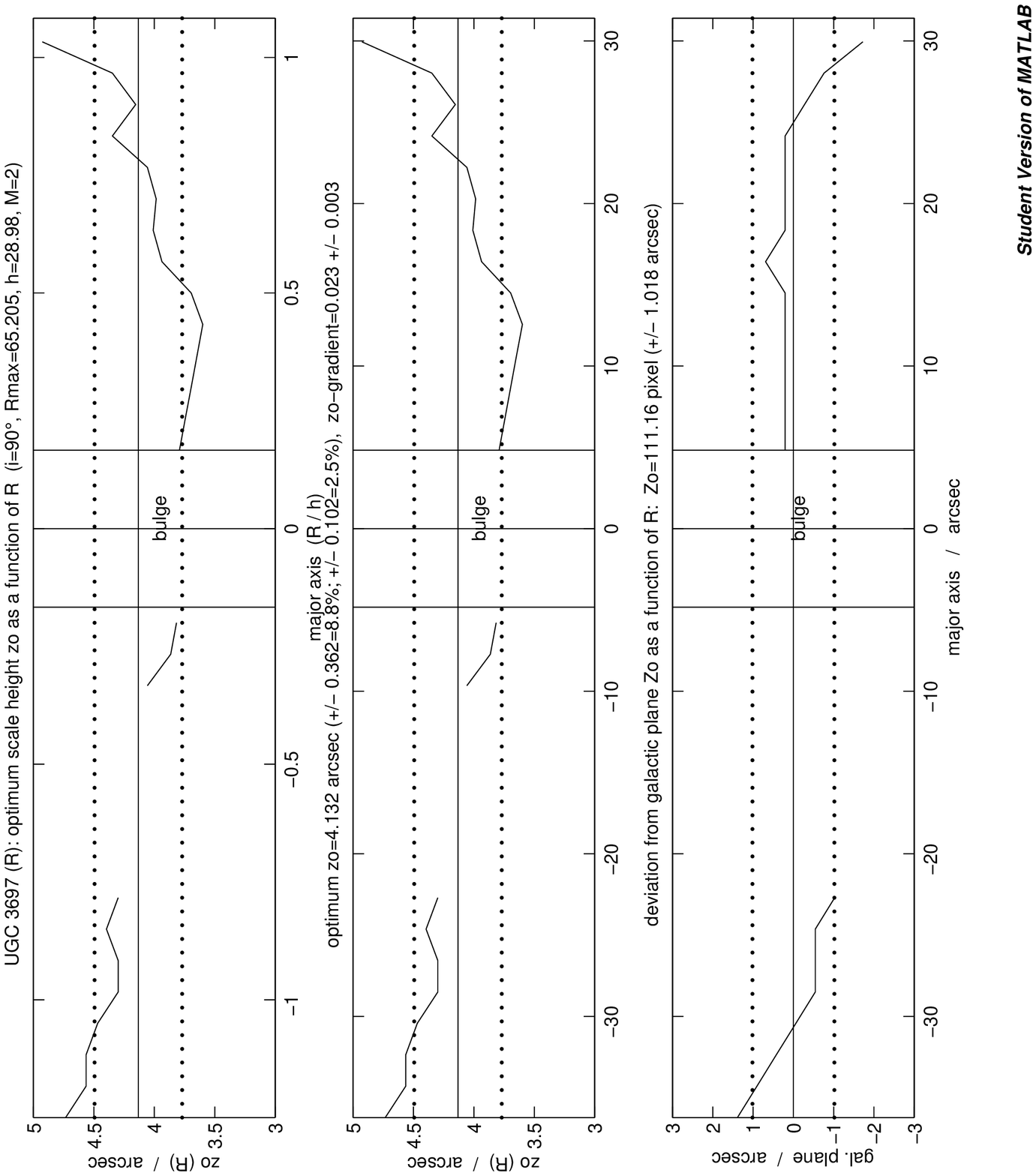}}
\end{picture}
\end{minipage}
\hfill
\begin{minipage}[b]{5.5cm}
\begin{picture}(3.0,3.0)
{\includegraphics[angle=180,viewport=40 50 400 730,clip,width=52mm]{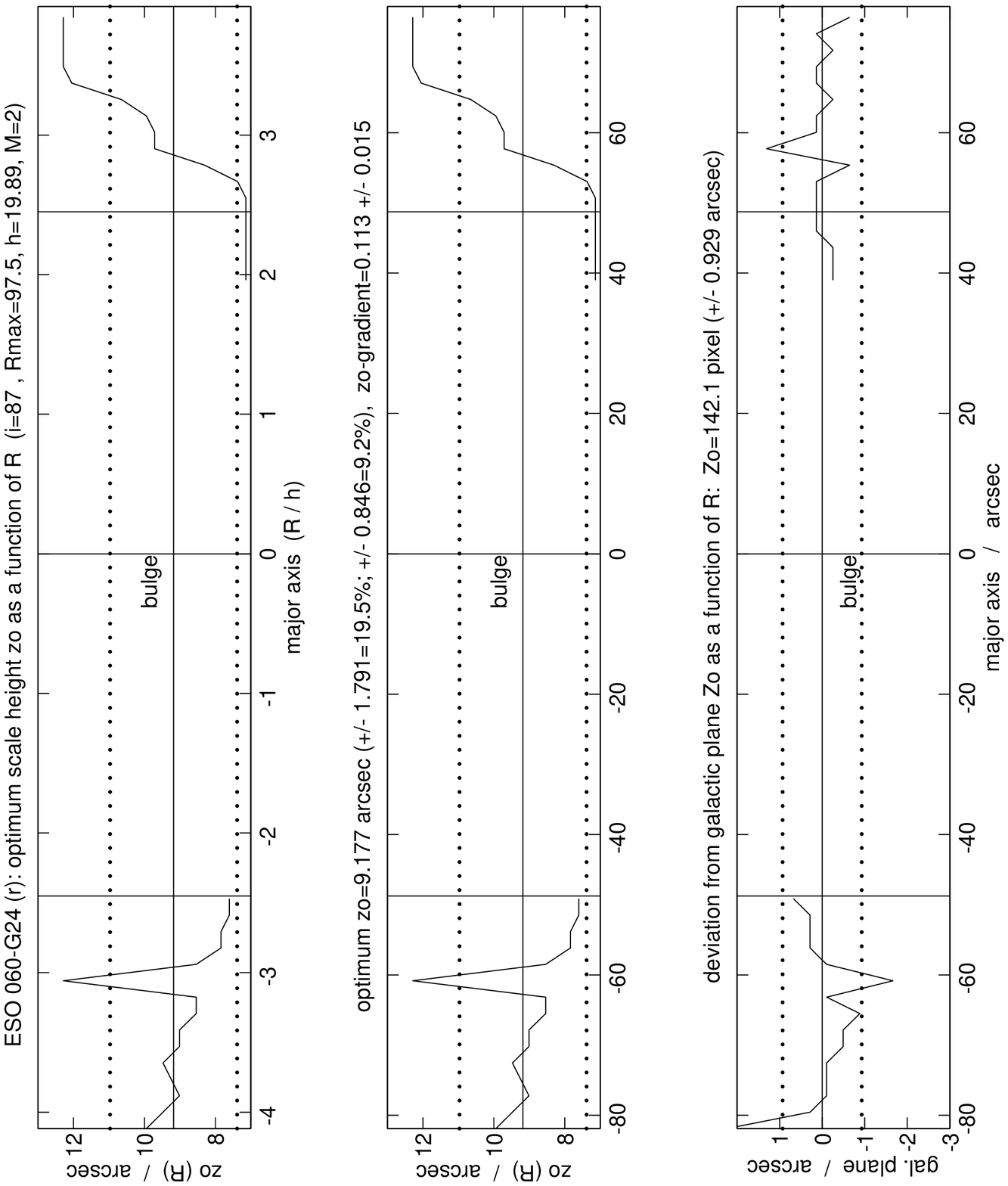}}
\end{picture}
\end{minipage}

\vspace*{98mm}

\hspace*{18mm}\parbox{165mm}{NGC 2188  \hspace{40mm}  UGC 3697  \hspace{46mm}  ESO 060-G24}

\vspace*{8mm}

\hspace*{8mm}\parbox{165mm}{
{\bf \noindent Appendix A.} (continued)
}
\end{figure*}

%%%%%%%%%%%%%%%%%%%%%%%%%%%%%%%%%%%%%%%%%%%%%%%%%%%%%%%%%%%%%%%%%%%

\clearpage

%%%%%%%%%%%%%%%%%%%%%%%%%%%%%  5  %%%%%%%%%%%%%%%%%%%%%%%%%%%%%%%%%

\begin{figure*}[t]
\vspace*{3mm}
\hspace*{5mm}
\begin{minipage}[b]{5.5cm}
\begin{picture}(3.0,3.0)
{\includegraphics[angle=180,viewport=40 50 400 730,clip,width=52mm]{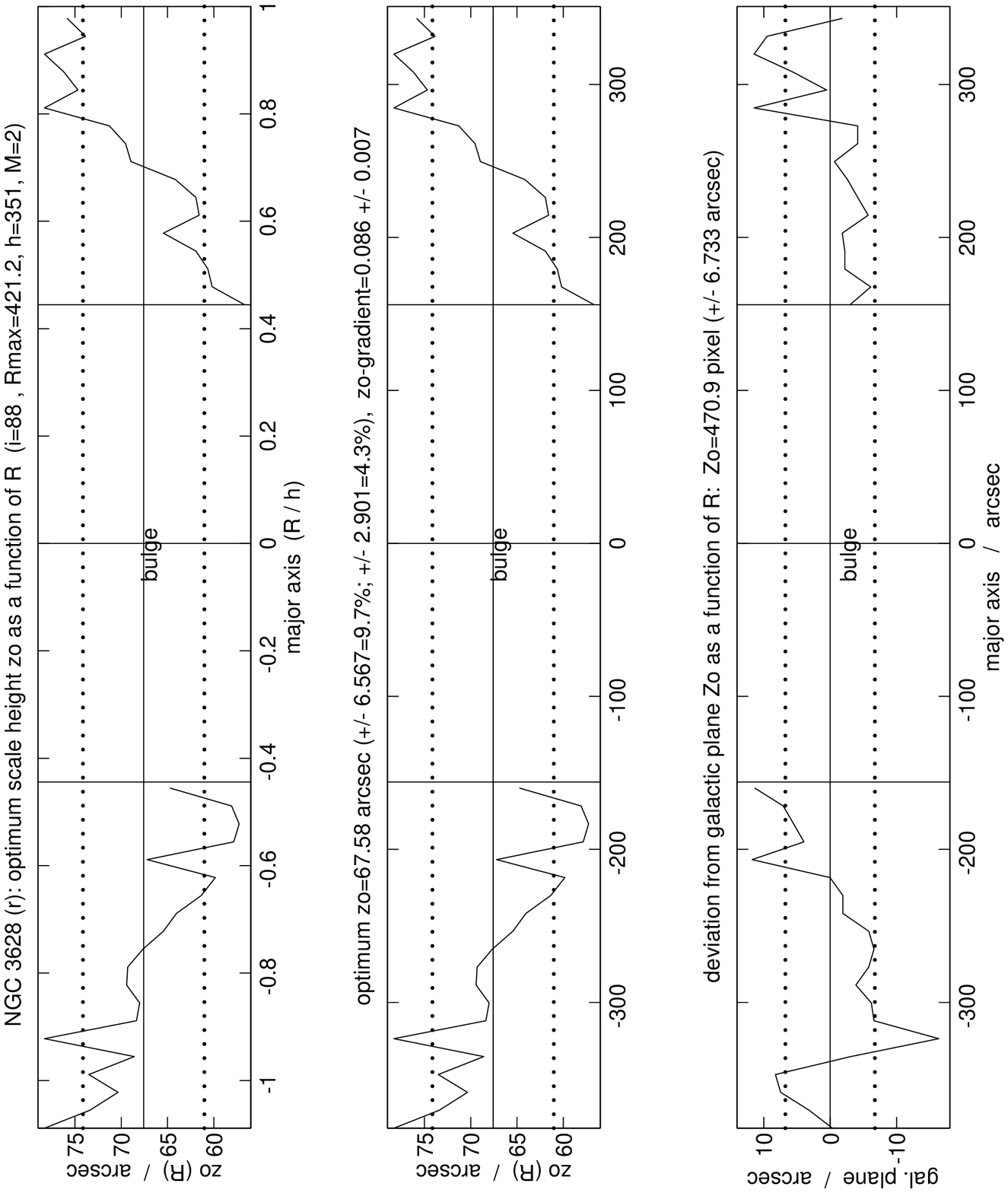}}
\end{picture}
\end{minipage}
\hfill
\begin{minipage}[b]{5.5cm}
\begin{picture}(3.0,3.0)
{\includegraphics[angle=180,viewport=40 50 400 730,clip,width=52mm]{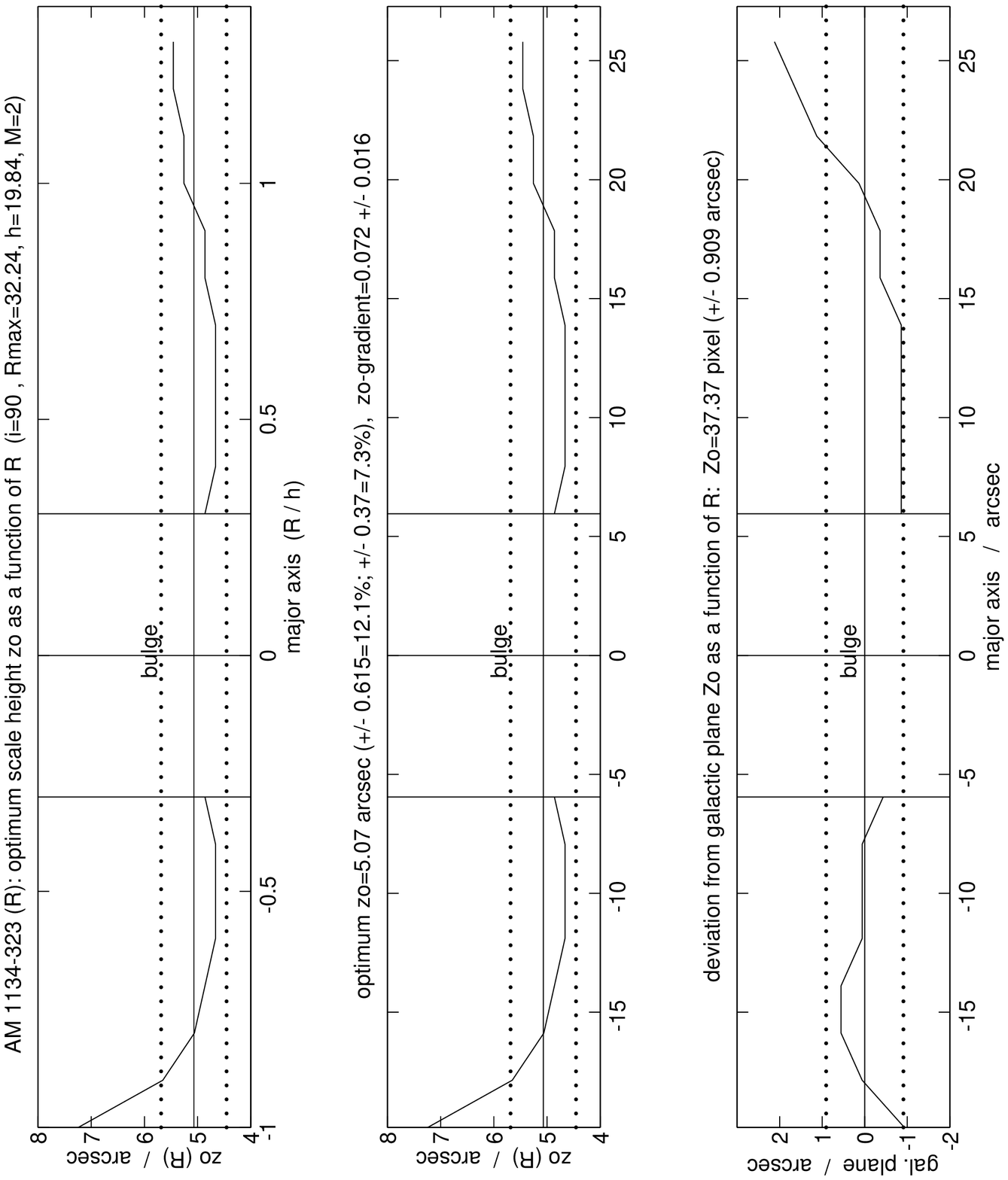}}
\end{picture}
\end{minipage}
\hfill
\begin{minipage}[b]{5.5cm}
\begin{picture}(3.0,3.0)
{\includegraphics[angle=180,viewport=40 50 400 730,clip,width=52mm]{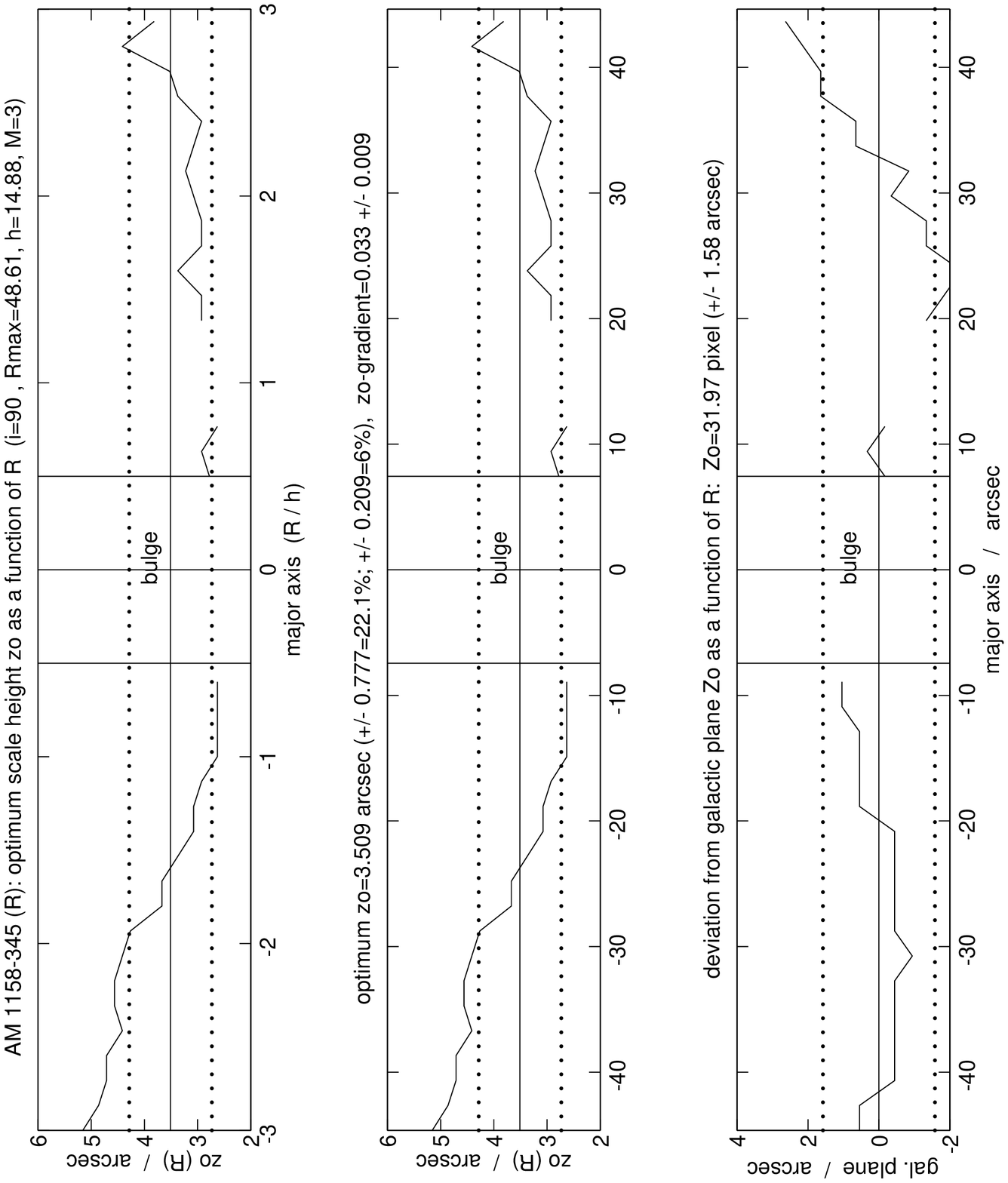}}
\end{picture}
\end{minipage}

\vspace*{98mm}

\hspace*{18mm}\parbox{165mm}{NGC 3628  \hspace{40mm}  ESO 378-G13  \hspace{40mm}  ESO 379-G20}

\vspace*{5mm}

\hspace*{5mm}
\begin{minipage}[b]{5.5cm}
\begin{picture}(3.0,3.0)
{\includegraphics[angle=180,viewport=40 50 400 730,clip,width=52mm]{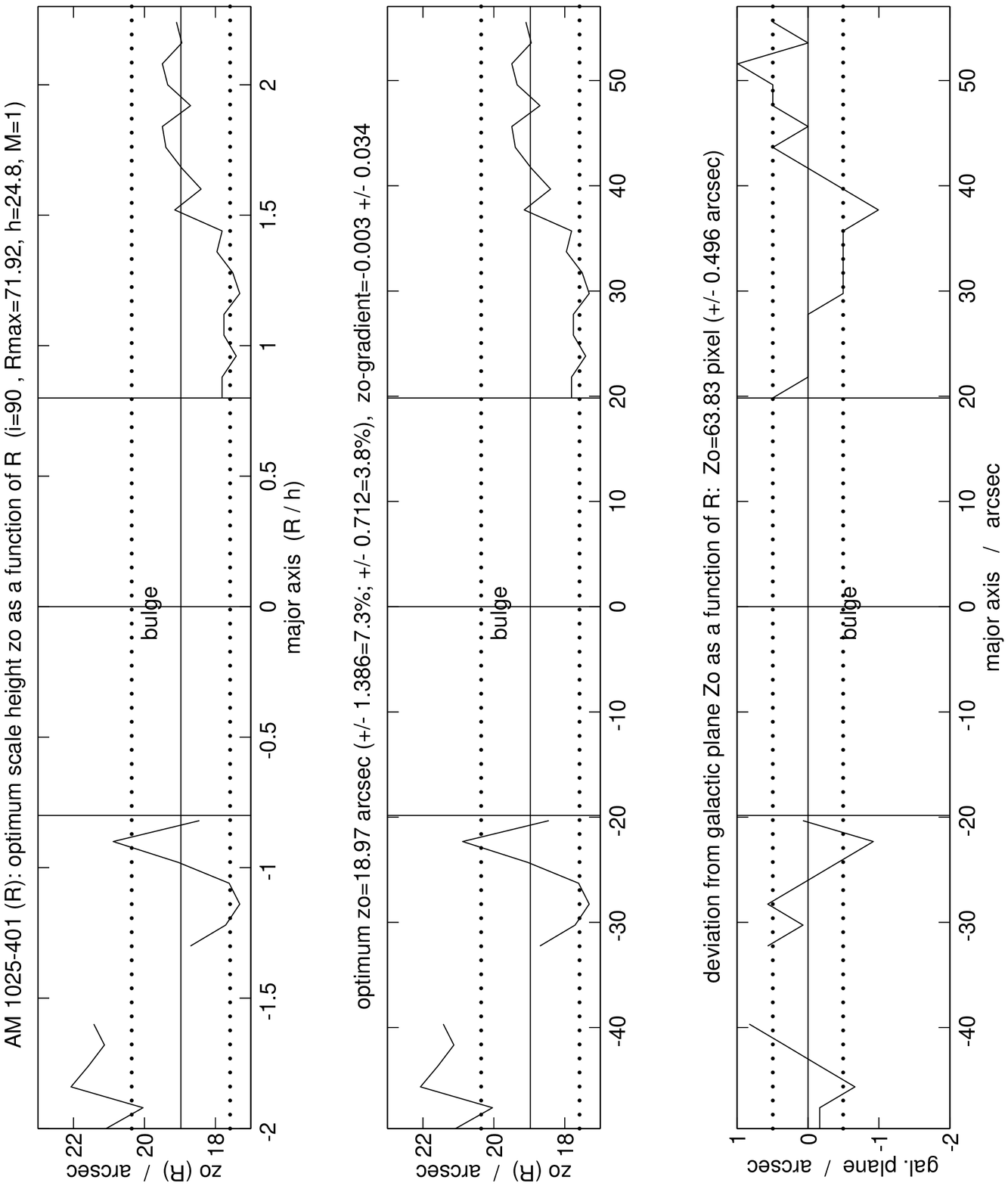}}
\end{picture}
\end{minipage}
\hfill
\begin{minipage}[b]{5.5cm}
\begin{picture}(3.0,3.0)
{\includegraphics[angle=180,viewport=40 50 400 730,clip,width=52mm]{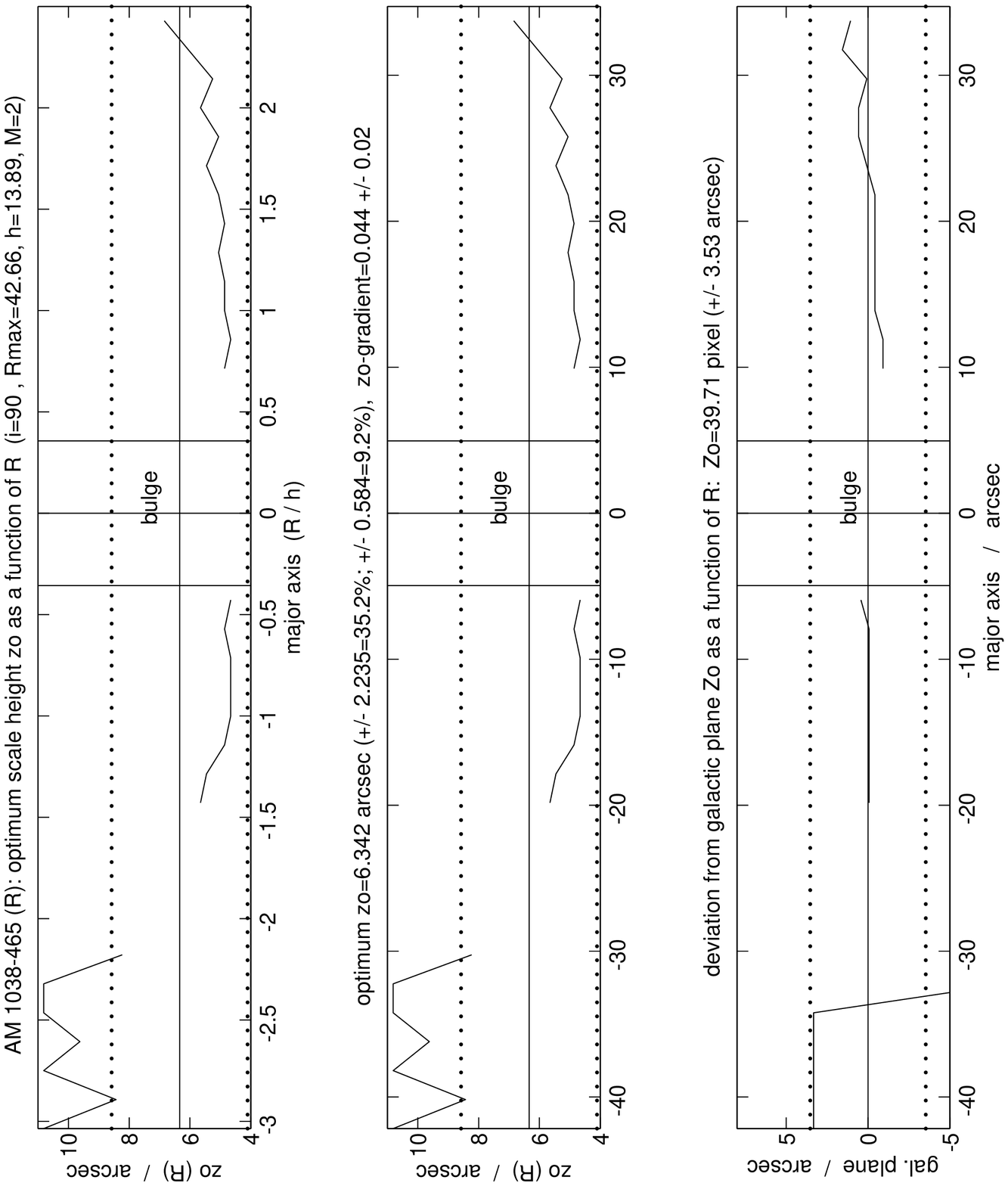}}
\end{picture}
\end{minipage}
\hfill
\begin{minipage}[b]{5.5cm}
\begin{picture}(3.0,3.0)
{\includegraphics[angle=180,viewport=40 50 400 730,clip,width=52mm]{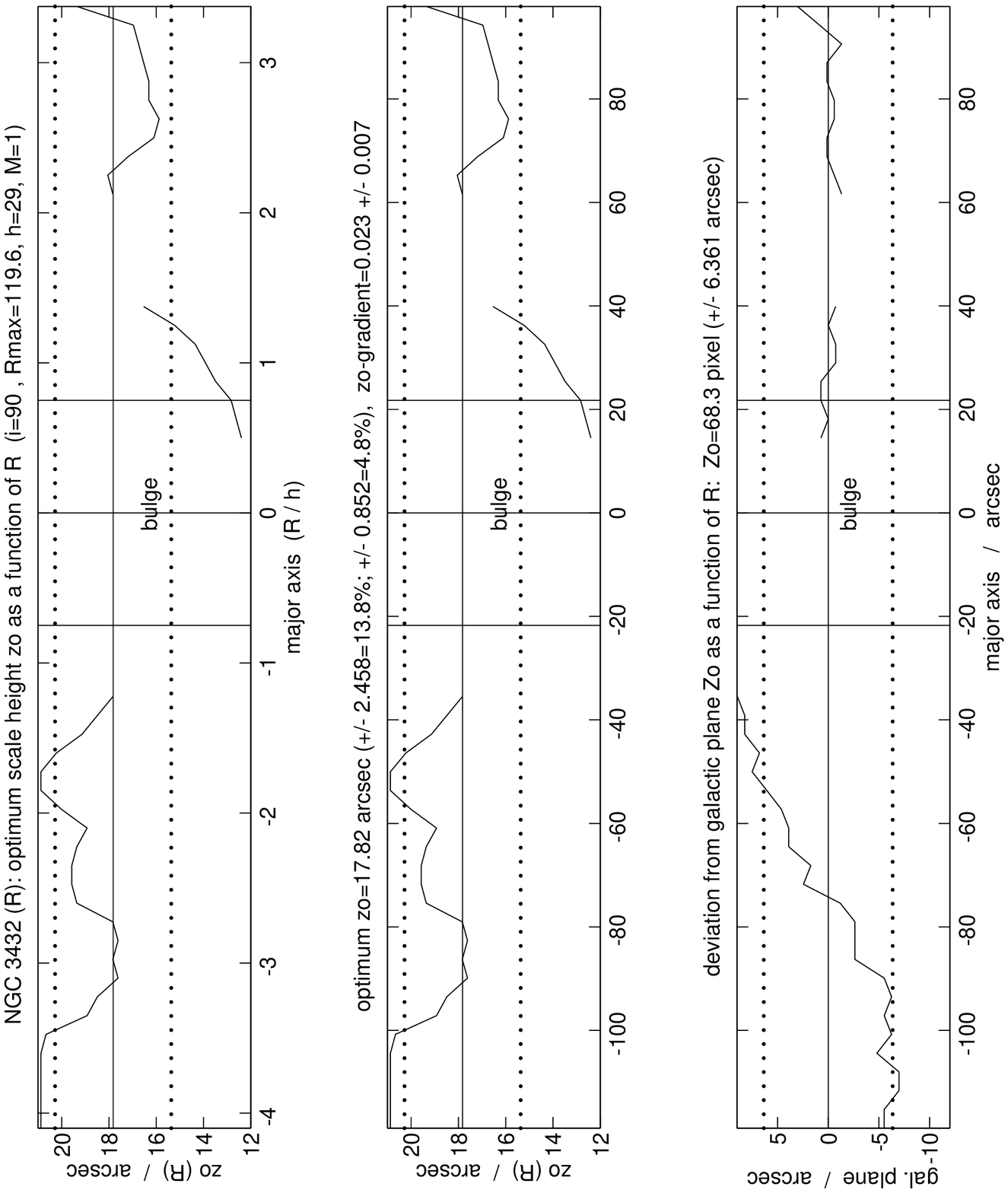}}
\end{picture}
\end{minipage}

\vspace*{98mm}

\hspace*{18mm}\parbox{165mm}{ESO 317-G29  \hspace{40mm}  ESO 264-G29  \hspace{40mm}  NGC 3432}

\vspace*{8mm}

\hspace*{8mm}\parbox{165mm}{
{\bf \noindent Appendix A.} (continued)
}
\end{figure*}

%%%%%%%%%%%%%%%%%%%%%%%%%%%%%%%%%%%%%%%%%%%%%%%%%%%%%%%%%%%%%%%%%%%

\clearpage

%%%%%%%%%%%%%%%%%%%%%%%%%%%%%  6  %%%%%%%%%%%%%%%%%%%%%%%%%%%%%%%%%

\begin{figure*}[t]
\vspace*{3mm}
\hspace*{5mm}
\begin{minipage}[b]{5.5cm}
\begin{picture}(3.0,3.0)
{\includegraphics[angle=180,viewport=40 50 400 730,clip,width=52mm]{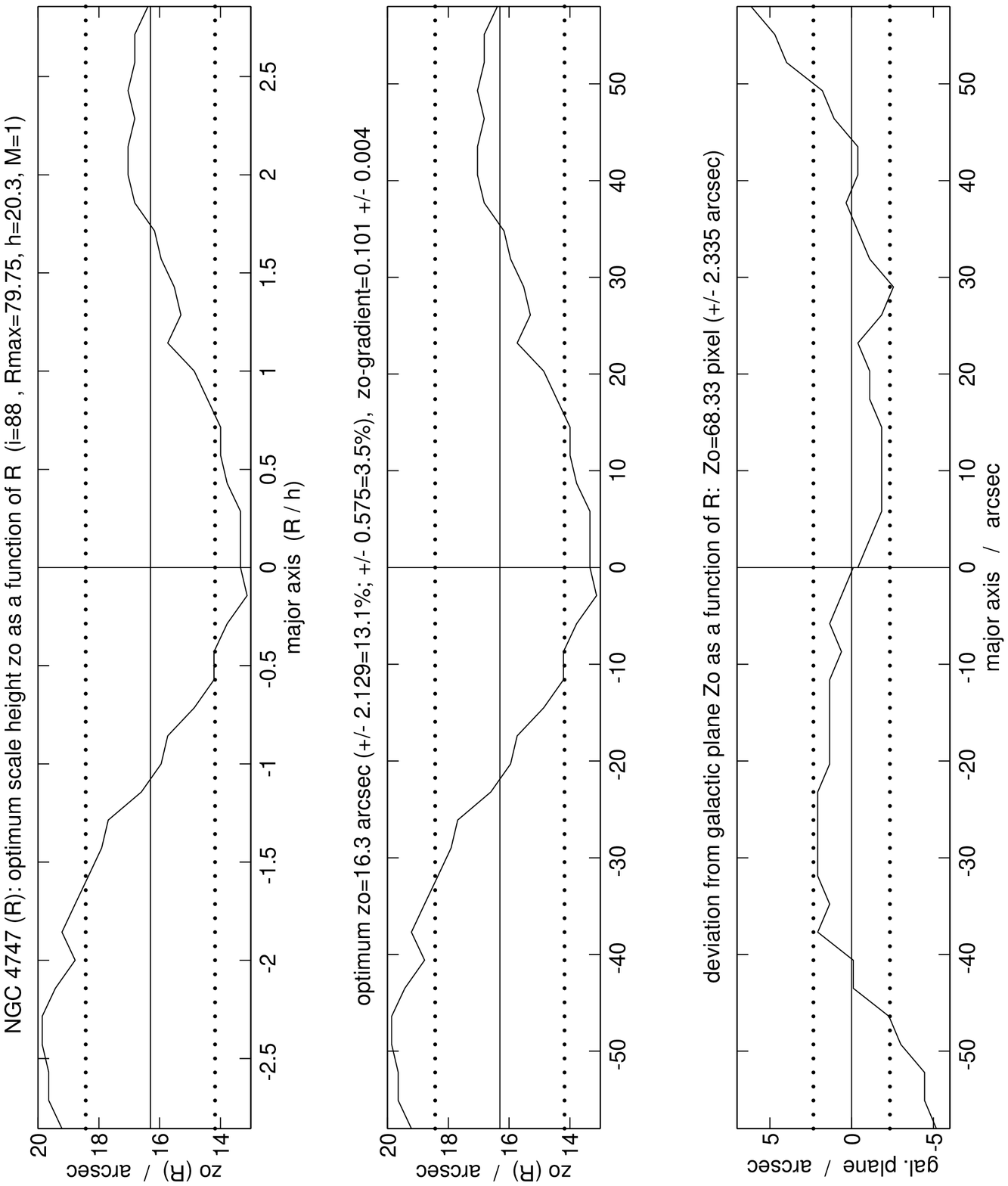}}
\end{picture}
\end{minipage}
\hfill
\begin{minipage}[b]{5.5cm}
\begin{picture}(3.0,3.0)
{\includegraphics[angle=180,viewport=00 -30 342 730,clip,width=50.5mm]{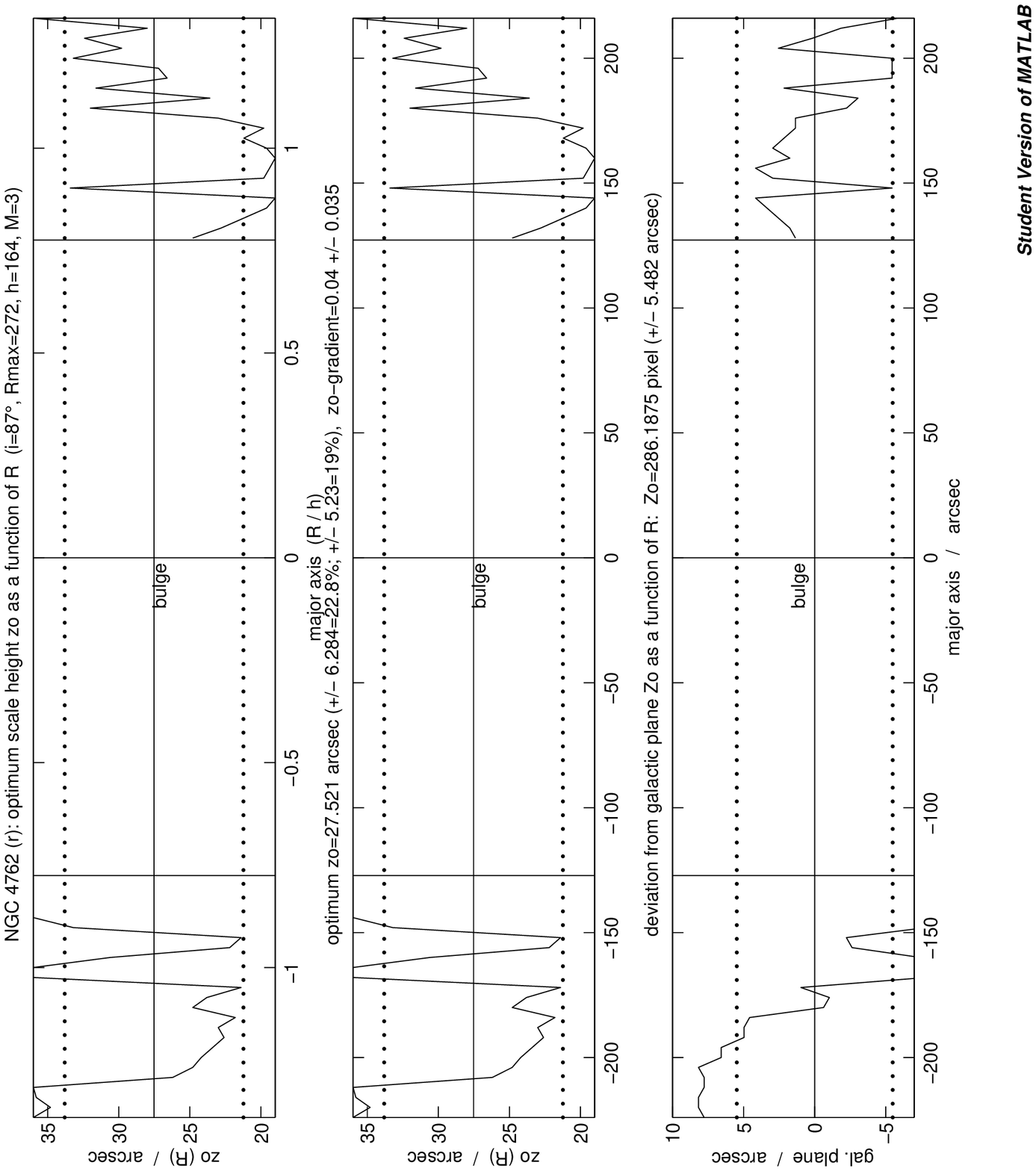}}
\end{picture}
\end{minipage}
\hfill
\begin{minipage}[b]{5.5cm}
\begin{picture}(3.0,3.0)
{\includegraphics[angle=180,viewport=40 50 400 730,clip,width=52mm]{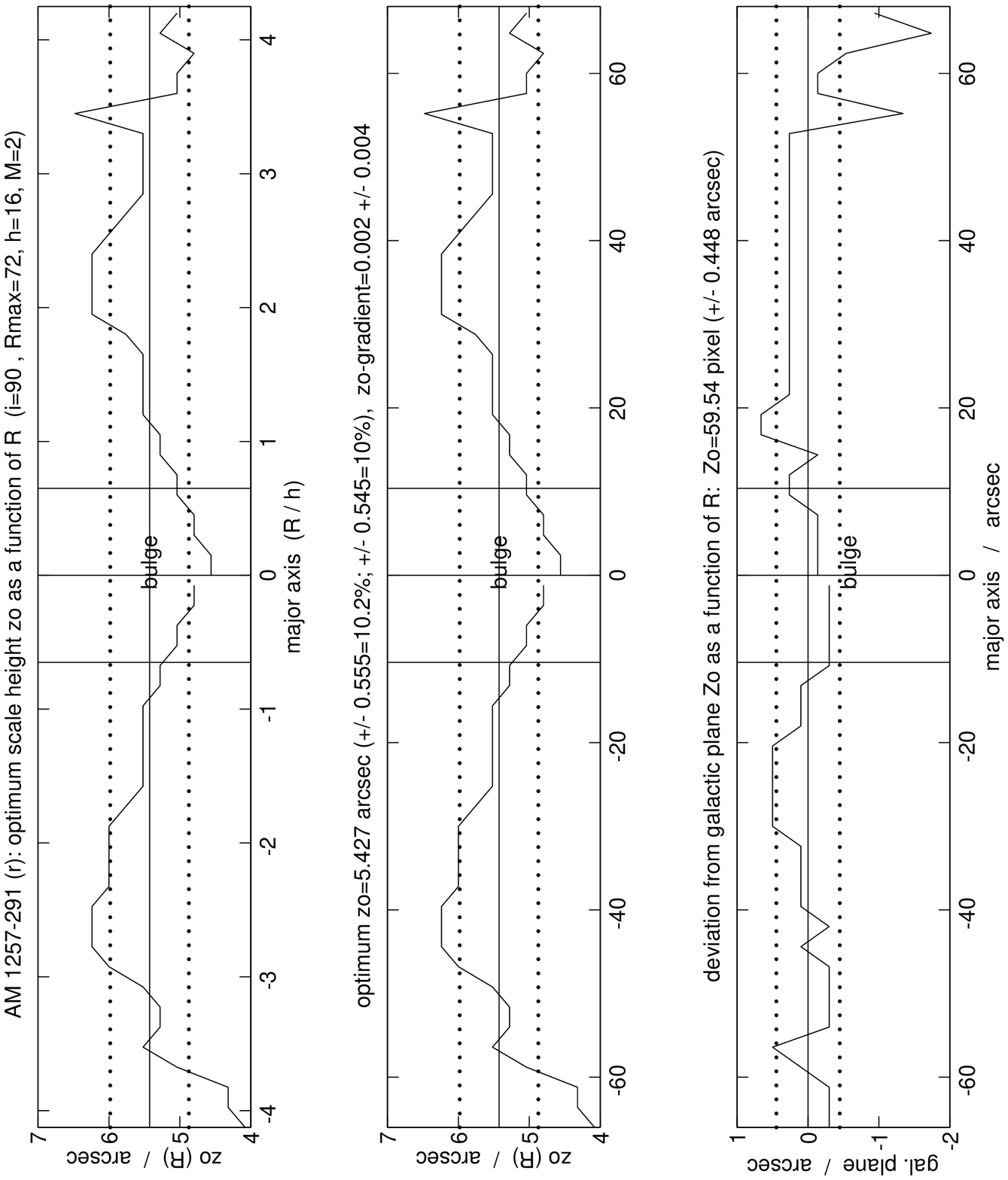}}
\end{picture}
\end{minipage}

\vspace*{98mm}

\hspace*{18mm}\parbox{165mm}{NGC 4747  \hspace{40mm}  NGC 4762  \hspace{44mm}  ESO 443-G21}

\vspace*{5mm}

\hspace*{5mm}
\begin{minipage}[b]{5.5cm}
\begin{picture}(3.0,3.0)
{\includegraphics[angle=180,viewport=40 50 400 730,clip,width=52mm]{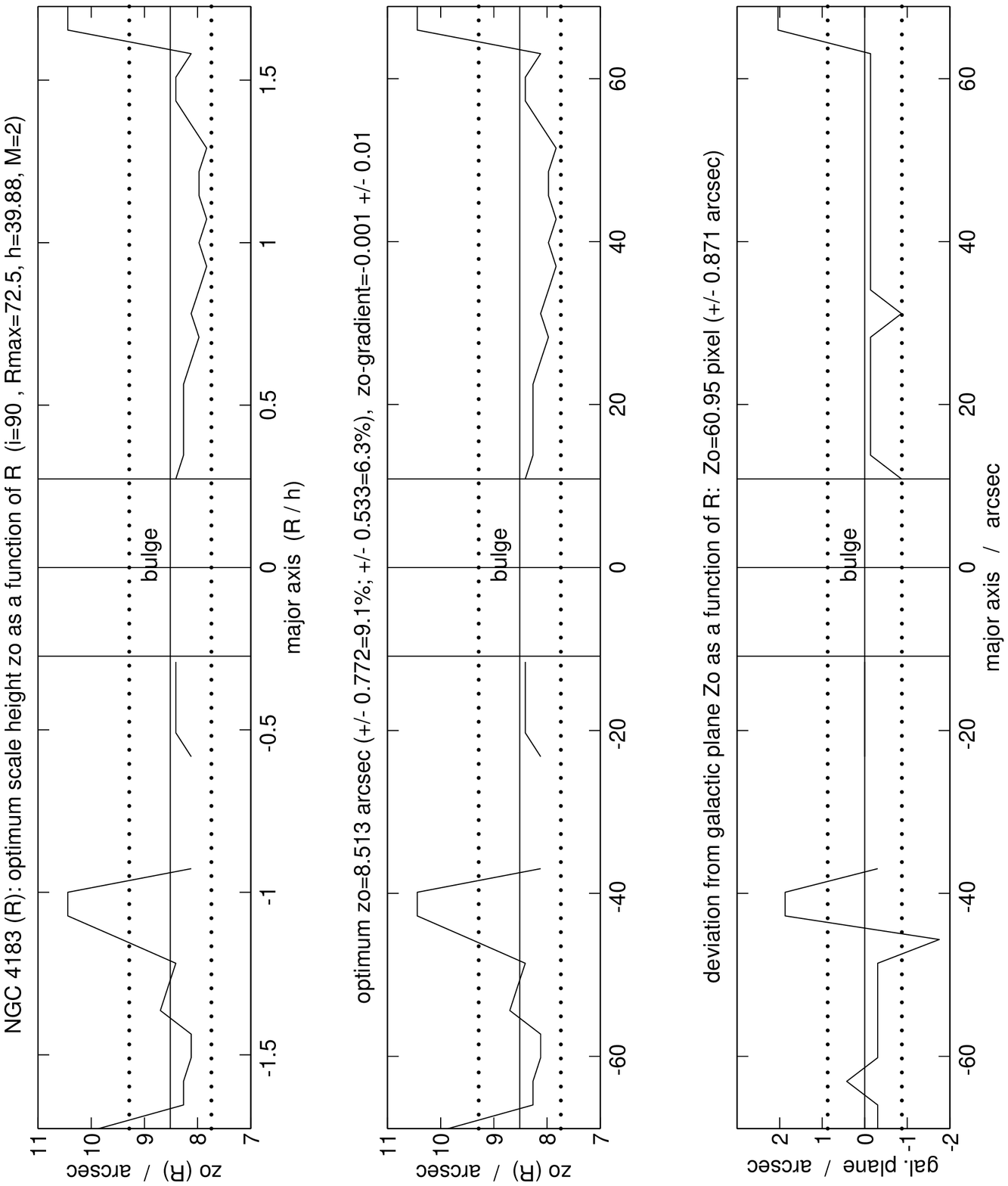}}
\end{picture}
\end{minipage}
\hfill
\begin{minipage}[b]{5.5cm}
\begin{picture}(3.0,3.0)
{\includegraphics[angle=180,viewport=40 50 400 730,clip,width=52mm]{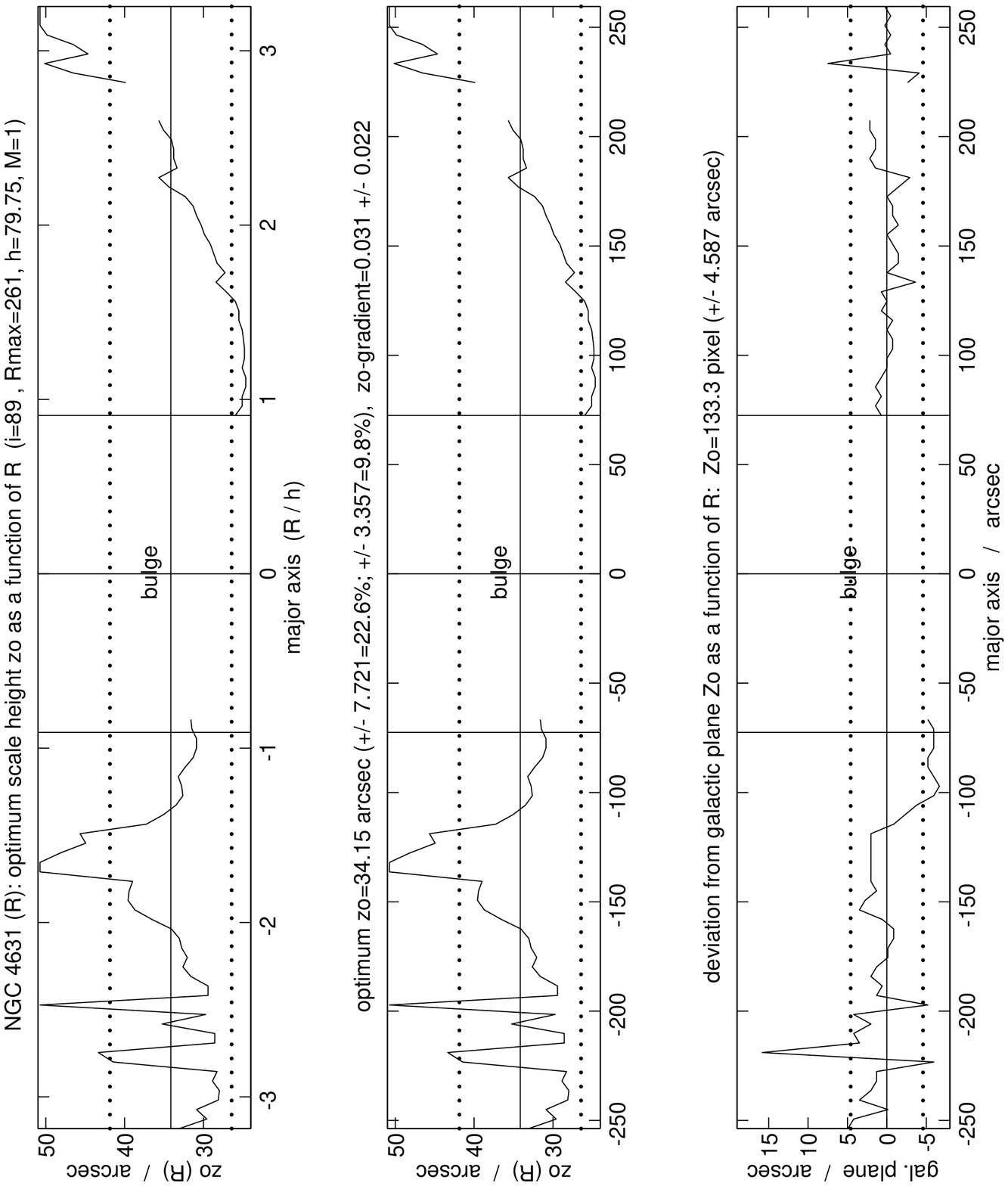}}
\end{picture}
\end{minipage}
\hfill
\begin{minipage}[b]{5.5cm}
\begin{picture}(3.0,3.0)
{\includegraphics[angle=180,viewport=40 50 400 730,clip,width=52mm]{ngc4634.ps}}
\end{picture}
\end{minipage}

\vspace*{98mm}

\hspace*{18mm}\parbox{165mm}{NGC 4183  \hspace{40mm}  NGC 4631  \hspace{48mm}  NGC 4634}

\vspace*{8mm}

\hspace*{8mm}\parbox{165mm}{
{\bf \noindent Appendix A.} (continued)
}
\end{figure*}

%%%%%%%%%%%%%%%%%%%%%%%%%%%%%%%%%%%%%%%%%%%%%%%%%%%%%%%%%%%%%%%%%%%

\clearpage

%%%%%%%%%%%%%%%%%%%%%%%%%%%%%  7  %%%%%%%%%%%%%%%%%%%%%%%%%%%%%%%%%

\begin{figure*}[t]
\vspace*{3mm}
\hspace*{5mm}
\begin{minipage}[b]{5.5cm}
\begin{picture}(3.0,3.0)
{\includegraphics[angle=180,viewport=40 50 400 730,clip,width=52mm]{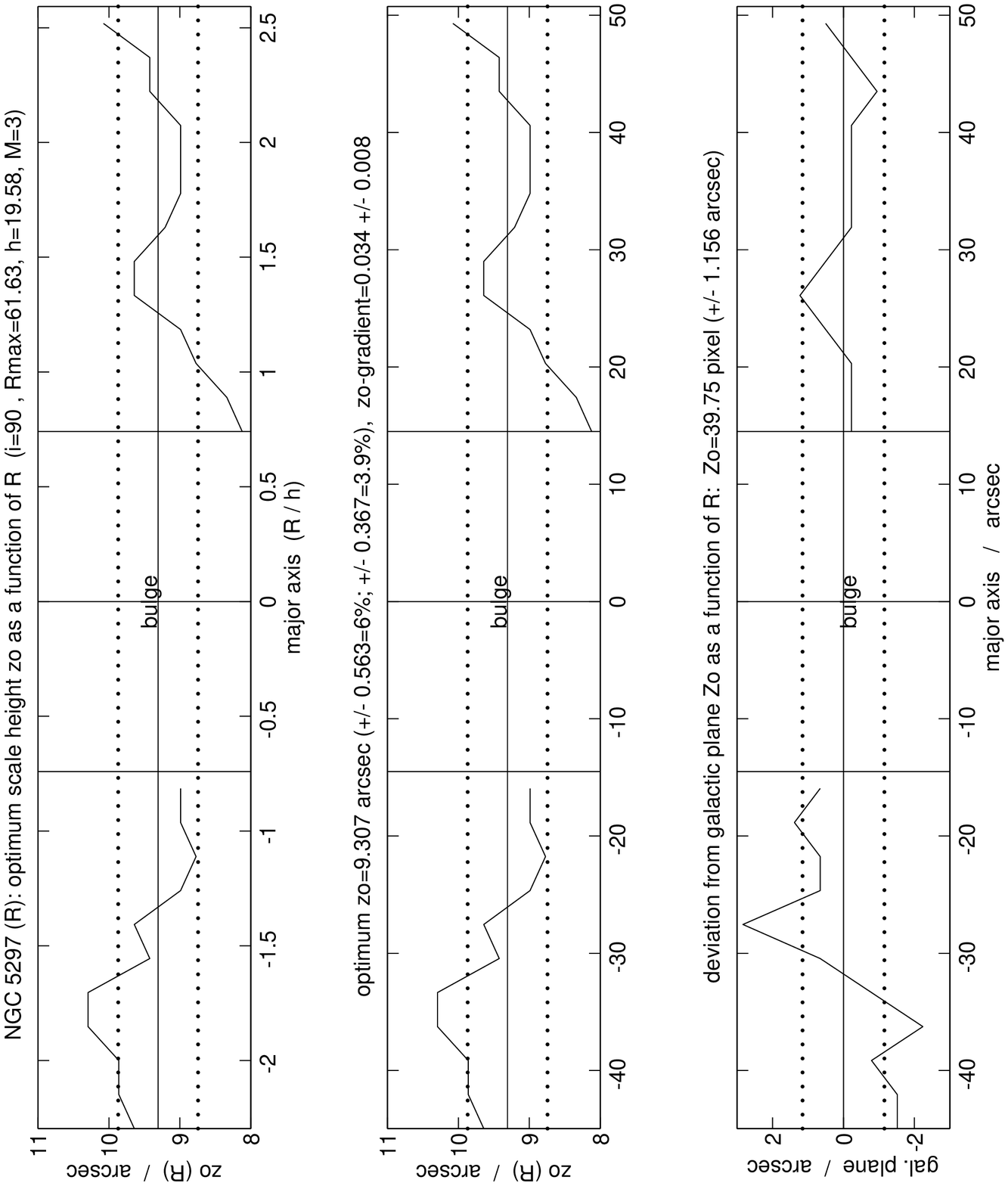}}
\end{picture}
\end{minipage}
\hfill
\begin{minipage}[b]{5.5cm}
\begin{picture}(3.0,3.0)
{\includegraphics[angle=180,viewport=40 50 400 730,clip,width=52mm]{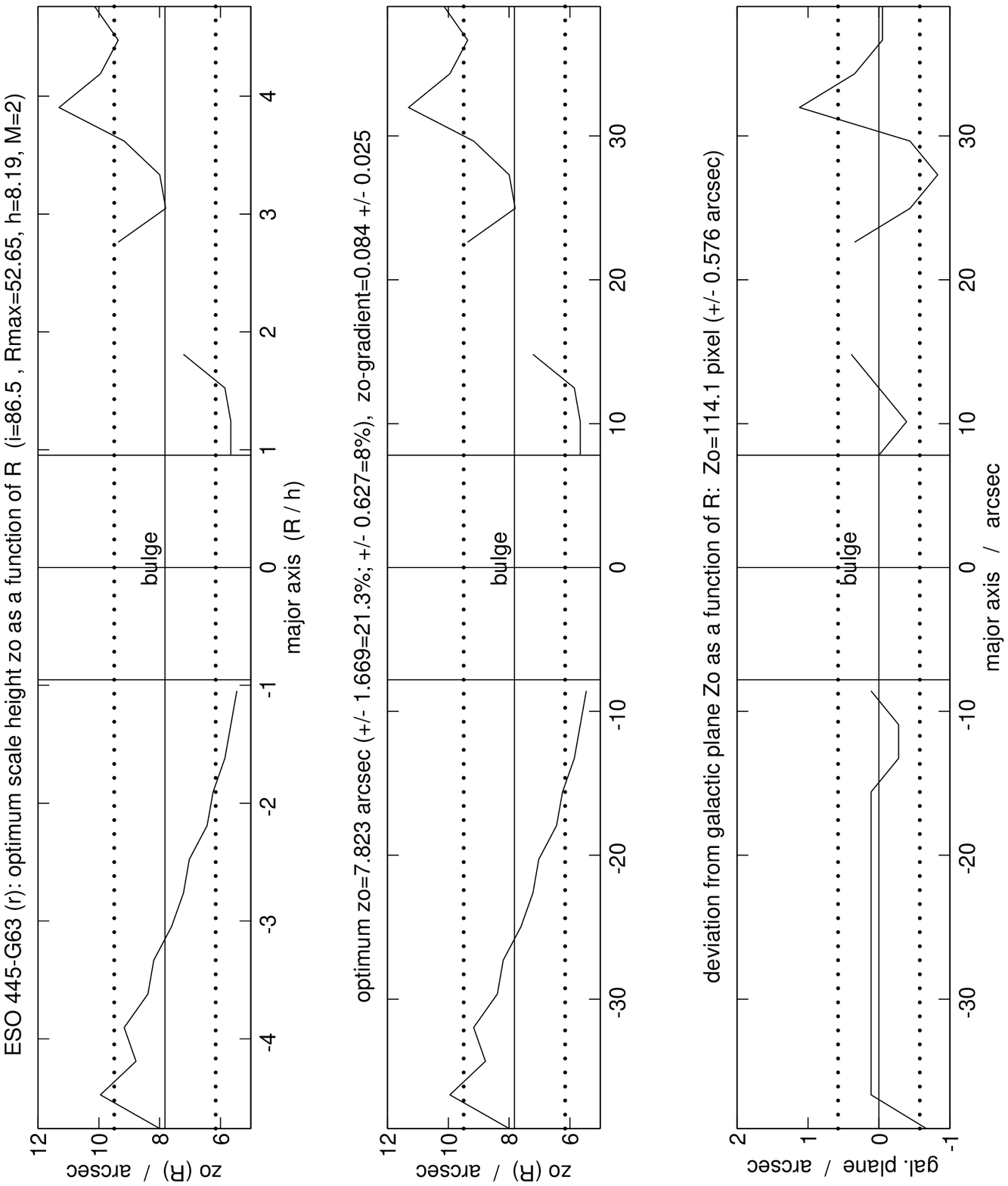}}
\end{picture}
\end{minipage}
\hfill
\begin{minipage}[b]{5.5cm}
\begin{picture}(3.0,3.0)
{\includegraphics[angle=180,viewport=00 -30 342 730,clip,width=50.5mm]{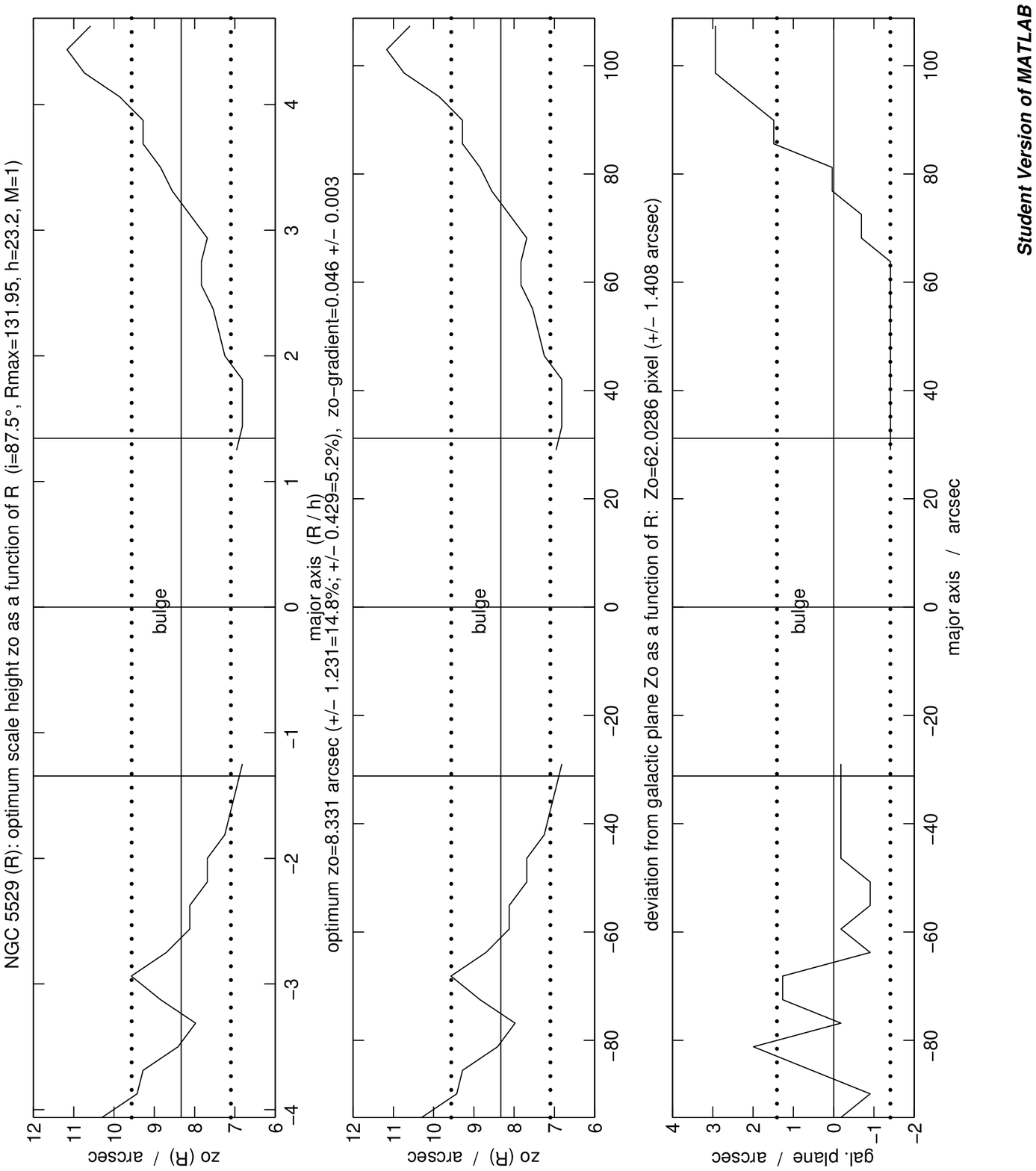}}
\end{picture}
\end{minipage}

\vspace*{98mm}

\hspace*{18mm}\parbox{165mm}{NGC 5297  \hspace{40mm}  ESO 445-G63  \hspace{40mm}  NGC 5529}

\vspace*{5mm}

\hspace*{5mm}
\begin{minipage}[b]{5.5cm}
\begin{picture}(3.0,3.0)
{\includegraphics[angle=180,viewport=40 50 400 730,clip,width=52mm]{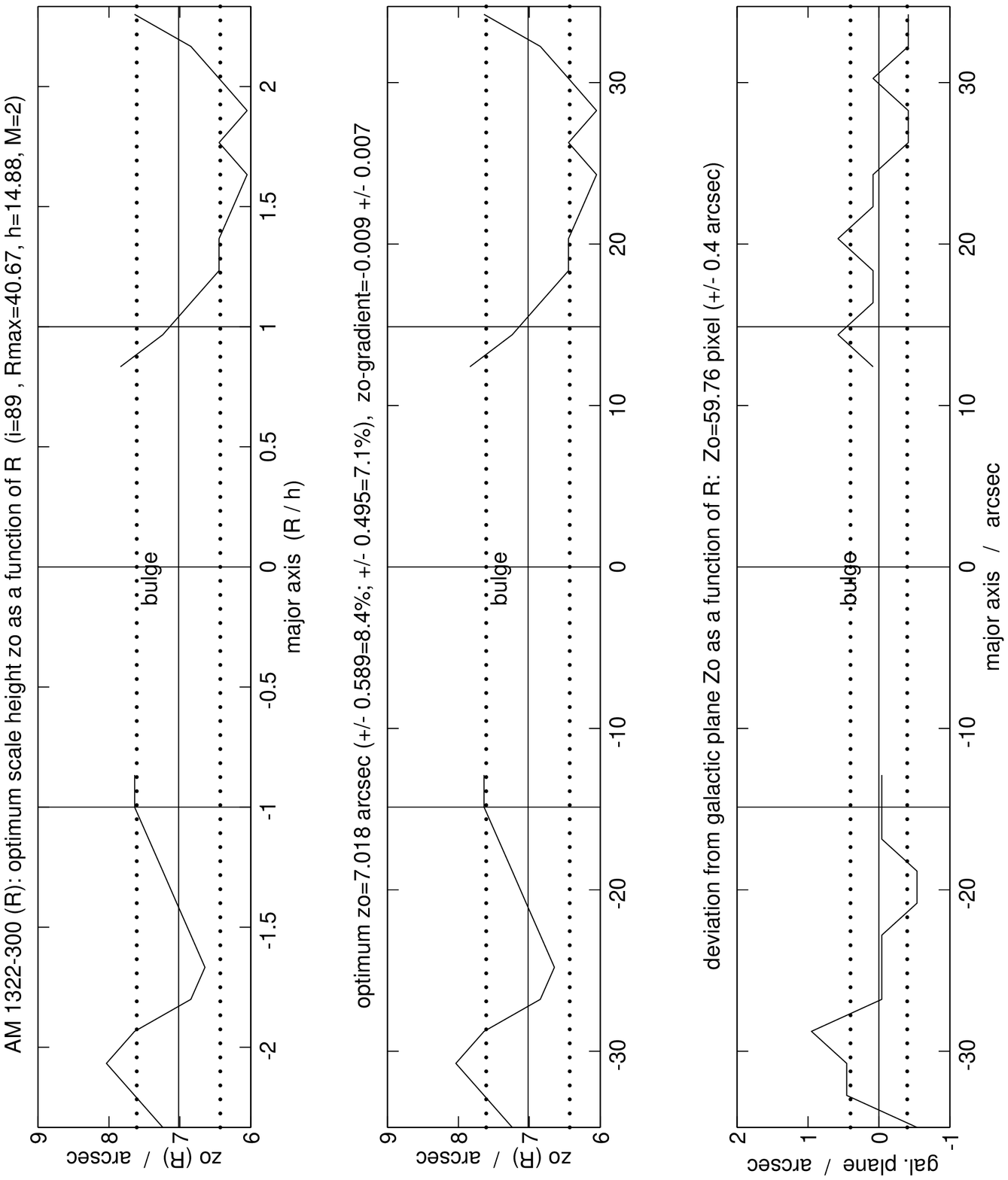}}
\end{picture}
\end{minipage}
\hfill
\begin{minipage}[b]{5.5cm}
\begin{picture}(3.0,3.0)
{\includegraphics[angle=180,viewport=40 50 400 730,clip,width=52mm]{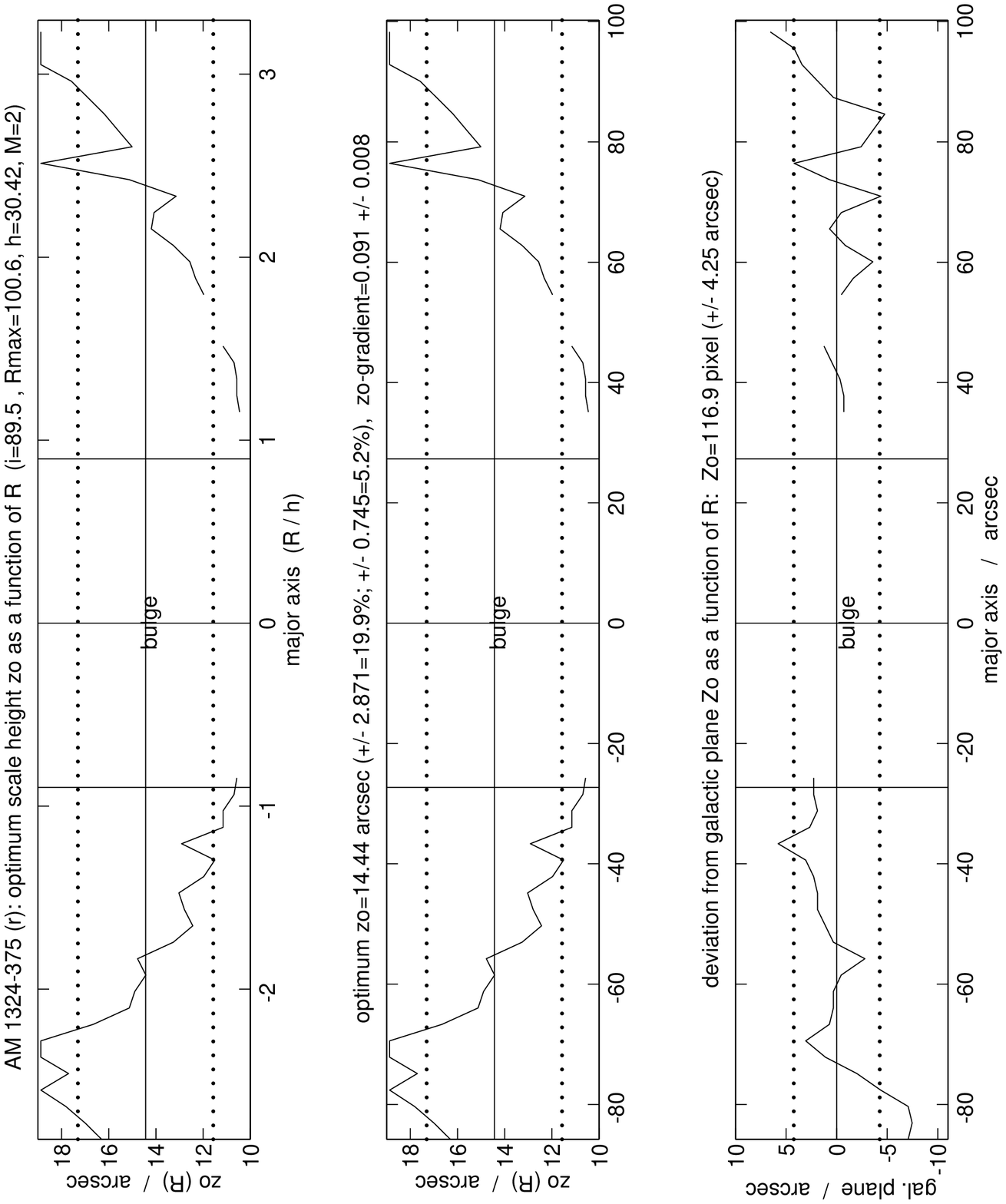}}
\end{picture}
\end{minipage}
\hfill
\begin{minipage}[b]{5.5cm}
\begin{picture}(3.0,3.0)
{\includegraphics[angle=180,viewport=40 50 400 730,clip,width=52mm]{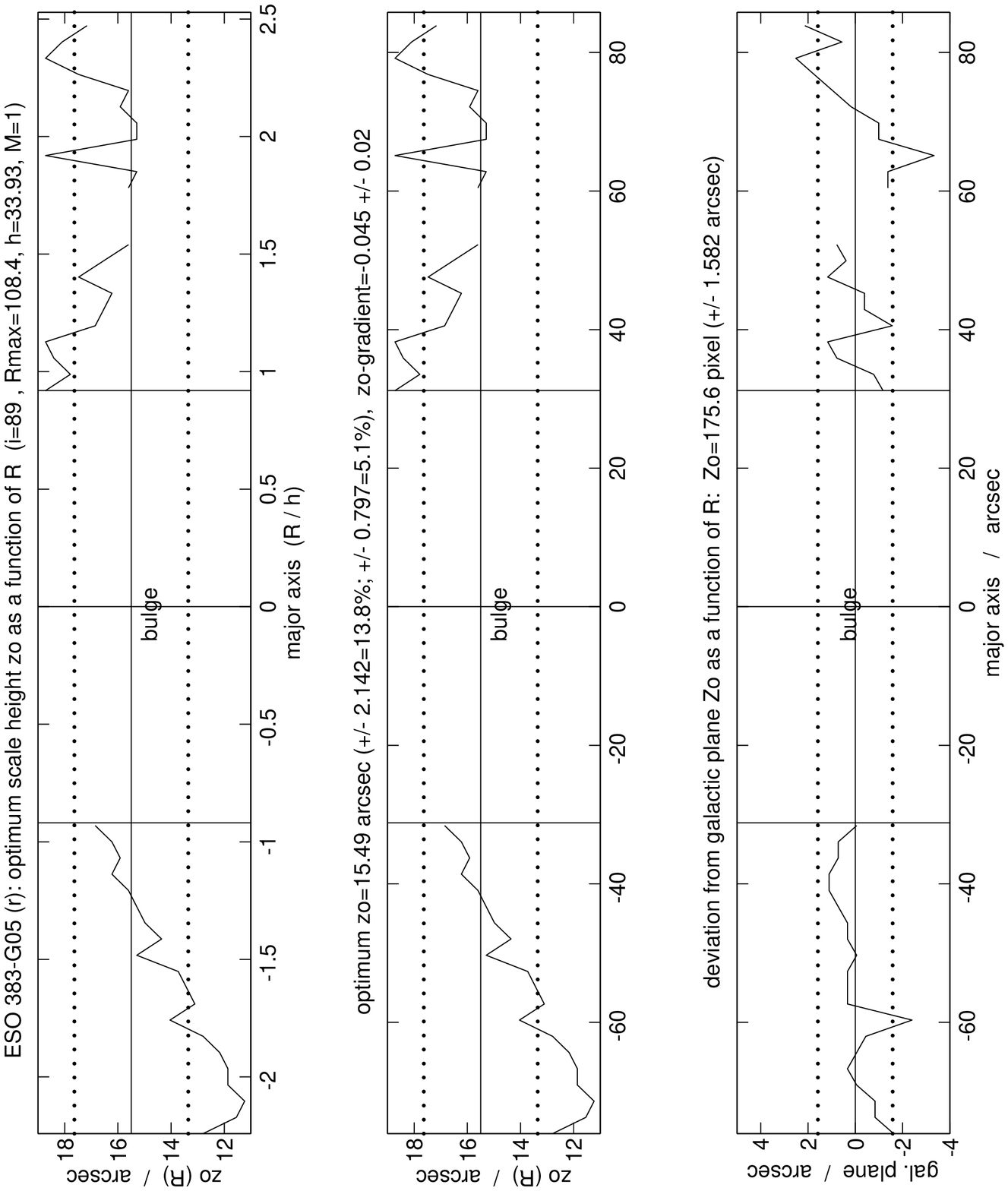}}
\end{picture}
\end{minipage}

\vspace*{98mm}

\hspace*{18mm}\parbox{165mm}{NGC 5126  \hspace{40mm}  ESO 324-G23  \hspace{40mm}  ESO 383-G05}

\vspace*{8mm}

\hspace*{8mm}\parbox{165mm}{
{\bf \noindent Appendix A.} (continued)
}
\end{figure*}

%%%%%%%%%%%%%%%%%%%%%%%%%%%%%%%%%%%%%%%%%%%%%%%%%%%%%%%%%%%%%%%%%%%

\clearpage

%%%%%%%%%%%%%%%%%%%%%%%%%%%%%  8  %%%%%%%%%%%%%%%%%%%%%%%%%%%%%%%%%

\begin{figure*}[t]
\vspace*{3mm}
\hspace*{5mm}
\begin{minipage}[b]{5.5cm}
\begin{picture}(3.0,3.0)
{\includegraphics[angle=180,viewport=40 50 400 730,clip,width=52mm]{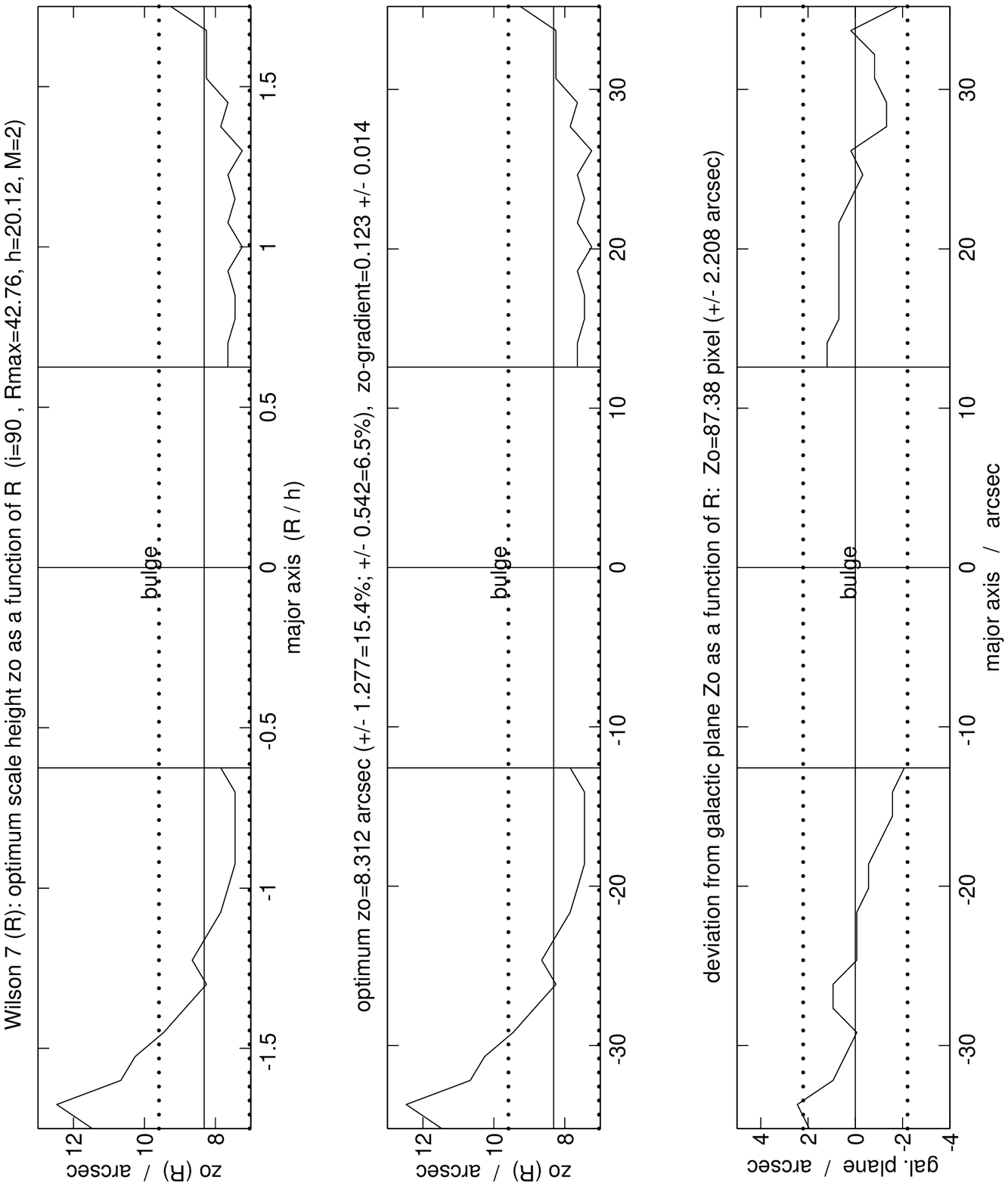}}
\end{picture}
\end{minipage}
\hfill
\begin{minipage}[b]{5.5cm}
\begin{picture}(3.0,3.0)
{\includegraphics[angle=180,viewport=40 50 400 730,clip,width=52mm]{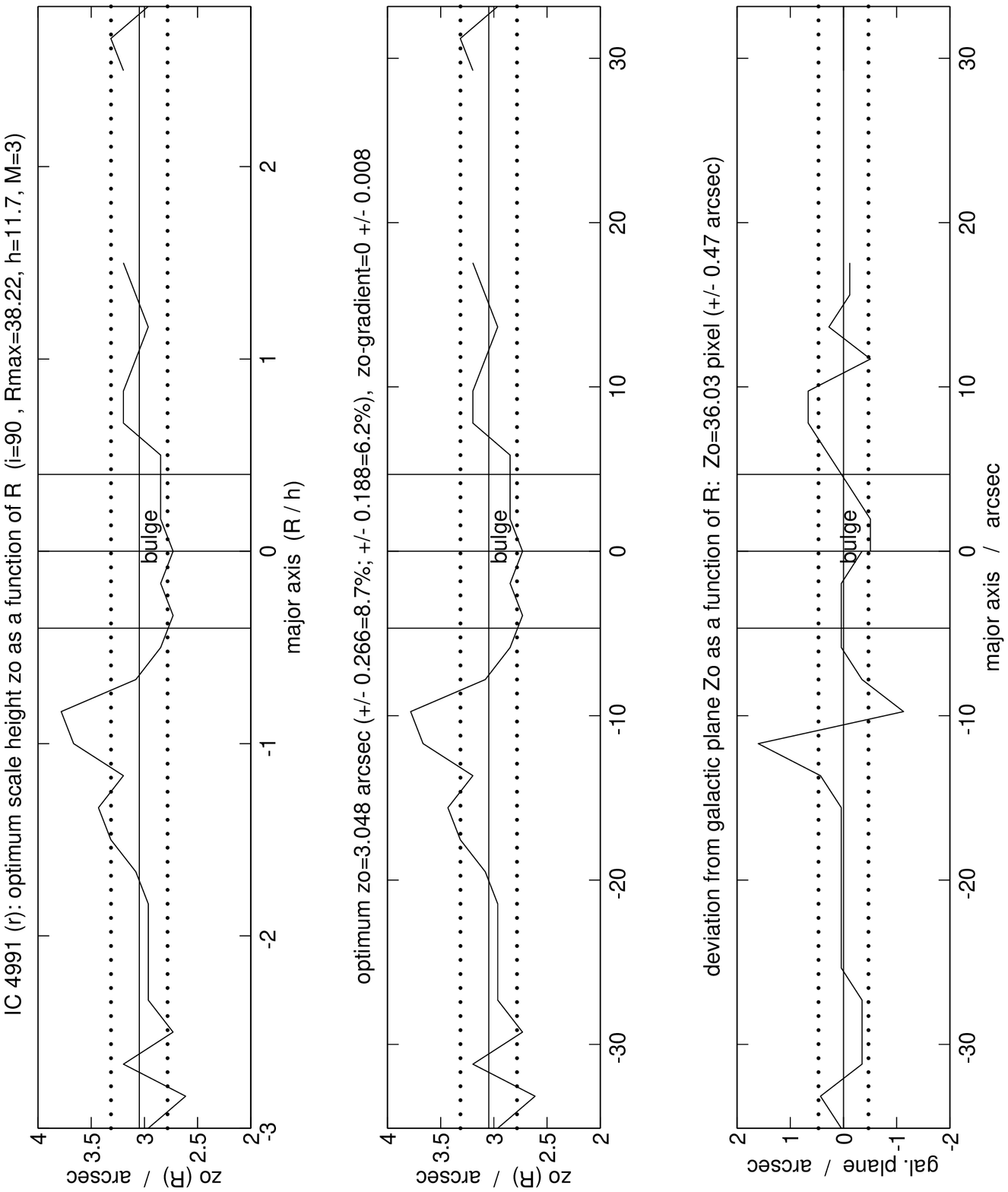}}
\end{picture}
\end{minipage}
\hfill
\begin{minipage}[b]{5.5cm}
\begin{picture}(3.0,3.0)
%{\includegraphics[angle=180,viewport=40 50 400 730,clip,width=52mm]{}}
\end{picture}
\end{minipage}

\vspace*{98mm}

\hspace*{22mm}\parbox{165mm}{Arp 121  \hspace{45mm}  IC 4991  \hspace{40mm}  }

\vspace*{5mm}

\hspace*{5mm}
\begin{minipage}[b]{5.5cm}
\begin{picture}(3.0,3.0)
{\includegraphics[angle=180,viewport=40 50 400 730,clip,width=52mm]{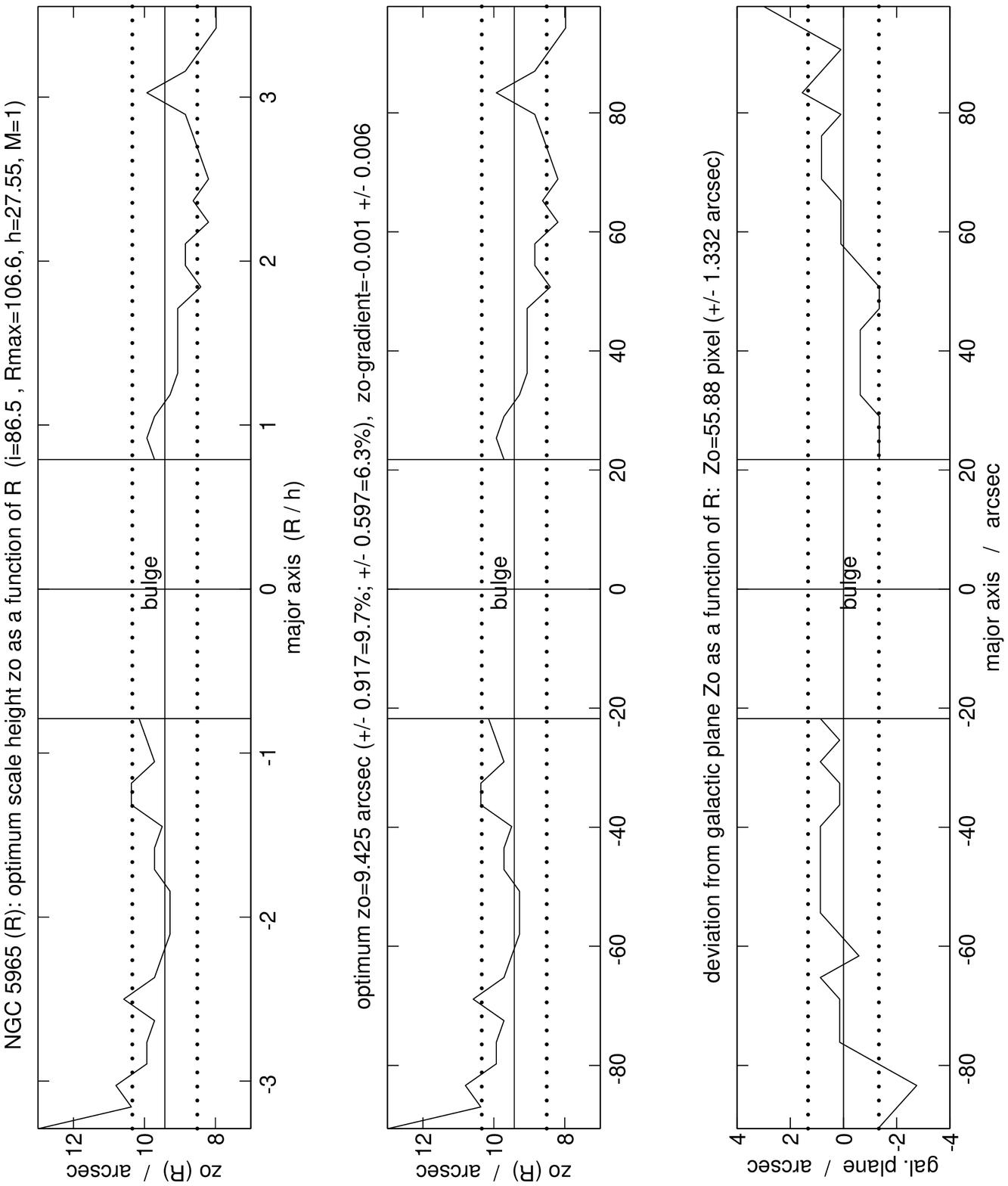}}
\end{picture}
\end{minipage}
\hfill
\begin{minipage}[b]{5.5cm}
\begin{picture}(3.0,3.0)
{\includegraphics[angle=180,viewport=40 50 400 730,clip,width=52mm]{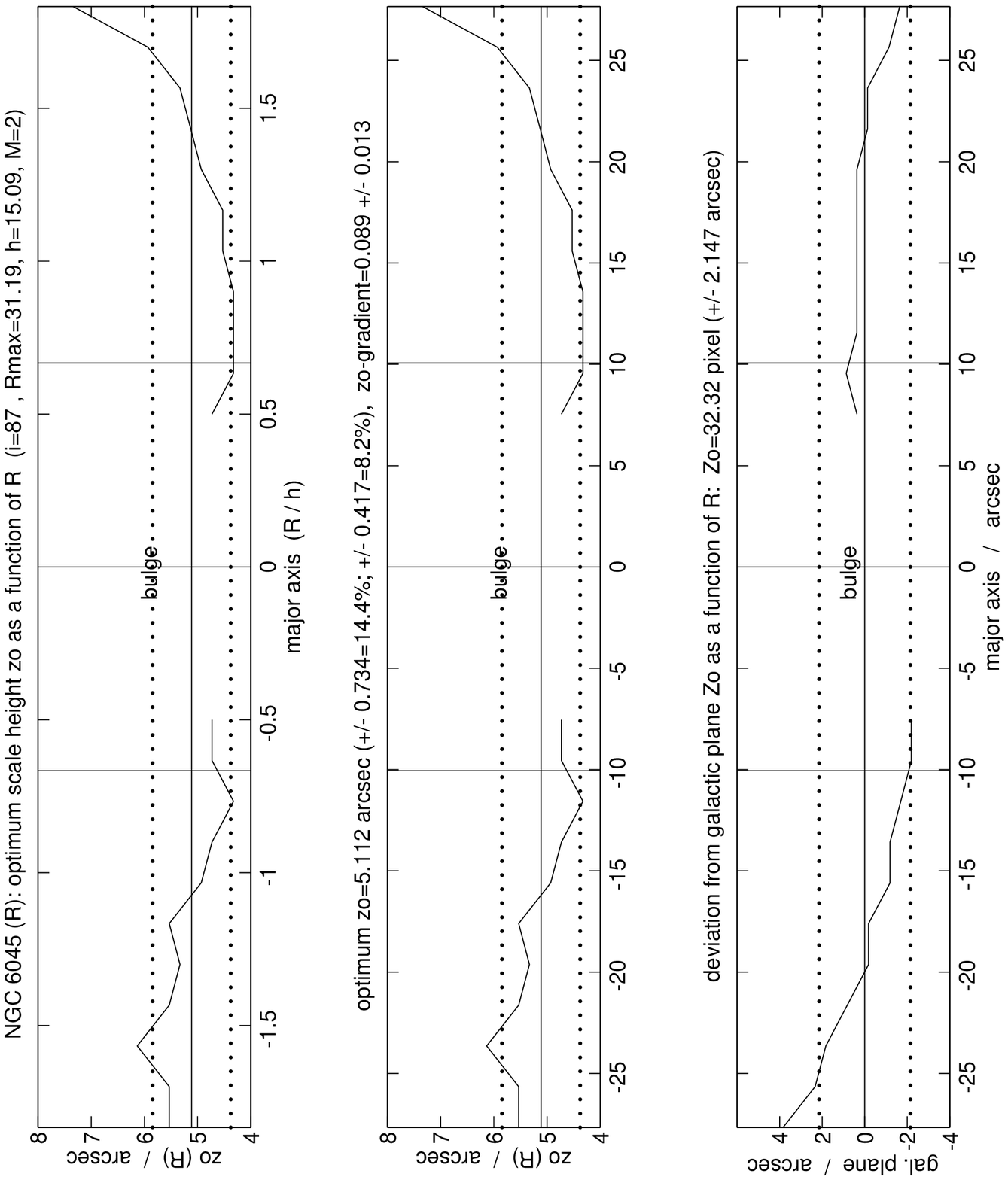}}
\end{picture}
\end{minipage}
\hfill
\begin{minipage}[b]{5.5cm}
\begin{picture}(3.0,3.0)
{\includegraphics[angle=180,viewport=40 50 400 730,clip,width=52mm]{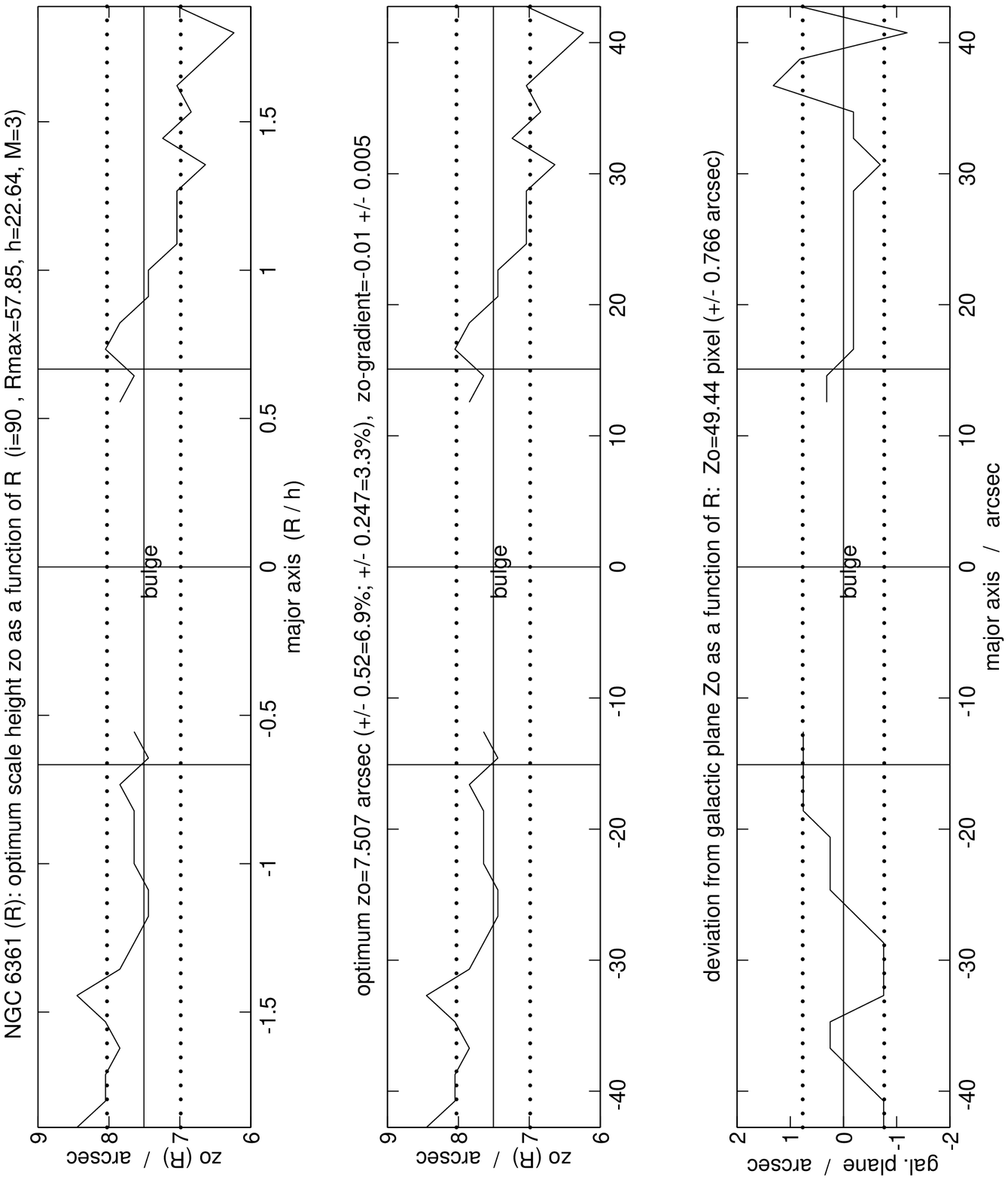}}
\end{picture}
\end{minipage}

\vspace*{98mm}

\hspace*{18mm}\parbox{165mm}{NGC 5965  \hspace{40mm}  NGC 6045  \hspace{45mm}  NGC 6361}

\vspace*{8mm}

\hspace*{8mm}\parbox{165mm}{
{\bf \noindent Appendix A.} (continued)
}
\end{figure*}

%%%%%%%%%%%%%%%%%%%%%%%%%%%%%%%%%%%%%%%%%%%%%%%%%%%%%%%%%%%%%%%%%%%

\clearpage

%%%%%%%%%%%%%%%%%%%%%%%%%%%%%  1  %%%%%%%%%%%%%%%%%%%%%%%%%%%%%%%%%

\begin{figure*}[t]
\vspace*{3mm}
\hspace*{5mm}
\begin{minipage}[b]{5.5cm}
\begin{picture}(3.0,3.0)
{\includegraphics[angle=180,viewport=40 50 400 730,clip,width=52mm]{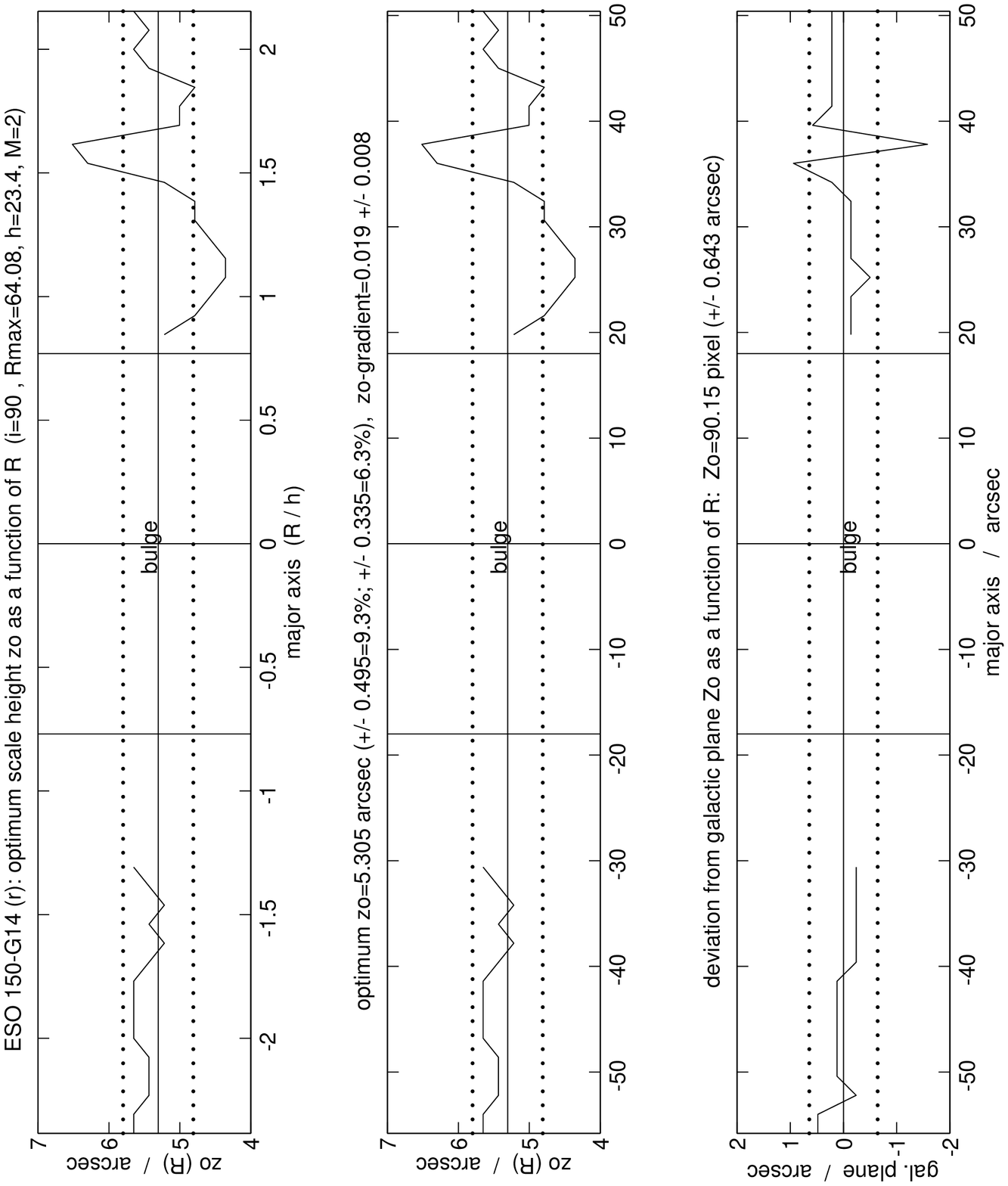}}
\end{picture}
\end{minipage}

\vspace*{-4mm}
\hspace*{66mm}
\begin{minipage}[b]{5.5cm}
\begin{picture}(3.0,3.0)
{\includegraphics[angle=180,viewport=00 -30 342 730,clip,width=50.5mm]{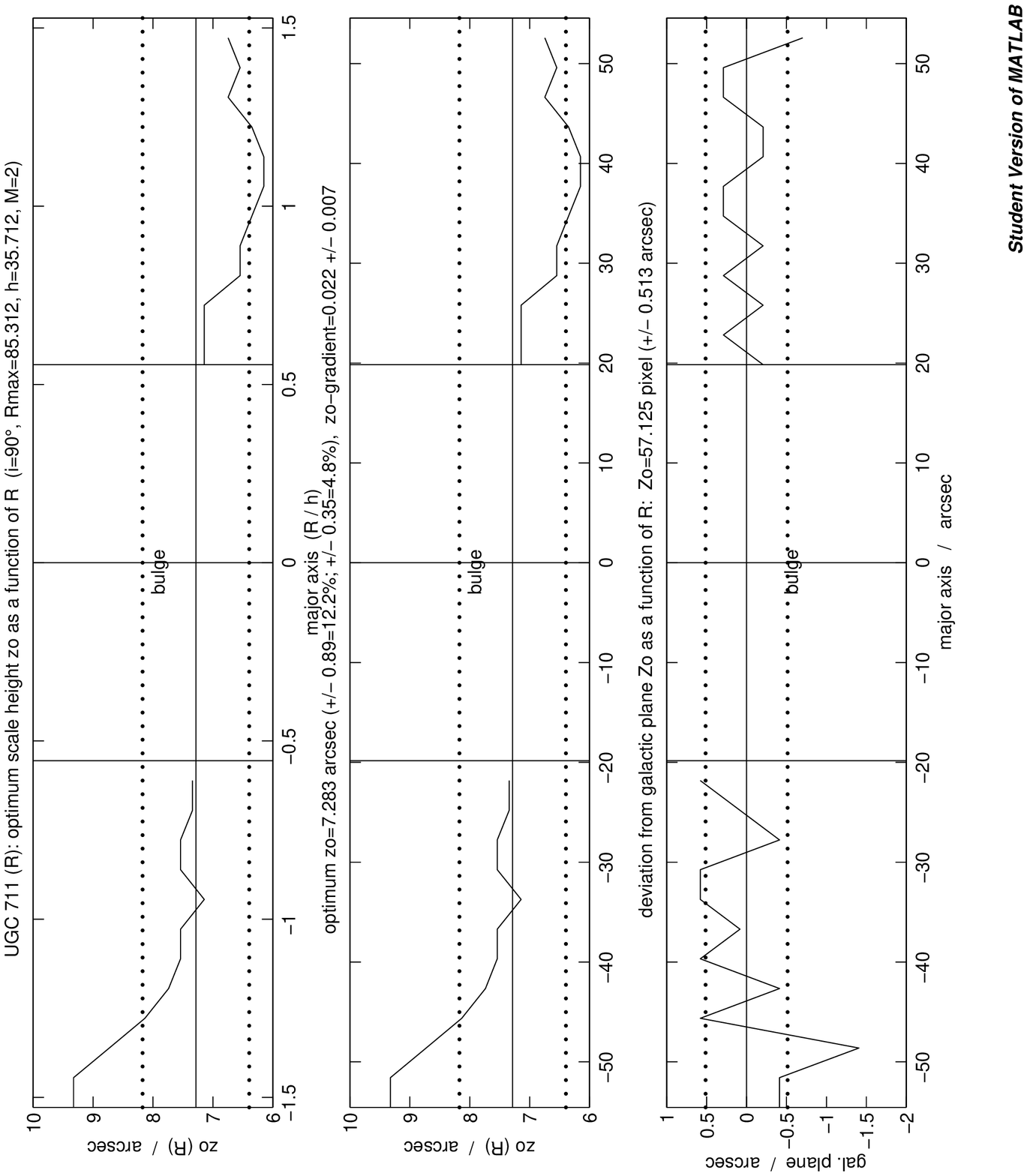}}
\end{picture}
\end{minipage}

\vspace*{-4mm}
\hspace*{125mm}
\begin{minipage}[b]{5.5cm}
\begin{picture}(3.0,3.0)
{\includegraphics[angle=180,viewport=40 50 400 730,clip,width=52mm]{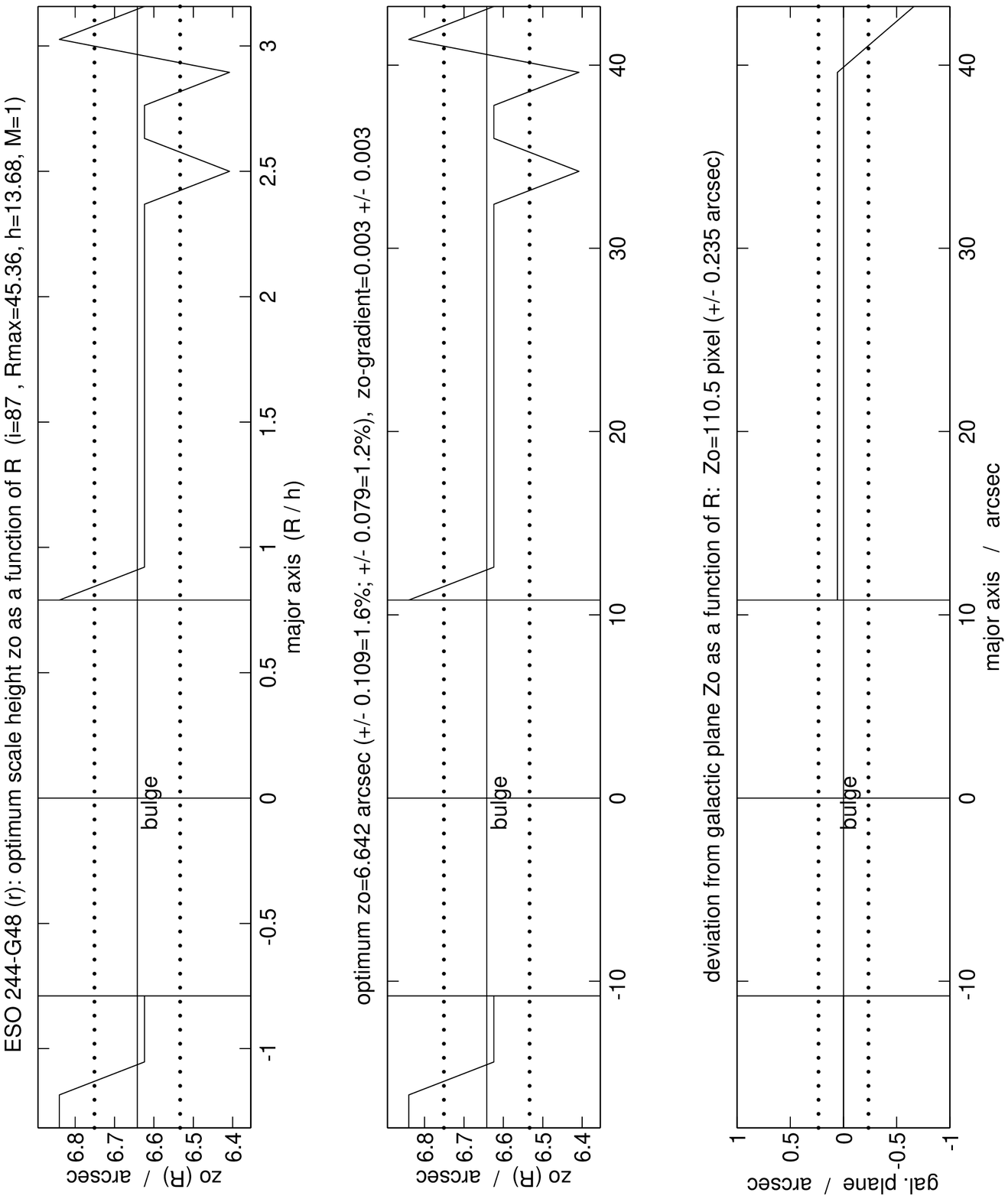}}
\end{picture}
\end{minipage}

\vspace*{98mm}

\hspace*{18mm}\parbox{165mm}{ESO 150-G14  \hspace{40mm}  UGC 711  \hspace{40mm}  ESO 244-G48}

\vspace*{5mm}

\vspace*{0mm}
\hspace*{6mm}
\begin{minipage}[b]{5.5cm}
\begin{picture}(3.0,3.0)
{\includegraphics[angle=180,viewport=00 -30 342 730,clip,width=50.5mm]{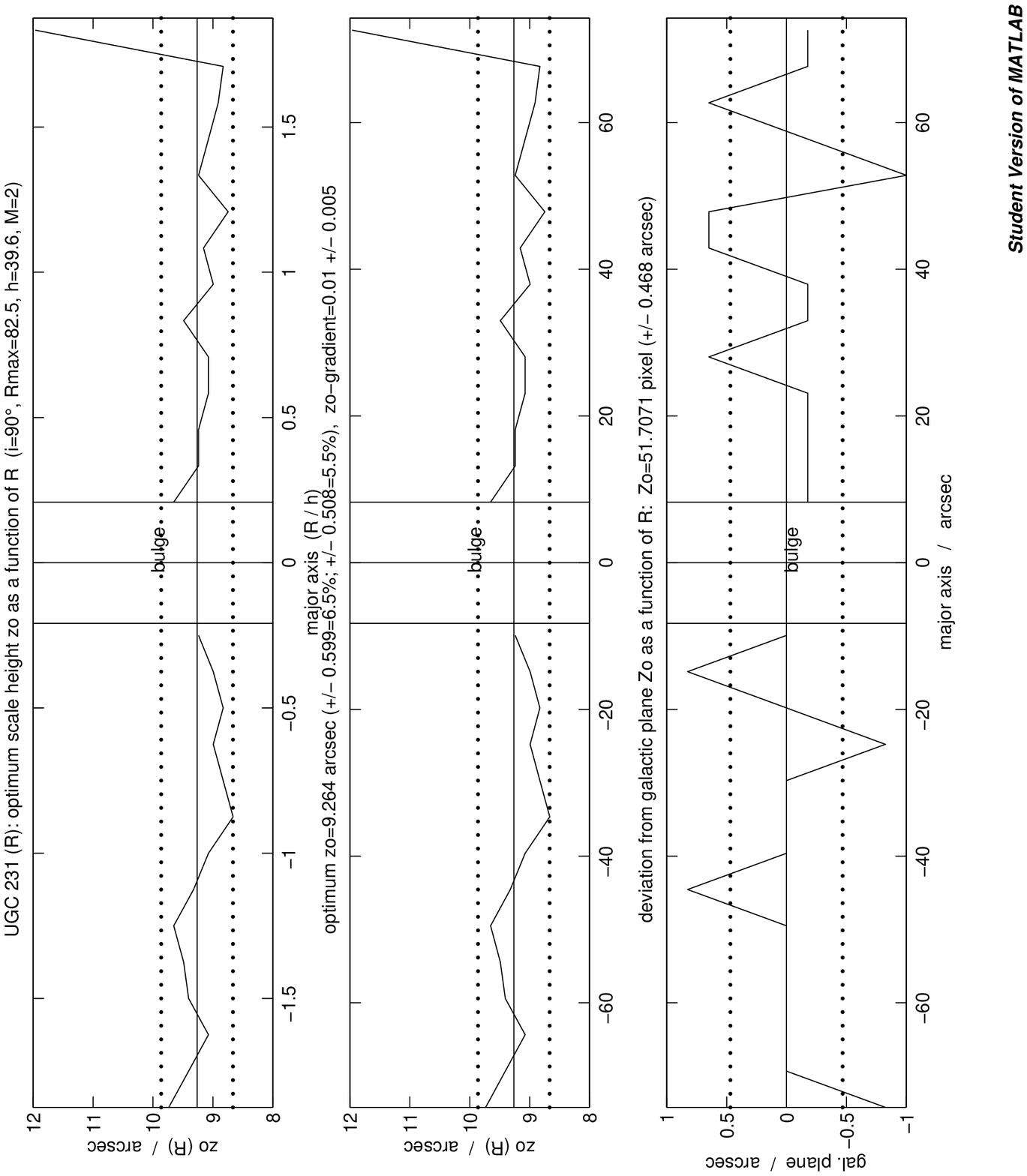}}
\end{picture}
\end{minipage}

\vspace*{-3mm}
\hspace*{64mm}
\begin{minipage}[b]{5.5cm}
\begin{picture}(3.0,3.0)
{\includegraphics[angle=180,viewport=40 50 400 730,clip,width=52mm]{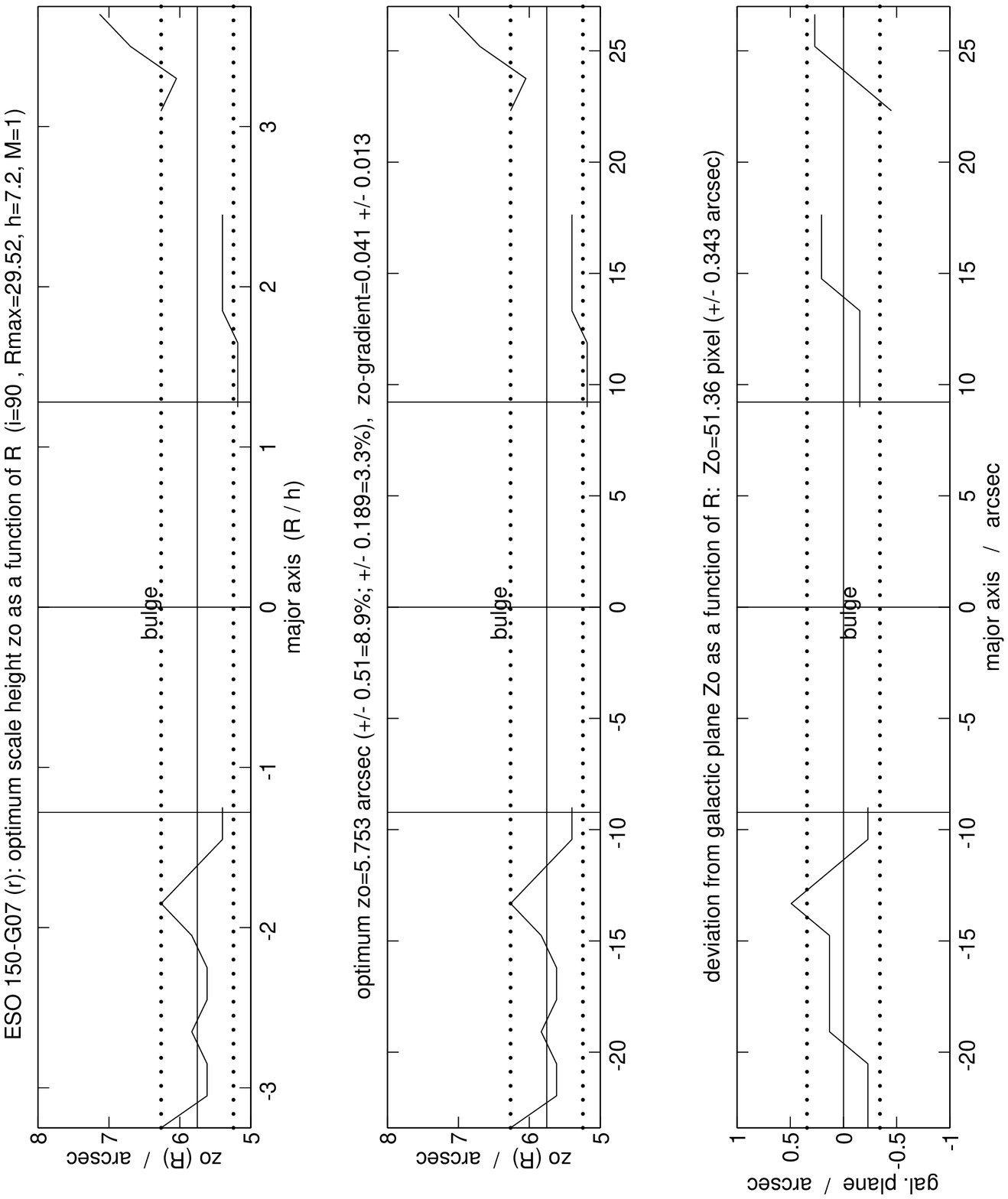}}
\end{picture}
\end{minipage}

\vspace*{-4mm}
\hspace*{125mm}
\begin{minipage}[b]{5.5cm}
\begin{picture}(3.0,3.0)
{\includegraphics[angle=180,viewport=40 50 400 730,clip,width=52mm]{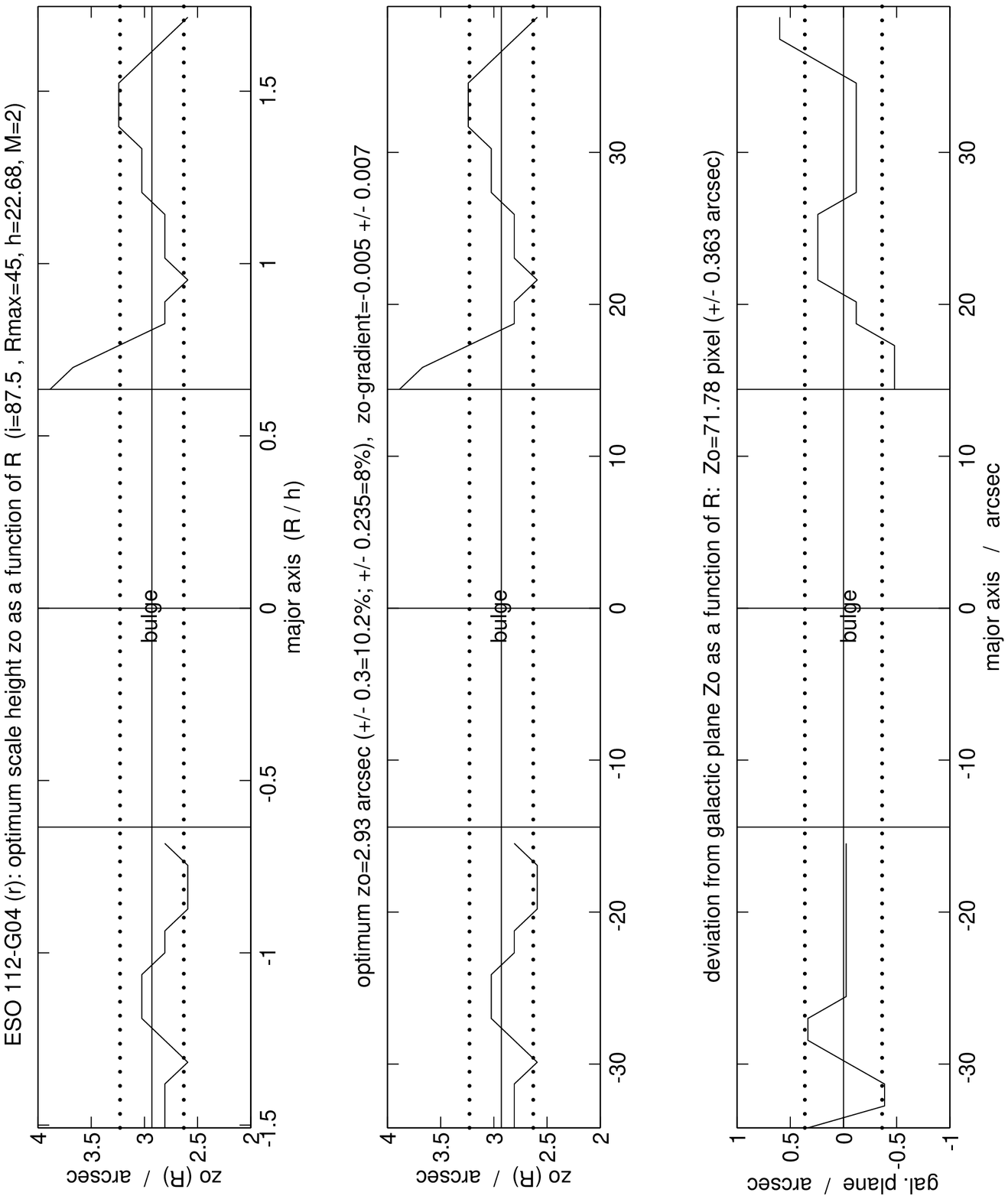}}
\end{picture}
\end{minipage}

\vspace*{98mm}

\hspace*{18mm}\parbox{165mm}{UGC 231  \hspace{40mm}  ESO 150-G07  \hspace{40mm}  ESO 112-G04}

\vspace*{8mm}

\hspace*{8mm}\parbox{165mm}{
{\bf \noindent Appendix B.} The measured behaviour of scale height (upper panels) and
mean galactic plane (lower panels) along the major axis of the disk, shown for each
galaxy in the interacting/merging sample. The main output parameters are indicated. For
details see Sect.\~3 and Table\~4 in this paper. Contour maps were given in Paper\~I,
Figs.\~4 and 5.
}

\end{figure*}

%%%%%%%%%%%%%%%%%%%%%%%%%%%%%%%%%%%%%%%%%%%%%%%%%%%%%%%%%%%%%%%%%%%

\clearpage

%%%%%%%%%%%%%%%%%%%%%%%%%%%%%  2  %%%%%%%%%%%%%%%%%%%%%%%%%%%%%%%%%

\begin{figure*}[t]
\vspace*{3mm}
\hspace*{6mm}
\begin{minipage}[b]{5.5cm}
\begin{picture}(3.0,3.0)
{\includegraphics[angle=180,viewport=00 35 342 730,clip,width=50.5mm]{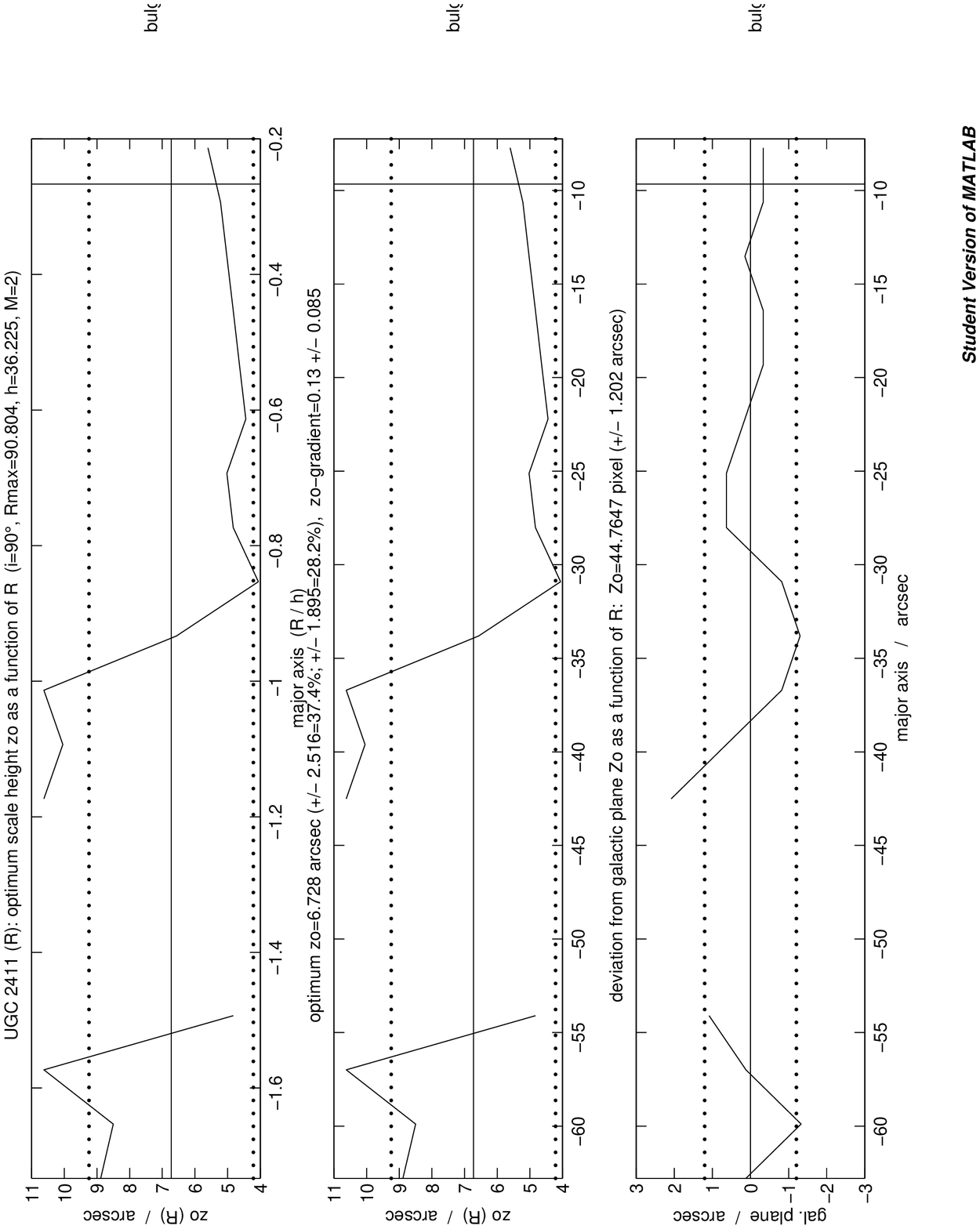}}
\end{picture}
\end{minipage}

\vspace*{-3mm}
\hspace*{66mm}
\begin{minipage}[b]{5.5cm}
\begin{picture}(3.0,3.0)
{\includegraphics[angle=180,viewport=00 -30 342 730,clip,width=50.5mm]{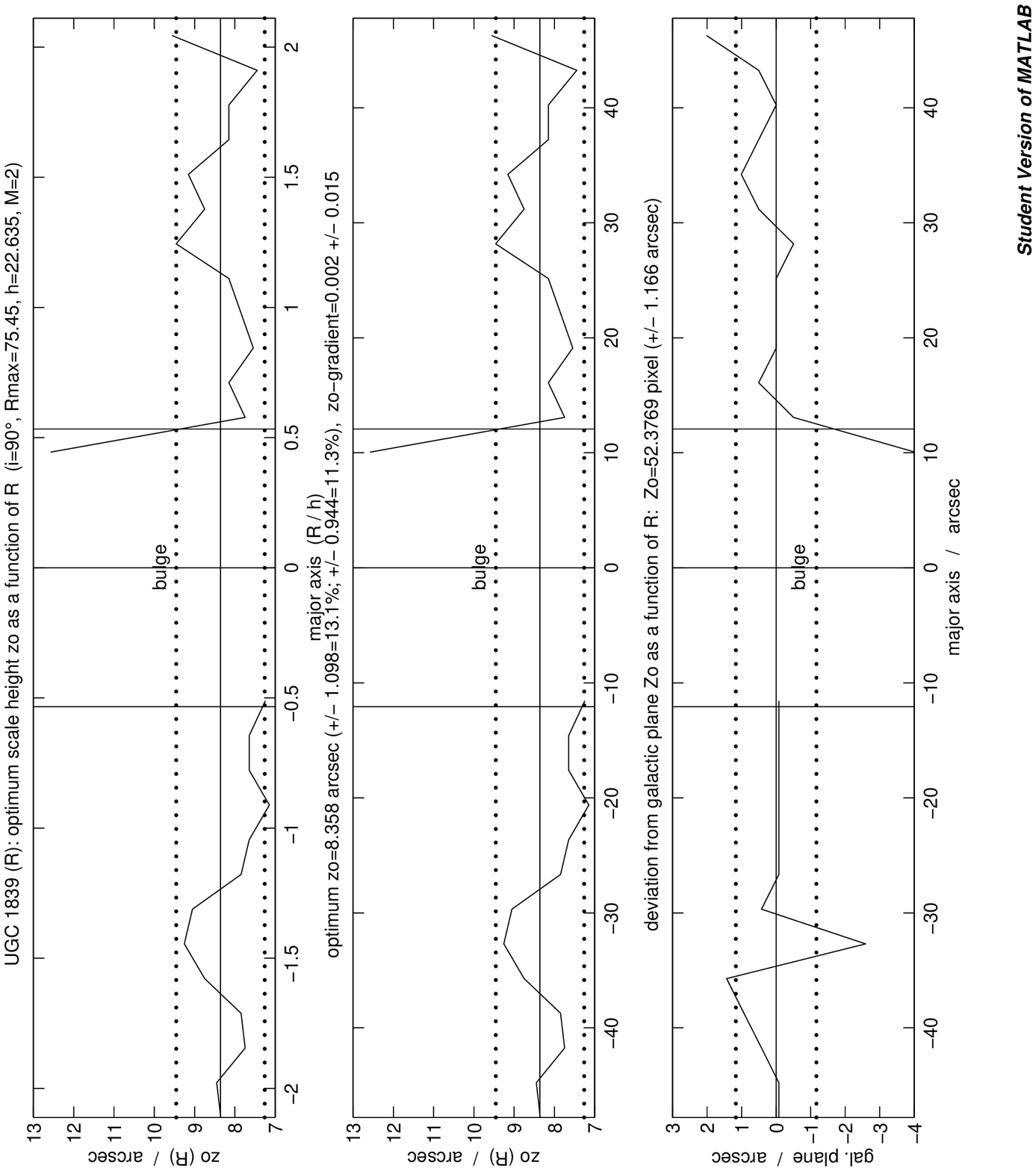}}
\end{picture}
\end{minipage}

\vspace*{93mm}
\hspace*{125mm}
\begin{minipage}[b]{5.5cm}
\begin{picture}(3.0,3.0)
{\includegraphics[angle=90,viewport=40 10 540 285,clip,width=53mm]{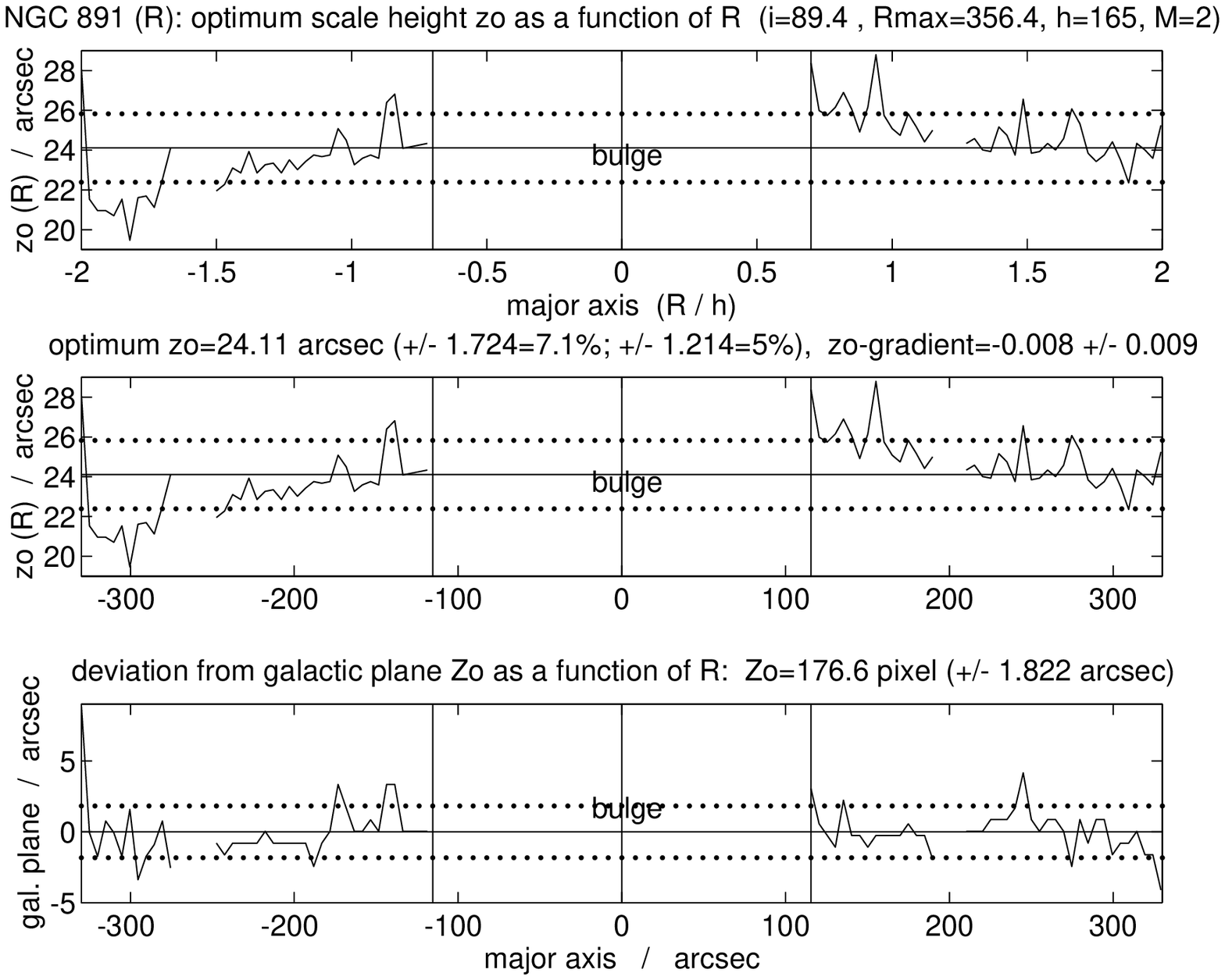}}
\end{picture}
\end{minipage}

\vspace*{0mm}

\hspace*{18mm}\parbox{165mm}{UGC 2411  \hspace{45mm}  IC 1877  \hspace{42mm}  ESO 201-G22}

\vspace*{2mm}

\vspace*{100mm}
\hspace*{6mm}
\begin{minipage}[b]{5.5cm}
\begin{picture}(3.0,3.0)
{\includegraphics[angle=90,viewport=40 10 540 285,clip,width=53mm]{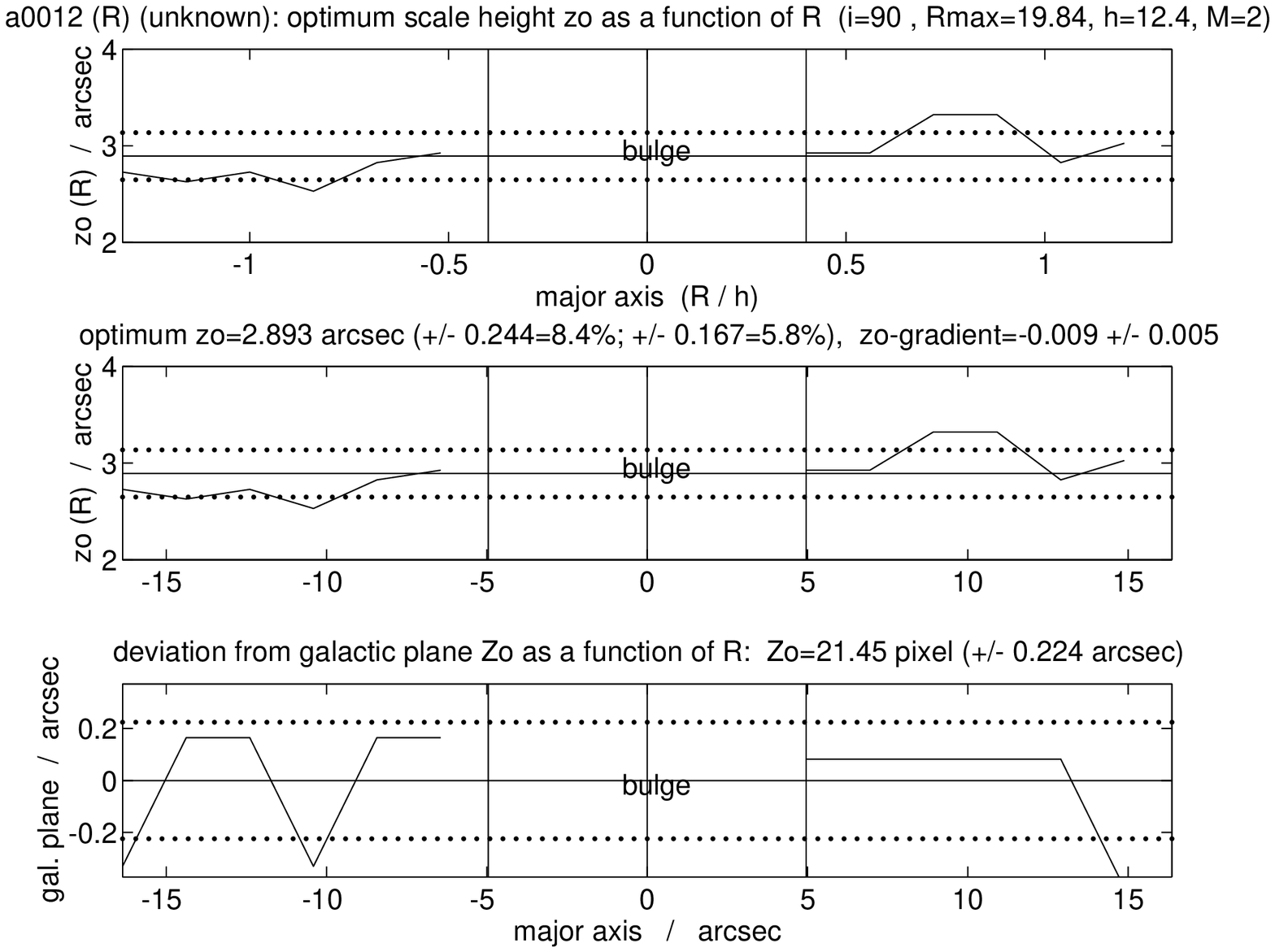}}
\end{picture}
\end{minipage}

\vspace*{-99mm}
\hspace*{60mm}
\hspace*{5mm}
\begin{minipage}[b]{5.5cm}
\begin{picture}(3.0,3.0)
{\includegraphics[angle=180,viewport=40 50 400 730,clip,width=50.5mm]{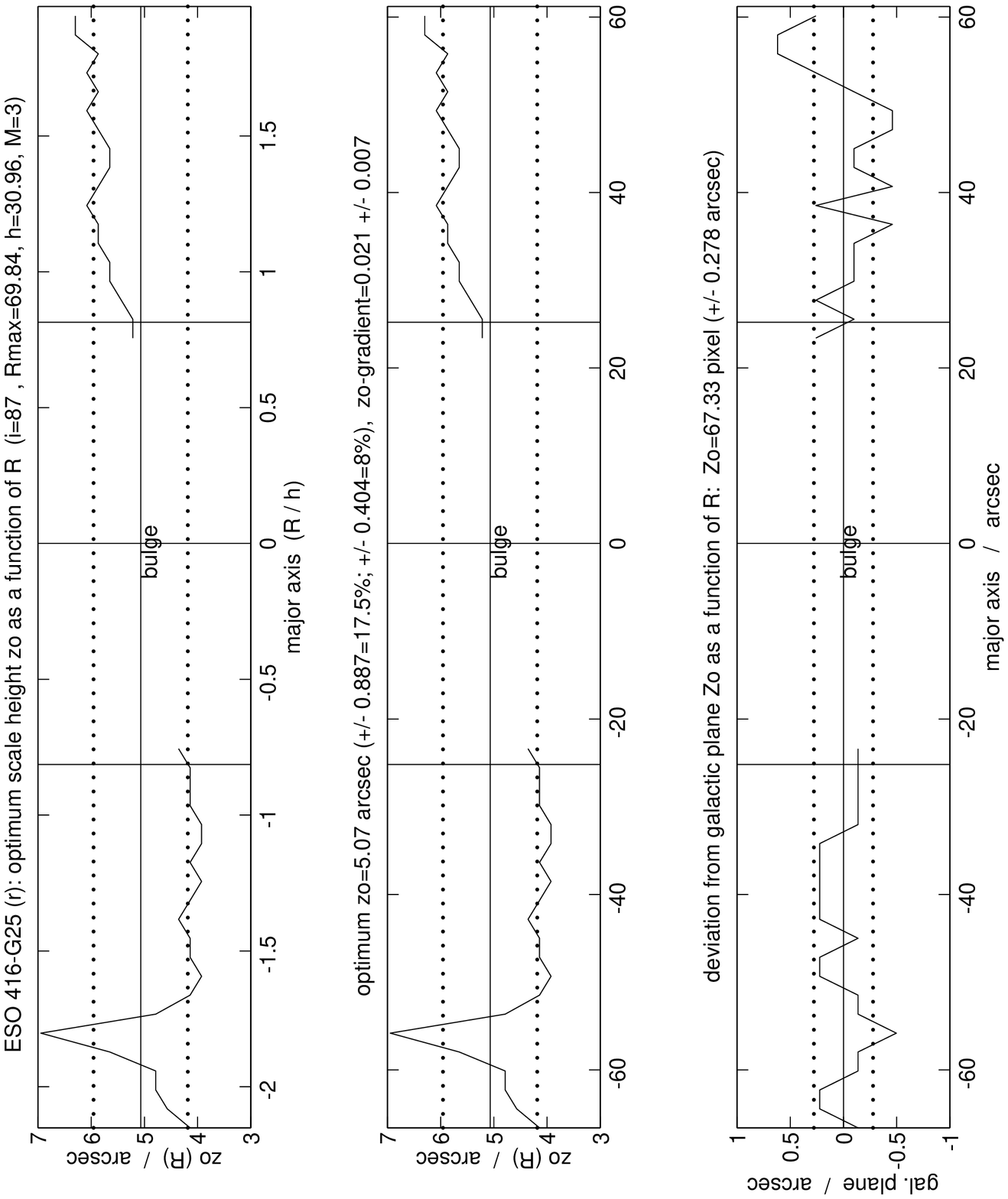}}
\end{picture}
\end{minipage}

\vspace*{-5mm}
\hspace*{125mm}
\begin{minipage}[b]{5.5cm}
\begin{picture}(3.0,3.0)
{\includegraphics[angle=180,viewport=00 -30 342 730,clip,width=50.5mm]{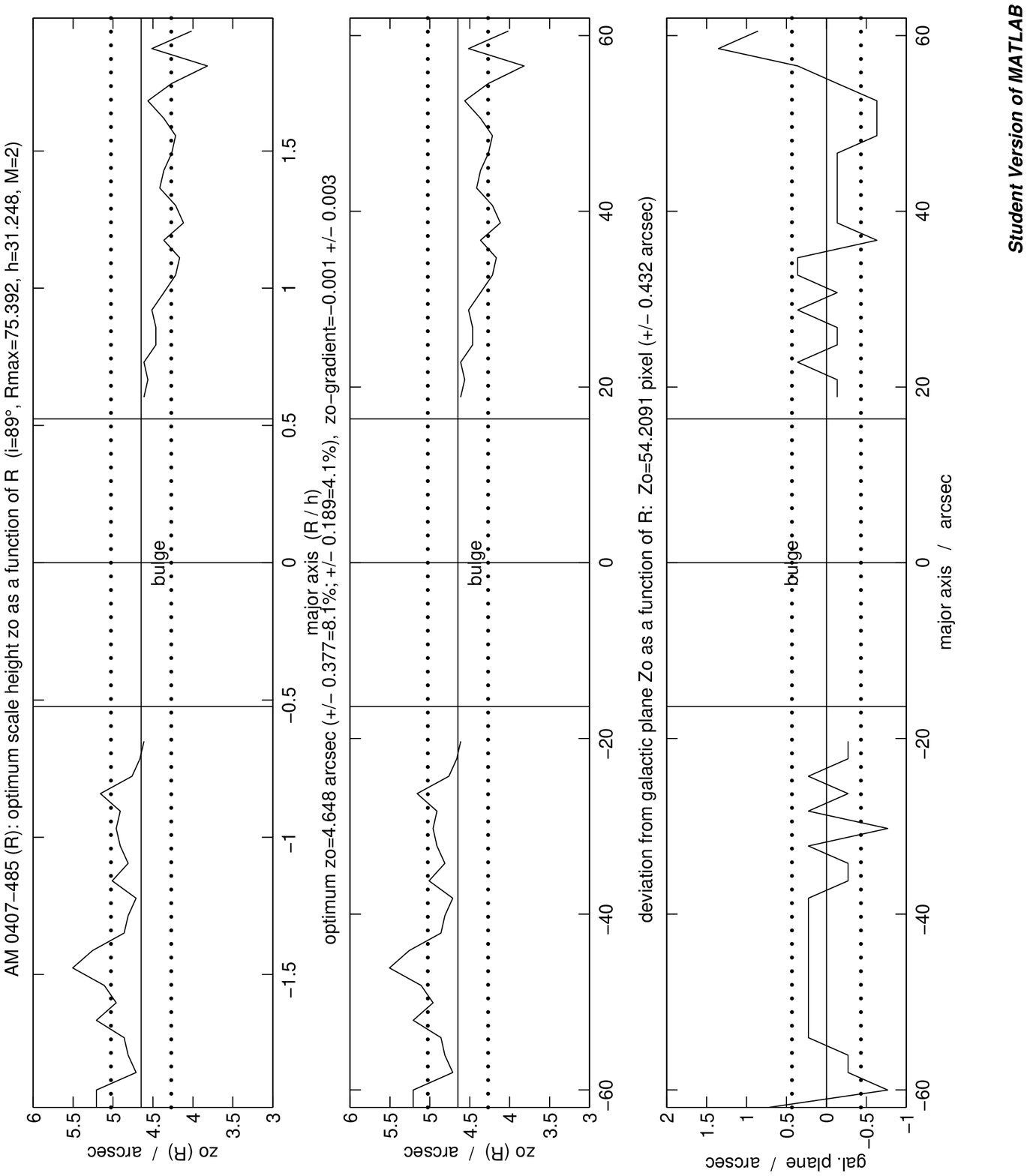}}
\end{picture}
\end{minipage}

\vspace*{98mm}

\hspace*{18mm}\parbox{165mm}{UGC 1839  \hspace{45mm}  NGC 891  \hspace{42mm}  ESO 416-G25}

\vspace*{8mm}

\hspace*{8mm}\parbox{165mm}{
{\bf \noindent Appendix B.} (continued)
}
\end{figure*}

%%%%%%%%%%%%%%%%%%%%%%%%%%%%%%%%%%%%%%%%%%%%%%%%%%%%%%%%%%%%%%%%%%%

\clearpage

%%%%%%%%%%%%%%%%%%%%%%%%%%%%%  3  %%%%%%%%%%%%%%%%%%%%%%%%%%%%%%%%%

\begin{figure*}[t]
\vspace*{3mm}
\hspace*{5mm}
\begin{minipage}[b]{5.5cm}
\begin{picture}(3.0,3.0)
{\includegraphics[angle=180,viewport=00 -30 342 730,clip,width=50.5mm]{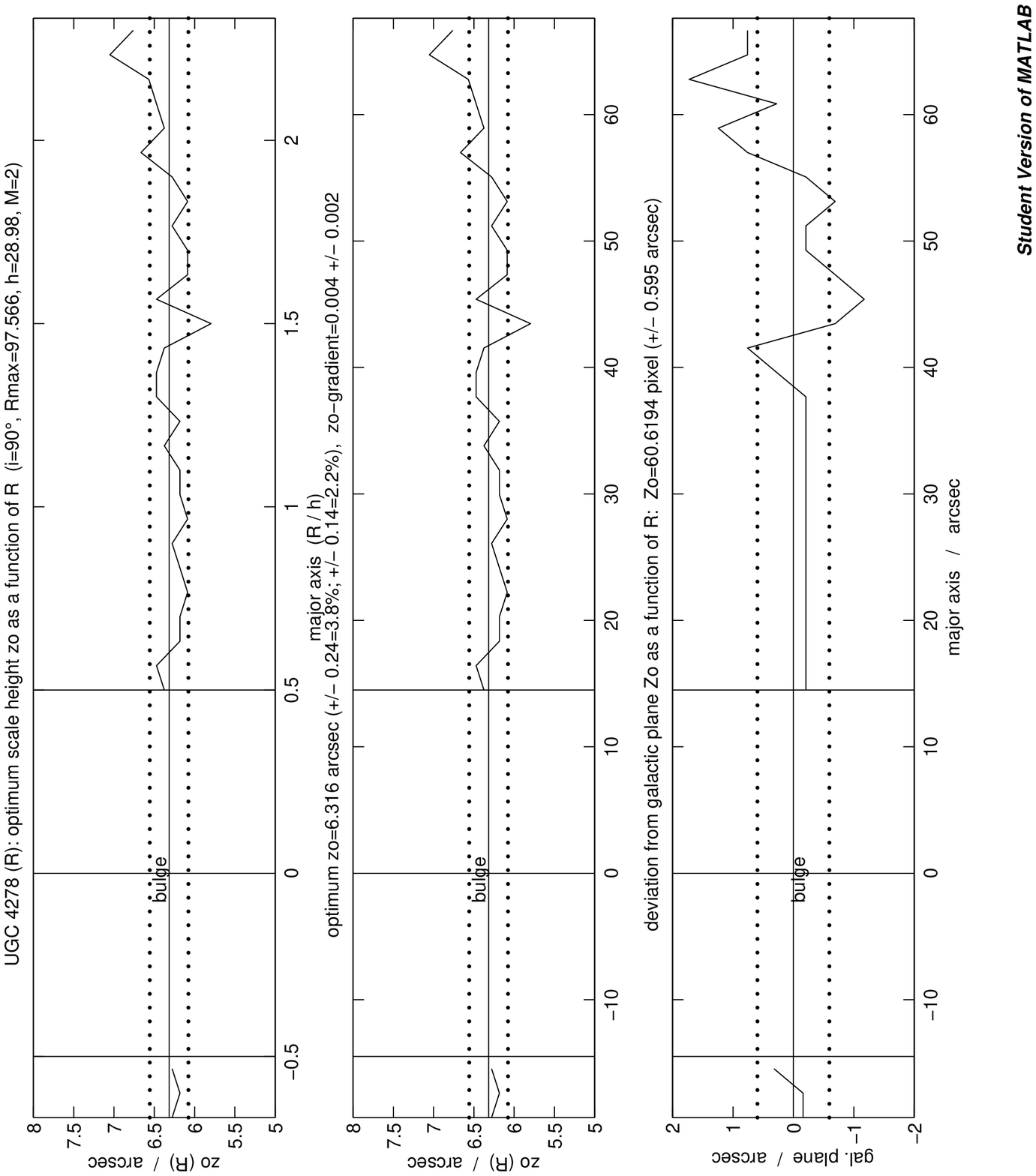}}
\end{picture}
\end{minipage}

\vspace*{-3mm}
\hspace*{65mm}
\begin{minipage}[b]{5.5cm}
\begin{picture}(3.0,3.0)
{\includegraphics[angle=180,viewport=40 50 400 730,clip,width=52mm]{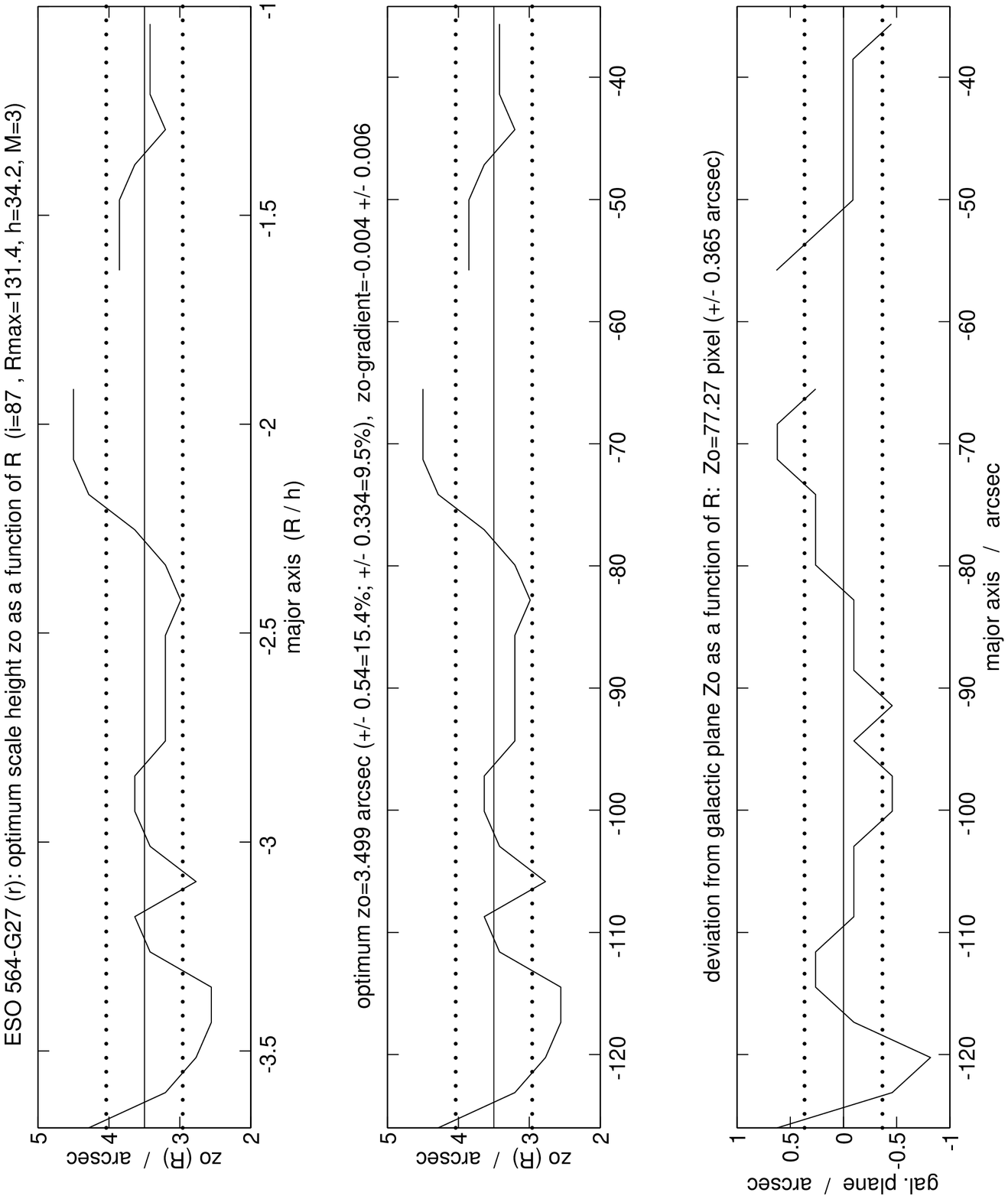}}
\end{picture}
\end{minipage}

\vspace*{-5mm}
\hspace*{125mm}
\begin{minipage}[b]{5.5cm}
\begin{picture}(3.0,3.0)
{\includegraphics[angle=180,viewport=00 -30 342 730,clip,width=50.5mm]{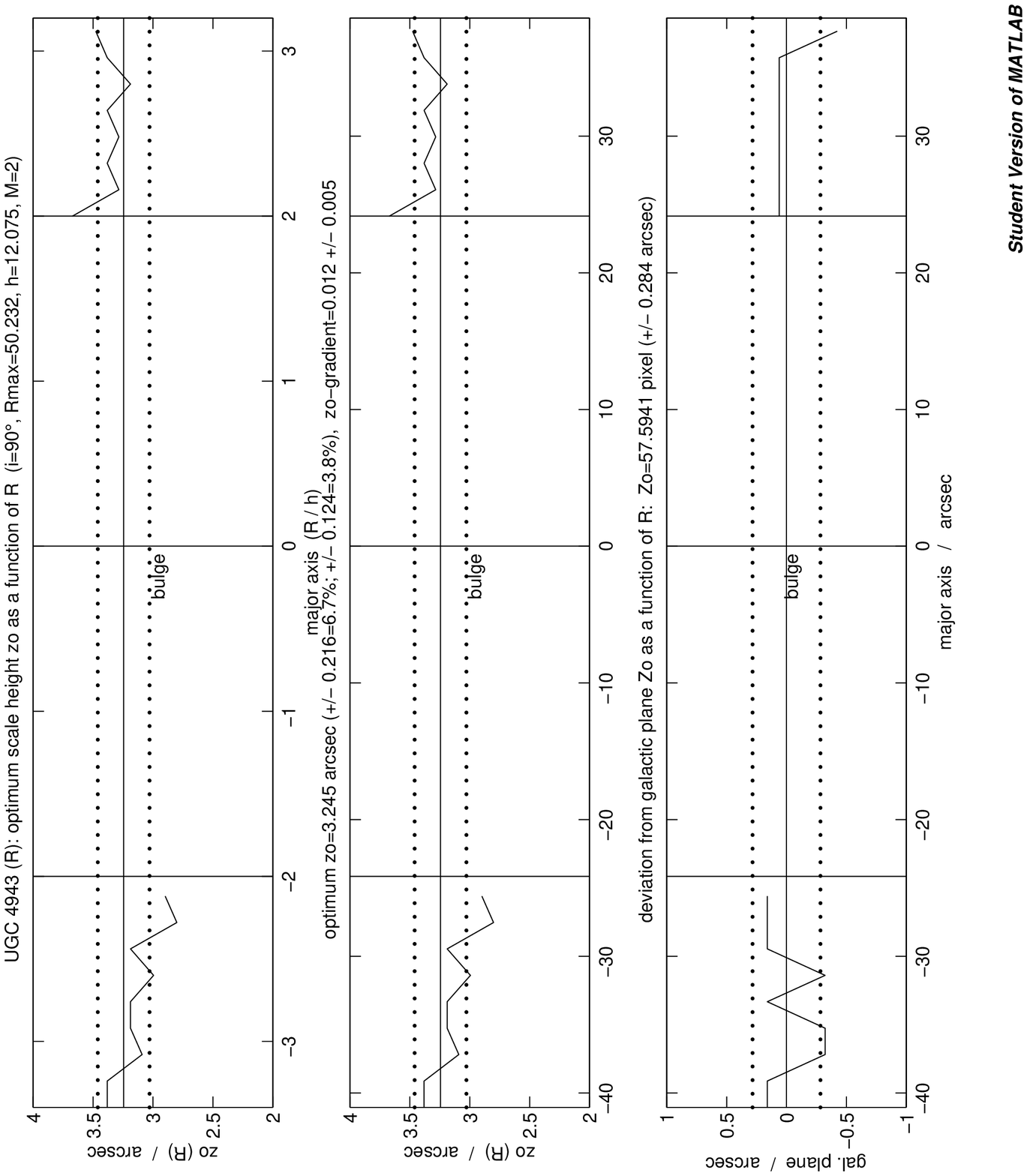}}
\end{picture}
\end{minipage}

\vspace*{98mm}

\hspace*{18mm}\parbox{165mm}{UGC 4278  \hspace{40mm}  ESO 564-G27  \hspace{40mm}  UGC 4943}

\vspace*{5mm}

\vspace*{98mm}
\hspace*{4mm}
\begin{minipage}[b]{5.5cm}
\begin{picture}(3.0,3.0)
{\includegraphics[angle=90,viewport=40 10 540 285,clip,width=53mm]{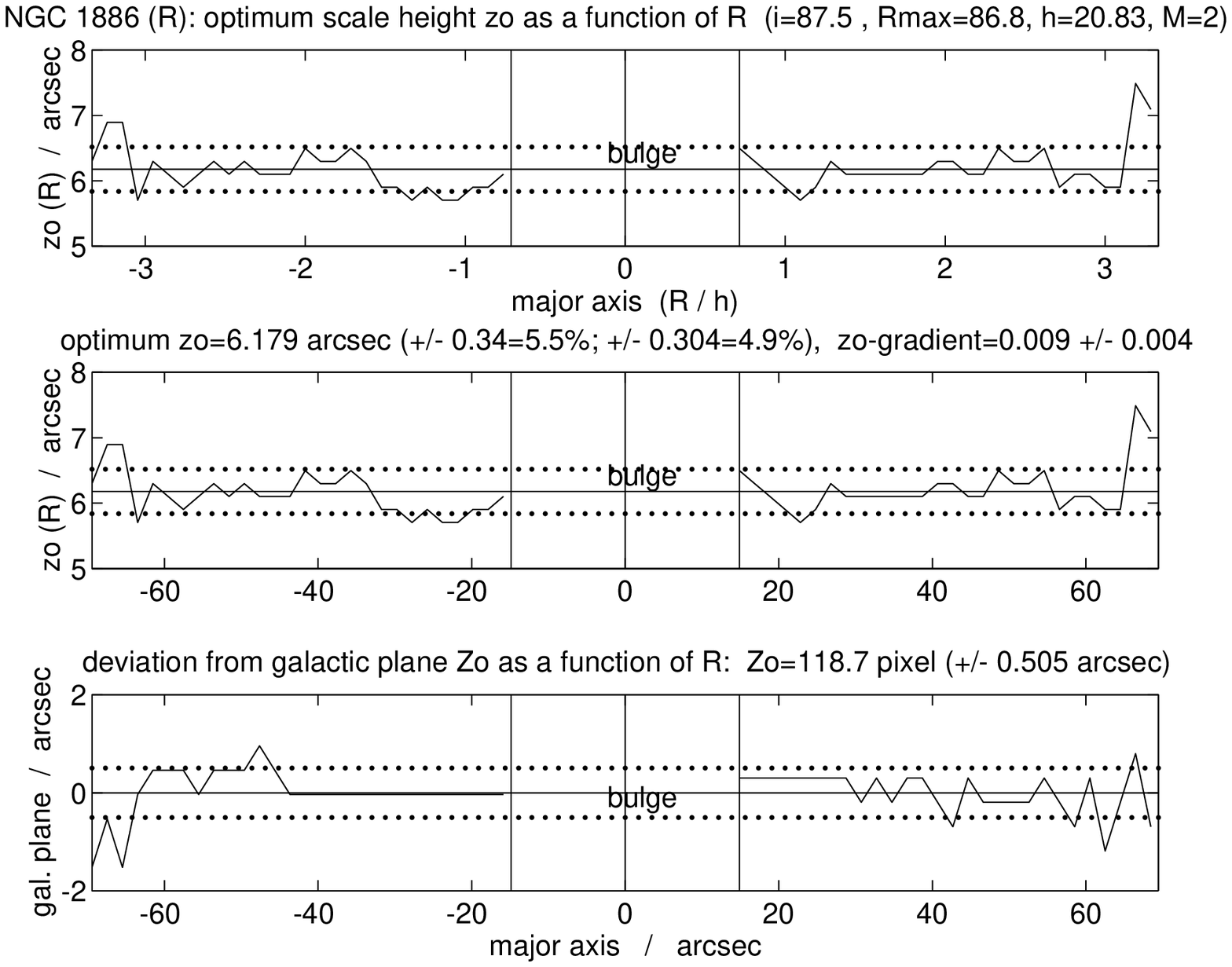}}
\end{picture}
\end{minipage}

\vspace*{-101mm}
\hspace*{65mm}
\begin{minipage}[b]{5.5cm}
\begin{picture}(3.0,3.0)
{\includegraphics[angle=180,viewport=00 -30 342 730,clip,width=50.5mm]{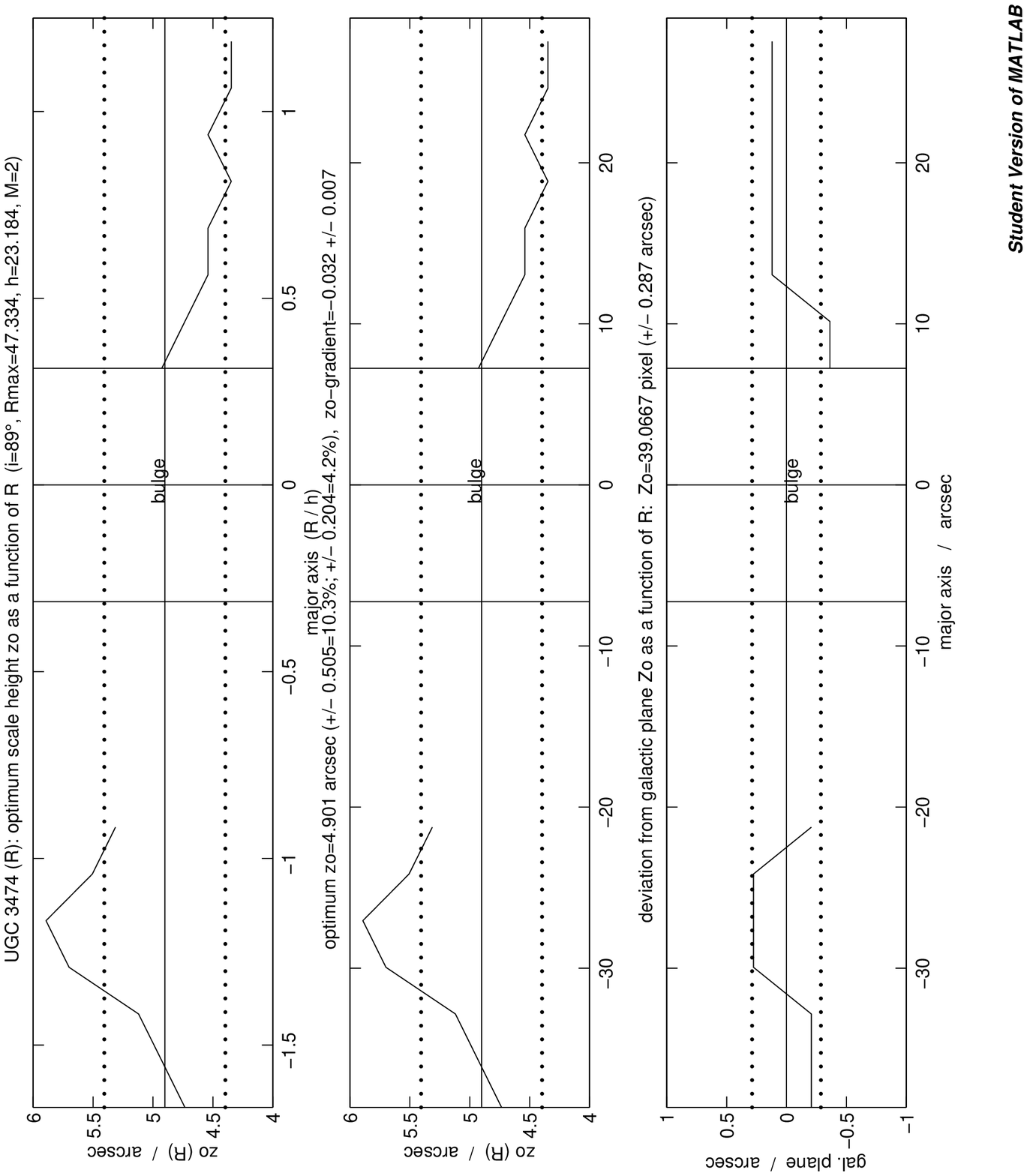}}
\end{picture}
\end{minipage}

\vspace*{94mm}
\hspace*{125mm}
\begin{minipage}[b]{5.5cm}
\begin{picture}(3.0,3.0)
{\includegraphics[angle=90,viewport=40 10 540 285,clip,width=53mm]{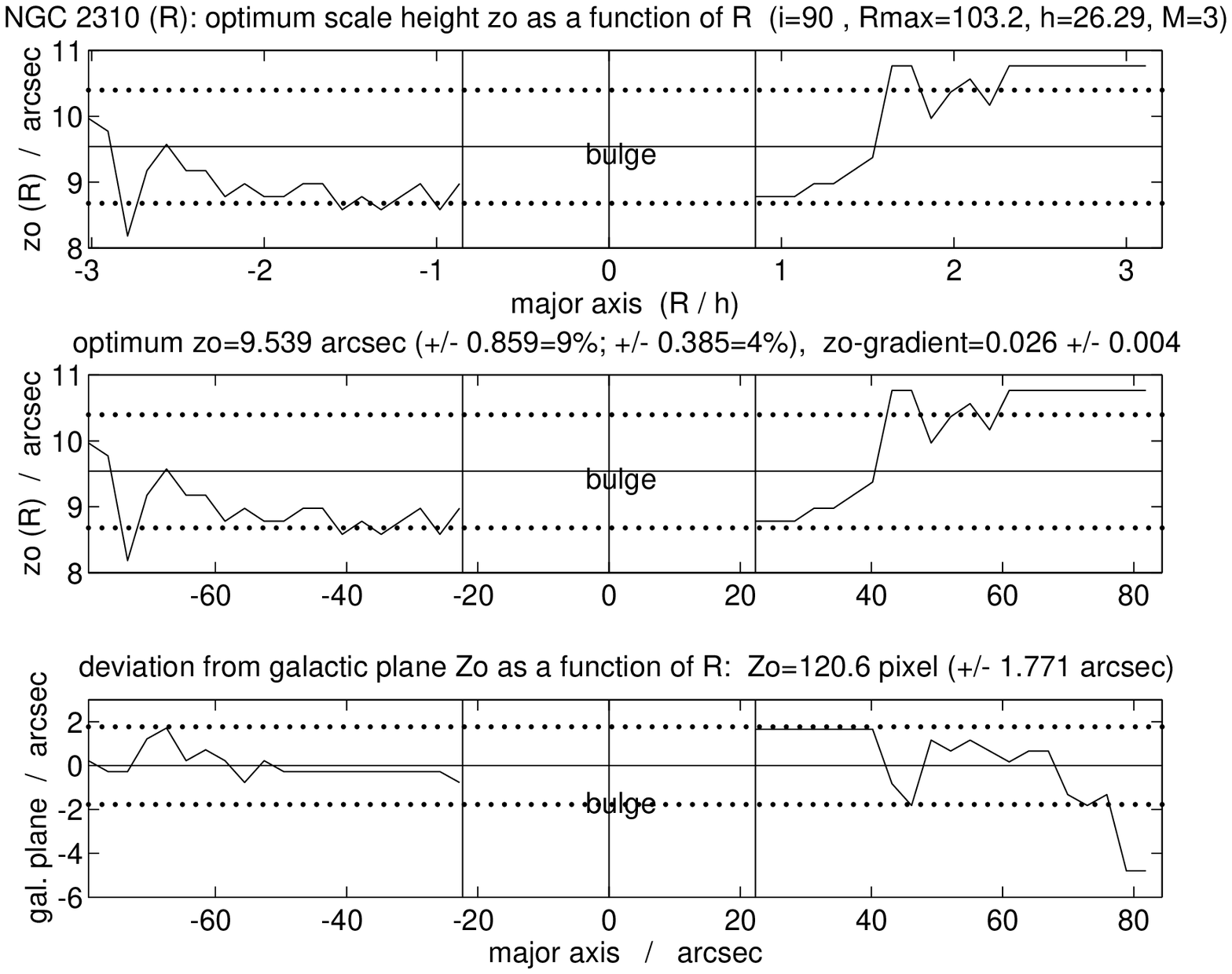}}
\end{picture}
\end{minipage}

\vspace*{0mm}

\hspace*{18mm}\parbox{165mm}{NGC 1886  \hspace{42mm} UGC 3474   \hspace{45mm}  NGC 2310}

\vspace*{8mm}

\hspace*{8mm}\parbox{165mm}{
{\bf \noindent Appendix B.} (continued)
}
\end{figure*}

%%%%%%%%%%%%%%%%%%%%%%%%%%%%%%%%%%%%%%%%%%%%%%%%%%%%%%%%%%%%%%%%%%%

\clearpage

%%%%%%%%%%%%%%%%%%%%%%%%%%%%%  4  %%%%%%%%%%%%%%%%%%%%%%%%%%%%%%%%%

\begin{figure*}[t]
\vspace*{3mm}
\hspace*{5mm}
\begin{minipage}[b]{5.5cm}
\begin{picture}(3.0,3.0)
{\includegraphics[angle=180,viewport=40 50 400 730,clip,width=52mm]{ngc3390.ps}}
\end{picture}
\end{minipage}
\hfill
\begin{minipage}[b]{5.5cm}
\begin{picture}(3.0,3.0)
{\includegraphics[angle=180,viewport=40 50 400 730,clip,width=52mm]{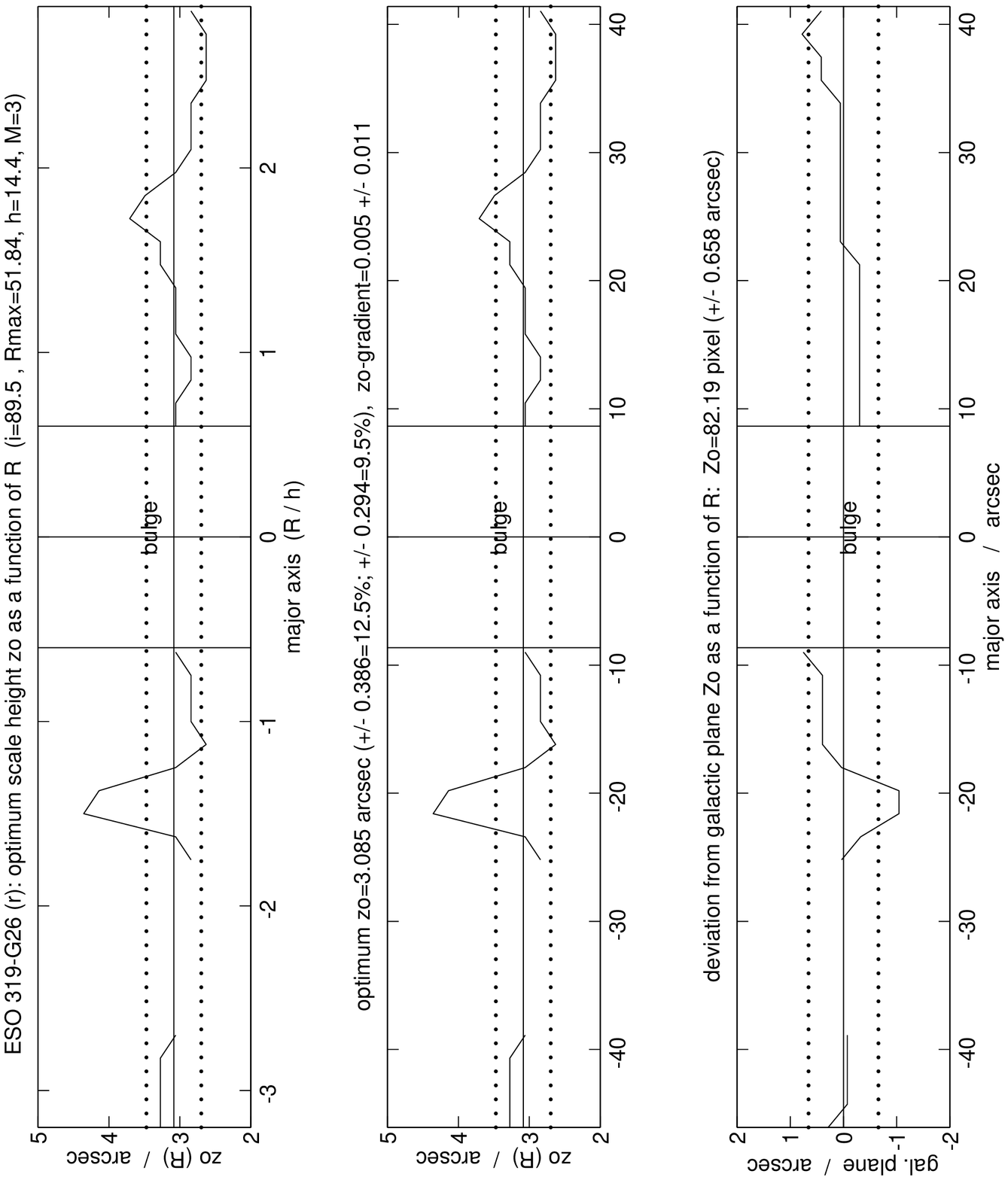}}
\end{picture}
\end{minipage}
\hfill
\begin{minipage}[b]{5.5cm}
\begin{picture}(3.0,3.0)
{\includegraphics[angle=180,viewport=40 50 400 730,clip,width=52mm]{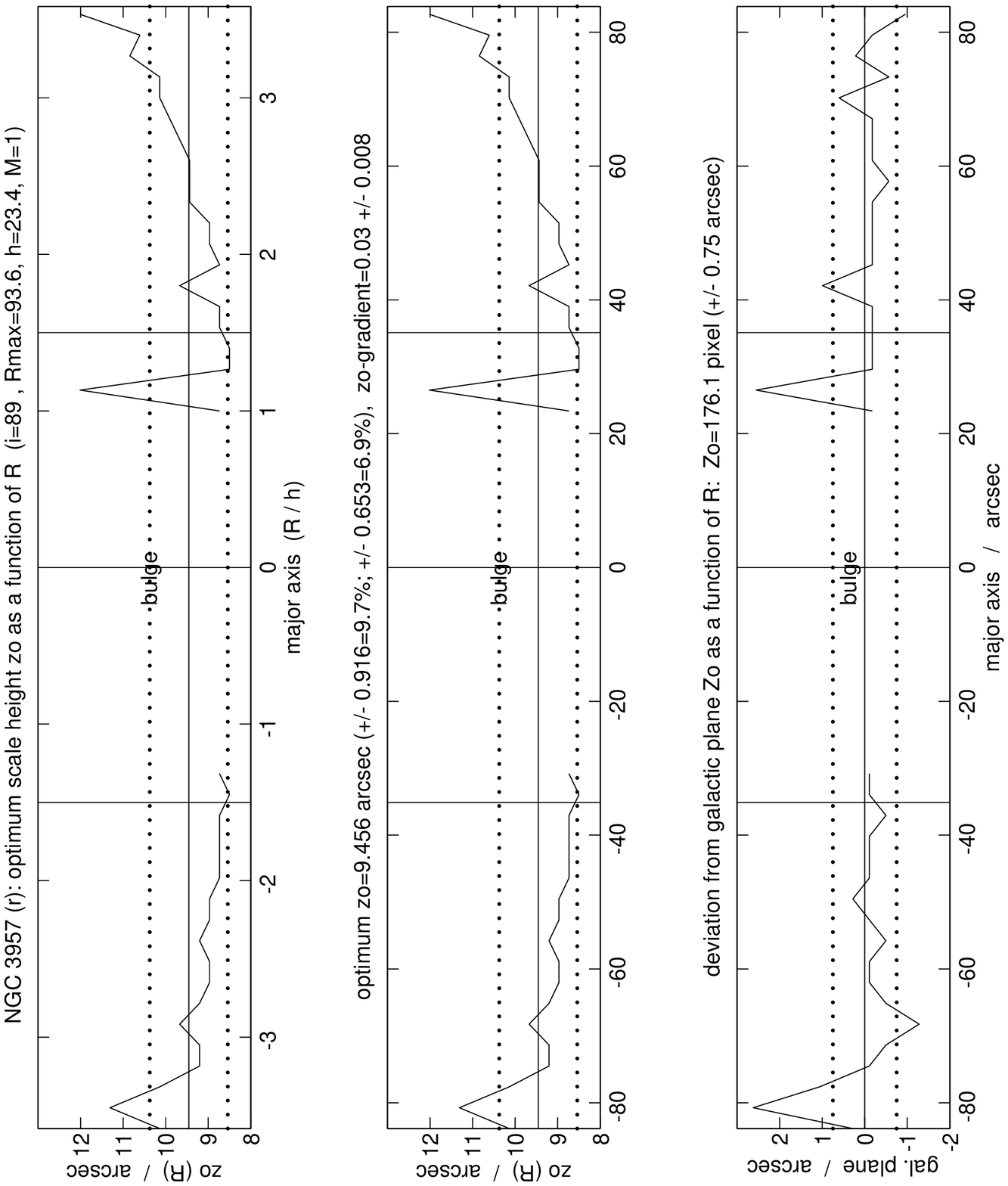}}
\end{picture}
\end{minipage}

\vspace*{98mm}

\hspace*{18mm}\parbox{165mm}{NGC 3390  \hspace{40mm}  ESO 319-G26  \hspace{40mm}  NGC 3957}

\vspace*{5mm}

\hspace*{5mm}
\begin{minipage}[b]{5.5cm}
\begin{picture}(3.0,3.0)
{\includegraphics[angle=180,viewport=40 50 400 730,clip,width=52mm]{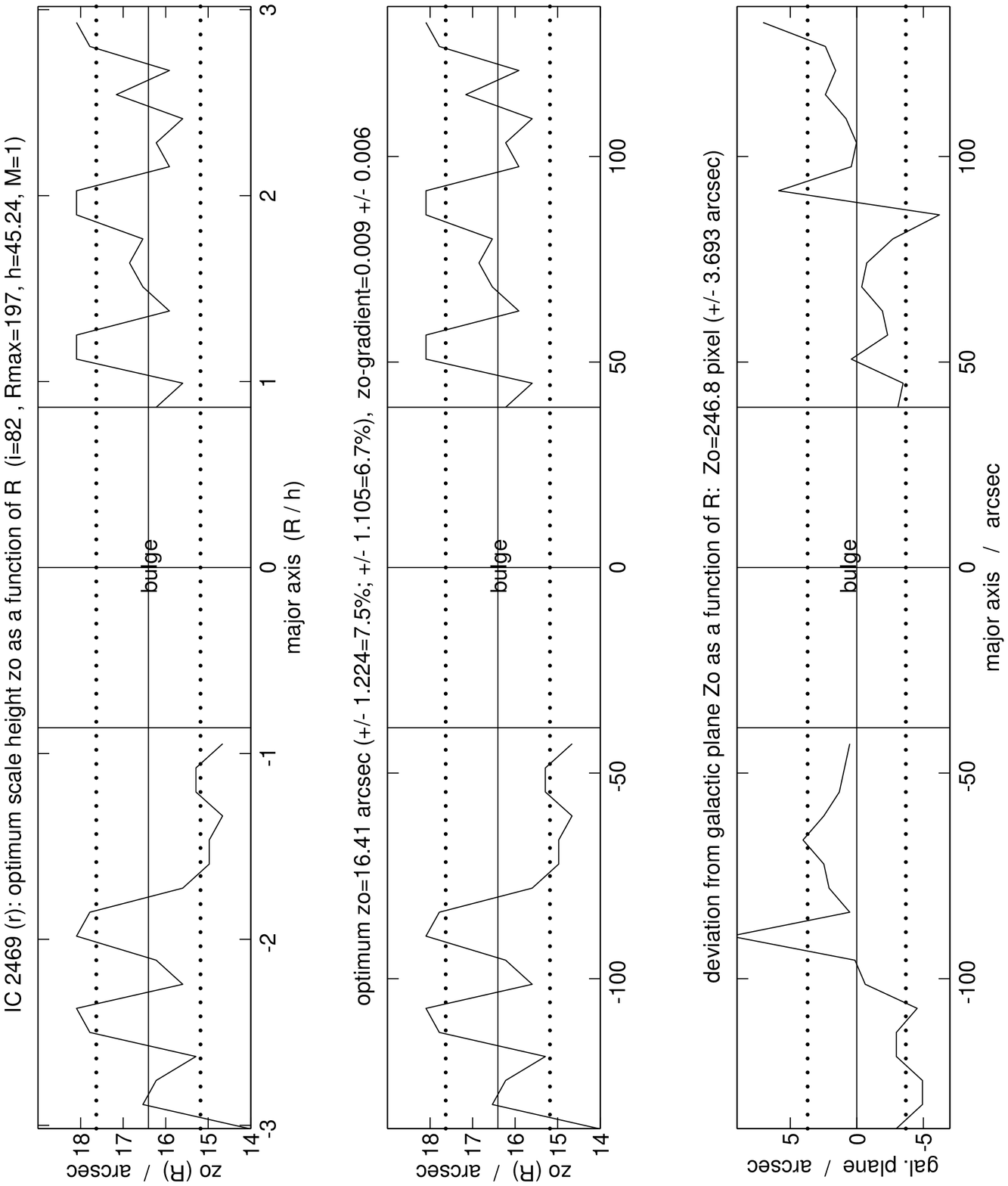}}
\end{picture}
\end{minipage}
\hfill
\begin{minipage}[b]{5.5cm}
\begin{picture}(3.0,3.0)
{\includegraphics[angle=180,viewport=00 -30 342 730,clip,width=50.5mm]{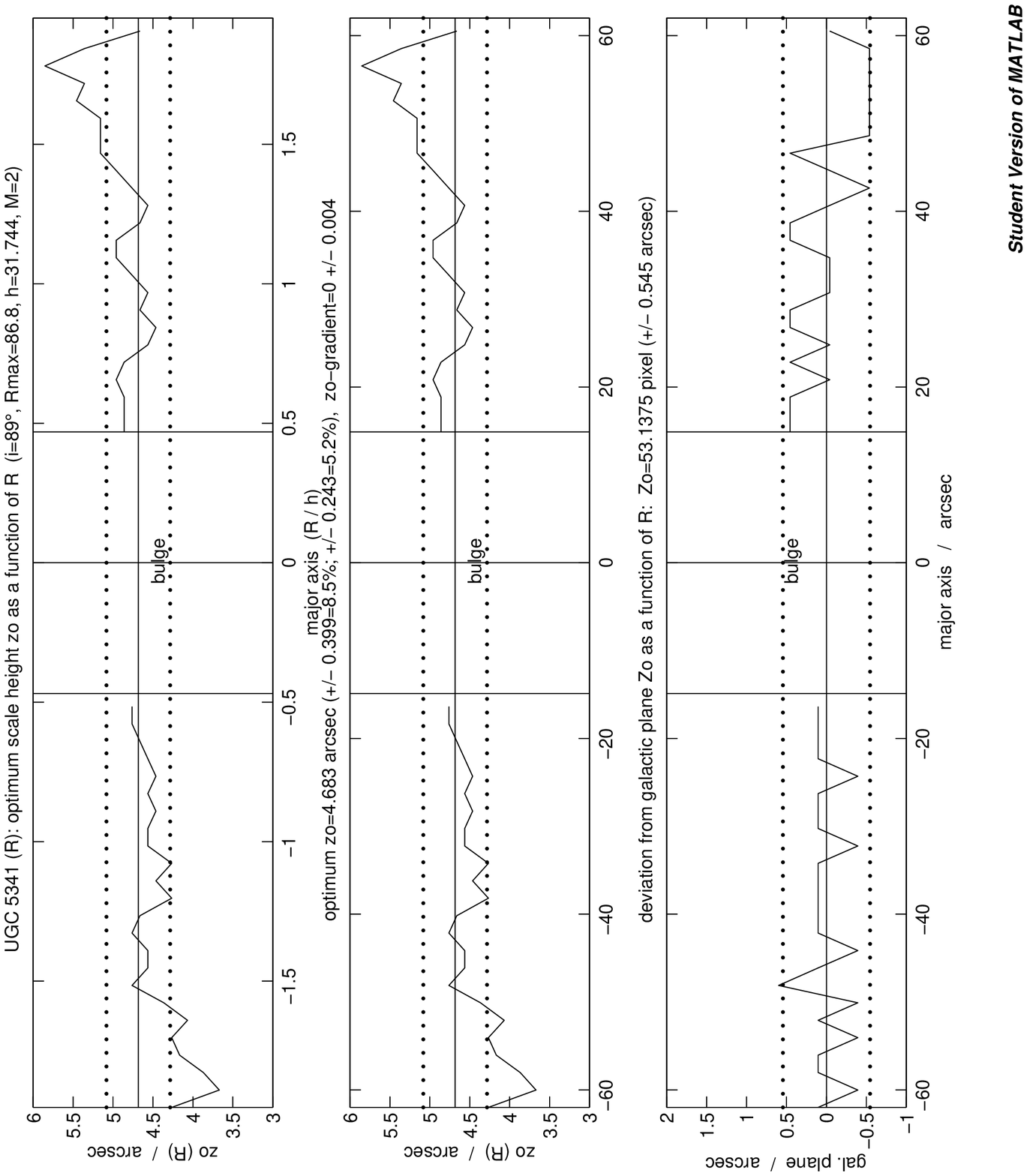}}
\end{picture}
\end{minipage}
\hfill
\begin{minipage}[b]{5.5cm}
\begin{picture}(3.0,3.0)
{\includegraphics[angle=180,viewport=00 -30 342 730,clip,width=50.5mm]{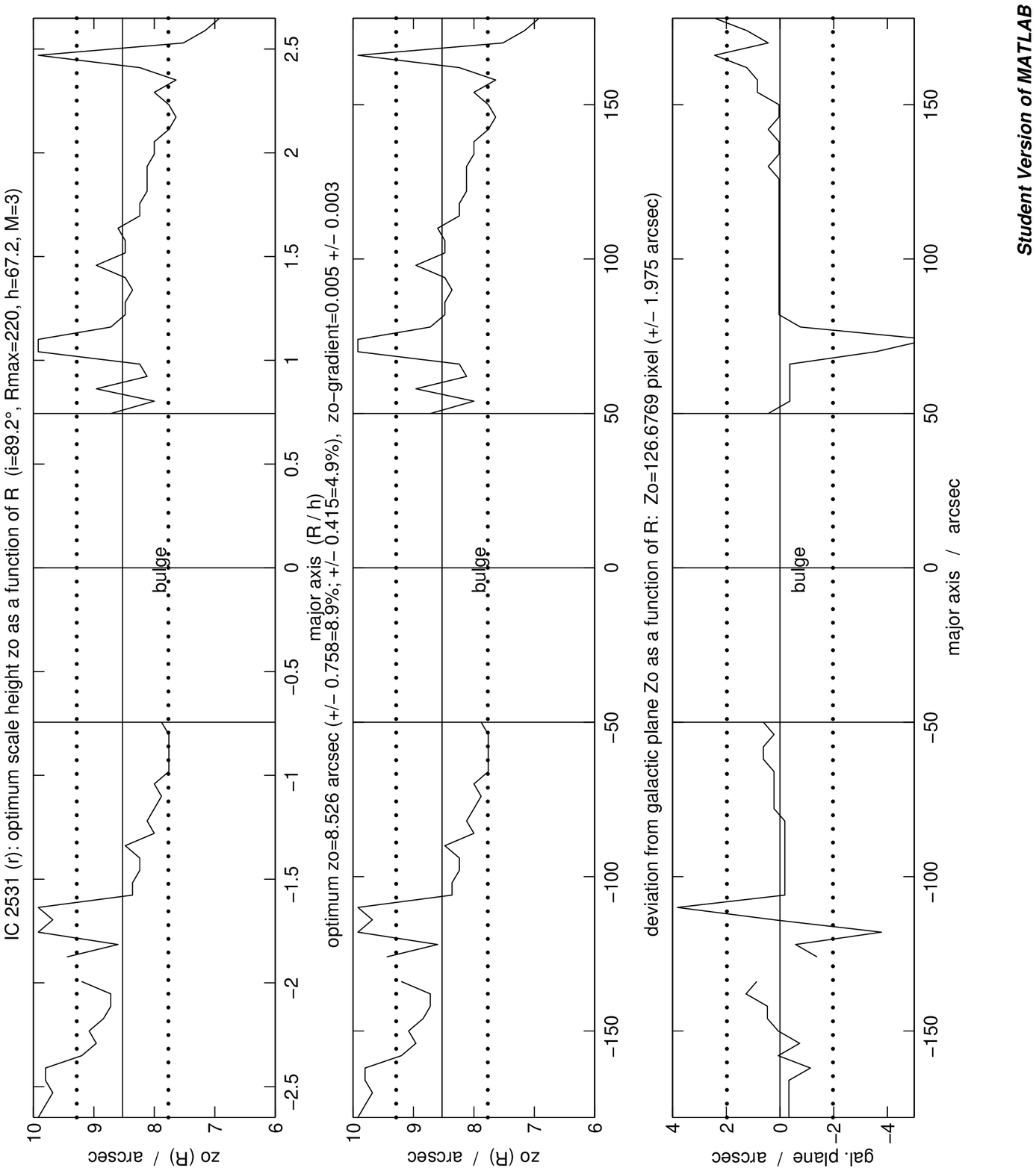}}
\end{picture}
\end{minipage}

\vspace*{98mm}

\hspace*{18mm}\parbox{165mm}{IC 2469  \hspace{45mm}  UGC 5341  \hspace{42mm}  IC 2531}

\vspace*{8mm}

\hspace*{8mm}\parbox{165mm}{
{\bf \noindent Appendix B.} (continued)
}
\end{figure*}

%%%%%%%%%%%%%%%%%%%%%%%%%%%%%%%%%%%%%%%%%%%%%%%%%%%%%%%%%%%%%%%%%%%

\clearpage

%%%%%%%%%%%%%%%%%%%%%%%%%%%%%  5  %%%%%%%%%%%%%%%%%%%%%%%%%%%%%%%%%

\begin{figure*}[t]
\vspace*{4mm}
\hspace*{5mm}
\begin{minipage}[b]{5.5cm}
\begin{picture}(3.0,3.0)
{\includegraphics[angle=180,viewport=40 50 400 730,clip,width=52mm]{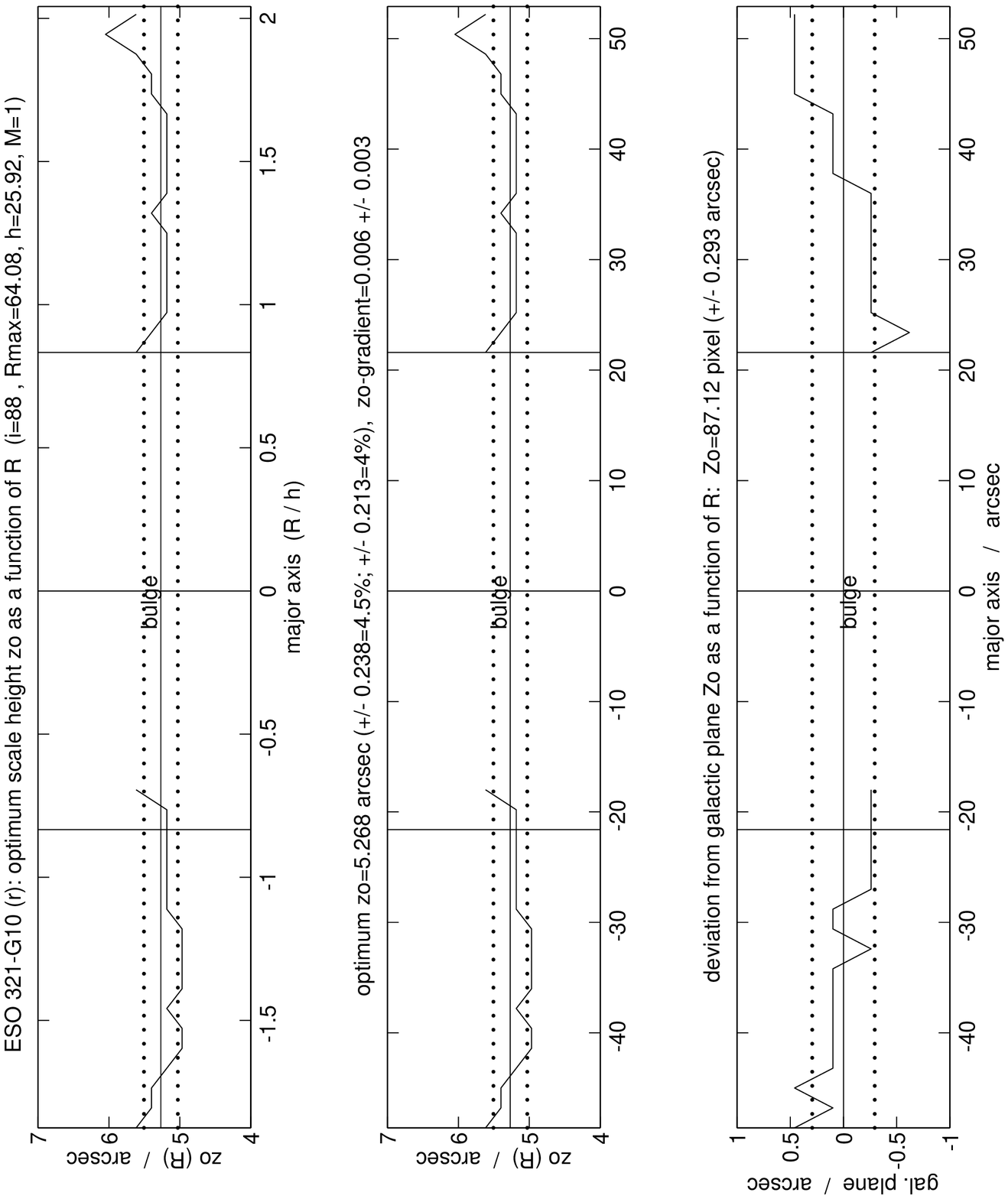}}
\end{picture}
\end{minipage}

\vspace*{95mm}
\hspace*{65mm}
\begin{minipage}[b]{5.5cm}
\begin{picture}(3.0,3.0)
{\includegraphics[angle=90,viewport=40 10 540 285,clip,width=54mm]{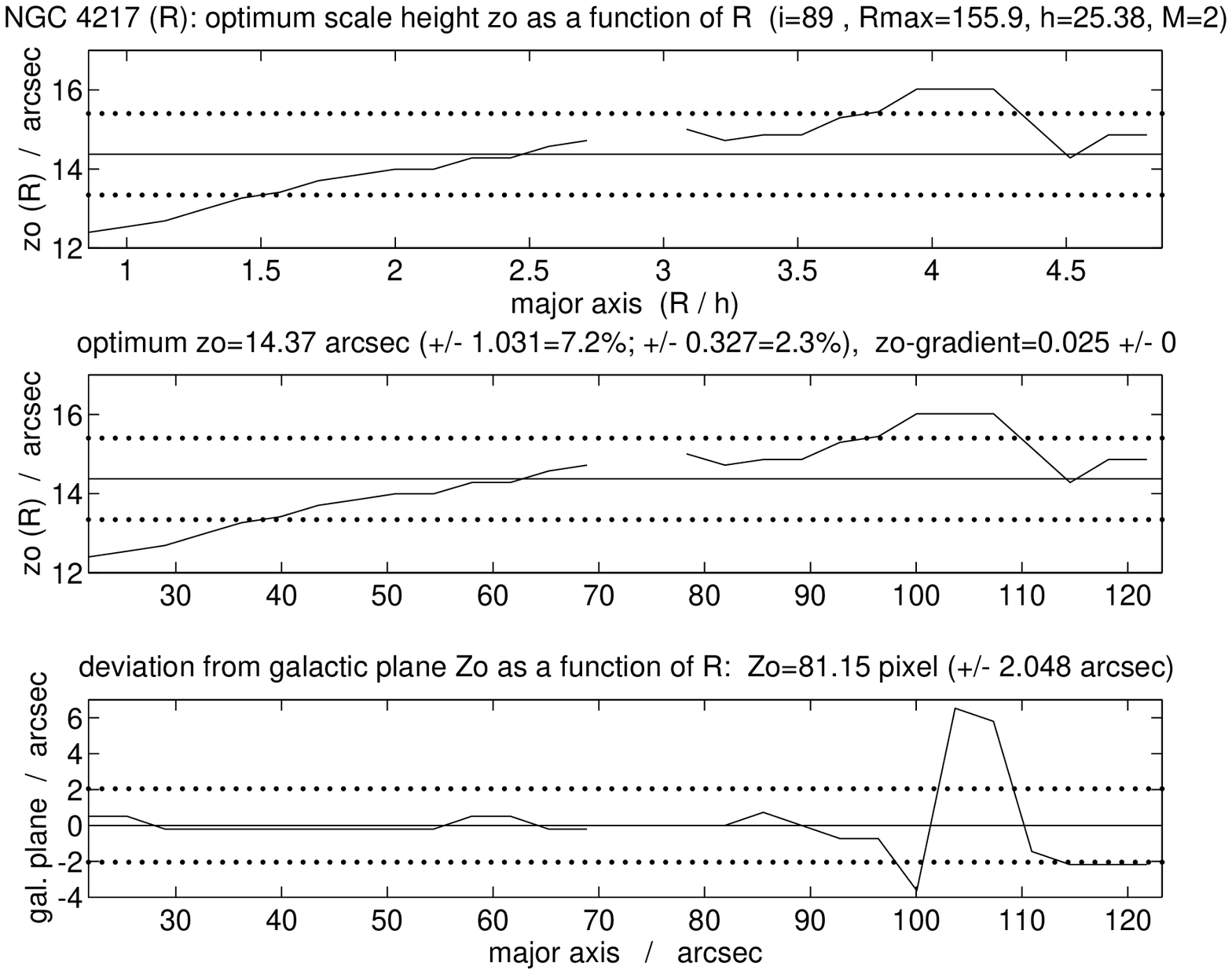}}
\end{picture}
\end{minipage}

\vspace*{-103mm}
\hspace*{125mm}
\begin{minipage}[b]{5.5cm}
\begin{picture}(3.0,3.0)
{\includegraphics[angle=180,viewport=00 -30 342 730,clip,width=50.5mm]{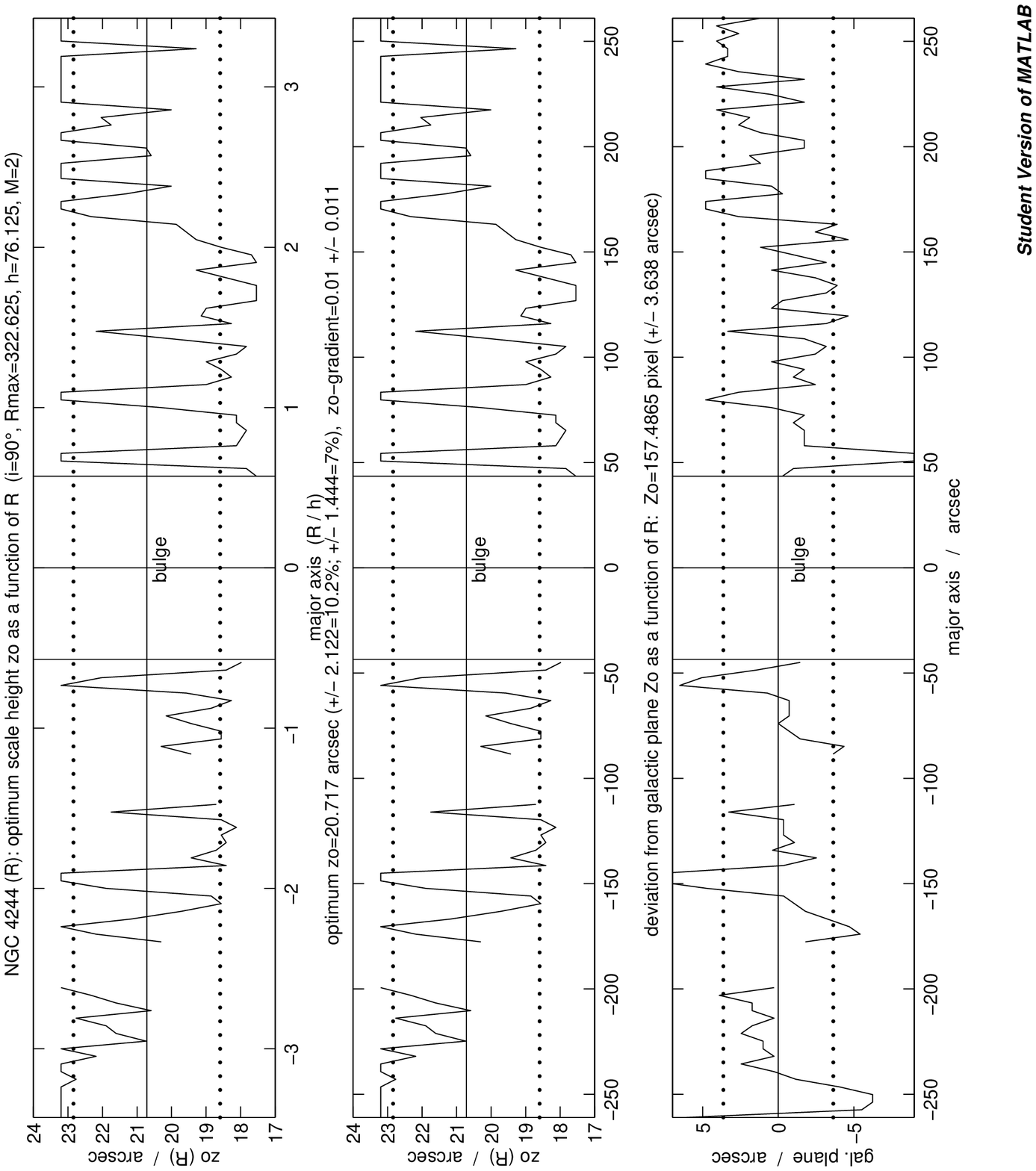}}
\end{picture}
\end{minipage}

\vspace*{99mm}

\hspace*{18mm}\parbox{165mm}{ESO 321-G10  \hspace{40mm}  NGC 4217  \hspace{40mm}  NGC 4244}

\vspace*{5mm}

\vspace*{96mm}
\hspace*{5mm}
\begin{minipage}[b]{5.5cm}
\begin{picture}(3.0,3.0)
{\includegraphics[angle=90,viewport=40 10 540 285,clip,width=54mm]{ngc4013.ps}}
\end{picture}
\end{minipage}

\vspace*{-101mm}
\hspace*{64mm}
\begin{minipage}[b]{5.5cm}
\begin{picture}(3.0,3.0)
{\includegraphics[angle=180,viewport=40 50 400 730,clip,width=52mm]{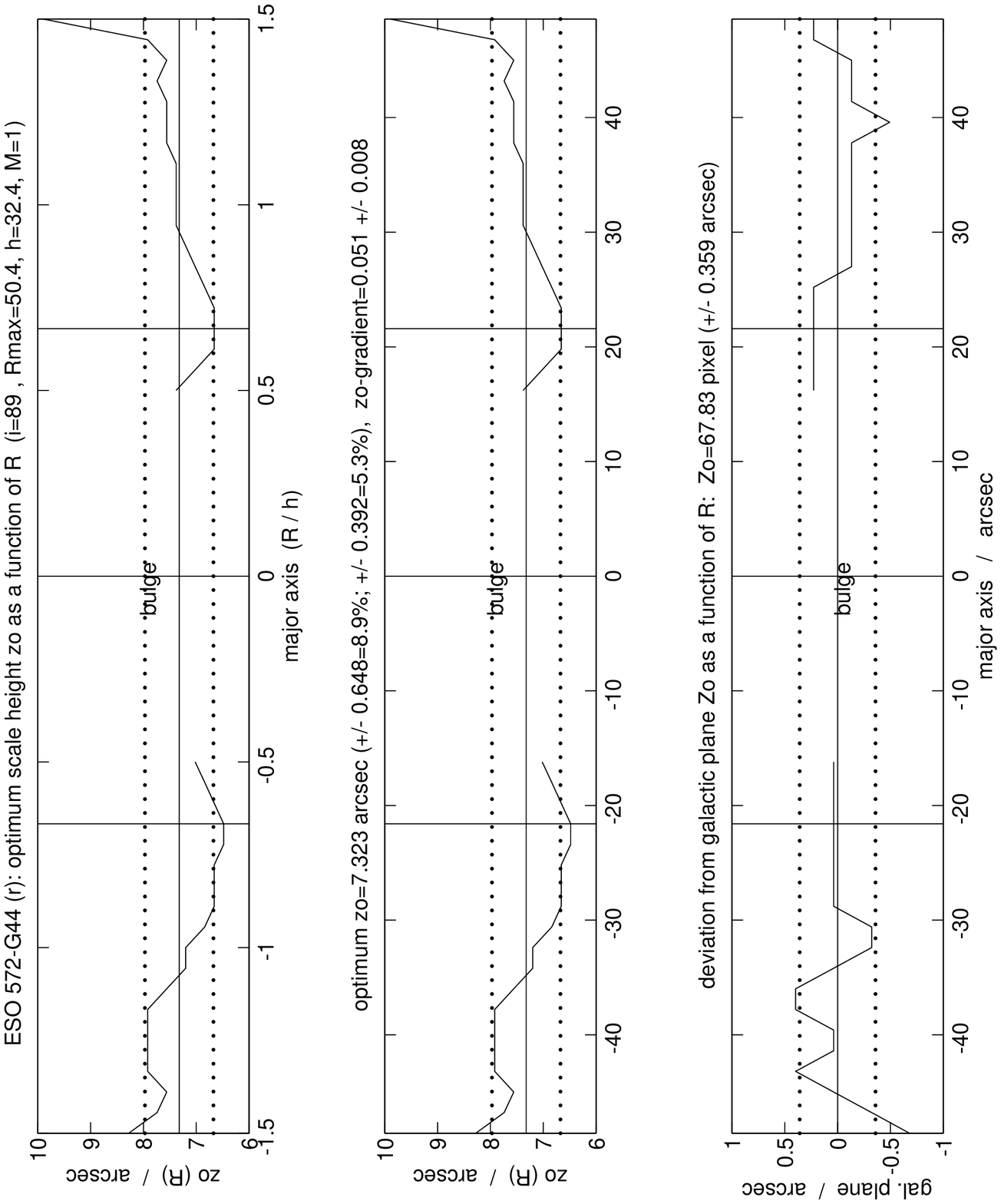}}
\end{picture}
\end{minipage}

\vspace*{-5mm}
\hspace*{125mm}
\begin{minipage}[b]{5.5cm}
\begin{picture}(3.0,3.0)
{\includegraphics[angle=180,viewport=00 -30 342 730,clip,width=50.5mm]{ugc7170.ps}}
\end{picture}
\end{minipage}

\vspace*{98mm}

\hspace*{18mm}\parbox{165mm}{NGC 4013  \hspace{40mm}  ESO 572-G44  \hspace{40mm}  UGC 7170}

\vspace*{8mm}

\hspace*{8mm}\parbox{165mm}{
{\bf \noindent Appendix B.} (continued)
}
\end{figure*}

%%%%%%%%%%%%%%%%%%%%%%%%%%%%%%%%%%%%%%%%%%%%%%%%%%%%%%%%%%%%%%%%%%%

\clearpage

%%%%%%%%%%%%%%%%%%%%%%%%%%%%%  6  %%%%%%%%%%%%%%%%%%%%%%%%%%%%%%%%%

\begin{figure*}[t]
\vspace*{100mm}
\hspace*{6mm}
\begin{minipage}[t]{5.5cm}
\begin{picture}(3.0,3.0)
{\includegraphics[angle=90,viewport=40 10 540 285,clip,width=53mm]{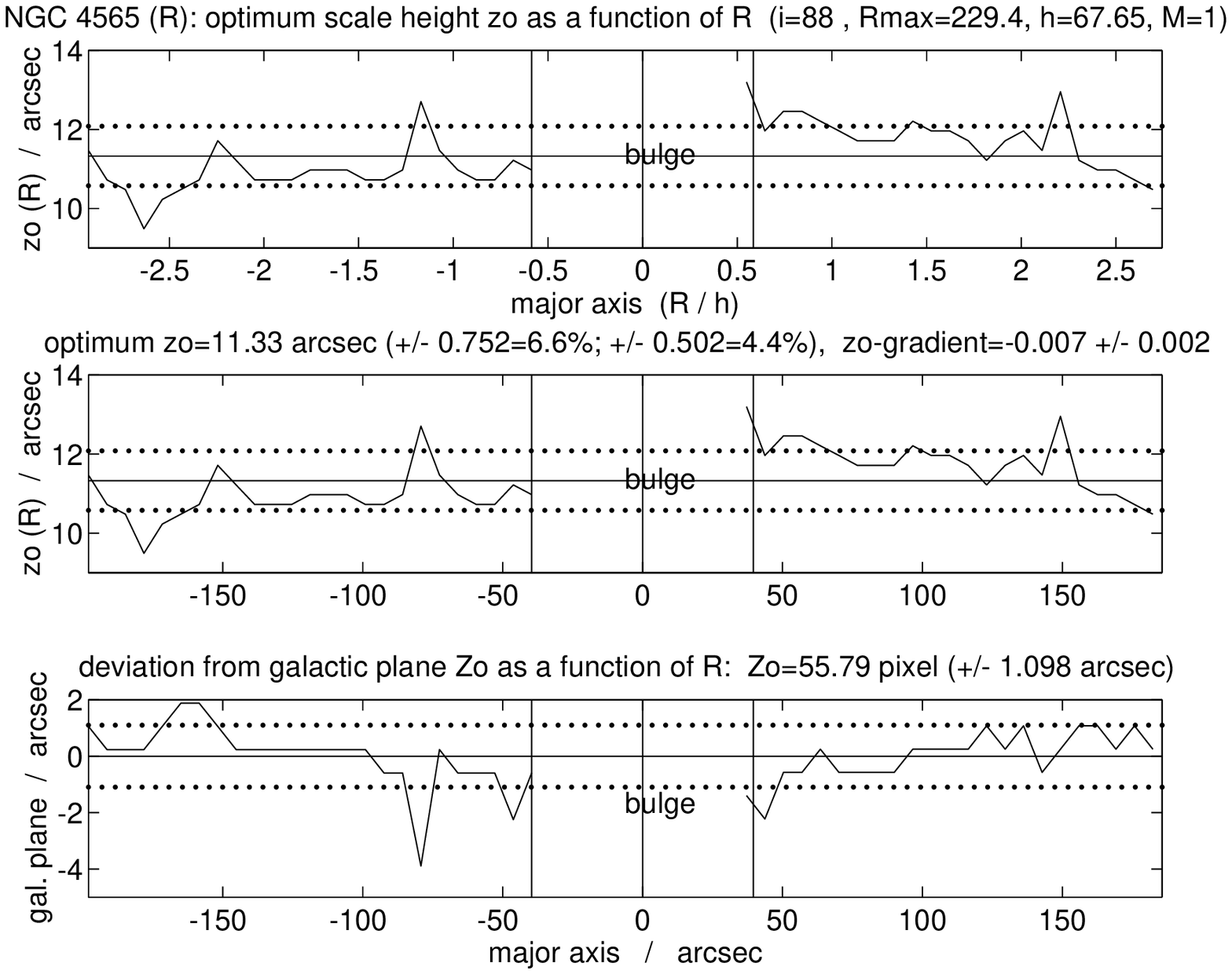}}
\end{picture}
\end{minipage}

\vspace*{-4mm}
\hspace*{65mm}
\begin{minipage}[t]{5.5cm}
\begin{picture}(3.0,3.0)
{\includegraphics[angle=90,viewport=40 10 540 285,clip,width=53mm]{ngc4710.ps}}
\end{picture}
\end{minipage}

\vspace*{-101mm}
\hspace*{125mm}
\begin{minipage}[t]{5.5cm}
\begin{picture}(3.0,3.0)
{\includegraphics[angle=180,viewport=00 -30 342 730,clip,width=50.5mm]{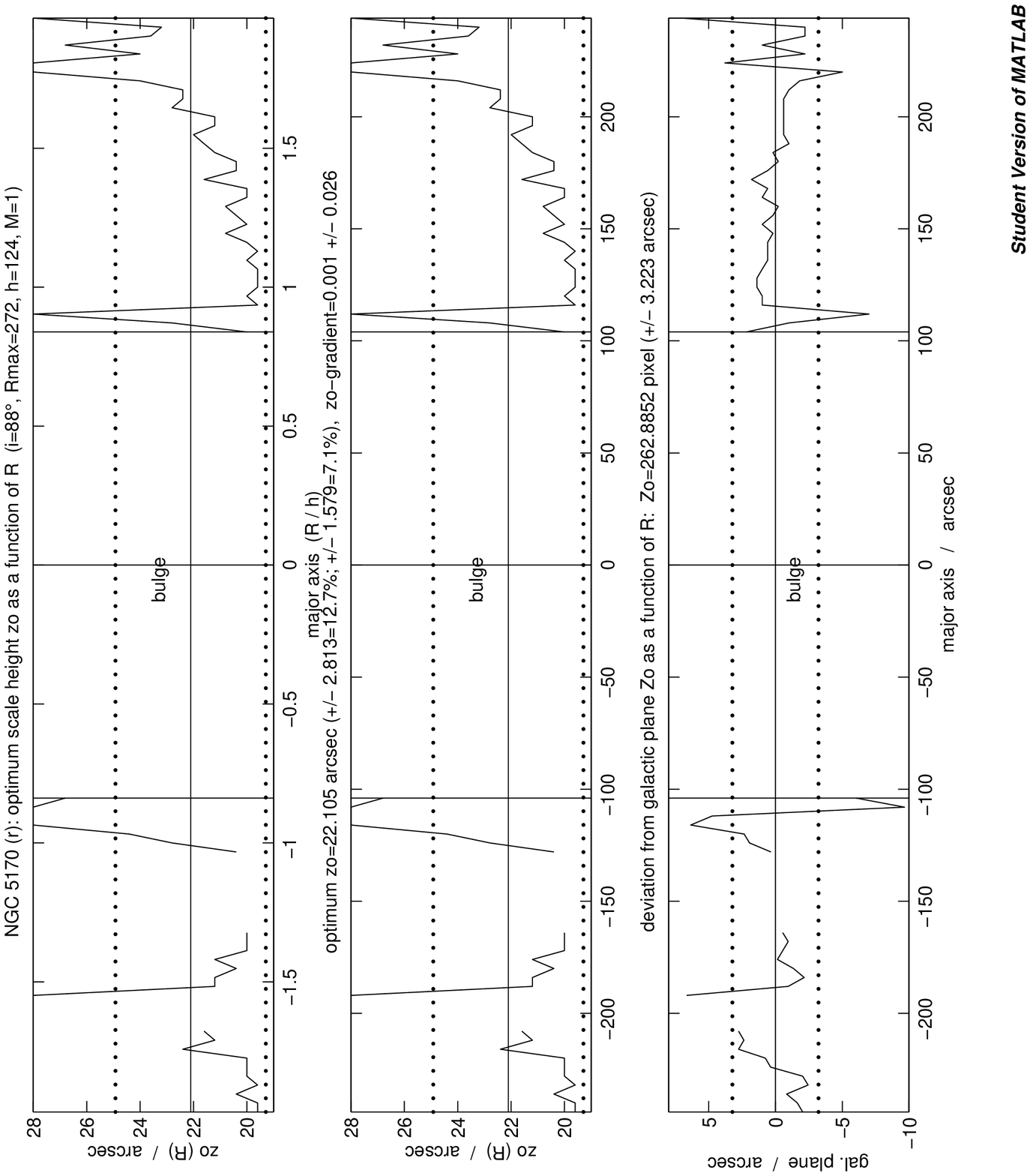}}
\end{picture}
\end{minipage}

\vspace*{98mm}

\hspace*{18mm}\parbox{165mm}{NGC 4565  \hspace{45mm}  NGC 4710  \hspace{42mm}  NGC 5170}

\vspace*{5mm}

\hspace*{7mm}
\begin{minipage}[b]{5.5cm}
\begin{picture}(3.0,3.0)
{\includegraphics[angle=180,viewport=00 -30 342 730,clip,width=50.5mm]{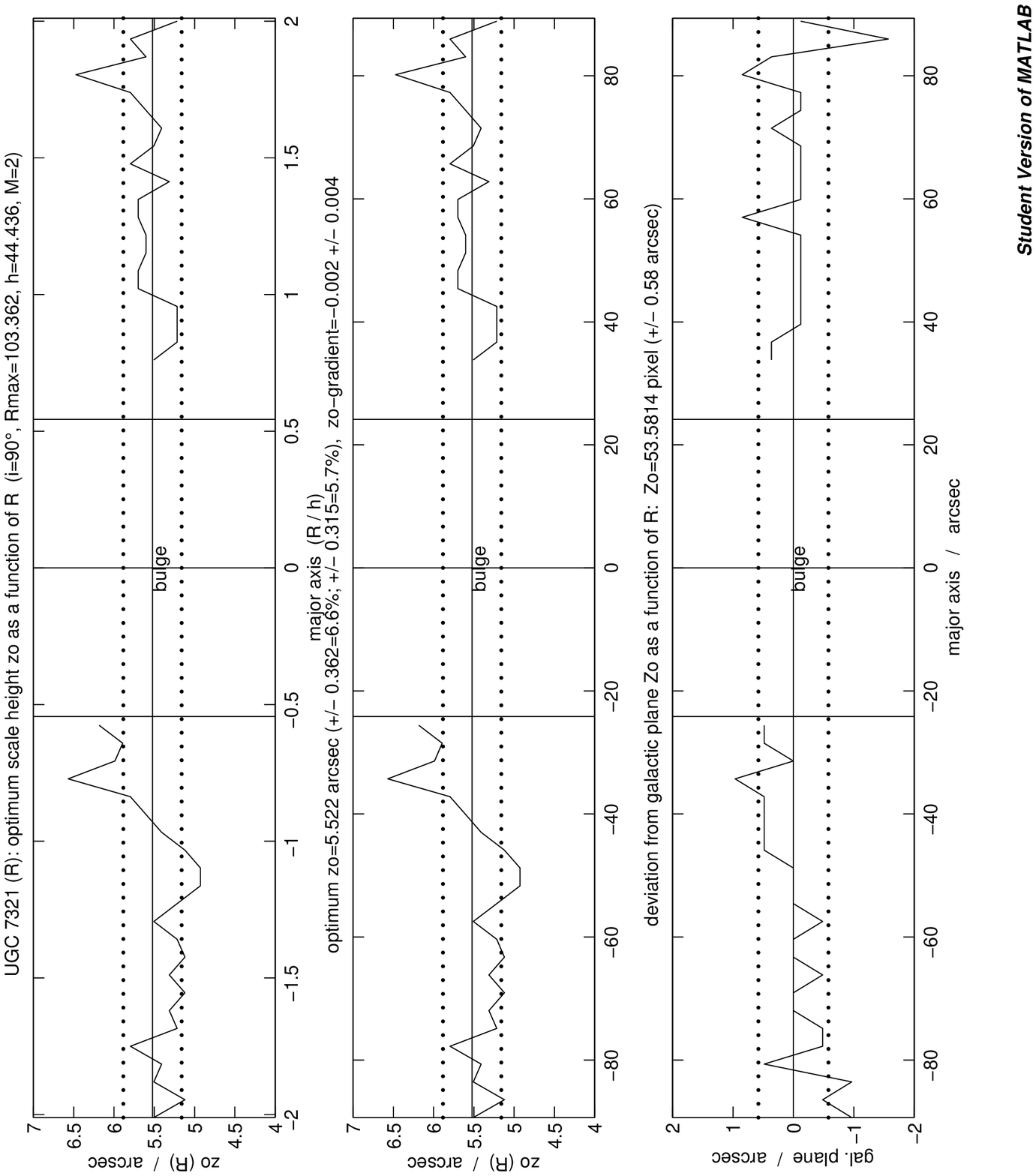}}
\end{picture}
\end{minipage}

\vspace*{95mm}
\hspace*{65mm}
\begin{minipage}[b]{5.5cm}
\begin{picture}(3.0,3.0)
{\includegraphics[angle=90,viewport=40 10 540 285,clip,width=53mm]{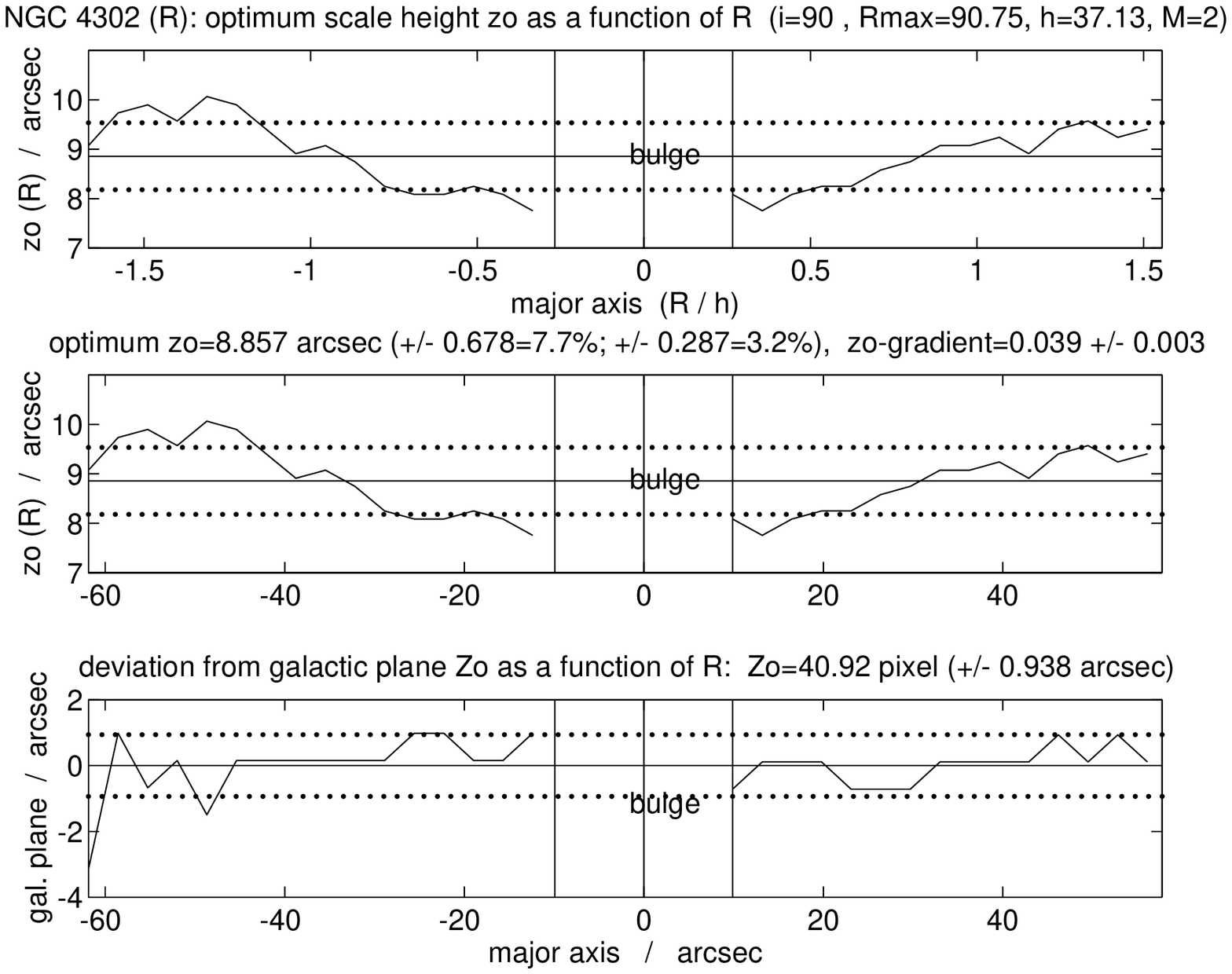}}
\end{picture}
\end{minipage}

\vspace*{-102mm}
\hspace*{125mm}
\begin{minipage}[b]{5.5cm}
\begin{picture}(3.0,3.0)
{\includegraphics[angle=180,viewport=00 -30 342 730,clip,width=50.5mm]{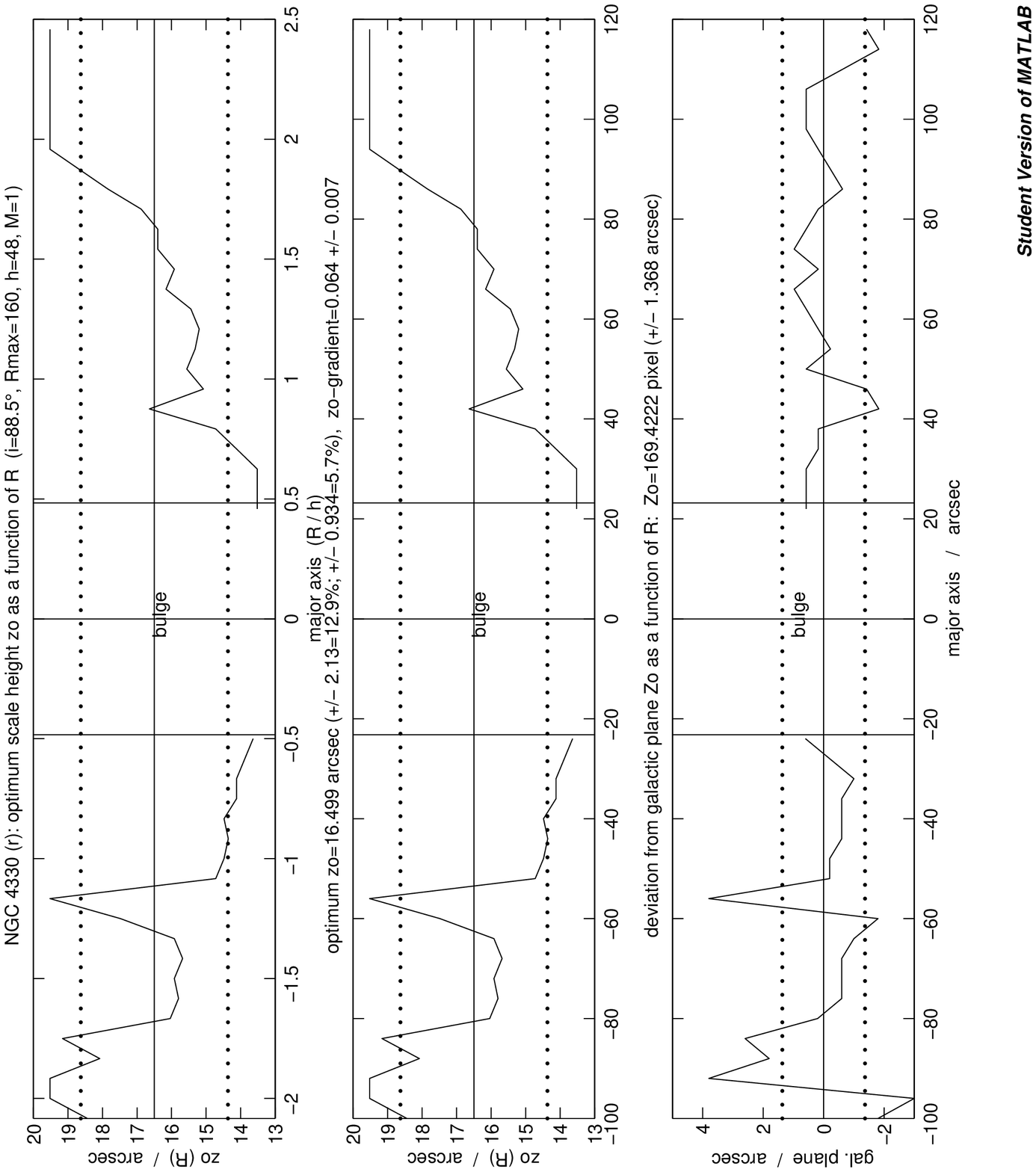}}
\end{picture}
\end{minipage}

\vspace*{98mm}

\hspace*{18mm}\parbox{165mm}{UGC 7321  \hspace{45mm}  NGC 4302  \hspace{42mm}  NGC 4330}

\vspace*{8mm}

\hspace*{8mm}\parbox{165mm}{
{\bf \noindent Appendix B.} (continued)
}
\end{figure*}

%%%%%%%%%%%%%%%%%%%%%%%%%%%%%%%%%%%%%%%%%%%%%%%%%%%%%%%%%%%%%%%%%%%

\clearpage

%%%%%%%%%%%%%%%%%%%%%%%%%%%%%  7  %%%%%%%%%%%%%%%%%%%%%%%%%%%%%%%%%

\begin{figure*}[t]
\vspace*{3mm}
\hspace*{5mm}
\begin{minipage}[b]{5.5cm}
\begin{picture}(3.0,3.0)
{\includegraphics[angle=180,viewport=00 -30 342 730,clip,width=50.5mm]{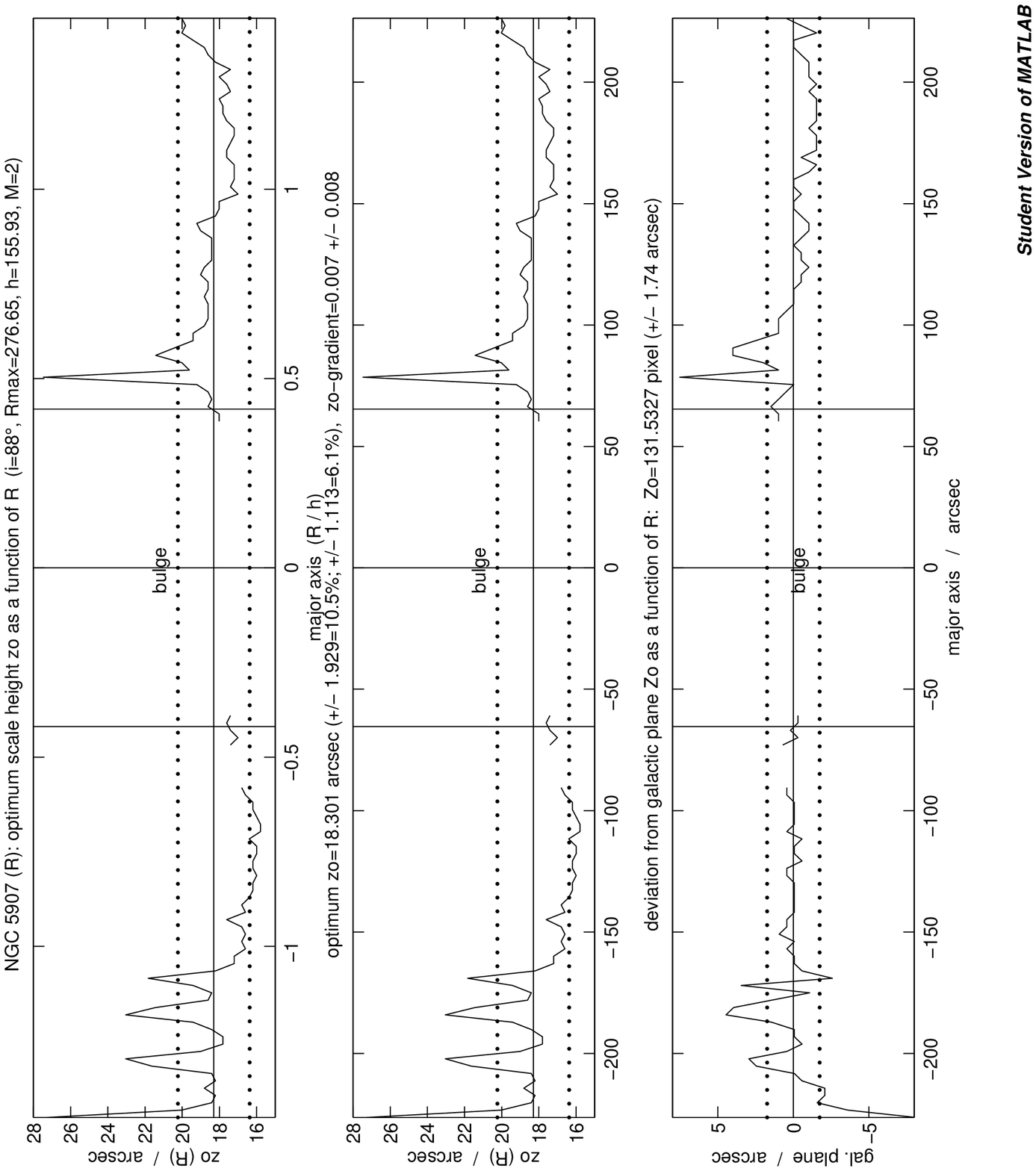}}
\end{picture}
\end{minipage}

\vspace*{95mm}
\hspace*{64mm}
\begin{minipage}[b]{5.5cm}
\begin{picture}(3.0,3.0)
{\includegraphics[angle=90,viewport=40 10 540 285,clip,width=53mm]{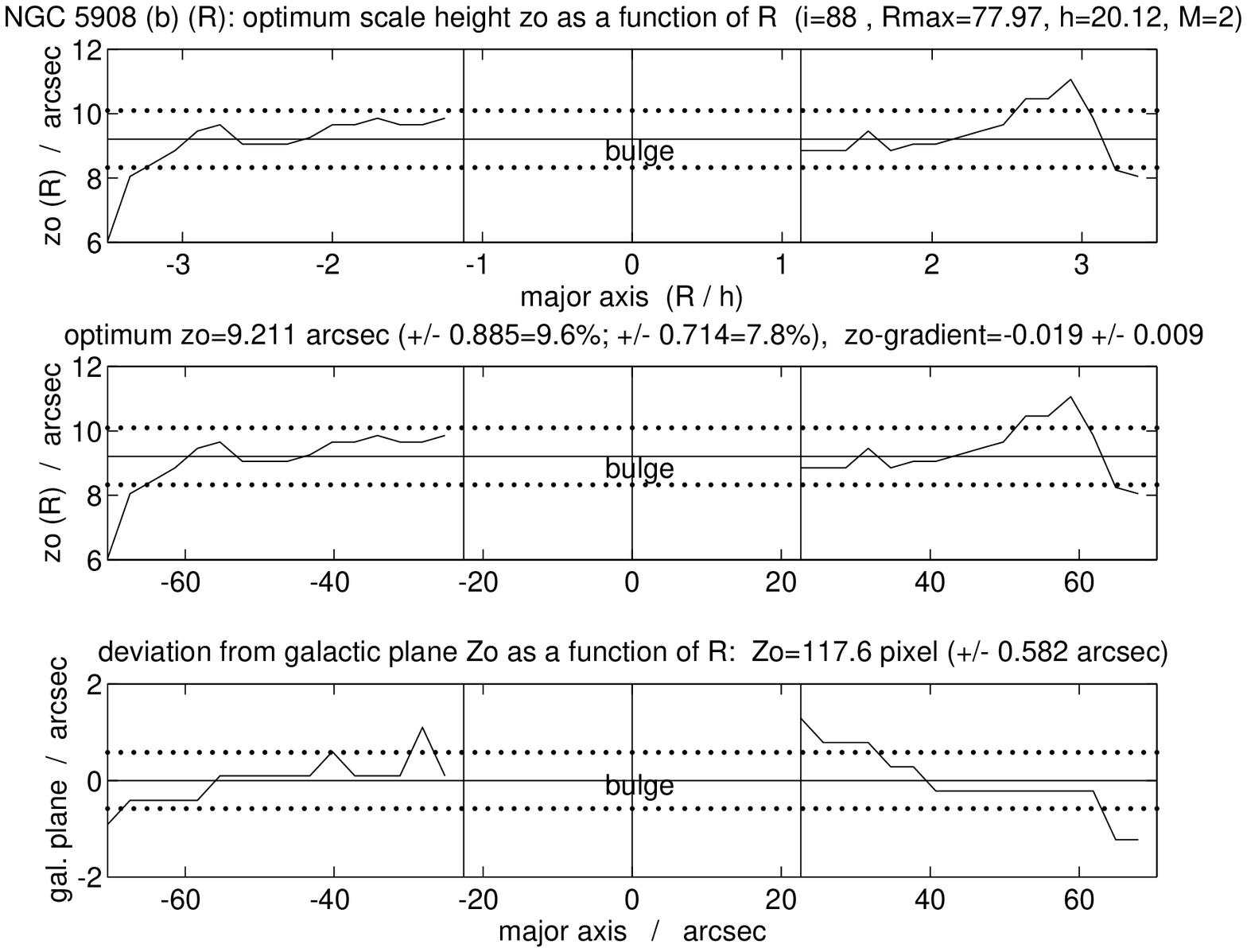}}
\end{picture}
\end{minipage}

\vspace*{-102mm}
\hspace*{125mm}
\begin{minipage}[b]{5.5cm}
\begin{picture}(3.0,3.0)
{\includegraphics[angle=180,viewport=40 50 400 730,clip,width=52mm]{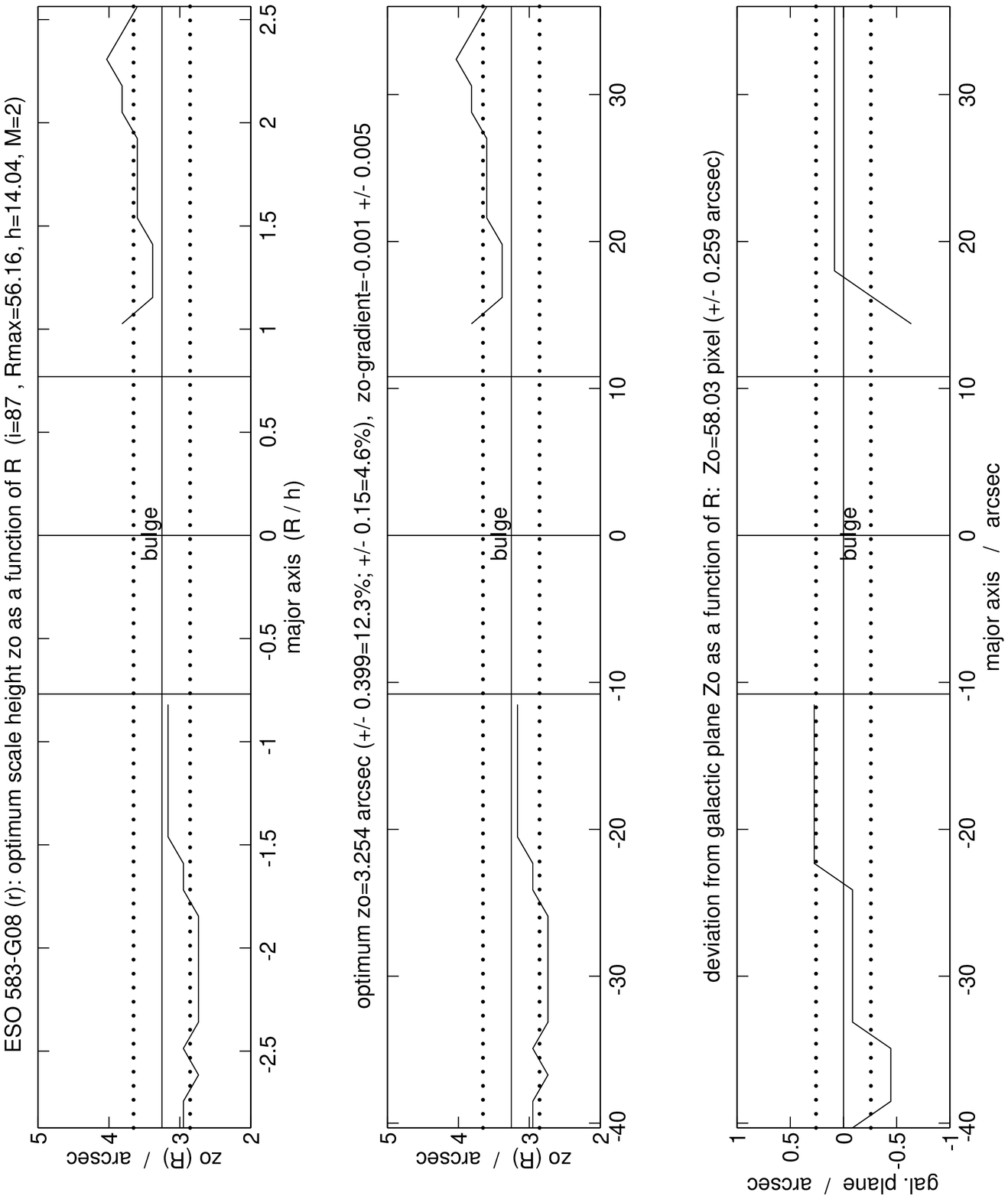}}
\end{picture}
\end{minipage}

\vspace*{98mm}

\hspace*{18mm}\parbox{165mm}{NGC 5907  \hspace{45mm}  NGC 5908  \hspace{42mm}  ESO 583-G08}

\vspace*{5mm}

\hspace*{5mm}
\begin{minipage}[b]{5.5cm}
\begin{picture}(3.0,3.0)
{\includegraphics[angle=180,viewport=40 50 400 730,clip,width=52mm]{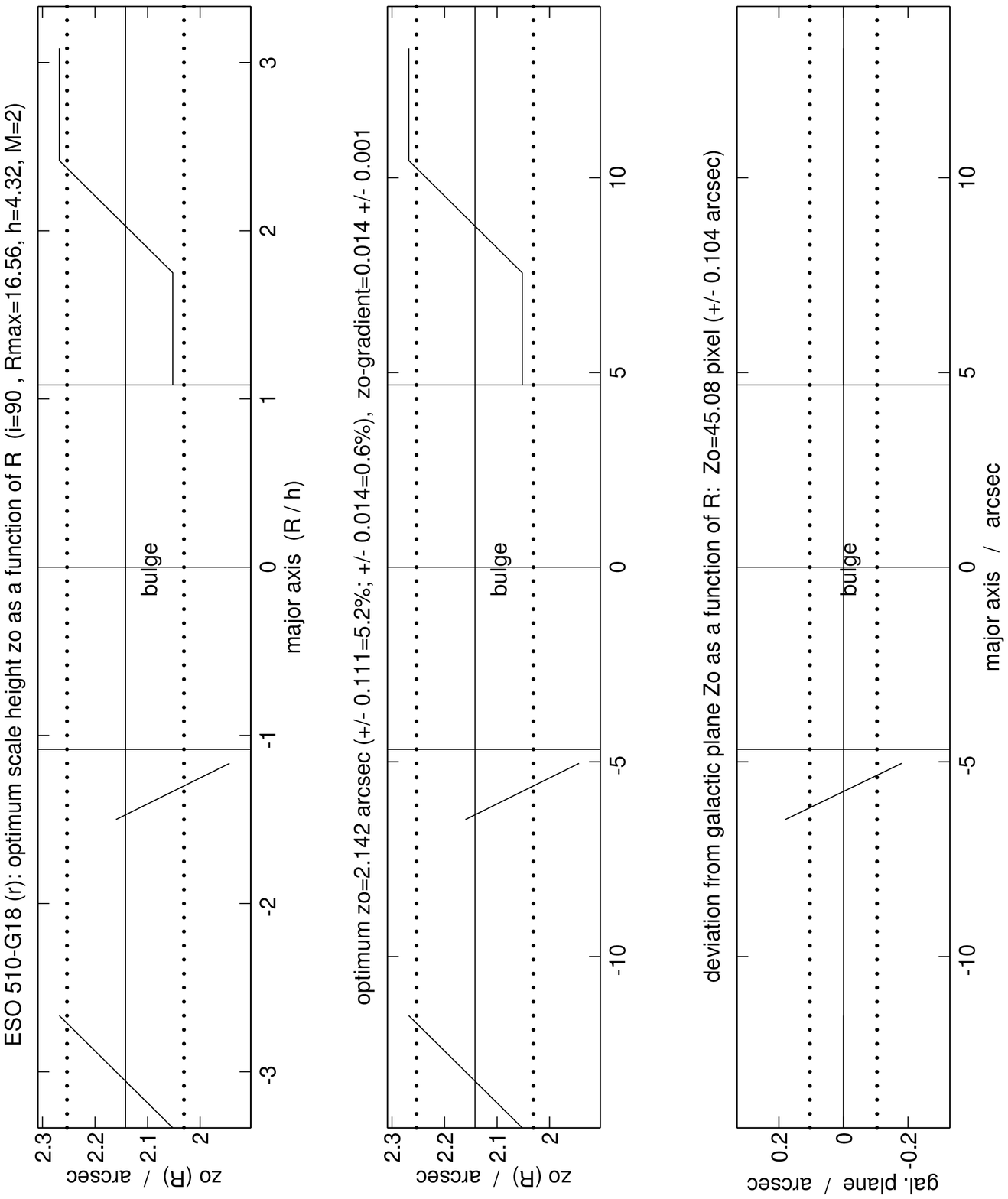}}
\end{picture}
\end{minipage}
\hfill
\begin{minipage}[b]{5.5cm}
\begin{picture}(3.0,3.0)
{\includegraphics[angle=180,viewport=00 -30 342 730,clip,width=50.5mm]{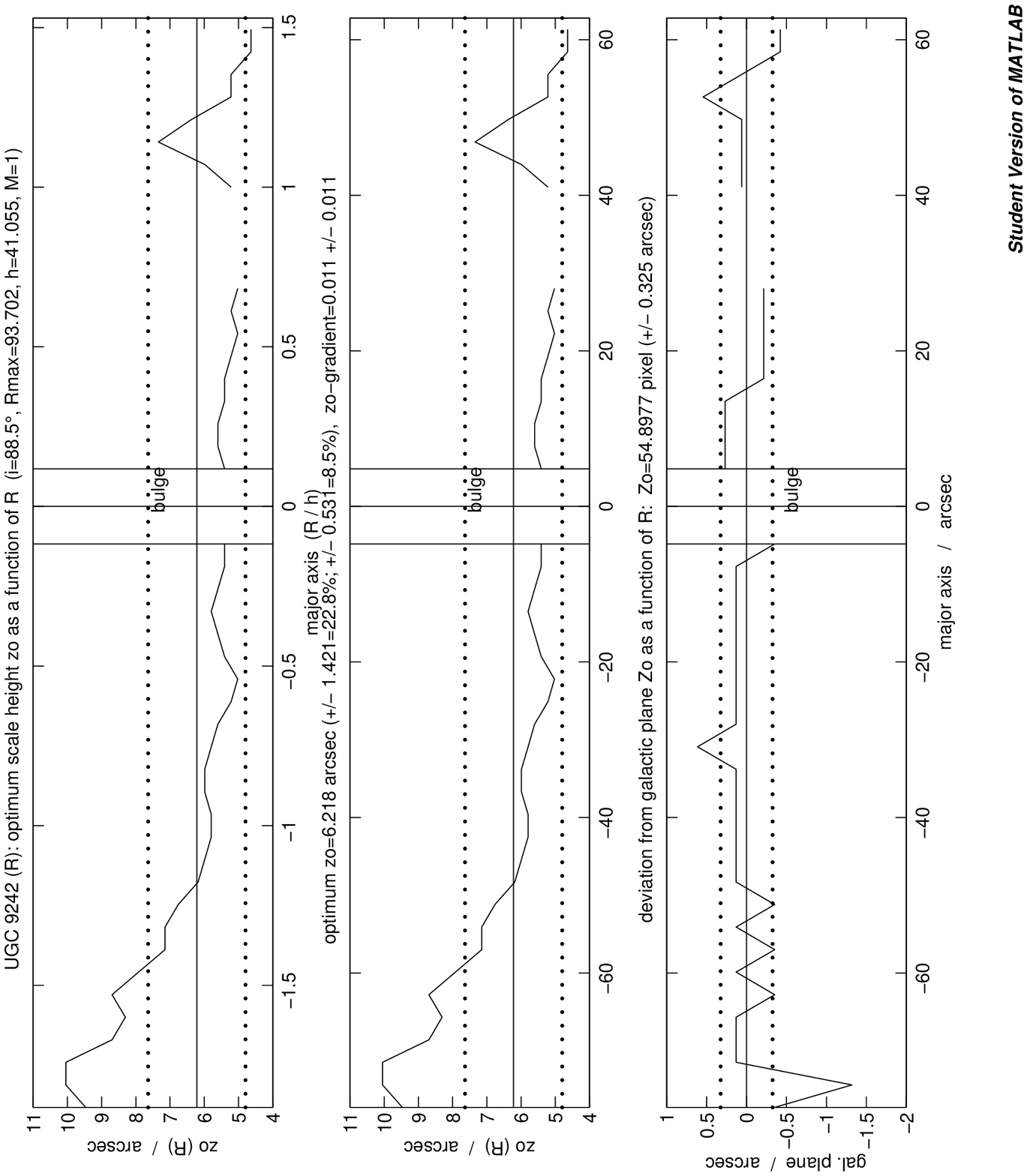}}
\end{picture}
\end{minipage}
\hfill
\begin{minipage}[b]{5.5cm}
\begin{picture}(3.0,3.0)
{\includegraphics[angle=180,viewport=40 50 400 730,clip,width=52mm]{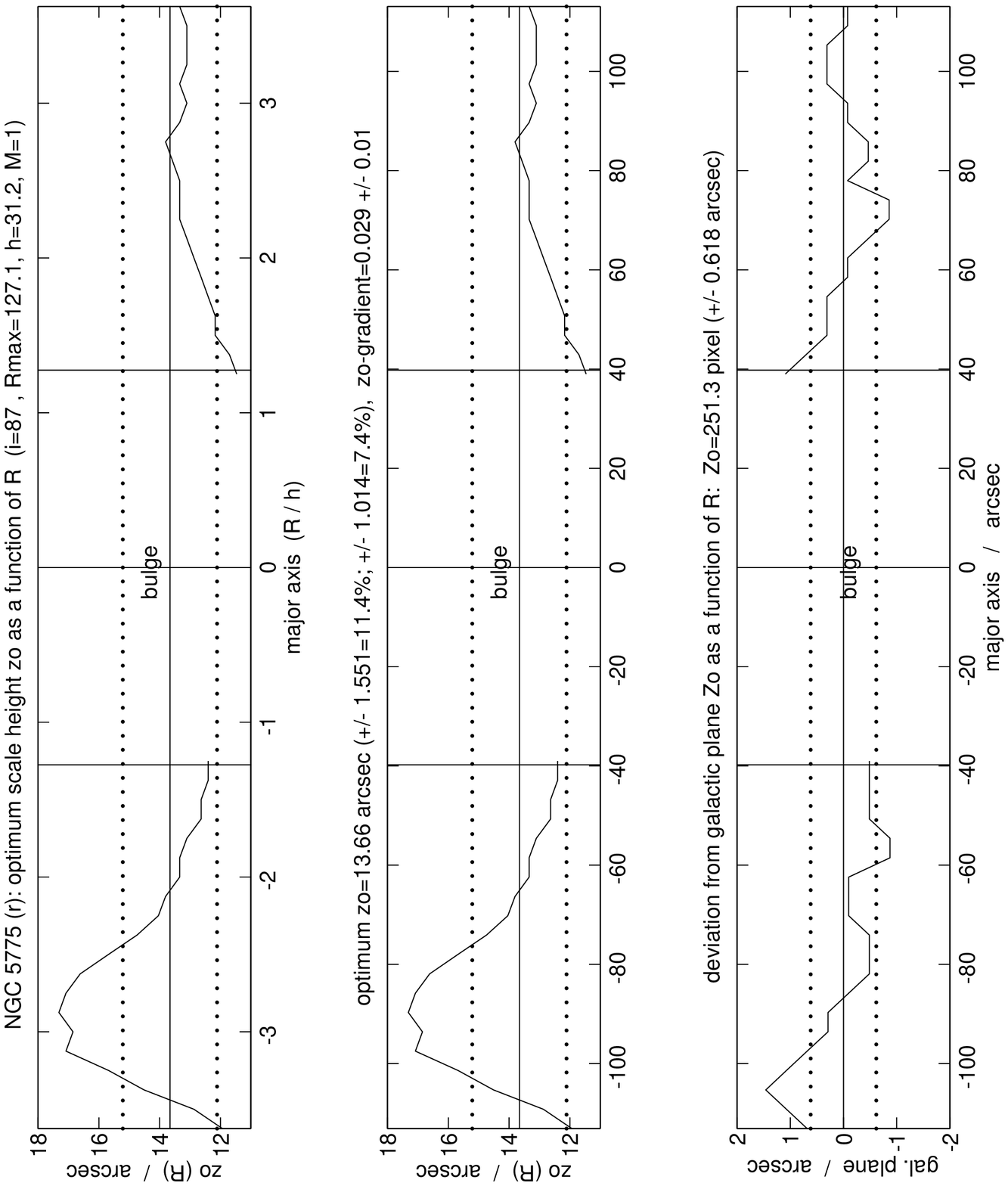}}
\end{picture}
\end{minipage}

\vspace*{98mm}

\hspace*{18mm}\parbox{165mm}{ESO 510-G18  \hspace{40mm}  UGC 9242  \hspace{40mm}  NGC 5775}

\vspace*{8mm}

\hspace*{8mm}\parbox{165mm}{
{\bf \noindent Appendix B.} (continued)
}
\end{figure*}

%%%%%%%%%%%%%%%%%%%%%%%%%%%%%%%%%%%%%%%%%%%%%%%%%%%%%%%%%%%%%%%%%%%

\clearpage

%%%%%%%%%%%%%%%%%%%%%%%%%%%%%  8  %%%%%%%%%%%%%%%%%%%%%%%%%%%%%%%%%

\begin{figure*}[t]
\vspace*{3mm}
\hspace*{5mm}
\begin{minipage}[b]{5.5cm}
\begin{picture}(3.0,3.0)
{\includegraphics[angle=180,viewport=40 50 400 730,clip,width=52mm]{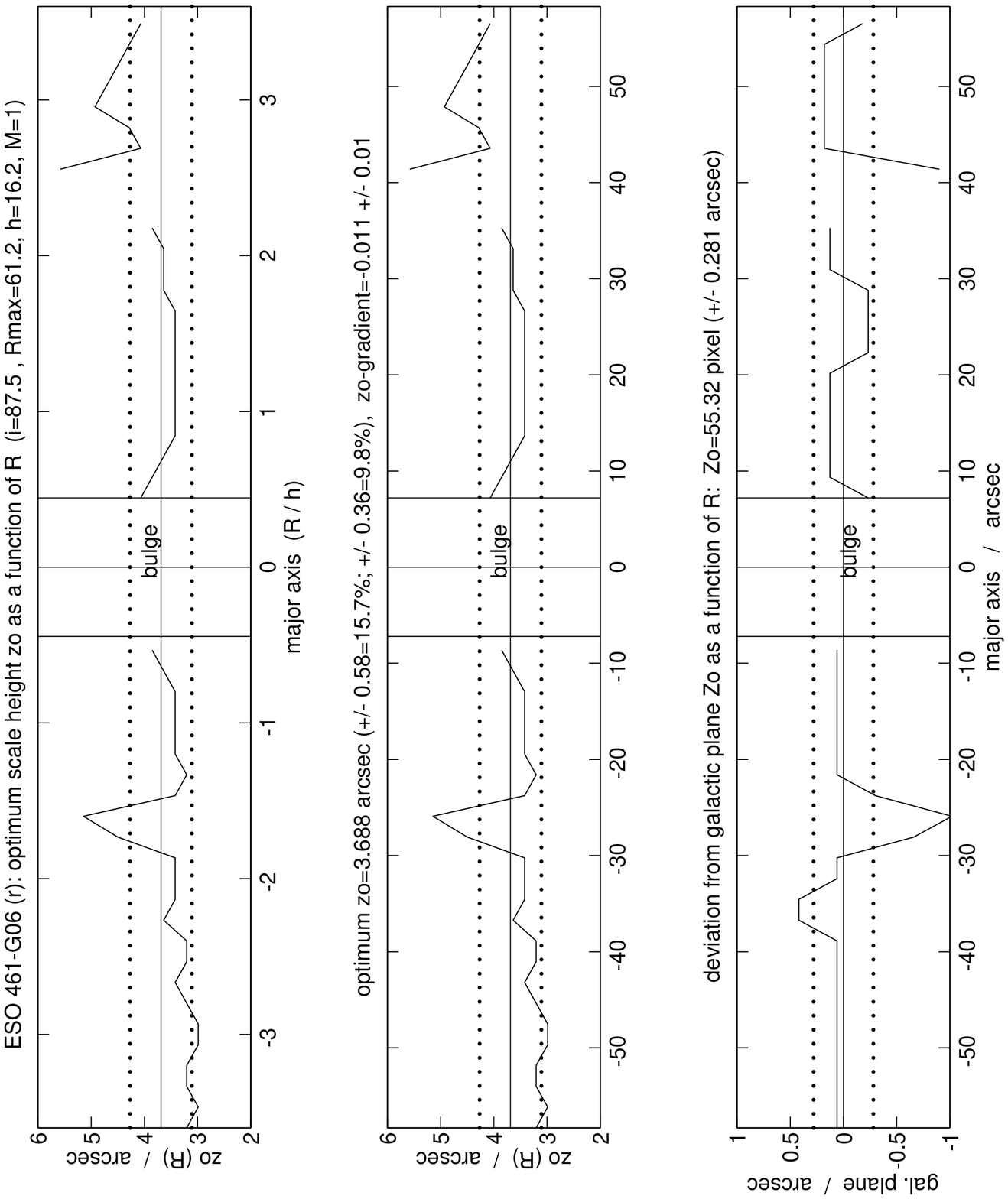}}
\end{picture}
\end{minipage}
\hfill
\begin{minipage}[b]{5.5cm}
\begin{picture}(3.0,3.0)
{\includegraphics[angle=180,viewport=40 50 400 730,clip,width=52mm]{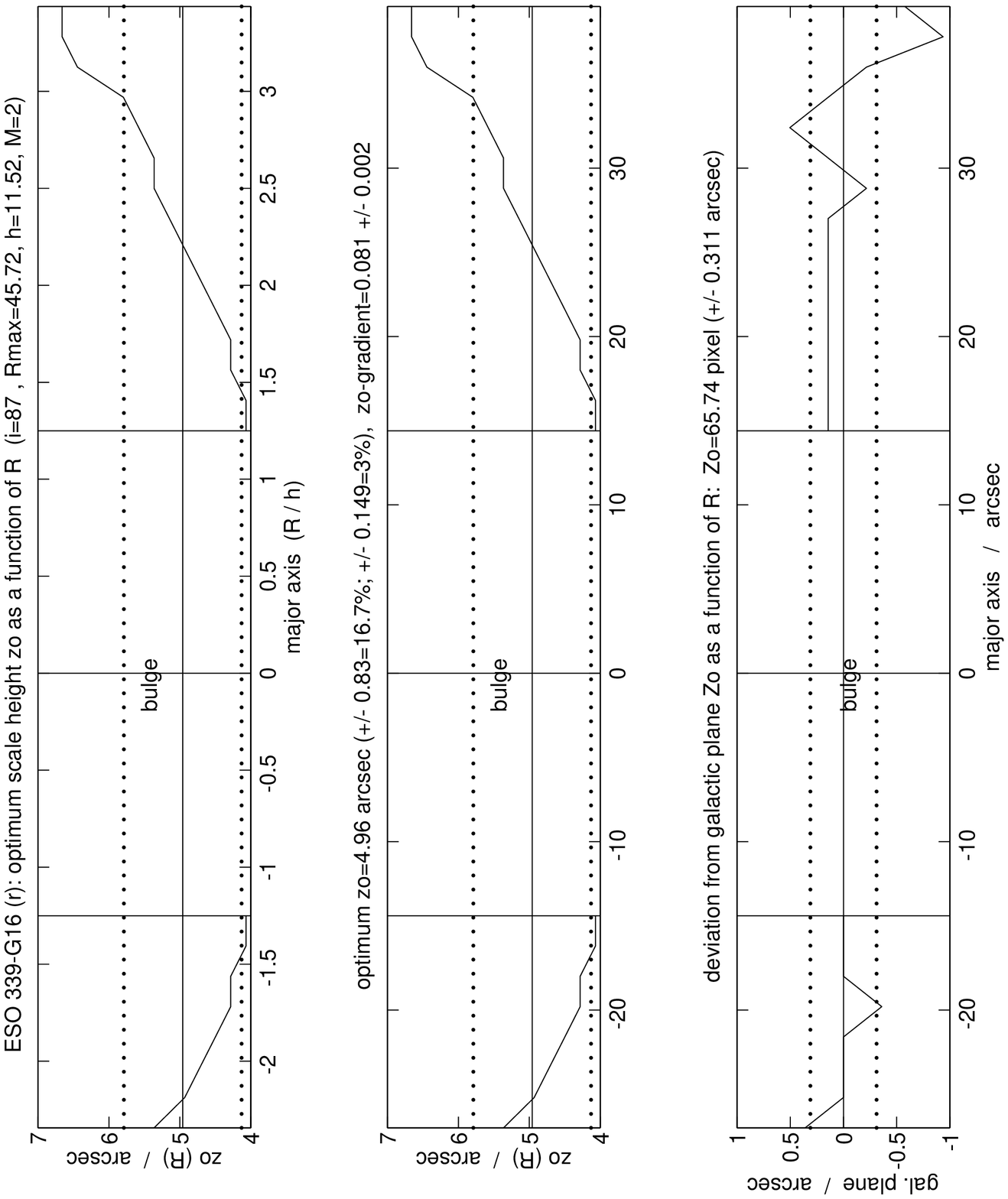}}
\end{picture}
\end{minipage}
\hfill
\begin{minipage}[b]{5.5cm}
\begin{picture}(3.0,3.0)
{\includegraphics[angle=180,viewport=40 50 400 730,clip,width=52mm]{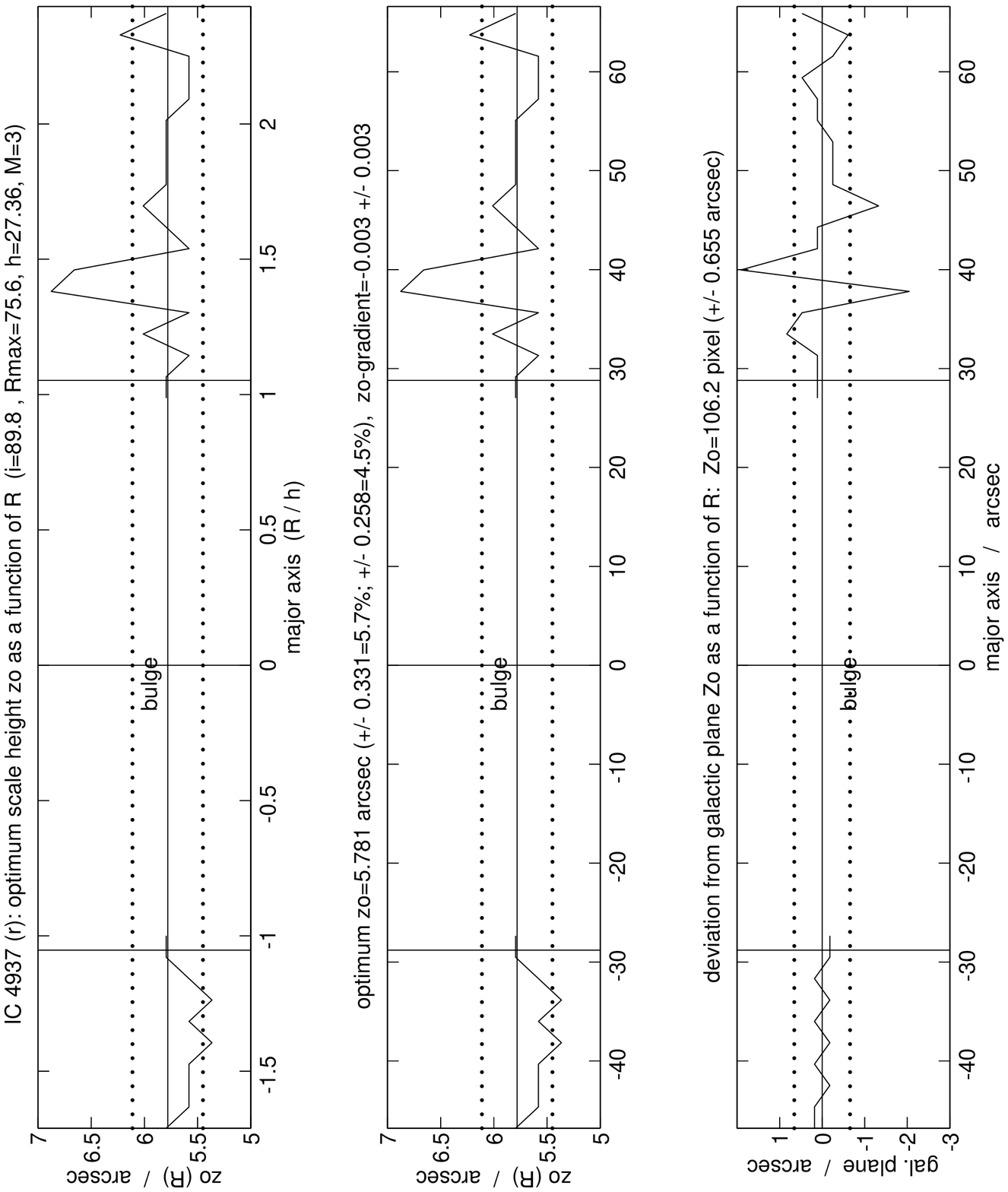}}
\end{picture}
\end{minipage}

\vspace*{98mm}

\hspace*{18mm}\parbox{165mm}{ESO 461-G06  \hspace{40mm}  ESO 339-G16  \hspace{40mm}  IC 4937}

\vspace*{102mm}
\hspace*{5mm}
\begin{minipage}[b]{5.5cm}
\begin{picture}(3.0,3.0)
{\includegraphics[angle=90,viewport=40 10 540 285,clip,width=54mm]{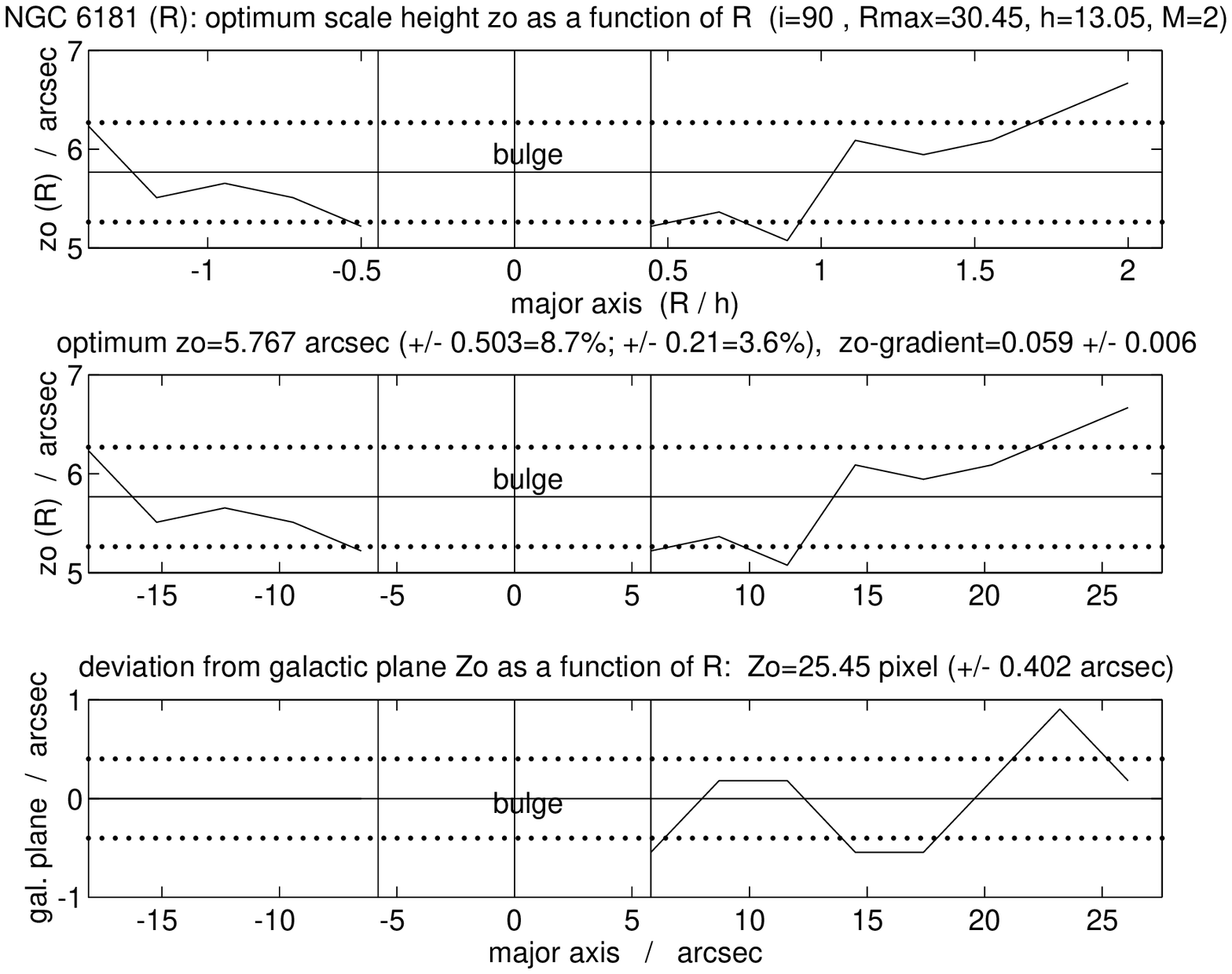}}
\end{picture}
\end{minipage}

\vspace*{-101mm}
\hspace*{64mm}
\begin{minipage}[b]{5.5cm}
\begin{picture}(3.0,3.0)
{\includegraphics[angle=180,viewport=40 50 400 730,clip,width=52mm]{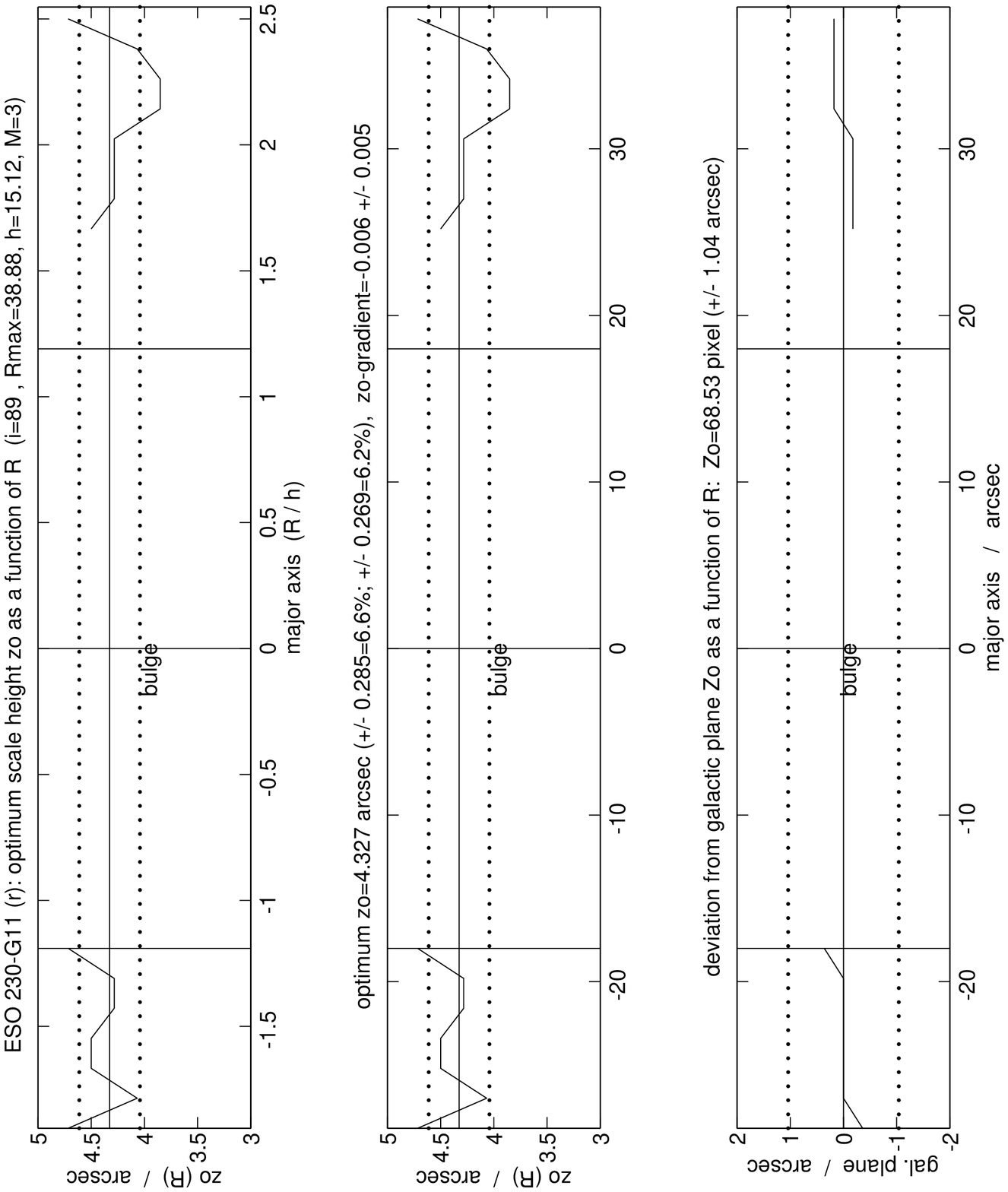}}
\end{picture}
\end{minipage}

\vspace*{-4mm}
\hspace*{125mm}
\begin{minipage}[b]{5.5cm}
\begin{picture}(3.0,3.0)
{\includegraphics[angle=180,viewport=40 50 400 730,clip,width=52mm]{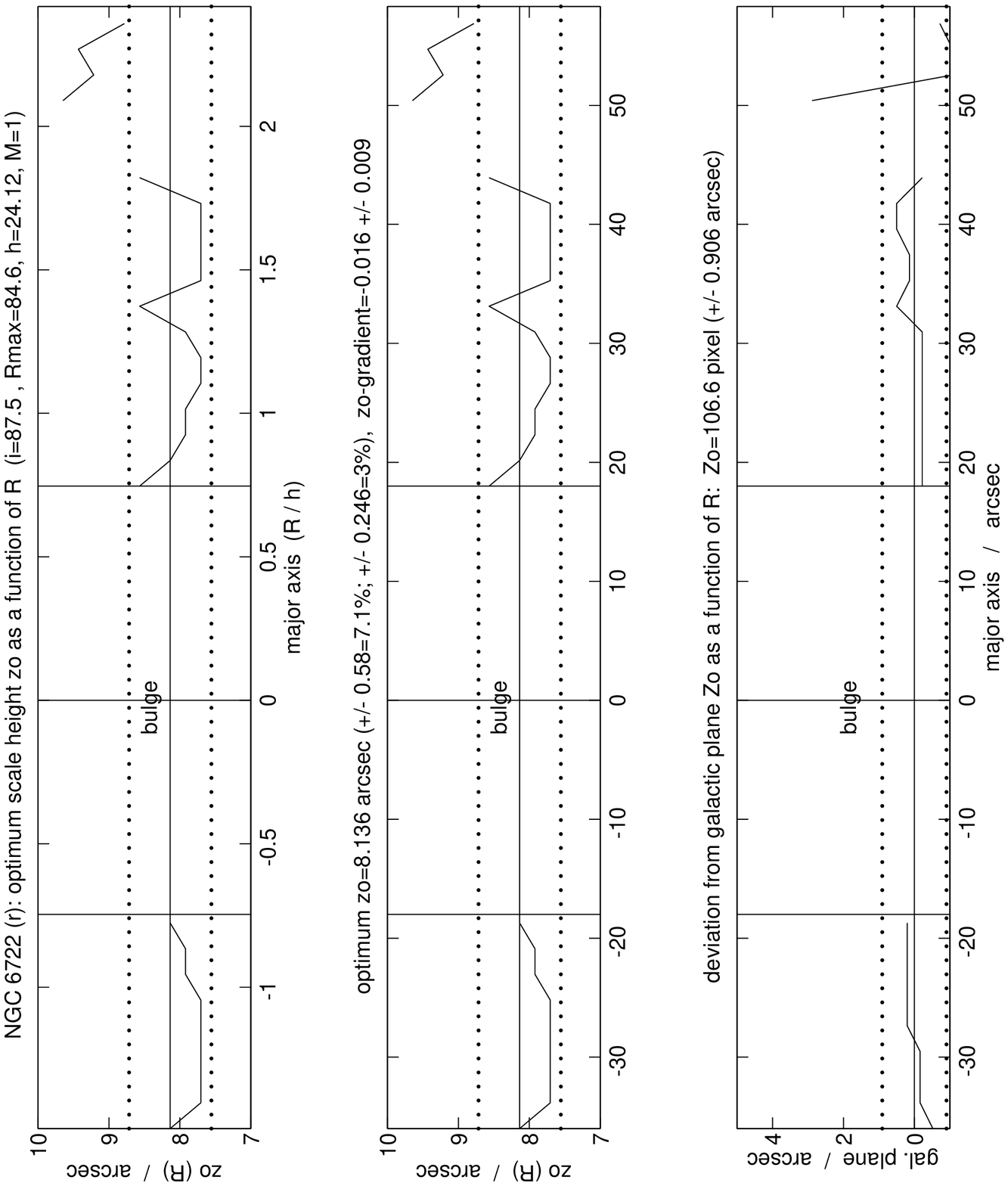}}
\end{picture}
\end{minipage}

\vspace*{98mm}

\hspace*{18mm}\parbox{165mm}{NGC 6181  \hspace{40mm}  ESO 230-G11  \hspace{40mm}  NGC 6722}

\vspace*{8mm}

\hspace*{8mm}\parbox{165mm}{
{\bf \noindent Appendix B.} (continued)
}
\end{figure*}

%%%%%%%%%%%%%%%%%%%%%%%%%%%%%%%%%%%%%%%%%%%%%%%%%%%%%%%%%%%%%%%%%%%

\clearpage

%%%%%%%%%%%%%%%%%%%%%%%%%%%%%  9  %%%%%%%%%%%%%%%%%%%%%%%%%%%%%%%%%

\begin{figure*}[t]
\vspace*{3mm}
\hspace*{5mm}
\begin{minipage}[b]{5.5cm}
\begin{picture}(3.0,3.0)
{\includegraphics[angle=180,viewport=40 50 400 730,clip,width=52mm]{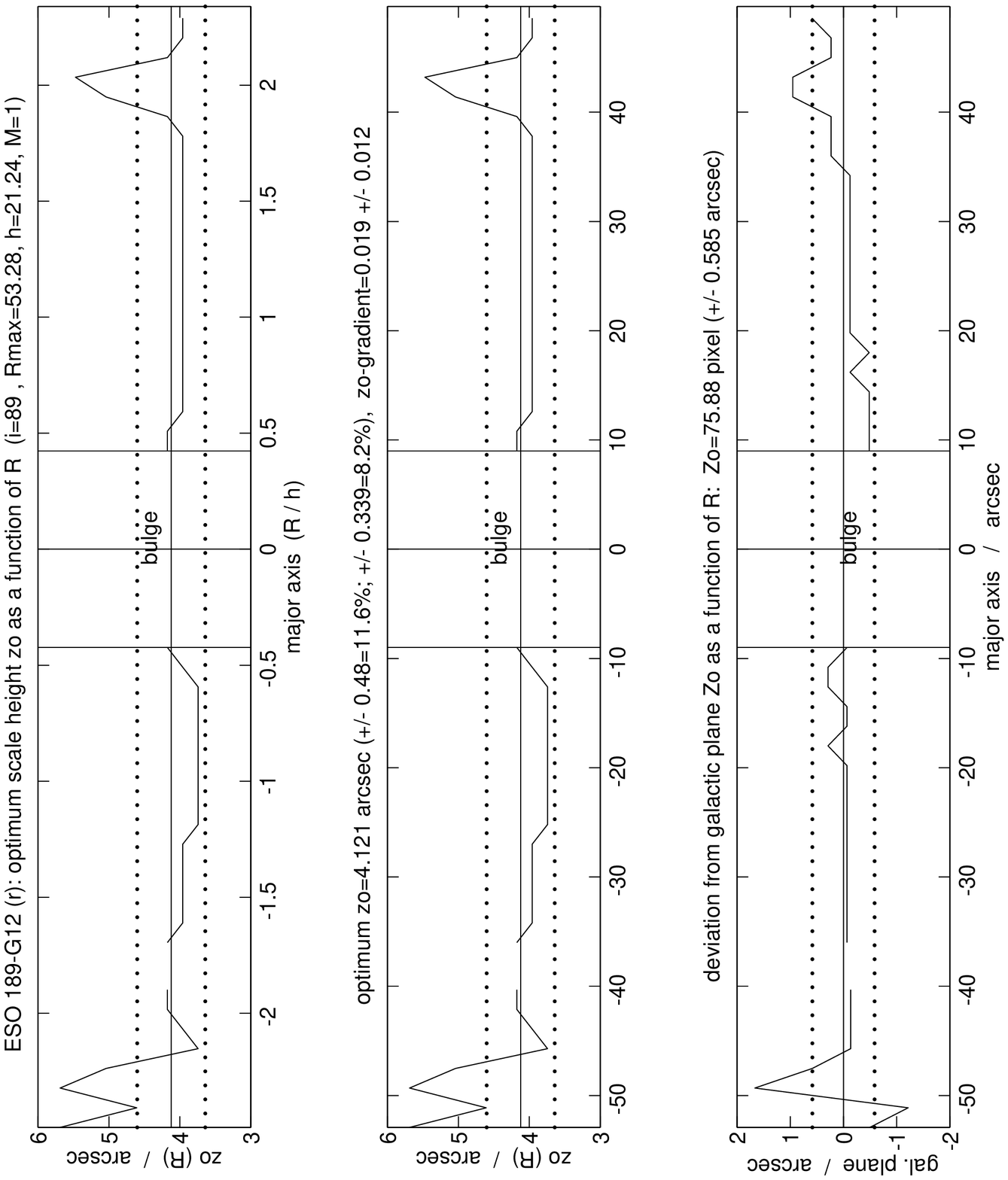}}
\end{picture}
\end{minipage}

\vspace*{-4mm}
\hspace*{65mm}
\begin{minipage}[b]{5.5cm}
\begin{picture}(3.0,3.0)
{\includegraphics[angle=180,viewport=00 -30 342 730,clip,width=50.5mm]{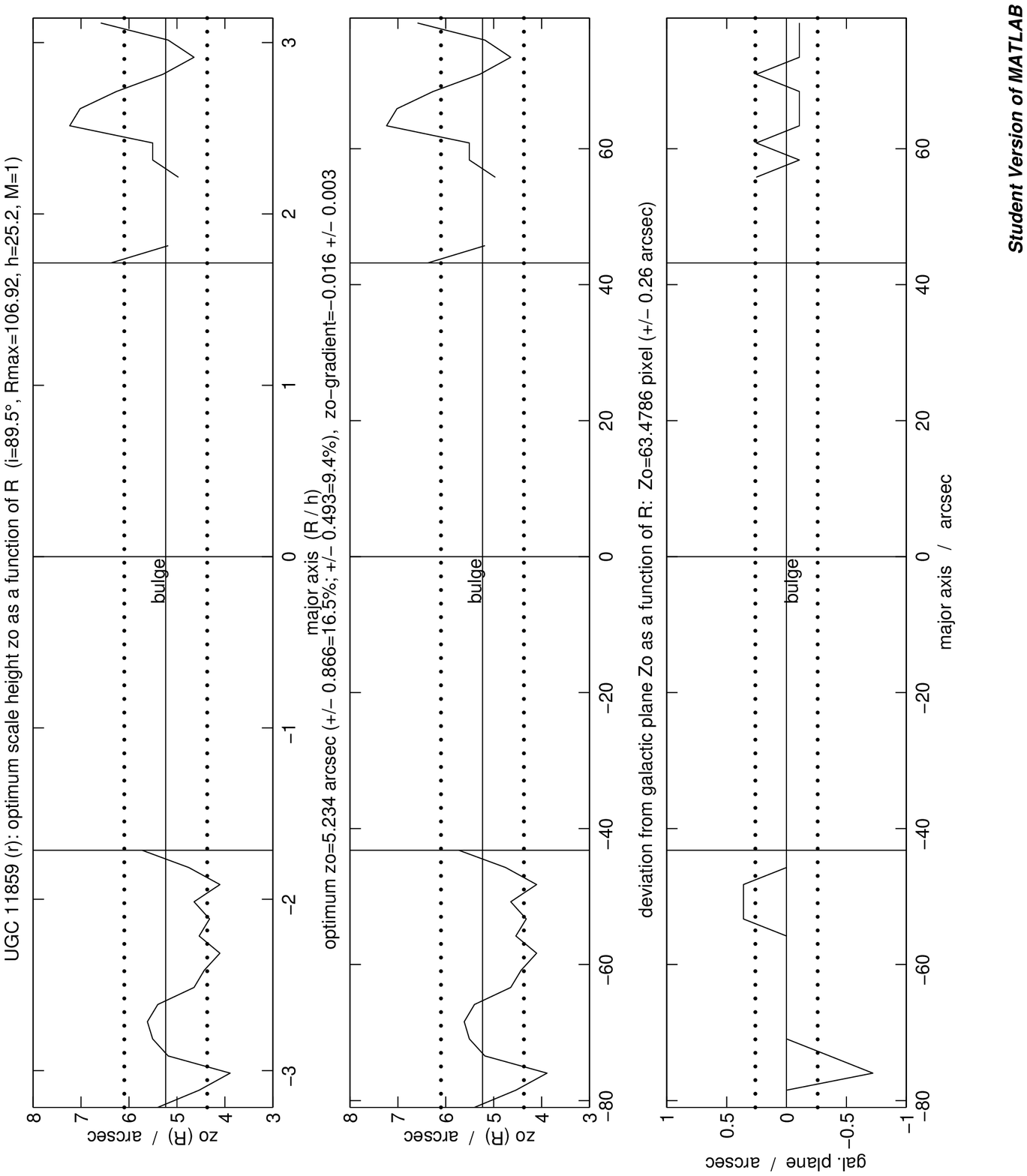}}
\end{picture}
\end{minipage}

\vspace*{94mm}
\hspace*{125mm}
\begin{minipage}[b]{5.5cm}
\begin{picture}(3.0,3.0)
{\includegraphics[angle=90,viewport=40 10 540 285,clip,width=53mm]{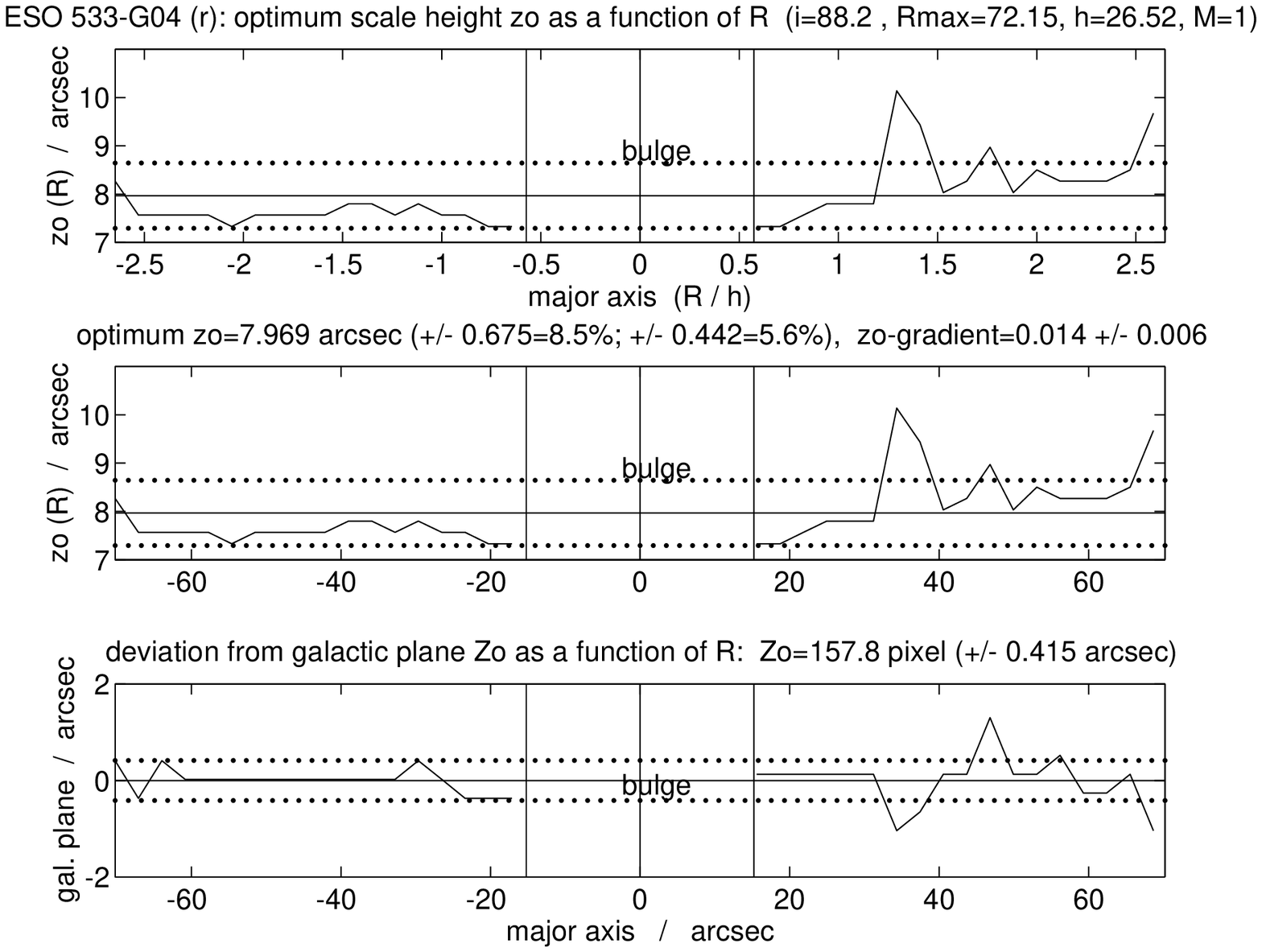}}
\end{picture}
\end{minipage}

\vspace*{1mm}

\hspace*{18mm}\parbox{165mm}{ESO 189-G12  \hspace{40mm}  UGC 11859  \hspace{40mm}  ESO 533-G04}

\vspace*{4mm}

\hspace*{5mm}
\begin{minipage}[b]{5.5cm}
\begin{picture}(3.0,3.0)
{\includegraphics[angle=180,viewport=40 50 400 730,clip,width=52mm]{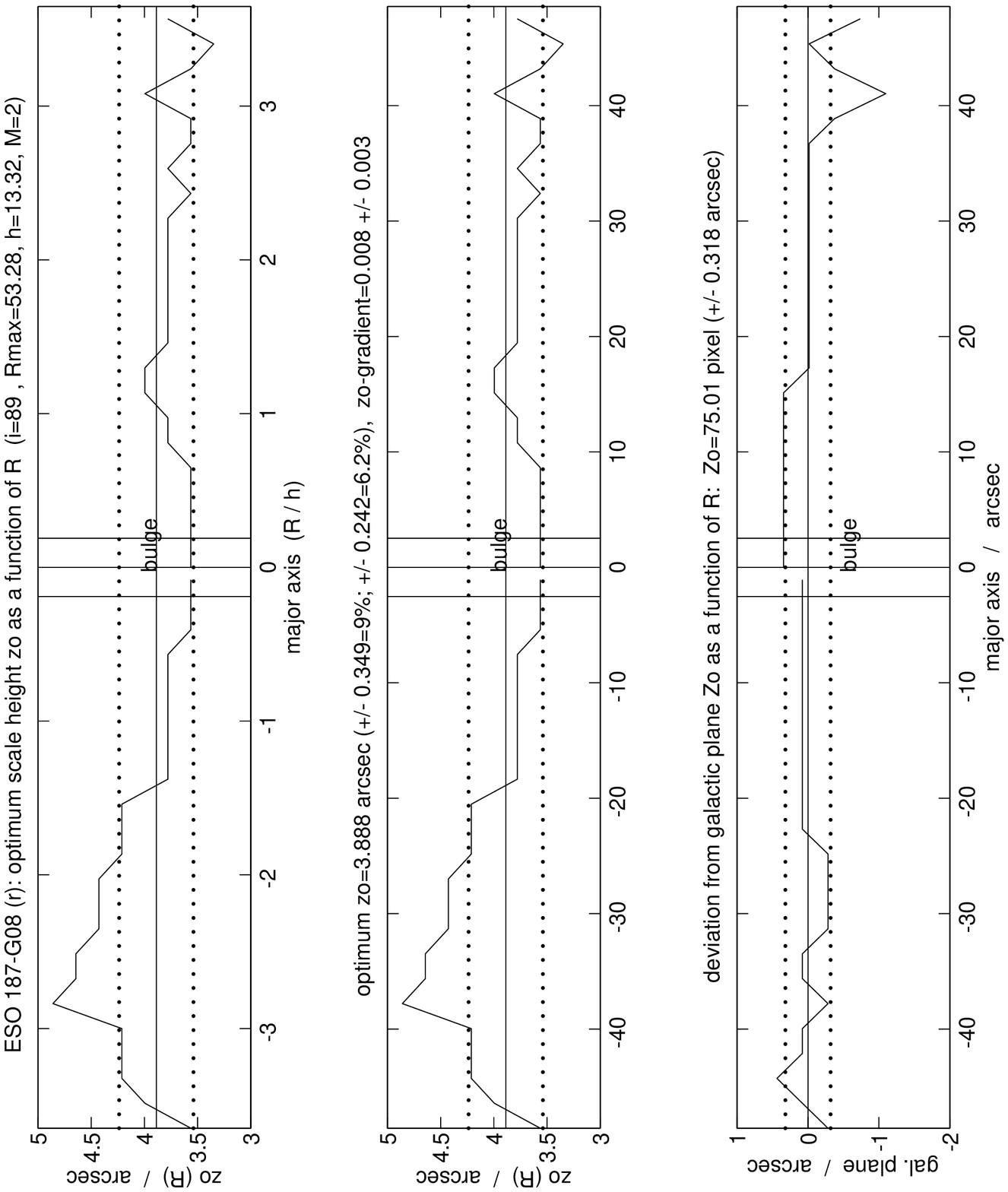}}
\end{picture}
\end{minipage}

\vspace*{95mm}
\hspace*{65mm}
\begin{minipage}[b]{5.5cm}
\begin{picture}(3.0,3.0)
{\includegraphics[angle=90,viewport=40 10 540 285,clip,width=54mm]{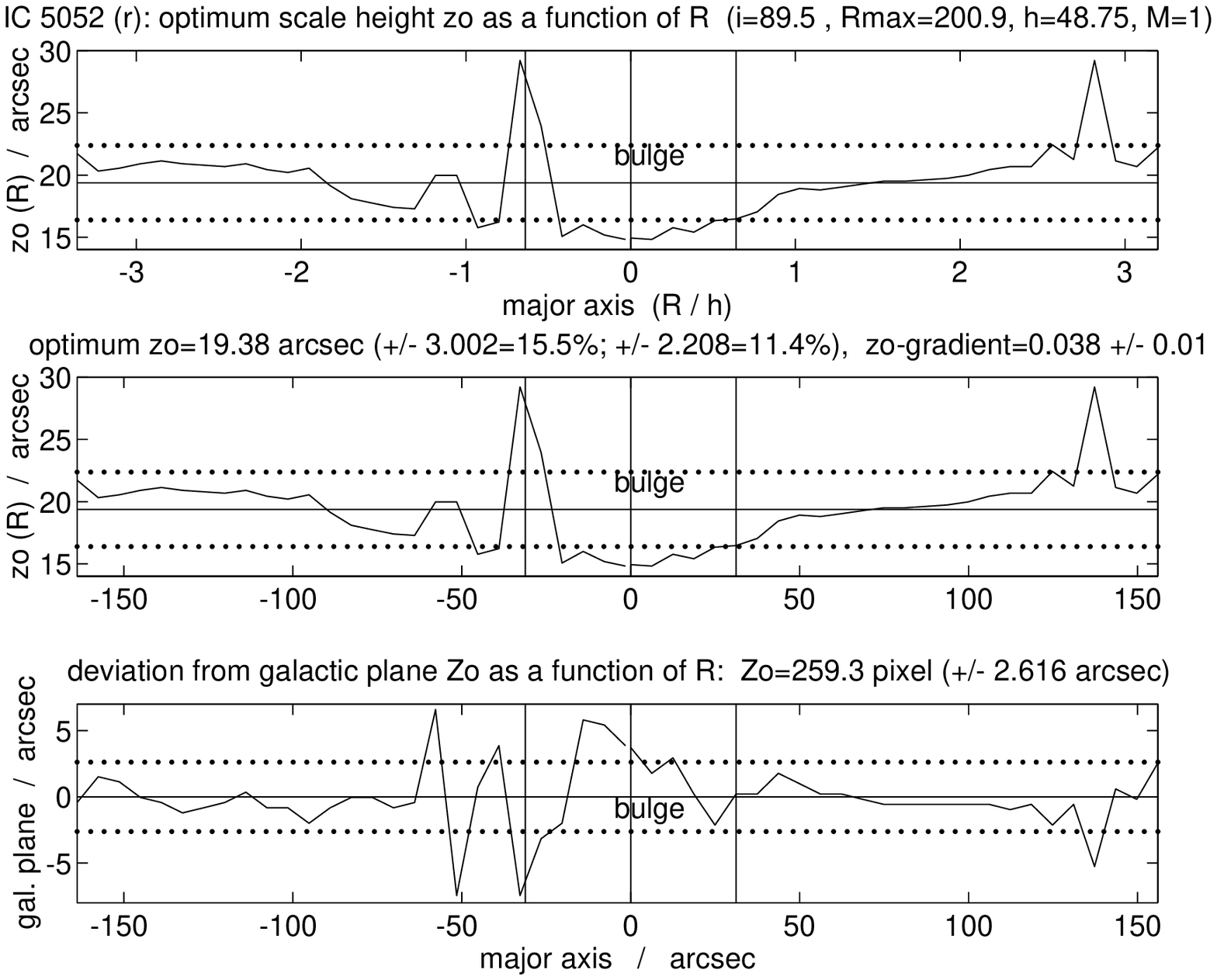}}
\end{picture}
\end{minipage}

\vspace*{-103mm}
\hspace*{125mm}
\begin{minipage}[b]{5.5cm}
\begin{picture}(3.0,3.0)
{\includegraphics[angle=180,viewport=40 50 400 730,clip,width=52mm]{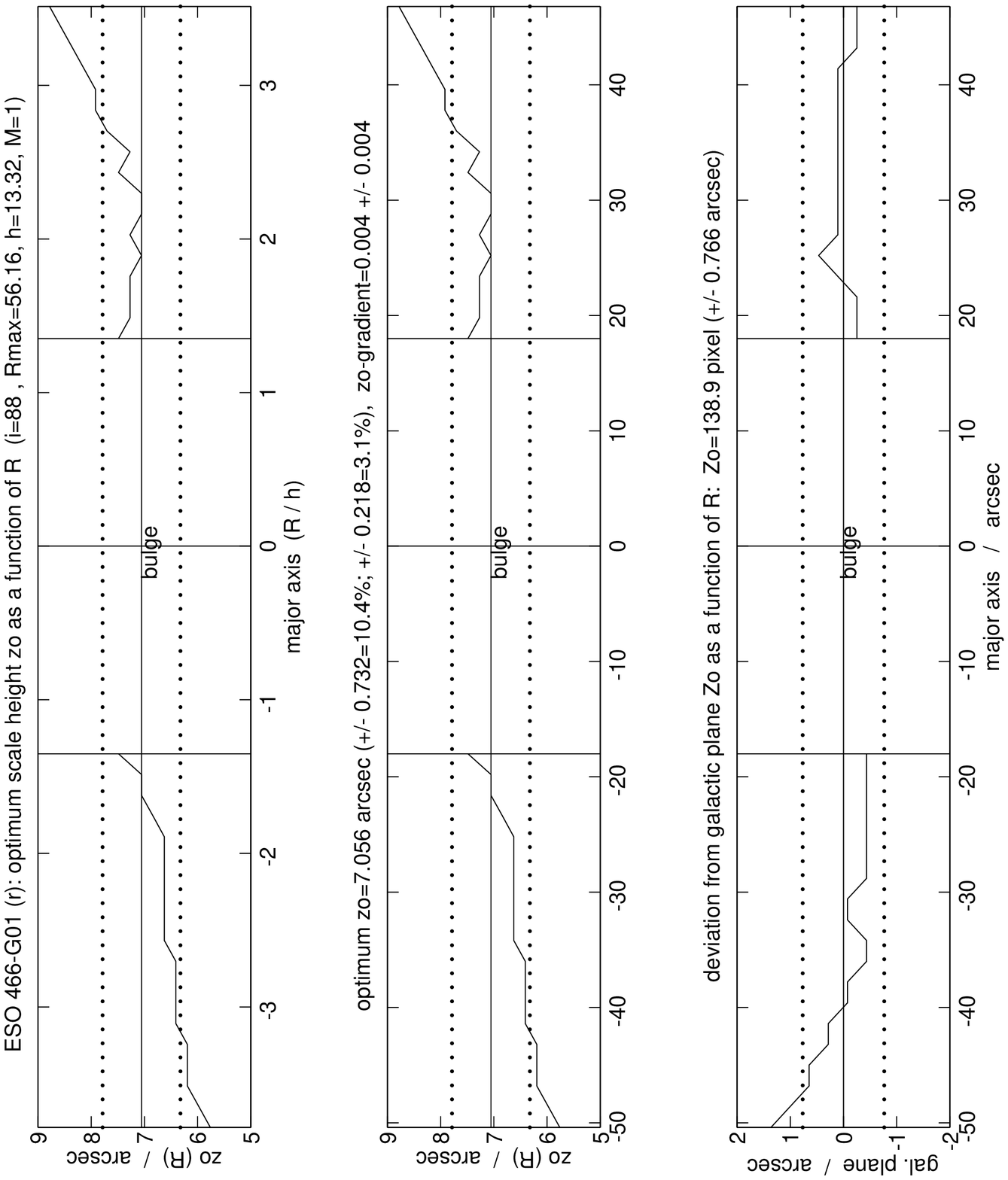}}
\end{picture}
\end{minipage}

\vspace*{98mm}

\hspace*{18mm}\parbox{165mm}{ESO 187-G08  \hspace{40mm}  IC 5052  \hspace{45mm}  ESO 466-G01}

\vspace*{8mm}

\hspace*{8mm}\parbox{165mm}{
{\bf \noindent Appendix B.} (continued)
}
\end{figure*}

%%%%%%%%%%%%%%%%%%%%%%%%%%%%%%%%%%%%%%%%%%%%%%%%%%%%%%%%%%%%%%%%%%%

\clearpage

%%%%%%%%%%%%%%%%%%%%%%%%%%%%%  10  %%%%%%%%%%%%%%%%%%%%%%%%%%%%%%%%%

\begin{figure*}[t]
\vspace*{3mm}
\hspace*{5mm}
\begin{minipage}[b]{5.5cm}
\begin{picture}(3.0,3.0)
{\includegraphics[angle=180,viewport=00 -30 342 730,clip,width=50.5mm]{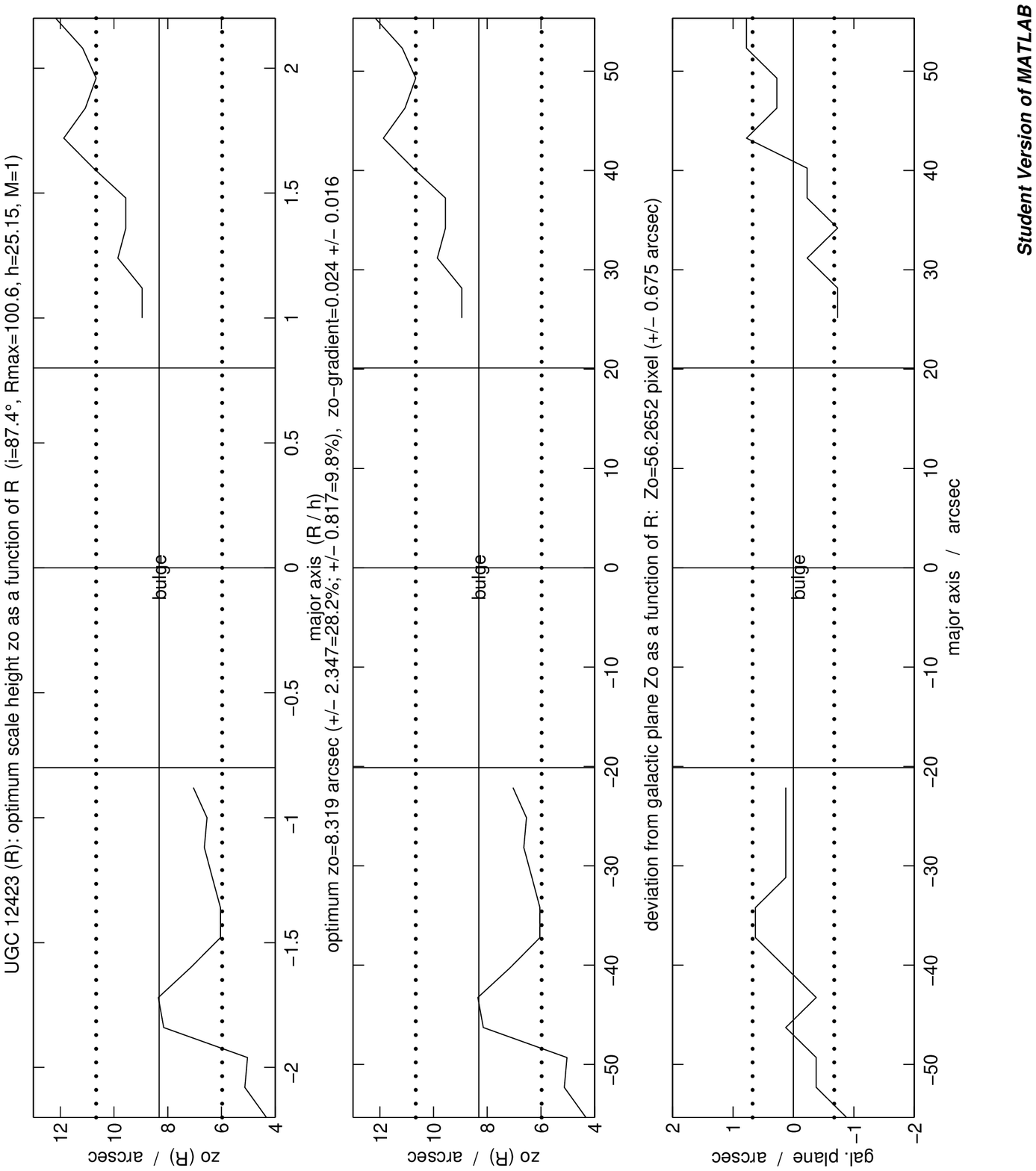}}
\end{picture}
\end{minipage}

\vspace*{94mm}
\hspace*{65mm}
\begin{minipage}[b]{5.5cm}
\begin{picture}(3.0,3.0)
{\includegraphics[angle=90,viewport=40 10 540 285,clip,width=53mm]{ngc7518.ps}}
\end{picture}
\end{minipage}

\vspace*{-100mm}
\hspace*{125mm}
\begin{minipage}[b]{5.5cm}
\begin{picture}(3.0,3.0)
{\includegraphics[angle=180,viewport=40 50 400 730,clip,width=51mm]{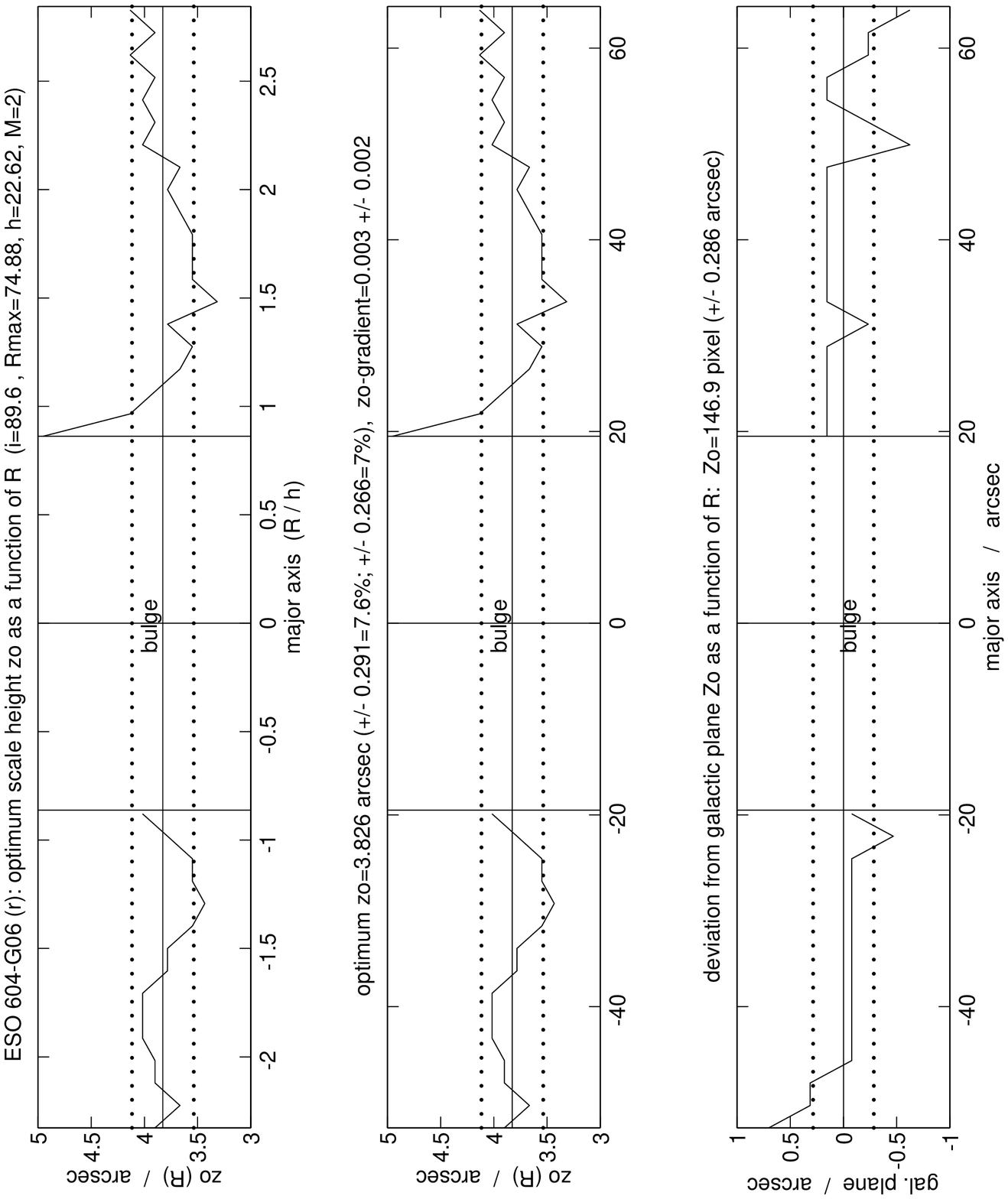}}
\end{picture}
\end{minipage}

\vspace*{98mm}

\hspace*{18mm}\parbox{165mm}{UGC 12423  \hspace{40mm}  NGC 7518  \hspace{45mm}  ESO 604-G06}

\vspace*{101mm}

\hspace*{6mm}
\begin{minipage}[b]{5.5cm}
 \begin{picture}(3.0,3.0)
{\includegraphics[angle=90,viewport=40 10 540 285,clip,width=54mm]{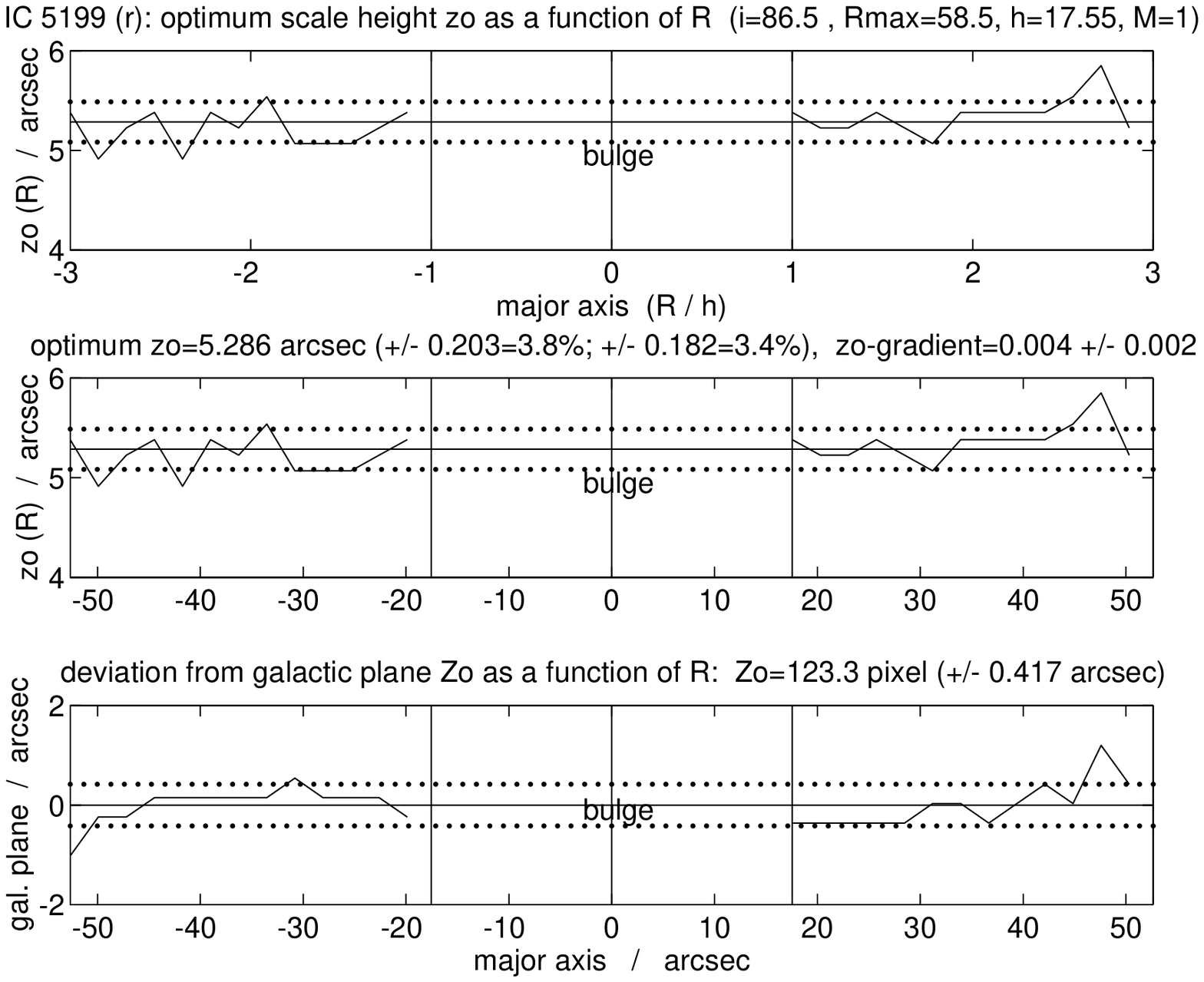}}
\end{picture}
\end{minipage}

\vspace*{-103mm}
\hspace*{65mm}
\begin{minipage}[b]{5.5cm}
\begin{picture}(3.0,3.0)
{\includegraphics[angle=180,viewport=00 -30 342 730,clip,width=50.5mm]{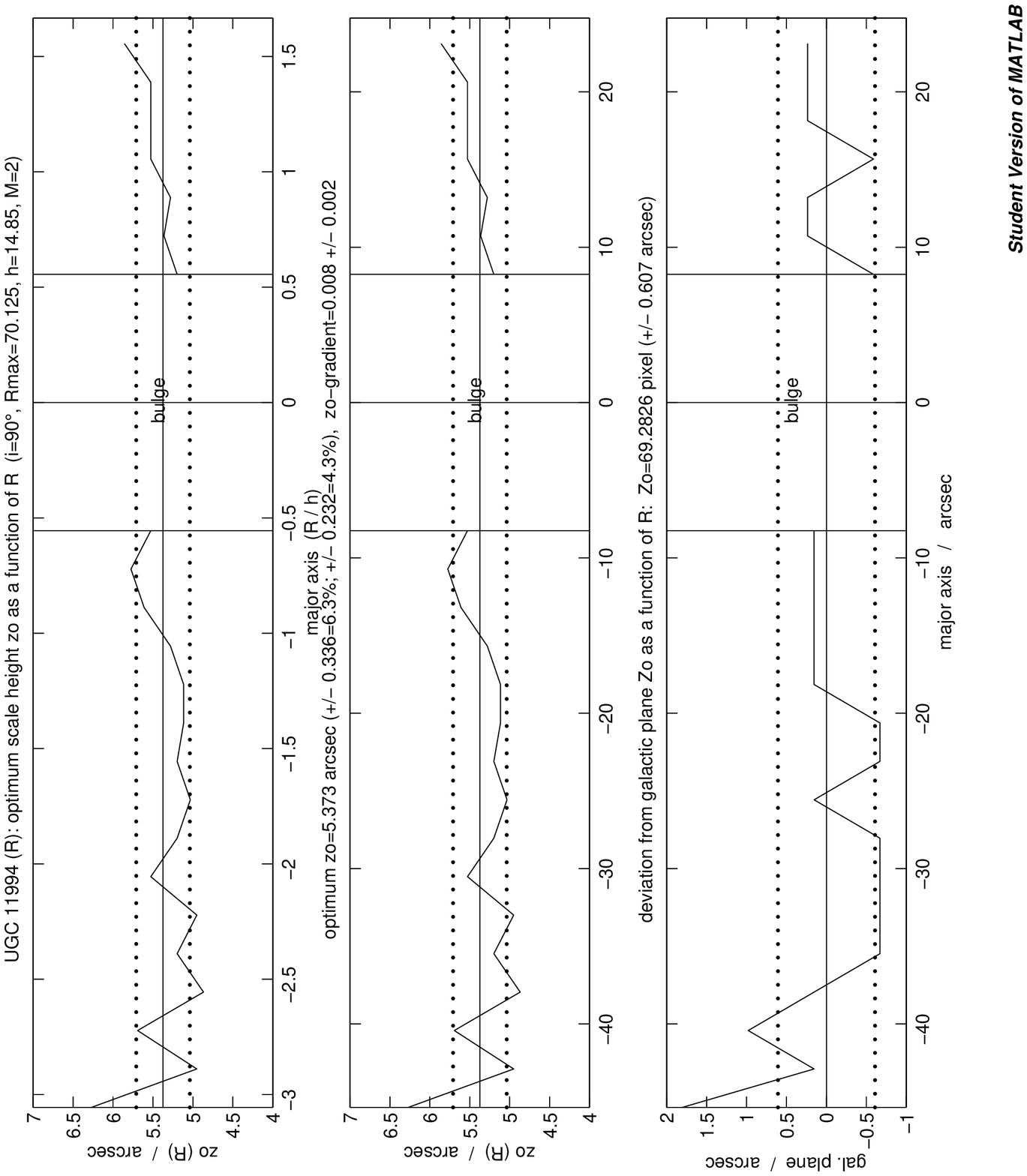}}
\end{picture}
\end{minipage}

\vspace*{-4mm}
\hspace*{125mm}
\begin{minipage}[b]{5.5cm}
\begin{picture}(3.0,3.0)
{\includegraphics[angle=180,viewport=00 -30 342 730,clip,width=50.5mm]{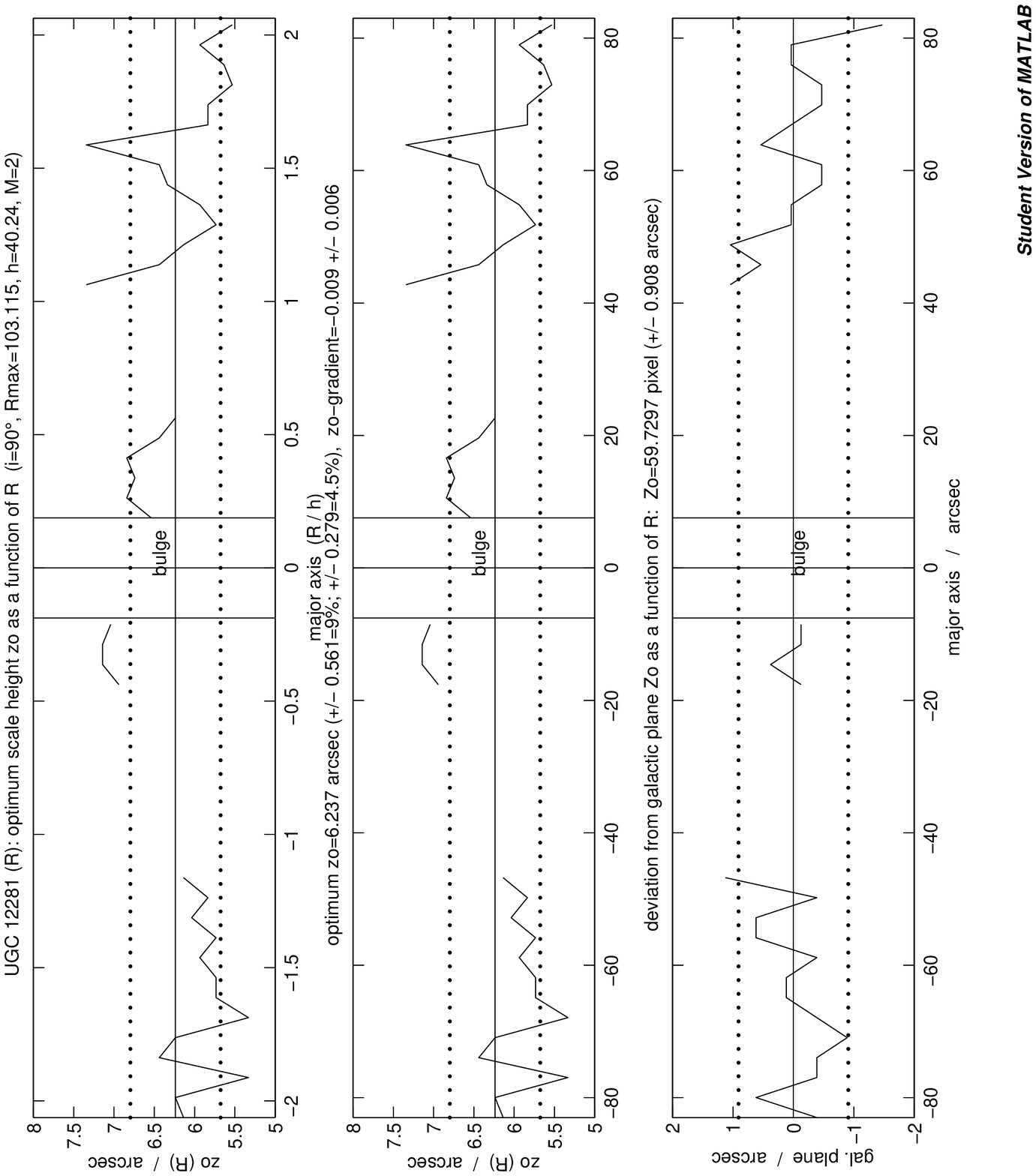}}
\end{picture}
\end{minipage}

\vspace*{98mm}

\hspace*{23mm}\parbox{165mm}{IC 5199  \hspace{42mm}  UGC 11994  \hspace{45mm}  UGC 12281}

\vspace*{8mm}

\hspace*{8mm}\parbox{165mm}{
{\bf \noindent Appendix B.} (continued)
}
\end{figure*}

%%%%%%%%%%%%%%%%%%%%%%%%%%%%%%%%%%%%%%%%%%%%%%%%%%%%%%%%%%%%%%%%%%%

\end{document}